\documentclass[usenatbib,usegraphicx,useAMS]{mn2e}

\usepackage{times}
\usepackage{amsmath}
\usepackage{amssymb}

\newcommand{\mbi}[1]{\mbox{\boldmath$#1$}}

\newcommand{\mat}[1]{\mbox{\rm\bf #1}}

\newfont{\itbf}{cmbxti10 at 9pt}
\newfont{\slbf}{cmbxsl10 at 9pt}

 \def\beqn{\vspace{2mm}
\begin{eqnarray}} \def\eeqn{\vspace{2mm} \end{eqnarray}}
 
  \def\la{\langle}

\begin{document}
\title[Bayesian reconstruction of the LSS]{Bayesian reconstruction of
  the cosmological large-scale structure:  {methodology, inverse algorithms and numerical optimization}}
\author[Kitaura  \& En\ss lin]{F.~S.~Kitaura\thanks{E-mail:
kitaura@mpa-garching.mpg.de} and T.~A.~En\ss lin\\
Max-Planck Institut f\"ur Astrophysik, D-85748 Garching, Germany}
\maketitle

\begin{abstract}
We address the inverse problem of cosmic large-scale structure reconstruction
from a Bayesian perspective.  For a linear data model, a number of known and novel reconstruction
schemes, which differ in terms of the underlying signal prior, data likelihood,
and numerical inverse extra-regularization schemes are derived and classified. 
The Bayesian methodology presented in this paper tries to unify and extend the following methods: Wiener-filtering, Tikhonov regularization, Ridge regression, Maximum Entropy, and inverse regularization techniques. The inverse techniques considered here are the asymptotic regularization, the Jacobi, Steepest Descent, Newton-Raphson, Landweber-Fridman, and both linear and non-linear Krylov methods based on Fletcher-Reeves, Polak-Ribi\`ere, and Hestenes-Stiefel Conjugate Gradients.  
The structures of the up-to-date highest-performing algorithms are presented, based on an operator scheme, which permits one to exploit the power of fast Fourier transforms. Using such an implementation of the generalized Wiener-filter in the novel \textsc{argo}-software package, the different numerical schemes are benchmarked with 1-, 2-, and 3-dimensional problems including structured white and Poissonian
noise, data windowing and blurring effects. A novel numerical Krylov scheme is shown to be superior in terms of performance and fidelity.
These fast inverse methods ultimately will enable the application of sampling techniques to explore complex joint posterior distributions.
We outline how the space of the dark-matter density field, the peculiar velocity field, and the power spectrum can jointly be investigated by a Gibbs-sampling
process. Such a method can be applied for the redshift distortions correction of the observed galaxies and for time-reversal reconstructions of the initial density field.
\end{abstract}



\section{Introduction}
\label{sec:intro}

According to our current picture of cosmogenesis, the galaxies, galaxy
clusters, galaxy filaments, and giant voids forming the cosmic large-scale
structure (LSS) are products of gravitational instability, which pulls
increasingly more matter onto the tiny primordial seed density fluctuations
generated at the very first epoch of inflation. The shape and size of the
cosmic matter distribution reflects the initial conditions set during or
shortly after Big Bang, as well as the interplay of the gravitational
self-attraction of matter and the diluting action of the Hubble  expansion of
cosmic space. Valuable information about the properties and the origin of the
cosmic inventory are encoded in the LSS, however, on small-scales, that information is being erased through dynamical non-linear processes.

 Our goal is to extract as much of this information as possible from astronomical measurements, which introduce uncertainties and, consequently, degeneracies. 
Therefore, we have to adapt an information-theoretical approach to solve the
reconstruction problem of cosmography. The Bayesian framework turns out to be
the most general approach as we will discuss later.   
In this paper we present the novel \textsc{argo}\footnote{{\bf A}lgorithm for the {\bf
    R}econstruction  of {\bf G}alaxy-traced {\bf O}ver-densities}-software
package, which reconstructs the three-dimensional density field from the
information provided by galaxy surveys with different Bayesian and inverse
methods. Here we focus our study on understanding the Bayesian theoretical
background and the required algorithmic aspects. Further extensions of the code
in which the power-spectrum and the peculiar velocities can be jointly sampled
    are presented and tested on mock galaxy catalogues. Some of the preliminary
    results are presented and future development is outlined.

The large number of telescopes performing galaxy surveys with increasing depth,
sky coverage, and accuracy in position and distance (or redshift) determination
provide us with superb data on the cosmic matter distribution at an
exponentially increasing rate. One problem is that the discrete objects these
instruments reveal to us, the galaxies, are the result of a complex non-linear
evolution of cosmic matter combined with complicated astrophysical processes
such as star formation. A translation of the galaxy data into the much better
understood large-scale dark matter (DM) distribution, which would be much
easier to analyze for imprints of cosmologically interesting effects, is far
from trivial. The discrete nature of galaxies introduces certain noise, usually modeled by shot noise. Moreover,
the partially understood galaxy-formation process inserts systematic
uncertainties. In addition, the limited volume of surveys adds complications beyond the problems of galaxy-distance determination being contaminated by observational and velocity redshift-distortions. All these complications have to be dealt with simultaneously and in a controlled fashion. Since it cannot be assumed that the correct or optimal values for the various degrees of freedom of the problem (bias factors, redshift-corrections, etc.) will be guessed a priory, repeated and iterative data analysis is mandatory in order to achieve a high-fidelity and well-understood cosmic map. For example, a correction of redshift-distortions of the galaxies requires the gravitational potential generated by the matter distribution to be reconstructed.

Repeated generation of cosmic matter maps increases the urge to face another
challenge, the scaling of the performance of the underlying map-generation
algorithms with the data size. Since the matter-density information displayed
at a location on a map may depend on all input data (galaxy positions), any
algorithm optimized to information theory scales super-linear\footnote{A map of
  galaxy counts can be generated by an algorithm with linear scaling to data
  size however, it is not an optimal representation of the underlying matter
  field.}. With increasing survey sizes, increasing requirements for spatial
resolution and volume coverage, and the need to frequently re-iterate the
map-generation step, the algorithm has to scale closely to linear with data
size, otherwise its application is strongly limited. Former applications in
cosmography suffered from such inconvenient performance-scaling, and an effort
has to be made to develop simultaneously high-performance and  accurate methods.

The work presented in this paper developes the general methodology of Bayesian
reconstruction of the cosmic matter distribution, based on the invaluable
pioneering work of many other scientists, which will be discussed below, and
extends this work to a series of new applications. Existing and novel map
making algorithms are summarized in terms of a classification of their Bayesian
likelihood and prior functions. The implementation, optimization, and
comparison of various numerical schemes are addressed in detail. This provides
a starting point for a correct information-theory approach to cosmography. Many
additional problems, not addressed in this paper, such as the galaxy bias, will
also have to be solved before accurate maps of the dark matter distribution in
our still mysterious Universe can be generated.

Such an undertaking would be highly rewarded in the short and long run. An
accurate map of the cosmic matter distribution would be valuable for a manifold
of direct scientific applications. These range from structure-formation
analysis, to cosmological parameter estimation via power-spectrum measurements,
dark energy studies, galaxy-cluster identification and galaxy-bias
studies. Accurate cosmic maps would help to determine weak signals associated
with the large-scale structure such as the integrated Sachs-Wolf (ISW) effect,
or the extended Sunyaev-Zel'dovich (SZ) effect, the detection of which relies on the construction of optimal statistical filters for these signals.

Finally, one could  argue that mapping the distribution of matter in the Universe represents a response to mankind's curiosity in its aim to discover {\it terra incognita} and find an orientation in space and time on cosmological scales and, therefore, should be a goal in itself.

In the remainder of this introduction we give the sources of uncertainties, we present an overview of existent and new Bayesian reconstruction methods, subsequently we briefly describe the algorithmic development presented in this paper, and in the final part we give a more detailed overview of the structure of this paper.  

\subsection{Classes of uncertainty}
\label{sec:uncert}

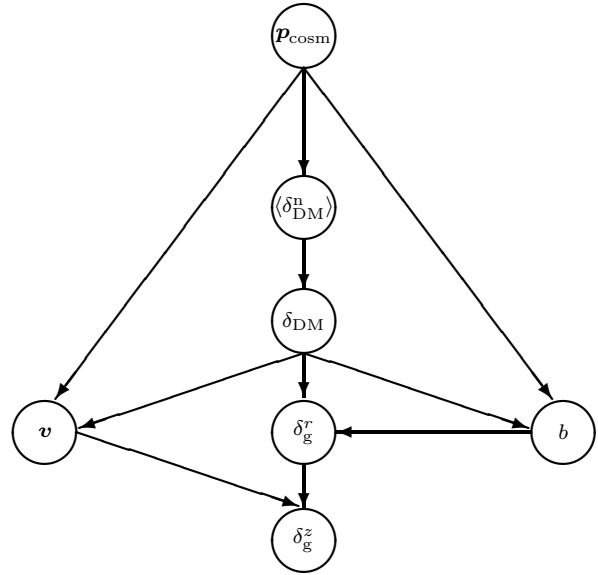
\begin{figure}
\begin{picture}(300,200)\thicklines
\put(120,200){\circle{25.0}\makebox(0,0){$\mbi p_{\rm cosm}$}}
\put(120,135){\circle{25.0}\makebox(0,0){$\langle\delta^{\rm n}_{\rm DM}\rangle$}}
\put(120,188){\vector(0,-1){41.0}}
\put(120,123){\vector(0,-1){19.0}}
\put(120,92){\circle{25.0}\makebox(0,0){$\delta_{\rm DM}$}}
\put(120,80){\vector(0,-1){17.0}}
\put(34,50){\vector(3,-1){84.0}}
\put(120,38){\vector(0,-1){17.0}}
\put(206,50){\vector(-1,0){74.0}}
\put(120,50){\circle{25.0}\makebox(0,0){$\delta_{\rm g}^r$}}
\put(120,80){\vector(-3,-1){85.0}}
\put(120,80){\vector(3,-1){85.0}}
\put(120,9){\circle{25.0}\makebox(0,0){$\delta_{\rm g}^z$}}
\put(22,50){\circle{25.0}\makebox(0,0){$\mbi v$}}
\put(120,188){\vector(-3,-4){94.0}}
\put(218,50){\circle{25.0}\makebox(0,0){$b$}}
\put(120,188){\vector(3,-4){94.0}}
\end{picture}
\caption{\label{fig:dag} The hierarchical Bayes model for a galaxy distribution in redshift space $\delta^z_{\rm g}$ is represented here in a directed acyclic graph (DAG). 
The cosmological parameters $\mbi p_{\rm cosm}$ govern the rest of the variables.
The initial density field coming from e.g.~inflationary scenarios can be statistically described by all its moments $\langle\delta^{\rm n}_{\rm DM}\rangle$. Here the power spectrum is usually taken, since the intitial perturbations are well described by a Gaussian realization of the initial seed fluctuations. 
The further evolution is described by nearly deterministic processes (given by structure and galaxy formation), which determine the later-time dark matter distribution $\delta_{\rm DM}$ with its peculiar velocity field $\mbi v$ and the bias function $b$ that relates the galaxy distribution to the dark matter density field. 
The dark matter distribution $\delta_{\rm DM}$ with the bias produces the galaxy distribution in real space $\delta_{\rm g}^r$. The peculiar velocities $\mbi v$ related to the density field through the continuity equation introduce the redshift distortion in $\delta_{\rm g}^r$ finally leading to the galaxy distribution in redshift space $\delta_{\rm g}^z$. }
\end{figure}
Several classes of uncertainties related to the density-field reconstruction
from galaxy surveys demand a statistical approach. Some of the uncertainties
are intrinsic to the nature of the underlying signal (the dark matter).  Other
uncertainties are intrinsic to the nature of the observable (the galaxies). And
finally there are uncertainties due to degeneracies which appear through the
observation and data mining process. 
\begin{enumerate}
\item {\bf Intrinsic stochastic character: cosmic variance}
%
%

In cosmology it is generally assumed that the structure of the Universe comes from some infinitesimal quantum fluctuations which were frozen out and stretched by an inflationary phase \citep[see][]{1981PhRvD..23..347G, 1982PhRvL..49.1110G, 1982PhLB..117..175S, 1982CMaPh..87..395H,  1982PhLB..108..389L, 1982PhRvL..48.1220A, 1983PhRvD..28..679B}, and later amplified by gravitational instability. 
According to this picture, the seed fluctuations would have an intrinsic stochastic character and are mainly Gaussian distributed.
However, the mechanisms that stretch the quantum fluctuations may also introduce deviations from Gaussianity which would then be imprinted in the seed fluctuations. In general all the moments of the initial fluctuations have to be considered $\langle\delta^{\rm n}_{\rm DM}\rangle$. 
Nevertheless, most of the inflationary scenarios predict the density field to
be very closely Gaussian distributed and it is generally sufficient to take the second order moment, the two-point correlation function, or the power-spectrum in Fourier-space. We will discuss below how to determine the power-spectrum and techniques to disentangle intrinsic non-Gaussianities within a Bayesian framework. 
Note, that there are alternative models to inflation in which e.g.~the seed fluctuations are identified with the topological defects that remain as relics of high-energy phase transitions \citep{1976JPhA....9.1387K}. Accurate reconstructions of the LSS could help to discriminate between the different models.

\item {\bf Physical uncertainties: galaxy bias}
%
%

The galaxy formation process is a complicated, non-linear and (probably) non-local process.  It is known that on large scales the galaxy power-spectrum fits well to the expected DM spectrum predicted from cosmic microwave background (CMB) observations, if some bias factor $b$ between the amplitude of the galaxy and DM fluctuations is assumed. Detailed studies show that the bias factor is not universal, but depends on galaxy type, galaxy formation time, redshift, etc.~\citep[see e.g.][and references therein]{cooray-2002-372}. 
For the purpose of  reconstructing the underlying density field, linear biases can  easily be tackled within the linear data model described below by including its effects in a selection function. Nevertheless, more complex biases have to be further investigated in a Bayesian framework.
Physical processes, which are not perfectly understood within galaxy formation
may  be treated in a statistical way, encoding the ignorance about certain
physical processes in probability distribution functions. Several works study
the  stochastic non-linear galaxy biasing \citep[see for
example][]{1998ApJ...504..601P,1999ApJ...520...24D,tegmark-1999-518}. Some of these
models could be implemented in the Bayesian reconstruction process. This issue
is out of scope in this paper, but should be further investigated in this framework.

\item {\bf Physical/observational uncertainties: redshift-distortions}

 The peculiar motion of galaxies with respect to the Hubble flow of the
 Universe: $\mbi v$, introduces  uncertainties in their redshift measurement, the
 so-called redshift-distortions \citep[see e.g.][for an introduction to this
 problem]{hamilton-1998}. The measured galaxy over-densities are thus said not
 to be in real-space $\delta^{ r}_{\rm g}$, but in redshift-space
 $\delta^{ z}_{\rm g}$. In the linear regime, where galaxies fall
 into the potential wells  of large scale structures, redshift-distortions cause a
 squashing of the linear over-densities in radial direction. However, in the
 non-linear regime, galaxies (e.g.~in a galaxy cluster) tend to behave like
 particles in a gas with randomized motions inside the clusters where the
 potentials are very high. This produces the so-called {\it finger-of-god} effect,
 a dispersion along the line of sight. The correction of these distortions is
 not trivial, since the process of structure formation partially erases the information about the initial
 fluctuations  after entering the non-linear regime. Consequently, determining
 the real position of galaxies poses a degenerate problem, which has in general many possible solutions.
Many efforts have been made  to correct for these distortions: in the linear
regime these efforts start with Kaiser's pioneering work
\citep[see][]{1987MNRAS.227....1K} and are followed by the linear redshift-distortions
operator \citep[for a detailed derivation see][]{hamilton-1998}. In the
non-linear regime, these efforts include a velocity dispersion factor (the {\it dispersion}-model) corresponding to an exponential pairwise velocity distribution function with no mean streaming \citep[see][]{1996MNRAS.282..877B}. \cite{2004PhRvD..70h3007S} presents an exact relationship between real-space and redshift-space two-point statistics through the pairwise velocity distribution function including all non-linearities.
More complex methods of correcting for redshift-distortions were classified by
\cite{1999AJ....118.1146S} into iterative methods, which uses the redshift-space
density to calculate a peculiar velocity field, and then iteratively corrects the density field distortions
\citep[][]{1991ApJ...372..380Y,1991ASPC...15..111K}. The other class decomposes the redshift-space
 density in radial and angular basis  functions
 from which the radial redshift-distortion is corrected \citep[see
 e.g.][]{1994ASPC...67..171L, 1994ApJ...421L...1N,1995MNRAS.272..885F,1999AJ....118.1146S} and more recently
 \cite{2005MNRAS.356.1168P}.
Below, we propose  a Bayesian method to correct for the linear and non-linear
 redshift-distortions in a statistical way (see section \ref{sec:MCMC}).

\item {\bf Observational uncertainties: measurements}
%
%

The action of measurement introduces uncertainties, either due to the
instruments, e.g.~blurring by the telescope, or due to the observational
strategy, which is included in the noise term, the selection function, and the
mask effects \citep[see][for a pioneering work in the LSS
field]{1995ApJ...449..446Z}. 
Ignoring selection functions, windowing, or blurring  will lead to strongly
biased reconstructions, which are far from the real signal, and thus allow only
very limited interpretation of the true physical picture.   A numerical implementation of these effects is presented in
section ({sec:operators}).  The influence of these effects will then be
analyzed separately and tested with our code.  The results are presented in
section (\ref{sec:codetesting}). Though \textsc{argo} demonstrates its capability to handle
these uncertainties, further work is required in order to apply it to real
data. Particular expressions for the selection function 
according to the redshift survey under study, as well, as masks, etc., have to be implemented.

\item {\bf Mathematical/numerical representation uncertainties: aliasing effects}

Some uncertainties are not intrinsic to the observable, but originate from
the mathematical representation one chooses. Treating galaxies as counts in
cells or with other mass-assignment schemes  will smear out the information about their measured
position for which one has to correct (see section~\ref{sec:respOp}) in order to derive other quantities, like the
power-spectrum (see section \ref{sec:PSEST}).

\end{enumerate}
From all the points mentioned above we conclude, that extracting the underlying
dark matter density field from the luminous matter distribution given by galaxy
redshift surveys poses a classical signal reconstruction problem.
A Bayesian network depicting the relation of these uncertainties is shown in fig.~(\ref{fig:dag}).

\subsection{Bayesian reconstruction methods}
%
%
Any Bayesian statistical approach requires the definition of a likelihood and a
prior. The former is the probability distribution function describing the
process generating the observational data. It can be interpreted as a distance
measure of the observed data to the underlying signal, as we will discuss
below. The prior stands for the distribution function modeling our prior
knowledge on the signal to be recovered. Mathematically it can be shown that it
regularizes the estimator in the presence of noise (see section
\ref{sec:baytik}).  Two kinds of priors have to be distinguished, informative
priors, in which the previous physical knowledge about the signal is encoded,
and non-informative priors, which try to give objective estimators for the
underlying signal based on purely information-theoretical arguments. Here, three non-informative priors are considered: flat priors (see section \ref{sec:flatprior}) with a constant probability distribution function (PDF), entropic priors based on Shannon's notion of information (see section \ref{sec:entr}), and Jeffrey's prior based on invariant statistical structures under transformation of variables (see section \ref{subsubsec:jeff}).
Finally, a maximization or sampling of the posterior distribution, which is proportional to the product of the likelihood and the prior, has to be done to complete the Bayesian estimation. 
The maximization of the posterior is called the maximum a posteriori method (MAP). The maximum likelihood (ML) and maximum entropy method (MEM) are particular cases of the MAP with flat priors and entropic priors, respectively.  
Complex posterior distribution functions may be sampled iteratively from conditional PDFs in a Markov Chain Monte Carlo fashion (MCMC), see section \ref{sec:MCMC}. 
 We show how different choices for these distribution functions together with the estimation procedure lead to different reconstruction algorithms, which consequently have distinct application fields (see table \ref{tab:rec}).
A review of existing methods is presented and new applications for the large-scale structure reconstruction, which naturally emerge within the Bayesian formalism, are developed.

%
%
In this work we consider Poissonian and Gaussian likelihoods for the galaxy
 distribution.  The former has been previously considered in image restoration
 especially for deconvolution purposes
 \citep[see][]{1972JOSA...62...55R,1974AJ.....79..745L}. For example, the Richardson-Lucy  algorithm can be derived as the ML of a Poissonian likelihood \citep[see][and appendix \ref{app:POISLIKEML}]{sheppvardi}. Here an image can be regarded as photon counts in cells represented by a Poissonian distribution.
 However, one should notice that this likelihood does not represent the galaxy-formation process. From a pure image reconstruction perspective, it can still be interesting for LSS estimations, because it naturally represents the discrete nature of a galaxy distribution.
The Gaussian likelihood  allows the incorporation of arbitrary noise structures through the variance. 
The CMB map-making algorithms, which aim to convert time-ordered data received from satellites into a map of the CMB signal on the sky as a projection on the sphere, usually use this likelihood.
 In this case, the ML leads to the simple COBE-filter first derived by \cite{1992issa.proc..391J}. Nevertheless, the complex scanning strategies and foreground removal can add unlimited complexity to these algorithms \citep[e.g.][]{natoli,mapcumba,maxima,madam,mirage}.

For the LSS the Gaussian prior arises as the natural informative prior due to the arguments discussed above. 
We propose a novel algorithm: GAPMAP, which maximizes the posterior with a Gaussian prior and a Poissonian likelihood (see section \ref{sec:gauspois} and appendix \ref{app:POISLIKE}). 
In contrast, the Gaussian likelihood with the Gaussian prior leads to the
well-known  Wiener-filter, which has been used for the LSS reconstruction
\citep[see][]{WienerFSL,1994ASPC...67..185H,1994ApJ...423L..93L,1994ASPC...67..171L,1995ApJ...449..446Z,1995MNRAS.272..885F,1997MNRAS.287..425W,1999ApJ...520..413Z,1999AJ....118.1146S,2004MNRAS.352..939E,2006MNRAS.373...45E}
and for  CMB-mapping \citep[see
e.g.][]{1994ApJ...432L..75B,1997ApJ...480L..87T}. It is also known to give
optimal results in terms of yielding the least square error, see the pioneering
work of \cite{1992ApJ...398..169R} and \cite{1995ApJ...449..446Z}. We present
in this paper a fast Wiener-filter extra-regularized with Krylov methods as we
will see below  (see table \ref{tab:beta} for a summary of different Krylov methods).

Intrinsic primordial non-Gaussianities  can be imprinted in the seed fluctuations depending on the particular theory responsible for the amplification of the fluctuations coming from the early Universe.
 To find such deviations, non-informative priors, which give non-linear
 estimates for the underlying signal are required. Entropic priors are well
 suited here, and have been previously applied for CMB studies. We extend this work for LSS reconstructions and develop the corresponding maximum entropy method for Gaussian and Poissonian likelihoods (see section \ref{sec:entr} and appendix \ref{app:MEM}).

Sampling methods have the advantage of determining the shape of distributions
and, thus, leading to a natural estimate of the uncertainty of the
estimator. Moreover, the mean can be calculated easily from the sample and is known to give more accurate results than the maximum in the case of asymmetric PDFs \citep[see e.g.][]{toolsstatinf}.   

As an example, \cite{2003MNRAS.338..765H} proposed a SZ-cluster detection algorithm using the Metropolis-Hasting algorithm method based on a Poissonian prior distribution, which is designed to find discrete objects. 
Recently \cite{2006ApJS..162..401S} developed a reconstruction method for radio-astronomy that samples from the multiplicity function (see eq.~\ref{eq:ent0}). 
Alternative approaches to the maximum likelihood for CMB-mapping algorithms try to jointly reconstruct the CMB-map with its power-spectrum using Gibbs-sampling techniques \citep{2004PhRvD..70h3511W,2004ApJ...617L..99O,2007ApJ...656..641E}. 
This approach is especially efficient with respect to other MCMC methods
because the transition probability matrix moves the system in each step of the
chain. For this special case the importance ratio is always one \citep[see e.g.][]{neal1993}. This MCMC method requires, however, the complete knowledge of the full conditional PDFs in order to sample from them.   
 Note, that the Gaussian prior for the signal simultaneously represents  the likelihood for the power-spectrum given the signal, which in this case is an inverse Gamma function for the power-spectrum (see section \ref{sec:PSEST}). This distribution naturally samples the power-spectrum, which strongly deviates from Gaussianity.  

With the aim of estimating the power-spectrum in an objective way,
non-informative priors are used. Usually a flat prior is  taken for the
power-spectrum. Alternatively, Jeffrey's prior,  for which we give a derivation
based on Fisher information (see appendix \ref{app:Jeff}), can be used. Alternatively, an entropic prior could also be taken.  

Other attempts have been made to estimate the power-spectrum from the LSS based on the distribution of galaxies. A modified Gaussian PDF with a log-normal mean has been used in this approach \citep[see][]{2005MNRAS.356.1168P}.
The same kind of concept, using a modified Gaussian distribution to sample deviations from Gaussianity, has been applied to SZ-cluster detection by \cite{2005AdSpR..36..757P}.

In this paper we propose  to apply a Gibbs-sampling algorithm to jointly
sample the underlying three-dimensional density field  with the power-spectrum
and the peculiar velocities, which can be used to correct for the
redshift-distortions.
Note, that the peculiar velocities can also be used to trace the initial density fluctuations back in time as we will discuss below.

\subsection{Algorithmic development}

 In this paper we focus our work on the numerical optimization of inverse techniques to show that a joint estimation of the LSS matter density field and its parameters is feasible (see sections \ref{sec:itinvreg} \& \ref{sec:codetesting}).

The calculation of the reconstructions, either through maximization or through
sampling, requires the inversion of certain matrices. For the Wiener-filter,
for instance, the reconstruction problem consists in one of its steps on the
inversion of the correlation matrix of the data. The methods used in this field
so far calculated this matrix and inverted it mainly using the Singular Value
Decomposition algorithm that scales as ${\cal O}(n^3)$ for a $n\times n$ matrix
\citep[see e.g.][]{1995ApJ...449..446Z}. However, this approach seems to be
hopeless in light of the overwhelming amounts of data coming from different
surveys and the possibility of combining them. We made special effort  to
implement an algorithm in which the involved matrices would not need to be
stored taking advantage of an operator formalism, which we worked out here for
different reconstruction methods (see table \ref{tab:oper} and section
\ref{sec:operators}).  Such a formalism also allows fast iterative numerical
methods that speed  the inverse step up to a scaling of ${\cal O}(n \log_2 n)$
thus reducing the main operations to fast Fourier transforms (FFTs). Some of
these numerical schemes have been used in CMB-mapping algorithms, but were
lacking a detailed comparison of the efficiency of the different methods. Such
a comparison is presented here. 
We derive  the different inverse methods in a unified way starting with a Bayesian motivation for iterative schemes (see appendix \ref{app:bayinv}) and following with a general formulation of the asymptotic regularization from which the Jacobi, the Steepest Descent, and the Krylov methods are derived. Moreover, non-linear inverse methods are discussed, like the Newton-Raphson, the Lanweber-Fridman and the non-linear Krylov methods. Preconditioning (see appendix \ref{app:prec}) was taken into account in all the derivations and the importance of such a treatment is tested in section (\ref{sec:codetesting}).
In addition, a previously not discussed Krylov method is derived (see formula \ref{eq:CG112}, section \ref{sec:itinvreg} and appendix \ref{app:CG}) and its superior efficiency is demonstrated (see section \ref{sec:codetesting}).

\subsection{Structure of the paper}

This paper is structured as follows: in section (\ref{sec:theory}) we state the
problem of signal reconstruction, then we define the data model. Subsequently,
we introduce a general statistical perspective within a Bayesian framework from
which different solutions to the reconstruction problem are presented,
including  Wiener-filtering, the COBE-filter, a novel GAPMAP algorithm with a
Poissonian likelihood and a Gaussian prior, Jeffrey's prior and the Maximum
Entropy method (MEM). Markov Chain Monte Carlo methods (MCMC) that sample the
global probability distribution function of the signal and all underlying
parameters are presented as the ideal approach to achieve a full Bayesian
solution of the reconstruction problem. In the numerical method section
(\ref{sec:itinvreg}), different iterative inverse schemes which have been
implemented in \textsc{argo} are presented, including a very efficient novel scheme. The
operator formalism is worked out for four novel algorithms in large-scale
structure reconstruction.  The efficiency of the different inverse
schemes is tested with the Wiener-filter under different reconstruction cases
with synthetic data, including structured noise, blurring, selection function
effects, and windowing in section (\ref{sec:codetesting}).
 Particular detailed derivations are presented in the appendix.

\section{Bayesian approach to signal reconstruction}
\label{sec:theory}

The reconstruction of a signal (here: DM distribution) given a set of
measurements (here: galaxy catalogues) is usually a highly degenerate problem,
as we have discussed above, where the signal is under-sampled and modified by systematic and intrinsic errors due to the nature of the observable. This is indeed the situation that we are facing, since most of the galaxy redshift surveys have partial sky coverage and the discrete nature of galaxies introduces shot noise. 

An expression for the data as a function of the real signal has to be modeled
in a first step. The reconstruction problem is classically seen as the inverse
of this functional dependence. The solution to this problem is far from being
trivial and essential issues, like solution existence, solution uniqueness, and
instability of the solving process, have to be considered. Regarding the
solution existence, there will be no model that exactly fits the data, since
the mathematical model of the physics of the system is approximate and the data
contain noise. That forces us to look for optimal solutions, rather than exact
solutions. We will have to deal especially with the last two points mentioned
above, uniqueness and stability, because an infinite set of possible solutions
can fit the data and because of the ill-conditioned  character of the system we
are treating. A regularization method that stabilizes the inverse process by
imposing additional constraints will be required. We show below how the
Bayesian framework permits us to do a regularization in a {\it natural} way and
furthermore to jointly estimate the signal and its parameters. The calculation
of the Bayesian estimators will require extra-regularization techniques, which
will be presented in section (\ref{sec:itinvreg}).
We will start posing the inverse problem by defining the model of the data.

\subsection{Data model}
\label{sec:datamodel}

 The galaxy formation process is  known to be a
complicated, non-linear and probably non-local process, as mentioned in the introduction. Thus, attempts to invert
the galaxy distribution into the original DM distribution suppose a great
challenge. It is known that, given some bias factor between the amplitude of the galaxy and the DM fluctuations, the galaxy power-spectrum on large scales fits well to the expected DM spectrum predicted from CMB observations. Detailed studies reveal that the bias factor is not universal, but
depends on galaxy type, galaxy formation time,
redshift, etc.
 The data model connecting the signal (DM distribution) to our observable (galaxy
counts) is in consequence complex, non-linear and non-local.
The main goal of this paper is to develop a Bayesian framework that permits
one to  split the
dependencies into separated problems, which can then be jointly tackled  with physical
and statistical techniques. 
In principle, also the bias of the galaxies can be sampled (see
discussion in the introduction).  However, this is out of the scope of this
paper. 

Here we present a linear data model which can easily be
extended to a simple non-linear data model by a non-linear weighting scheme
(e.g.~by weighting the
 galaxies according to their
apparent luminosity). 
Nevertheless, many of the uncertainties we are facing,
such as the convolution effects due to the blurring of a telescope, the
pixelization scheme, the mask effects due to the observation strategy, or the
selection effects due to the limited sensitivity of the detectors, can be described with a linear model. 
This linear model will contain non-linear information in the noise term.

\subsubsection{Linear data model}
\label{sec:datamodel}

The general linear reconstruction problem formally can be written as the inverse problem of recovering the signal $\mbi s$ from the observations $\mbi d$ related in the following way
\begin{equation}
{ d}(\mbi x)=\int{\rm d} \mbi y{ R}(\mbi x,\mbi y){ s}^\epsilon(\mbi y){,}
\label{eq:fred}
\end{equation}
where $R$ represents the kernel of the Fredholm integral equation of the first kind defined by (\ref{eq:fred}), with noise on the signal $s$ being expressed by the superscript $\epsilon$.
Discretizing eq.~(\ref{eq:fred}) and assuming additive noise, we can formulate the signal degradation model by
\begin{equation}
{\mbi d} = {\mat R}{\mbi s} + {\mbi \epsilon}{.}
\label{eq:data}
\end{equation}
where the $m\times 1$ vector ${\mbi d}$ represents the data points resulting from the measurements (here: galaxy counts), the statistical noise and the underlying signal are a $m\times 1$ vector ${\mbi \epsilon}$,  and a  $n\times 1$ vector ${\mbi s}$ respectively.
The object that operates on the signal is $\mat R$ a  $m\times n$ matrix which commonly describes blurring effects caused by the atmosphere, the point-spread function (PSF) of the telescope or the response function of the detectors of the instrument. 

Let us denote the physical observation process encoded in the $\mat R$-matrix as $\mat R_{\rm P}$. 
We are interested in the selection function of the survey $f_{\rm S}$ with the corresponding masks $f_{\rm M}$, which can also be included in $\mat R$.
One has to be careful with the data model defined in eq.~\ref{eq:data}. As
several authors point out, there is a correlation between the underlying signal
$\mbi s$ and the level of shot noise produced by the discrete distribution of
galaxies \citep[see e.g.][]{1998ApJ...503..492S}. Since, by definition,
additive noise assumes no correlation with the signal --otherwise we would have
signal content in the noise -- we define the effective noise $\mbi\epsilon$ as the product of  a structure function $f_{\rm SF}$, which could be correlated with the signal, with a random noise component ($\epsilon_{\rm N}$) that is uncorrelated with the signal. Given the above definitions, the effective noise $\epsilon$ is uncorrelated with the signal.   
We may then rewrite eq.~(\ref{eq:data}) in continuous representation as
\begin{equation}
{ d}(\mbi x) = \int{\rm d} \mbi y{ R}_{\rm P}(\mbi x,\mbi y){ f}_{\rm S}(\mbi y){ f}_{\rm M}(\mbi y){ s}(\mbi y) + { f}_{\rm SF}(s(\mbi x)){ \epsilon}_{\rm N}(\mbi x){,}
\label{eq:data1}
\end{equation} 
where $ R(\mbi x,\mbi y)={ R}_{\rm P}(\mbi x,\mbi y){ f}_{\rm S}(\mbi y){ f}_{\rm M}(\mbi y)$ and $\epsilon(\mbi x)={ f}_{\rm SF}(\mbi s(\mbi x)){ \epsilon}_{\rm N}(\mbi x)$. In practice, we will assume white noise (i.e.~constant noise in Fourier space), $\epsilon_{\rm N}=\epsilon_{\rm WN}$. However, none of the presented techniques in this paper depend on this simplification.
Some of the previous studies of large-scale structure reconstruction also included the inverse of the linear redshift-distortions operator as a matrix multiplying ${\mat R}$ \citep[see e.g.][]{1994ApJ...423L..93L}. Such an operator cannot easily be found for the non-linear regime. Earlier works try to correct the non-linear redshift-distortions with an additional factor in the power-spectrum analogous to Kaiser's factor \citep[see][]{1987MNRAS.227....1K, 1996MNRAS.282..877B, 2004MNRAS.352..939E}. Here, we propose a Bayesian solution to the signal reconstruction problem as it will be discussed later.

In most cases, the signal will be strongly under-constrained due to
under-sampling, i.e.~$n\gg m$, which is nearly unavoidable due to partial sky coverage of surveys. The linear equation (eq.~\ref{eq:data}) to be inverted is a rank-deficient system. Such systems are characterized by non-uniqueness, since the matrix $\mat R$ has a nontrivial null space. By superposition, any linear combination of the null space models (models ${\mbi s_0}$ that satisfy ${\mat R}{\mbi s_0}=0$) can be added to a particular solution leading to infinite solutions. Consequently, we cannot discriminate between situations where the solution is truly zero \citep[see for example][]{paramest}.
As is well known, a direct inversion of eq.~(\ref{eq:data}) (${\mat
  R}^{-1}{\mbi d}$) will amplify the statistical noise and lead to an unstable
solution \citep[see e.g.][]{1995ApJ...449..446Z}. Instead, a regularization
method, which often follows several steps, has to be applied . The first step consists of finding an expression for an estimator of the signal $\mbi s$ that approximately satisfies the data model (eq.~\ref{eq:data}) and copes with the noise. Further regularization methods are usually required in a second step to actually calculate the estimator. This happens whenever some ill-posed linear or non-linear operators have to be inverted. We shall distinguish between noise regularization and inverse regularization according to the first and the second step, respectively.  
As \cite{1995ApJ...449..446Z} pointed out, using a mean variance estimator
alone does not completely solve the inverse problem. Therefore, they proposed
the singular value decomposition algorithm (SVD) to extra-regularize these
problems. However, this method requires one to calculate the correlation matrix of the data implying a slow algorithm, scaling as ${\cal O}(n^3)$, and needs large storage facilities. 
We will show that a Bayesian approach is a {\it natural} regularizer for the noise, which then can be regularized further for the inverse purpose  with efficient methods that scale as ${\cal O}(n \log_2 n)$ (see section \ref{sec:itinvreg}).
Let us address the problem of signal reconstruction from a statistical inference perspective.  

\subsection{Inversion via statistical estimator}
\label{sec:stat}

In parametric modeling  it is assumed that observational data have been generated by random processes with probability density distributions, depending on the model parameters \citep[see for example][]{bayeschoice}. 
Statistical analysis in this context is essentially an inverse method, which aims at retrieving the causes (here reduced to the parameters of the probabilistic generating mechanism) from the effects (here summarized by the observations). 

 Traditionally, one tries to find a way where the available information is optimally used and a unique estimator is selected from an infinite set of solutions.
 One of the classical approaches consists of minimizing the variance of the
 residuals, which is the variance of the discrepancy between the estimator and
 the set of possible realizations consistent with the data
 \citep[see][]{1992ApJ...398..169R}. This conjecture is reasonable because the
 least deviation from the set of {\it true} signals is searched. The estimator
 obtained in this way is called the least squares quadratic (LSQ)
 estimator. However, a transparent statement of the statistical assumptions is
 missing in this method, contrary to the Bayesian approach used in this work as
 will be shown below. Moreover, Bayesian statistics allows sampling the PDF of
 the system under consideration in a natural way. Strictly speaking, one does not look for a unique estimator in this framework. Nevertheless, a summary of the PDF can be given by the mean of the sample (see section \ref{sec:MCMC}).

The most general approach to determine an estimator, however, should be based on the global (joint) PDF over all relevant quantities, like the signal ${\mbi s}$ and all model parameters ${\mbi p}$, without neglecting any possible dependences. 
Let us assume that $P({\mbi s}, {\mbi p}\mid {\mbi d})$, the joint PDF of the system under consideration, depends on the signal ${\mbi s}$ and a series of additional parameters ${\mbi p}$, given the observations ${\mbi d}$. 
One  solution  would then be to calculate the expectation of the signal over the joint PDF space 
\begin{equation}
{\rm E}_{\rm joint}({\mbi s})\equiv \int  {\rm d}{\mbi s}\, {\rm d}{\mbi p}\,  \Big[ P({\mbi s}, {\mbi p}\mid {\mbi d}) \,{\mbi s} \Big]\equiv\langle\mbi s\rangle_{(\mbi s,\mbi p|\mbi d)} {,}
\label{eq:joint}
\end{equation}
where we have introduced the ensemble average $\langle\rangle_{(\mbi s,\mbi p|\mbi d)}$  with the subscript representing the PDF over which the integral is done $P({\mbi s}, {\mbi p}\mid {\mbi d})\rightarrow({\mbi s}, {\mbi p}\mid {\mbi d})$\footnote{Sometimes, however, the ensemble angles will denote the estimator of some signal or parameter in a more general sense, like the maximum likelihood or the maximum a posteriori (see sections \ref{sec:likel} and \ref{sec:prior}, respectively). Note that a bracket formalism could be introduced at this point, in which eq.~(\ref{eq:joint}) would be represented in the following way: $(\mbi s|\mbi s|\mbi p,\mbi d)$.}. Expression (\ref{eq:joint}) can consequently be read as the ensemble average over all possible signals and parameters.
The joint PDF is unfortunately quite hard to calculate directly, and the
integral in eq.~(\ref{eq:joint}) is  computationally too expensive for
realistic cases as it involves many parameters and a large amount of data. 
To disentangle the uncertainties in parameter and signal spaces, let us apply the product rule of statistics\footnote{$P({\mbi s}, {\mbi p}\mid {\mbi d})=P({\mbi s}\mid \mbi p, {\mbi d})P({\mbi p}\mid\mbi d)$} to eq.~(\ref{eq:joint}) 
\begin{eqnarray}
{\rm E}_{\rm joint}({\mbi s})&=& \int {\rm d}{\mbi p}\, P({\mbi p}\mid\mbi d) \left[\int  {\rm d}{\mbi s}\, \Big[ P({\mbi s}\mid \mbi p, {\mbi d})\,{\mbi s}  \Big]\right]\nonumber \\
&=& {\rm E}_{\mbi p}\Big[{\rm E}_{\mbi s}\left({\mbi s}\mid {\mbi p},{\mbi d}\right)\mid \mbi d\Big]=\langle\langle\mbi s\rangle_{(\mbi s|\mbi p,\mbi d)}\rangle_{(\mbi p|\mbi d)} {.}
\label{eq:eq3}
\end{eqnarray}
This means that the expectation of the signal ${\mbi s}$  corresponds to 
the average of the conditional mean of $\mbi s$ over the marginal distribution of $\mbi p$ \citep[see for example][]{bayesdataanal}, where the conditional mean is given by
\begin{equation}
{\rm E}_{\rm cond}({\mbi s})= {\rm E}_{\mbi s}({\mbi s}\mid{\mbi p,\mbi d}) = \int  {\rm d}{\mbi s}\, \Big[ {P({\mbi s}\mid {\mbi p, \mbi d}) \,\mbi s} \Big]=\langle\mbi s\rangle_{(\mbi s|\mbi p,\mbi d)}{.}
\label{eq:cond}
\end{equation}
Traditionally, the conditional PDF has been used to determine the estimator of the signal assuming that all the parameters are known \citep[e.g.~][]{1995ApJ...449..446Z}.

As the reconstruction step of the density field is computationally expensive, a joint estimation of the parameters is out of scope. Therefore, the reduced approach of basing the estimators on conditional PDFs provides a computationally more feasible way to tackle problems of this kind. 
In particular, we will demonstrate that an operator formalism allows efficient sampling of the conditional PDFs, enabling us to  sample the joint PDF in a Bayesian framework.   

\subsection{Bayesian approach}
\label{sec:bay}

Given a data model, one can usually find an expression for the sampling distribution, i.e.~the probability of obtaining the data given the signal and some additional parameters ${\mbi p}$, $P({\mbi d}\mid {\mbi s, \mbi p})$. This is much less difficult than a direct calculation of the posterior $
P({\mbi s}\mid {\mbi d, \mbi p})$. We need an expression which relates both the sampling and the posterior distribution given by Bayes theorem. The derivation of Bayes theorem is straightforward from the joint PDF of the signal and the data, using the product rule and the fact that the joint PDF is invariant under permutations of its arguments\footnote{
\begin{eqnarray}
P({\mbi s}, {\mbi d, \mbi p}, I)&=&P({\mbi s}\mid {\mbi d, \mbi p}, I)P({\mbi d \mid \mbi p}, I)=\nonumber\\
P({\mbi d}, {\mbi s, \mbi p}, I)&=&P({\mbi d}\mid {\mbi s, \mbi p}, I)P({\mbi s \mid \mbi p}, I)\nonumber
\end{eqnarray}
}. Bayes theorem can be expressed by the following equation
\begin{equation}
P({\mbi s}\mid {\mbi d, \mbi p}, I)= \frac{P({\mbi d}\mid {\mbi s, \mbi p}, I)P({\mbi s\mid \mbi p}, I)}{P({\mbi d \mid \mbi p}, I)}{,}
\label{eq:bayes}
\end{equation}
where $P({\mbi s \mid \mbi p}, I)$ represents the prior knowledge about the signal, as it models the signal before any observations occur. 
The PDF given by $P({\mbi d \mid \mbi p}, I)$ stands for the so-called evidence that is treated as the normalization of the posterior 
\begin{equation}
P({\mbi d}\mid {\mbi p}, I)= \int{\rm d}{\mbi s}\,P({\mbi d}\mid {\mbi s}, {\mbi p}, I)P({\mbi s}\mid {\mbi p}, I) {.}
\label{eq:evidence}
\end{equation}
It is worth mentioning that all the probabilities are conditional to the
underlying physical picture, or prior information $I$. This has to be
explicitly considered in case of model comparisons.
In the following sections, we will present the steps for completing a Bayesian
analysis, starting with the likelihood, then discussing the importance of the
prior, and finishing with sampling through the joint signal and parameter
space. Note that different choices for these three components (likelihood,
prior, and sampling) lead to different classes of reconstruction algorithms. An
overview  of the different reconstruction scheme implementations based on this classification can be found in table (\ref{tab:rec}).

\subsection{The likelihood}
\label{sec:likel}

The likelihood function is formally any function of the parameters
${\mbi\theta}$ proportional to the sample density \cite[see][]{toolsstatinf}
\begin{equation}
{\cal L}({\mbi \theta}\mid{\mbi d})\propto P({\mbi d}\mid{\mbi \theta}){.}
\label{eq:like}
\end{equation}
Many inference approaches are based on the likelihood function, justified by
the likelihood principle, which states that the information obtained by an
observation ${\mbi d}$ about ${\mbi \theta}$ is entirely contained in the
likelihood function ${\cal L}({\mbi \theta}\mid{\mbi d})$.   
To be specific, if $d_1$ and $d_2$ are two observations depending on the same
parameter $\theta$ such that there exists a constant $c$ satisfying ${\cal
  L}_1(\theta\mid d_1)=c{\cal L}_2(\theta\mid d_2)$ for every $\theta$, $d_1$
and $d_2$ then bring the same information about $\theta$ and must hence lead to
identical inferences \cite[see][]{bayeschoice}. 
 
Maximum likelihood (ML) methods, for example, rely on the likelihood principle with an estimator of the parameters given by
\begin{equation}
\langle{\mbi\theta}\rangle_{\rm ML}={\rm arg}\,{\rm sup}_{\mbi\theta}\,{\cal L} ({\mbi \theta}\mid{\mbi d}){,} 
\label{eq:ML}
\end{equation}
i.e., the value of ${\mbi\theta}$ that maximizes the probability density at ${\mbi d}$. 
Bayesian methods take also advantage of the likelihood principle incorporating the decision-related requirement of the inferential problem through the definition of a prior distribution (see section \ref{sec:prior}).
The definition of the likelihood is the first step in a Bayesian framework to determine the posterior distribution (see eq.~\ref{eq:bayes}).
 In using galaxy redshift surveys to trace the matter distribution, we have to deal with the discrete nature of the data sample. Thus the likelihood may be derived here for Poissonian statistics.

\subsubsection{Poissonian likelihood}
\label{sec:pois}

The likelihood of our galaxy distribution may be approximately represented by a Poissonian distribution (the real statistics should describe the much more complex galaxy formation process). Under the assumption of independent and identically distributed ({\it iid}) observations, this yields
\begin{eqnarray}
{\cal L}({\mbi s}\mid{\mbi d},{\mbi p})
& \propto &\\
 P({\mbi d}\mid{\mbi s},{\mbi p})&=& \prod_{i=1}^m {\rm exp}\left(-\left[({\mat
       R}{\mbi s}')_i+c_i \right]\right)\frac{[({\mat R}{\mbi s}')_i+c_i]^{{d}'_{i}}}{{d}'_{i}!}\nonumber{,}
\label{eq:likepois}
\end{eqnarray}
where ${d}'_{i}$ are the galaxy counts per cell $i$ and the real, positive signal of the expectation value of the number of galaxies is given by ${\mbi s}'_{i}=\overline{n_{\rm g}}(1+b{\mbi s}_i)$, with $s_i=\delta_{\rho i}=\frac{\rho_i-\overline{\rho}}{\overline{\rho}}$ the DM over-density, our target signal. 
The quantity $n_{\rm g}$ stands for the mean number of galaxies, $\overline{\rho}$ represents the mean density and $b$ the bias factor. All these quantities are redshift-dependent.
The additional parameters $\mbi p$ in this case would be represented by some background $c_i$ and would enter into the operator $\mat R$ that modifies the signal $\mbi s$.

For a similar application in astronomy see \cite{1989MNRAS.240..753L} and
\cite{maxentbaymeth}.  If ${d}'_{i}$ is not an integer, e.g.~due to some
interpolation process, a Gamma function may be used instead of the factorial, ${d}'_{i}! \rightarrow \Gamma({d}'_{i}+1 )$.

\subsubsection{Gaussian likelihood}
\label{sec:gauslike}

When the number of counts is large the Poisson distribution can be approximated
by the normal distribution. In that case, the likelihood can be given by a Gaussian distributed noise
\begin{eqnarray}
{\cal L}({\mbi s}\mid{\mbi d},\mbi p) 
& \propto & \nonumber\\
P({\mbi d\mid\mbi s},\mbi p)&=&\frac{1}{[(2\pi)^m{\rm det}({\mat N})]^{1/2}}{\rm exp}\left(-\frac{1}{2}{\mbi \epsilon}^{\dagger}{\mat N}^{-1}{\mbi \epsilon}\right) \nonumber\\
&\propto & {\rm exp}\left[-\frac{1}{2}\chi^2({\mbi s})\right]{,}
\label{eq:likegaus}
\end{eqnarray}
where ${\mat N}\equiv\langle\mbi\epsilon\mbi\epsilon^{\dagger}\rangle_{(\mbi\epsilon|\mbi p)}$ is the covariance matrix of the noise ${\mbi\epsilon\equiv \mbi d-\mat R\mbi s}$, and
\begin{equation}
\chi^2({\mbi s})=({\mbi d}-{\mat R}{\mbi s})^{\dagger}{\mat N}^{-1}({\mbi d}-{\mat R}{\mbi s}){.}
\label{eq:chi2}
\end{equation}
The parameters $\mbi p$ determine the structure of the noise $\mbi\epsilon$ (and therefore the structure of the covariance matrix $\mat N$), and also enter into the operator $\mat R$. We give different expressions for the noise covariance matrix $\mat N$ in section (\ref{sec:operators}).

Note that $\chi^2$ coincides with the square of the Mahalanobis
distance\footnote{We introduce here a generalized definition of the Mahalanobis
  distance as: ${ D}^2_{\rm Mah}(\mbi x,\mbi y)_{\mat M}=(\mbi x- \mbi
  y)^\dagger\mat M(\mbi x-\mbi y)$, with $\mbi x$ and $\mbi y$ being two
  vectors in the $N$-dimensional space and $\mat M$ a $N\times N$ matrix.}
between $\mbi d$ and ${\mat R}{\mbi s}$, and also coincides with the squared $\mat N^{-1}$-norm of the error
\begin{equation}
\chi^2({\mbi s})={ D}^2_{\rm Mah}({\mbi d},{\mat R}{\mbi s})_{\mat N^{-1}}=||\mbi\epsilon||^2_{\mat N^{-1}}{.}
\label{eq:mah}
\end{equation}
In this case, the ML will correspond to the least squares of
the error. It will minimize the $\chi^2({\mbi s})$ and hence minimize the
Mahalanobis distance between the data and the noise-free data model. Therefore,
the ML is equivalent to searching the estimator that fits the data better without constraining the model for the signal. Let us study the prior that precisely sets constraints on the signal $\mbi s$.  

\begin{table*}
\rotatebox[]{90}{
\begin{tabular}{|c|c|c|c|c|} 
\multicolumn{5}{|c|}{ \hspace{3cm}{\bf Non-informative priors}  \hspace{5.5cm}  {\bf Informative priors (MAP)} } \\
    &  \hspace{-2.cm}{\bf Prior}\hspace{1.cm} {\bf Flat} ({\bf ML}) &  {\bf Entropic} ({\bf MEM}) & {\bf Gaussian}              &  {\bf Poissonian} \\ 
 {\bf Likelihood} &  &&\\ \hline
 {\bf Gaussian} &   &  & WIENER (Tikhonov, Ridge)      &        \\ 
 {\bf --Radio} &  &  \cite{2006ApJS..162..401S}$^\#$  &       &       \\ 
      {\bf --CMB}           &  COBE:  \cite{1992issa.proc..391J}   & \cite{1997MNRAS.290..313M}  &    \cite{1995ApJ...446...49B}          &    \cite{2003MNRAS.338..765H}$^{\#}$               \\   
     &   Tegmark (1997)    &      \cite{1998MNRAS.300....1H}   &     Tegmark (1997)        &          \\
        &   ROMA: \cite{natoli}     &           &             &       \\
        &  MAPCUMBA: \cite{mapcumba}    &      &          &                \\
        & MAXIMA: \cite{maxima}   &        &              &             \\
        & MAGIC$^{\#}$: \cite{2004PhRvD..70h3511W}  &        &  MAGIC$^{\#}$: \cite{2004PhRvD..70h3511W}     &\\        
&  MIRAGE:  \cite{mirage}     &       &   \cite{2004ApJ...617L..99O}$^{\#}$           &          \\
&  MADAM:  \cite{madam}    &    &    \cite{2007ApJ...656..641E}$^{\#}$               &          \\    
 & &   &\cite{2007ApJ...656..653L}$^{\#}$ &\\
 {\bf  --LSS}        &     &  &  \cite{WienerFSL}   &                \\
        &    &      &  \cite{1994ASPC...67..185H}                           \\
        &    &     &    \cite{1994ApJ...423L..93L}, \cite{1994ASPC...67..171L}&                            \\
        &      &  &     \cite{1995ApJ...449..446Z}                     \\
        &   &     & \cite{1995MNRAS.272..885F}            &                 \\
        &      &     &  \cite{1997MNRAS.287..425W}            &                \\
        &      &     &  \cite{1999ApJ...520..413Z}            &                \\
        &      &     &   \cite{1999AJ....118.1146S}                         \\
        &    &     &    \citeauthor{2004MNRAS.352..939E} (2004,2006) \nocite{2006MNRAS.373...45E}                        \\
         &       & \textsc{argo}: MEMG$^*$ &   \textsc{argo}: WIENER$^{**\#}$   &            \\
 & & (section \ref{sec:entr} and appendix \ref{app:MEM})         &(sections \ref{sec:WF}, \ref{sec:MCMC}, \ref{sec:codetesting} and appendix \ref{sec:mapeq}) & \\\hline
 {\bf Poissonian} & \cite{1972JOSA...62...55R}   & \textsc{argo}: MEMP$^*$ & \textsc{argo}: GAPMAP$^*$ &  \\
        &  \cite{1974AJ.....79..745L}    &  (section \ref{sec:entr} and appendix \ref{app:MEM})      &  (section \ref{sec:gauspois} and appendix \ref{app:POISLIKE})     &                    \\
\hline
{\bf Inverse Gamma} &  & &  &\\
{\bf --CMB} & MAGIC$^{\#}$: \cite{2004PhRvD..70h3511W} & &  &\\
  &\cite{2004ApJ...617L..99O}$^{\#}$ & &  &\\
  &\cite{2007ApJ...656..653L}$^{\#}$ & &  &\\
 &\cite{2007ApJ...656..641E}$^{\#}$ & &  &\\
{\bf --LSS} & \textsc{argo}$^{*\#}$ & &  &\\
 & (section \ref{sec:PSEST}) & &  &\\
\hline
{\bf Modified Gaussian} &  & &  &\\
{\bf --CMB}  & \cite{2005AdSpR..36..757P}$^{\#}$ & &  &\\
{\bf --LSS} & \cite{2005MNRAS.356.1168P}$^{\#}$ & &  &\\
\hline
\multicolumn{5}{|c|}{$^{*}$developed and presented in this paper; $^{**}$developed, tested and presented in this paper; $^{\#}$able to sample PDFs}\\
\hline
\multicolumn{5}{|c|}{We have left out the reconstruction methods that are focused on the cosmological initial conditions, since they address a different problem and, in general,}\\
\multicolumn{5}{|c|}{cannot be classified in terms of the PDFs listed in this table. Neither can other reconstruction algorithms based on geometrical arguments, }\\
\multicolumn{5}{|c|}{like Voronoi, Delaunay tessellations, {\it friends-of-friends} schemes or {\it cloud-in-cell} interpolation schemes, be classified here. }\\
\hline
\end{tabular}
}
\caption{ \label{tab:rec} Classification of reconstruction methods in
  astrophysics based on the prior (columns) and likelihood (rows). Note that
  most of the reconstruction algorithms in other research areas, such as
  tomography, where Tikhonov-regularization is widely used, or the algebraic
  reconstruction technique (ART), which is based on the asymptotic
  regularization, fall into the class of Wiener-filtering schemes as we show in
  section (\ref{sec:baytik}) and appendix \ref{app:bayinv}.  The differences in
  the ML CMB-map-making algorithms reside mainly in the modeling of the complex
  noise structure that arises  due to the scanning strategies of the satellites
  and in the various foreground removal methods. The LSS Wiener-filtering
  methods on the other hand present improvements in the redshift distortions
  treatment, or  are based on the different input data, either galaxy-positions
  or peculiar velocities. The discrete object detection
  \citep{2003MNRAS.338..765H} algorithm was developed to find
  Sunyaev-Zel'dovich clusters. This is also the case for the modified Gaussian
  by Pierpaoli \& Anthoine (2005). The reconstruction of the power-spectrum is also listed here. In CMB the joint map and power-spectrum estimation is done by MAGIC. \citet{2005MNRAS.356.1168P} samples the power-spectrum with a modified Gaussian likelihood given by a log-normal mean.
We propose to follow the steps done in CMB and sample the density field
  and the power-spectrum jointly (see section \ref{sec:PSEST}). This paper covers three new areas in LSS (GAPMAP, MEMG, MEMP) and presents four novel algorithms with which reconstructions can be done very fast. 
}
\end{table*}


\subsection{The prior}
\label{sec:prior}

A second step in Bayesian analysis is to specify the prior distribution for the
signal, which contains the prior knowledge about the signal before the
measurements were carried out. For little informative data it can strongly
affect the posterior distribution and thus modify any inference based on
it. For this reason, frequentists criticize Bayesian methods as being
subjective. Other definitions of probability, like the frequentist, however,
can be shown in most of the situations to be  particular cases of the Bayesian
approach \citep[see e.g.][]{toolsstatinf}, implying the use of an implicit
prior. The advantage of defining the prior knowledge about the system under
consideration is that the interpretation of the results is straightforward,
especially because assumptions flowing into the inference procedure are clearly stated. 
Once the prior is defined, we can obtain the maximum a posteriori (MAP) estimator, by maximizing the posterior distribution, which is proportional to the likelihood multiplied by the prior, 
\begin{equation}
\langle{\mbi\theta}\rangle_{\rm MAP}={\rm arg}\,{\rm sup}_{\mbi\theta}\,{P} ({\mbi \theta}\mid{\mbi d}){.} 
\label{eq:MAP}
\end{equation}
Note that there is a crucial difference to the maximum likelihood estimator (eq.~\ref{eq:ML}) due to the incorporation of the prior information.

\subsubsection{Bayes and regularization methods: the prior as a regularizer}
\label{sec:baytik}

Looking at the $\log$-probabilities, we see that the MAP estimator maximizes the following quantity using Bayes theorem (${\log P(\mbi \theta\mid\mbi d)\propto\log(P(\mbi d\mid\mbi\theta)P(\mbi\theta))}$)
\begin{equation}
Q=\log{P} ({\mbi d}\mid{\mbi \theta})+\log{P} ({\mbi \theta}){.} 
\label{eq:Q}
\end{equation}
 If we assume that the error is Gaussian distributed, (which is a fair assumption if there is no prior information about the noise), and we parameterize the prior of the parameter, say the signal $\mbi s$,  we can rewrite eq.~(\ref{eq:Q}) as ($2Q\rightarrow Q$) 
\begin{equation}
Q=-\chi^2({\mbi s})+\alpha f_{\rm p}({\mbi s}) {,} 
\label{eq:Q2}
\end{equation}
where we absorbed the factor $2$ in the Lagrangian multiplier $\alpha$, and
$f_{\rm p}$ represents the penalty function that obliges the estimator to
fulfill some constraint on the parameter $\mbi s$, to the detriment of the
$\chi^2({\mbi s})$ that strongly relies on the data. If we further assume that
$\mat N^{-1}=\mat I$ (say we have white noise), the Mahalanobis distance
reduces to the Euclidean distance \\(${ D}^2_{\rm Mah}({\mbi d},{\mat R}{\mbi
  s})|_{\mat N^{-1}=\mat I}={ D}^2_{\rm Euc}({\mbi d},{\mat R}{\mbi s})$), and the quantity one wants to minimize reads
\begin{equation}
||\mbi\epsilon||^2+\alpha f_{\rm p}({\mbi s}) {,} 
\label{eq:Q3}
\end{equation}
where we have absorbed the minus sign in $\alpha$.
Expression (\ref{eq:Q3}) is equivalent to least squares with a regularization term, and belongs to Ridge-regression problems \citep{ridgeregression,ridgeregres}.
Assuming that the penalty function takes the following form $f_{\rm p}({\mbi s})=||\mbi s||^2$, we can write expression (\ref{eq:Q3}) as
\begin{equation}
||\mbi\epsilon||^2+\alpha ||{\mbi s}||^2 {,} 
\label{eq:Q4}
\end{equation}
which then becomes the Tikhonov regularization method \citep{tikhonov}. The
parameter $\alpha$ is called the regularization parameter. These methods lead
to linear filters and are essentially identical to Wiener-filtering
\citep{wienerkolmogorov}, which will be presented in the next section. Note
that Tikhonov regularization is equivalent to MAP of a Gaussian likelihood with
noise covariance matrix ${\mat N=\mat I}$ and Gaussian prior, with signal
covariance matrix ${\mat S=\alpha^{-1}\mat I}$. Nevertheless, the penalty
function $f_{\rm p}$ in general can be a non-linear function of the parameter to be estimated (say the signal $\mbi s$) leading to non-linear estimators. We will introduce MEM as such an example. Tikhonov regularization can also be generalized to non-linear problems by introducing a non-linear kernel operator $\mat R(\mbi s)$.
 
Summarizing the exposed theory of signal reconstruction, we might interpret  the likelihood as some distance measure between the data and the noise-free model of the data, and the prior as some constraint that tightens the estimator to the model of the signal. 
We have shown here that the classical methods of signal reconstruction, like the Tikhonov regularization,  are particular cases of the Bayesian approach.
The inclusion of a prior can be regarded as a {\it natural} regularization, in
the sense that the regularization term is provided by a (physical) model of the
{\it true} signal. In appendix \ref{app:bayinv}, we discuss the relation between other regularization methods and the Bayesian approach.
In the following sections we introduce different priors that are relevant for large-scale structure reconstruction and are implemented in \textsc{argo}.

\subsubsection{Gaussian prior}
\label{gaussprior}

The distribution of the primordial density field should be very close to
Gaussianity according to most of the inflationary scenarios
\citep{1981PhRvD..23..347G, 1982PhLB..108..389L, 1982PhRvL..48.1220A}. In fact,
the measurements of the CMB show very small deviations from Gaussianity
\citep[see e.g.][]{2003ApJS..148..119K}. Non-Gaussianities in the matter
distribution arose mainly from non-linear gravitational collapse. The
non-linear regime of structure formation is responsible for the strong radial
redshift-distortions, the {\it finger-of-god} effect, limiting the accuracy of
reconstructions. Previous attempts to correct for these distortions have
modified the power-spectrum by introducing a Lorentzian factor \citep[see
e.g.][]{1996MNRAS.282..877B, 2004MNRAS.352..939E}. In section (\ref{sec:MCMC}) we propose an alternative way to do this in a Bayesian framework, where peculiar velocities are sampled together with the three dimensional map of the matter distribution. For the underlying DM density fluctuation we will assume a Gaussian prior. This is a crude approximation for the density field at the present epoch of the Universe, especially on small-scales. It is, however, a valid description on large-scales and allows to incorporate non-linear corrections in a MCMC fashion, as will be discussed in section (\ref{sec:MCMC}). Following \citet{1986ApJ...304...15B} we may thus write the PDF of the signal as a multivariate Gaussian distribution
\begin{equation}
P({\mbi s}\mid \mbi p)=\frac{1}{[(2\pi)^n{\rm det}({\mat S})]^{1/2}}{\rm exp}\left(-\frac{1}{2} {\mbi s}^{\dagger}{\mat S}^{-1}{\mbi s}\right){,}
\label{eq:priorgaus}
\end{equation}
with ${\mat S}$ being the covariance matrix of the signal (${\mat S=\mat S(\mbi p)\equiv\langle\mbi s\mbi s^\dagger \rangle_{(\mbi s|\mbi p)}}$).
This formula emphasizes the high dimensional character of the problem (n dimensions of the signal reconstruction, with n being typically between $10^3$ and $10^9$).

\subsubsection{Gaussian prior and Gaussian likelihood: the Wiener-filter}
\label{sec:WF}

The Gaussian prior together with the Gaussian likelihood lead to the
Wiener-filter, completing the square for the signal in the exponent of the
posterior distribution (see \cite{1995ApJ...449..446Z} and appendix~\ref{app:WF}),
\begin{eqnarray}
\lefteqn{P({\mbi s}\mid{\mbi d},\mbi p)}\nonumber\\
&&\propto {\rm exp}\left(-\frac{1}{2}\left[ {\mbi s}^{\dagger}{\mat S}^{-1}{\mbi s}+({\mbi d-\mat R\mbi s})^{\dagger}{\mat N}^{-1}({\mbi d-\mat R \mbi s})\right]\right)\nonumber\\
 &&\propto {\rm exp}\left(-\frac{1}{2}\left[ ({\mbi s-\langle{\mbi s}\rangle_{\rm WF}})^{\dagger}{(\mbi\sigma_{\rm WF}}^2)^{-1}({\mbi s-\langle{\mbi s}\rangle_{\rm WF}}) \right]\right){,}\hspace{0.5cm}
\label{eq:WFBAY}
\end{eqnarray}
where the Wiener-filter used to calculate the estimator from the data $\langle{\mbi s}\rangle_{\rm WF}=\mat F_{\rm WF}\mbi d$ is given by 
\begin{equation}
{\mat F}_{\rm WF}={(\mat S^{-1} +\mat R^\dagger \mat N^{-1}\mat R)^{-1}\mat R^\dagger\mat N^{-1}}{,}
\label{eq:WF2}
\end{equation}
 and the corresponding covariance is 
\begin{equation}
{\mbi\sigma}^2_{\rm WF}=\langle\mbi r\mbi r^\dagger \rangle_{\rm WF}=({\mat S^{-1}+\mat R^\dagger\mat N^{-1}\mat R})^{-1}{,}
\label{eq:WFvar}
\end{equation}
with ${\mbi r}={\mbi s}-\langle{\mbi s}\rangle_{\rm WF}$ being the residual.
A similar filter to the Wiener-filter can be obtained by the LSQ estimation \footnote{Note
  that in this case, the least squares are referred to the residuals $\mbi r$,
  i.e.~the difference between the real signal $\mbi s$ and the estimated signal
  $\langle\mbi s\rangle_{\rm LSQ}$: ${||\mbi r||^2=||{\mbi s}-\langle{\mbi
      s}\rangle_{\rm LSQ}||^2}$, where the prior on $\mbi s$ is given in a more
  implicit way by assuming a linear relation between the estimator and the data
  and statistical homogeneity.} \citep[for an explicit derivation see][and
appendix \ref{sec:mapeq}]{1995ApJ...449..446Z} leading to the following
expression
\begin{equation}
\langle\mbi s\rangle_{\rm LSQ} =\langle{\mbi s}
{\mbi d}^\dagger\rangle \langle{\mbi d}{\mbi d}^\dagger\rangle^{-1}\mbi d {,}
\label{eq:WF3}
\end{equation}
where the correlation matrix of the signal and the data ($\langle{\mbi s}
{\mbi d}^\dagger\rangle$) is multiplied by the inverse of the autocorrelation matrix of the
data ($\langle{\mbi d}{\mbi d}^\dagger\rangle^{-1}$).
Given that the signal and the noise are uncorrelated ($\langle\mbi
s\mbi\epsilon^\dagger\rangle=0$), the correlation matrix of the signal and the
data reduces to: $\langle\mbi s\mbi d^\dagger\rangle= \mat S \mat
R^\dagger$. Thus, the filter in eq.~(\ref{eq:WF3}) can be reformulated as
\begin{equation}
{\mat F}_{\rm LSQ}={\mat S \mat R^\dagger(\mat R \mat S \mat
  R^\dagger+\langle\mat N\rangle_{(\mbi s|\mbi p)})^{-1}}{.}
\label{eq:LSQ}
\end{equation}
The noise covariance matrix for the LSQ estimator will differ from the one in
the likelihood, if there is a signal dependence in the structure function of
the noise term as it is the
case for a Poissonian-like distribution.

From the structure of the LSQ filter $\mat F_{\rm LSQ}$ (eq.~\ref{eq:LSQ}), one could postulate another expression for the Wiener-filter given by:
\begin{equation}
{\mat F}_{\rm WF}={\mat S \mat R^\dagger(\mat R \mat S \mat R^\dagger+\mat N)^{-1}}{.}
\label{eq:WF}
\end{equation}
We show in appendix~(\ref{app:WIENER}) that both expressions for the
Wiener-filter (eqs.~\ref{eq:WF2} and \ref{eq:WF}) are equivalent.
From now on, we will call eq.~(\ref{eq:WF}) the data-space representation of the
Wiener-filter, and eq.~(\ref{eq:WF2}) the signal-space representation of the Wiener-filter.
Note, that the LSQ estimator will coincide with the Wiener-filter  after performing an ensemble
average over all possible signal realizations: $\langle\mbi s\rangle_{\rm LSQ} =\langle\langle\mbi s\rangle_{\rm
  WF}\rangle_{(\mbi s|\mbi p)}$.

The following notation can be introduced for the posterior PDF 
\begin{equation}
{P({\mbi s}\mid{\mbi d},\mbi p)\propto G(\mbi s-\langle{\mbi s}\rangle_{\rm
 WF},\mbi\sigma^2_{\rm WF})}{,}
\label{eq:PWF}
\end{equation}
 i.e.~given a dataset ${\mbi d}$ derived from a Gaussian process, the possible signals are Gaussian distributed around the Wiener-filter reconstruction $\langle{\mbi s}\rangle_{\rm WF}$ with a covariance ${\mbi \sigma}_{\rm WF}^{2}$. 
The parameters $\mbi p$ enter the operator $\mat R$, including also the
cosmological parameters that determine the signal covariance matrix $\mat S$.
We will discuss in section (\ref{sec:MCMC}) how to sample $\mat S$ and to determine cosmological parameters.

A remarkable characteristic of the Wiener-filter is that it suppresses the
signal in the presence of a high noise level resulting in the null estimator and gives just the deblurred data when noise is negligible. In this sense it is a  biased estimator, since its covariance matrix has less power than the original one. Some attempts have been made to derive an equivalent unbiased estimator \citep[see][]{2002MNRAS.331..901Z}. However, one might be especially interested in obtaining a conservative estimator. Sampling the joint PDF will fill the missing modes \citep[see e.g.][]{2004PhRvD..70h3511W} and in this way complete the signal in regions where it is under-sampled or the signal to noise ratio is low.  
It is interesting to note that the Wiener Filter coincides with the MAP
estimator in the case of a Gaussian prior on ${\mbi s}$ and a Gaussian
likelihood ($\langle\mbi s\rangle_{\rm WF}=\langle\mbi s\rangle_{\rm
  MAP}$). Performing the integral of the
conditional PDF (see eq.~\ref{eq:cond}) one obtains the same estimator again,
thus $\langle\mbi s\rangle_{\rm WF}=\langle\mbi s\rangle_{(\mbi s|\mbi d,\mbi
  p)}$. This is a very important result, since it permits one to sample the conditional PDF. We propose to exploit this property for the joint estimation of the signal and its power-spectrum as is done in the CMB (see \cite{2004PhRvD..70h3511W} and section \ref{sec:PSEST}).

\subsubsection{Gaussian prior and Poissonian likelihood: the GAPMAP estimator}
\label{sec:gauspois}

The Gaussian likelihood constitutes a valid approximation when the Poissonian character of the distribution is appropriately modeled in the noise correlation matrix ${\mat N}$. However, one would rather describe a discrete sampling process like a galaxy survey with a Poissonian likelihood. Unfortunately, there is no filter available for such a case. Thus, we present a novel iterative equation for the MAP estimator with a Gaussian prior and a Poissonian likelihood, which we call GAPMAP (see appendix \ref{app:POISLIKE} for a derivation) 
\begin{equation}
{\mbi s}^{j+1}={\mat S}{\mat R}^{\dagger}b\overline{n_{\rm g}}\left(-\vec{1}+{\rm diag}\left({\mat R}\overline{n_{\rm g}}(\vec{1}+b{\mbi s}^j)+{\mbi c}\right)^{-1}{\mbi d}'\right)
{.}
\label{eq:likelpois}
\end{equation}


\subsubsection{Flat prior} 
\label{sec:flatprior}

 With the aim of deriving objective posterior distributions, non-informative prior distributions are introduced. A non-informative prior would suggest that any value is reasonable. 
Flat priors where the probability distribution is assumed to be constant ${P}({\mbi s})= {\rm const}$ are thus very often applied. Note, however, that these are improper priors, since the integral of these distributions diverges to infinity. In this case, the posterior is proportional to the likelihood. The maximum likelihood solution coincides in this way with the MAP estimator assuming a flat prior ($\langle\mbi s\rangle_{\rm ML}=\langle\mbi s\rangle_{\rm MAP}|_{\rm flat}$). 

\subsubsection{Flat prior and Gaussian likelihood: the COBE-filter} 
\label{sec:flatgaus}

In CMB map-making algorithms it is common to use the so-called COBE-filter
\citep[see][]{1992issa.proc..391J, 1997ApJ...480L..87T}, which can easily be
derived by maximizing the likelihood given in eq.~(\ref{eq:likegaus})
\begin{equation}
{\mat F}_{\rm COBE}=({\mat R^\dagger\mat N^{-1}\mat R})^{-1}\mat R^\dagger\mat N^{-1}{.}
\label{eq:COBE}
\end{equation}
This filter has the property that among all unbiased linear estimators (with a
noise of zero mean), it leads to the minimum variance \citep{natoli}. Here unbiased means that the statistical mean of the estimator is equal to the {\it true} signal. This is, however, only fulfilled when the inverse of $\mat R^\dagger\mat N^{-1}\mat R$ exists (see appendix~\ref{app:COBE}). 
 The covariance for the COBE-filter can found to be 
\begin{equation}
{\mbi\sigma}^2_{\rm COBE}=\langle\mbi r\mbi r^\dagger \rangle_{\rm COBE}=({\mat R^\dagger\mat N^{-1}\mat R})^{-1}{.}
\label{eq:COBEvar}
\end{equation}
Note that, in general, the following relation holds: ${\mbi\sigma}^2_{\rm WF}\le{\mbi\sigma}^2_{\rm COBE}$, as a comparison to eq.~(\ref{eq:WFvar}) shows.

\cite{1997ApJ...480L..87T} claims that several linear filters like the COBE or
the Wiener-filter conserve information by comparing the Fisher information
matrix corresponding to the filtered signal to the one of the un-filtered time
ordered data. This property apparently permits one to perform cosmological parameter estimation from the reconstructed signal after filtering the data. However, linear filters conserve information only if they are invertible, which is not provided for realistic cases as we show in appendix \ref{app:FISHER}. 
A consistent estimation of cosmological parameters has to be done in a full Bayesian framework by estimating the joint PDF of the signal and the parameters, as we will see in section (\ref{sec:MCMC}) \citep{2004PhRvD..70h3511W}.

\subsubsection{Flat prior and Poissonian likelihood: the Richardson-Lucy algorithm} 
\label{sec:flatpois}

A widely used deblurring algorithm in astronomy and medical tomography is the
Richardson-Lucy algorithm \citep{1972JOSA...62...55R, 1974AJ.....79..745L},
which was shown to be the maximum likelihood solution with a Poissonian
likelihood by \cite{sheppvardi}. We show the derivation in appendix
\ref{app:POISLIKEML}, as a simplified case with respect to
eq.~(\ref{eq:likelpois}). The Richardson-Lucy algorithm cannot prevent serious
noise amplifications in the restoration process \citep[see
e.g.][]{carasso}. This is a natural consequence when a prior  that regularizes
the solution is missing. A toy application is presented in section (\ref{sec:blurring}).  

\subsubsection{Jeffrey's prior} 
\label{subsubsec:jeff}

Other non-informative priors have been suggested based on invariant statistical structures under transformation of variables in a Bayesian formalism. Considering a one-to-one transformation in the one-dimensional case of the parameter: $\phi=f(\theta)$, the equivalence between the respective prior densities is expressed by
\begin{equation}
P({ \phi})=P({ \theta})\left|\frac{{\rm d}\theta}{{\rm d}\phi}\right|=P({ \theta})\left|f'(\theta)\right|^{-1}{.}
\label{eq:Jef}
\end{equation}
  This relation is satisfied by Jeffrey's prior ${P(\theta)\propto[J(\theta)]^{1/2}}$, where $J(\theta)$ is the Fisher information\footnote{The generalization to the multidimensional case leads to the following matrix form: $\mat J_{ij}({ \mbi\theta})\equiv\langle\frac{{\partial}\log P({\mbi d}|{\mbi \theta})}{{\partial}\mbi\theta_i}\frac{{\partial}\log P({\mbi d}|{\mbi \theta})}{{\partial}\mbi\theta_j}\rangle_{(\mbi d|\mbi \theta)}$ (see appendix \ref{app:FISHER}).} 
\begin{equation}
J({ \theta})\equiv\langle\left(\frac{{\partial}\log P({ d}|{ \theta})}{{\partial}\theta}\right)^2\rangle_{( d|\theta)}=-\langle\frac{{\partial}^2\log P({ d}|{ \theta})}{{\partial}\theta^2}\rangle_{( d|\theta)}{,}
\label{eq:Jef2}
\end{equation}
and where we have assumed the following regularity condition ${ \int {\rm d}d\frac{\partial^2}{\partial\theta^2}P(d\mid\theta)=0}$.
Relation (\ref{eq:Jef}) can be proved easily by doing the evaluation $J({ \phi})=-\langle\frac{{\partial}^2\log P({ d}|{ \phi})}{{\partial}\phi^2}\rangle_{( d|\theta)}=J({ \theta})\left|\frac{{\rm d}\theta}{{\rm d}\phi}\right|^2$
\citep[see e.g.][]{bayesdataanal}. 
Note, however, that in the multidimensional case, Jeffrey's prior may lead to incoherences or even paradoxes \citep[see e.g.][]{bergerbernardo, bayeschoice}. Jeffrey's prior is applied adequately, when not even the order of magnitude of the parameter to be estimated is known a priori. 
We derive Jeffrey's ignorance prior for the 3-D power-spectrum ($\mat S={\rm
  diag} ({P_{\rm S}}(\mbi k))$)\footnote{Here the autocorrelation matrix $\mat S$ is represented in k-space. We will discuss this in further detail in section (\ref{sec:operators}).} in appendix \ref{app:Jeff} (see section \ref{sec:PSEST} for an application of this prior).

\subsubsection{Entropic prior and Maximum Entropy method} 
\label{sec:entr}

Another approach searches the least informative model compatible with the data using a prior based on Boltzmann's definition of entropy $S^{\rm E}$ \footnote{Not to be confused by the signal autocorrelation ${\mat S}$.} \citep[or equivalently, Shannon's notion of information, see][]{Shannon1948},
\begin{equation}
P({\mbi s}\mid \mbi p)={\rm exp}(\alpha S^{\rm E}){,}
\label{eq:priorent}
\end{equation}
 and maximizing the resulting posterior distribution, being $\alpha$ some constant, and ${\mbi s}$ the so-called hidden image (or signal).  This inference procedure is called the  Maximum Entropy method (MEM) \citep{jaynes1963, jaynes1968, 1972JOSA...62..511F, 1978Natur.272..686G, gull1989, 1989mebm.conf.....S, 1997MNRAS.290..313M, 1998MNRAS.300....1H}. For a review see \cite{1986ARA&A..24..127N}.  From now on we will represent the underlying signal by $\mbi s$ in the framework of MEM. 
The MEM can be considered as MAP estimation with an entropic prior.

The particular expression for the entropy depends on the statistical formulation of the non-informative prior. 
Let us think of a positive signal as a grid with $q$ cells, with each cell $i$ having a certain intensity value $s_i$, $i=1,\dots,q$, with an uncertainty on each value given by $\pm\alpha^{-1}$.
Then we define some discrete {\it quanta} $n_i$ on each cell related to the intensity through the uncertainty: $n_i=\alpha s_i$. The signal can be guessed by distributing the $n_i$ {\it quantas} in the grid. 
In this way, the image is modeled in this way analogously to the energy configuration space of a thermo-dynamical system.
If we further demand each cell to be {\it iid}, the number of ways this object can occur is given by the multiplicity 
\begin{equation}
W=\frac{N_{\rm q}!}{n_1!n_2!\dots n_q!}{,}
\label{eq:ent0}
\end{equation}
with $N_{\rm q}$ being the total amount of {\it quantas} to be distributed in all cells ($N_{\rm q}=\sum_in_i$).
The probability of any particular result is then given by the multinomial distribution 
\begin{equation}
P({\mbi s}'\mid \mbi p)=Wq^{-N_{\rm q}}{.}
\label{eq:ent1}
\end{equation}
\cite{2006ApJS..162..401S} propose to sample from the multiplicity function directly to perform reconstructions in radioastronomy.
By using Stirling's formula for the factorials ($n!\sim n^ne^{-n}$) we can write\begin{equation}
\log P({\mbi s}'\mid \mbi p)=-\alpha\sum_{i}s'_{i}\log s'_{i} +const{.}
\label{eq:ent1}
\end{equation}
Comparing this expression with eq.~(\ref{eq:priorent}), we recover Shannon's definition of entropy ($S^{\rm E}_+=\sum_{i}s'_{i}\log s'_{i} $)\footnote{The ``+'' symbol in $S^{\rm E}_+$ denotes that the definition is only valid for positive signals $\mbi s'$.}.
The expression that is commonly used for the entropy is a generalization of Shannon's formula by Skilling that can be derived based only on consistency arguments within probabilistic information theory for positive and additive distributions (PADs) \citep{1989mebm.conf.....S}.  

This generalization implies the definition of a Lebesgue measure ($\mbi m$) for the integral of some function of the hidden image to represent the entropy
\begin{equation}
S^{\rm E}_+({\mbi s}'\mid \mbi p)=\sum_i \Big[s'_{i}-m_i- s'_{i}\log\left(s'_{i}/m_i\right)\Big]{,}
\label{eq:entpos}
\end{equation}
here in its discretized form.
Skilling's expression for the entropy can also be derived by considering a {\it team of monkeys} throwing balls at $q$ cells at random with Poissonian expectation $\mu_i$: ${P(\mbi n|\mu)=\prod_i\mu_i^{n_i}e^{-\mu_i}/n_i!}$, where ${n_i=\alpha s_i}$ and ${\mu=\alpha m_i}$ \citep{1989mebm.conf.....S}. 
For a review on further expressions for the entropy see \cite{imagerest}. 

The global maximum of $S^{\rm E}$ over $\mbi s$ in the absence of further constraints is found to be ${\mbi s'=\mbi m}$. Consequently, $\mbi m$ can also be thought of as a prior model for the image.
However, this expression for the entropy will allow reconstructing positive
signals only. \cite{1995ApJ...449..446Z} propose to define ${\mbi s'=\mbi\rho}$ and ${\mbi m=\mbi \rho_0}$, to avoid the possibility of having a negative distribution for $\mbi s$. 

According to \cite{gullskilling} the MEM can be extended to reconstruct
distributions, which can be either positive or negative, as in the case of density fluctuations. Such distributions can be described as the difference between two subsidiary positive distributions (PADs)
\begin{equation}
\mbi s=\mbi u-\mbi v {,}
\label{eq:ME0}
\end{equation}
relative to a common model $\mbi m$ \footnote{The ``$\pm$'' symbol in $S^{\rm E}_\pm$ denotes that the definition is valid for positive and negative signals $\mbi s$.}
\begin{eqnarray}
S^{\rm E}_\pm({\mbi u,\mbi v}\mid \mbi p)&=&\sum_i \Big[u_{i}-2 m_i- u_{i}\log(u_{i}/m_i)\Big]\nonumber\\
&+&\sum_i \Big[v_{i}-2 m_i- v_{i}\log(v_{i}/m_i)\Big]{.}
\label{eq:ME1}
\end{eqnarray}
One can see from eq.~(\ref{eq:ME0}) that ${\partial S^{\rm E}_\pm/\partial\mbi
  u=-\partial S^{\rm E}_\pm/\partial\mbi v}$, hence yielding 
\begin{equation}
\mbi u\mbi v= \mbi m^2{.}
\label{eq:ME2}
\end{equation}
From the relations given by eqs.~(\ref{eq:ME0}) and (\ref{eq:ME2}), it is easy to derive 
\begin{equation}
\mbi u=\frac{1}{2}(\mbi w+ \mbi s){,}
\label{eq:ME3}
\end{equation}
\begin{equation}
\mbi v=\frac{1}{2}(\mbi w- \mbi s){,}
\label{eq:ME4}
\end{equation}
with ${w_i=( s_{i}^2+4 m_i^2)^{1/2}}$.
Using these expressions, the total entropy can be rewritten as 
\begin{equation}
S^{\rm E}_\pm({\mbi s}\mid \mbi p)=\sum_i \Big[w_i-2 m_i- s_i\log\Big((w_i+ s_i)/2 m_i\Big)\Big]{.}
\label{eq:entposneg}
\end{equation}
The Maximum Entropy method gives a non-linear estimator of the underlying signal that one wants to reconstruct. This method is especially interesting to study deviations from Gaussianity \citep{1997MNRAS.290..313M, 1998MNRAS.300....1H}. 
It is equivalent to maximize $\chi^2$ with a Lagrangian multiplier, which
includes a penalty function given by the entropy. Maximum Entropy in this
context searches the hidden image that adds the least additional information to the data. 

The quantity we need to maximize is given by
\begin{equation}
Q^{\rm E}({\mbi s}\mid \mbi p)=\alpha S^{\rm E}({\mbi s}\mid \mbi p) + \log{\cal L}({\mbi s}\mid{\mbi d},\mbi p){,}
\label{eq:ME5}
\end{equation}
where the $\log{\cal L}$ is given by eq.~(\ref{eq:chi2}) or eq.~(\ref{ap:pois1}). The equation we want to solve is
\begin{equation}
\nabla Q^{\rm E}({\mbi s}\mid \mbi p)=0{.}
\label{eq:ME6}
\end{equation}
In section (\ref{sec:nonlinear}), different iterative algorithms to solve this non-linear problem will be discussed. The required expressions for the gradient of $Q^{\rm E}$ and its curvature for positive and positive/negative expressions of the entropy (eqs.~\ref{eq:entpos} and \ref{eq:entposneg}) and for both Gaussian and Poissonian likelihoods are presented in appendix \ref{app:MEM}.

Note that in the limit of low density fluctuations, i.e.~in the linear regime,
the expression of the entropy reduces to the quadratic entropy (eventually with
an offset of the origin of $\mbi s$), ${S^{\rm E}(\mbi s\mid \mbi
  p)\simeq-\sum_i s_i^2/2m_i}$. This expression is very similar to a Gaussian
prior for the signal with a variance given by ${\mbi m}$. In that case Maximum Entropy leads to the Wiener-filter.

\subsection{Markov Chain Monte Carlo: sampling the joint PDF}
\label{sec:MCMC}

The drawback of the maximization methods hitherto mentioned, is that they find a unique estimator that is most probably subject to the chosen values for the required parameters.
As already mentioned, the complete characterization of a system is contained in
the joint PDF in the product space of possible signals and parameters. Thus, it would be desirable to sample from this PDF to find the region of highest confidence for our estimator.
This is possible using Markov Chain Monte Carlo (MCMC). The importance of sampling from the joint PDF and the viability of doing that with MCMCs has already been discussed in other contexts in astronomy \citep{2003MNRAS.338..765H, 2004ApJ...609....1J, 2004PhRvD..70h3511W}.
With the MCMC method, the whole system can be moved in its configuration space by updating all variables successively in a Monte Carlo fashion, until the system relaxes ({\it burns-in}) and reaches the highest density region. 

The expectation of the $i$-th parameter ($\theta_i$) can be calculated by the so-called ergodic average, which is given by the mean of the sample 
\begin{equation}
\langle{\theta}_i\rangle_{(\mbi\theta|\mbi d)}\simeq\frac{1}{N_{\rm b}}\sum_{t=0}^{N_{\rm b}-1}{\theta_i}^t{,}
\label{eq:MCMC1}
\end{equation}
with $N_{\rm b}$ being the size of the sample drawn once the Markov Chain has {\it burned-in}.
In general, the mean estimator is more reliable than the maximum of the
distribution, especially in cases with deviations from Gaussianity \citep[see
e.g.][]{bayesdataanal}. The MCMC method permits one to approximately solve the integral in eq.~(\ref{eq:joint}) through expression (\ref{eq:MCMC1}).

\subsubsection{Gibbs sampling}
\label{sec:gibbs}

The most straightforward MCMC method is the Gibbs sampler \citep{gibbsamp}, also known as the {\it heatbath} algorithm. The Gibbs algorithm samples from the joint PDF by repeatedly replacing each component with a value drawn from its distribution conditional on the current values of all other components.
This process can be seen as a Markov Chain with transition probabilities $\mbi\pi_k$ for ${k=1,...,n}$,
\begin{equation}
{\mbi \pi_k}({\mbi\theta},{\mbi\theta'})= P(\theta'_k\mid\{\theta_i:i\neq k\})\cdot\prod_{i\neq k}\delta_{\rm K}(\theta_i,\theta_i'){,} 
\label{eq:MCMC2}
\end{equation}
where ${\{\theta_i:i\neq k\}=(\theta_1,...,\theta_{k-1},\theta_{k+1},...,\theta_{n})}$ \citep[see e.g.][]{neal1993} and $\delta_{\rm K}$ is the Kroenecker delta-function.
The Gibbs sampler starts with some initial values ${\mbi\theta^{(0)}=(\theta_1^{(0)}, ...,\theta_n^{(0)})}$ and obtains new updates ${\mbi\theta^{(j)}=(\theta_1^{(j)}, ...,\theta_n^{(j)})}$ from the previous step  ${\mbi\theta^{(j-1)}}$ through successive generation of values
\begin{eqnarray}
{\theta_1^{(j)}}&\sim& P({\theta_1}\mid\{\theta_i^{(j-1)}: i\neq 1\}) \nonumber\\
{\theta_2^{(j)}}&\sim& P({\theta_2}\mid\theta_1^{(j)},\{\theta_i^{(j-1)}: i> 2\}) \nonumber\\
&\vdots&\nonumber\\
{\theta_n^{(j)}}&\sim& P({\theta_n}\mid\{\theta_i^{(j)}: i\neq n\}) 
\label{eq:MCMC3}
\end{eqnarray}
In this way a random walk on the vector $\mbi\theta$ is performed by making subsequent steps in low-dimensional subspaces, which span the full product space. This is similar to individual collisions of particles in a mechanical system that drives a many-body system to an equilibrium distribution for all degrees of freedom.
We are especially interested in this sampling method because of its efficiency that permits us to tackle large dimensional problems in contrast to other algorithms, which include acceptance and rejection rules. 
See \cite{2004PhRvD..70h3511W} for applications in CMB-mapping and power-spectrum estimation. However, in the case where the particular distribution function is unknown or cannot be explicitely expressed rejection sampling methods will be necessary (see section \ref{sec:reds}), like the Metropolis-Hastings algorithm \citep{metroplis, hastings}. 

The MCMC method can be applied to perform simultaneously the reconstruction of the density field and the estimation of other parameters, such as the power-spectrum, the peculiar velocities, the bias, or the comological parameters (see fig.~\ref{fig:dag}).
We propose in the next sections two novel applications of this method to power-spectrum estimation from a galaxy redshift survey
and redshift-distortion corrections, which can also be used in a joint
algorithm. 
Note, that a higher degree of complexity can be achieved in the schemes we
present here by going beyond linear perturbation theory or considering higher moments of the density field.

\subsubsection{Joint signal and power-spectrum estimation: sampling the cosmic
  variance with data augmentation}
\label{sec:PSEST}

The joint PDF considered here is given by the joint PDF of the signal and the power-spectrum ${P(\mbi s, \mat S|\mbi d)}$.
For the initial guess either an expression for the power-spectrum can be applied \citep[see e.g.][]{1992MNRAS.258P...1E,1994MNRAS.267.1020P,1998MNRAS.297..910S,1999ApJ...511....5E}, or the power-spectrum of the CMB can be taken and calculated for the required redshifts with some transfer functions \citep[see e.g.][]{1999ApJ...511....5E}. Then the following sampling processes are iterated until the chain {\it burns-in}
\begin{equation}
{\mbi s^{(j+1)}}\sim P({\mbi s}\mid\mat S^{(j)},\mbi d) {,}
\end{equation}
\begin{equation}
{\mat S^{(j+1)}}\sim P({\mat S}\mid\mbi s^{(j+1)}) {,}
\end{equation}
The DM signal is sampled with the following PDF (see section \ref{gaussprior})
\begin{equation}
P({\mbi s}\mid\mat S^{(j)},\mbi d) \propto G\left(\mbi s-\mat F_{\rm WF}(\mat S^{(j)})\mbi d,\mbi \sigma^2_{\rm WF}(\mat S^{(j)})\right){.}
\end{equation}
The Wiener reconstruction is known to give a biased estimator, which attenuates
the power especially for the modes where the noise becomes important, as
discussed in section~(\ref{sec:WF}). This filtering effect has to be compensated by
adding a fluctuating term with statistics according to the correct covariance \citep[see][]{2004PhRvD..70h3511W} 
\begin{equation}
{\mbi s^{(j)}} = \langle{\mbi s}^{(j)}\rangle_{\rm WF}+{\mbi  y}^{(j)}_{\sigma_{\rm WF}}{.}
\end{equation}
To generate the data augmentation $\mbi y^{(j)}_{\sigma_{\rm WF}}$ one has to solve the following set of equations \citep[see][]{2007ApJ...656..641E}
\begin{eqnarray}
\label{eq:Wsamp}
 \lefteqn{{\mbi y}^{(j)}_{\sigma_{\rm WF}} =}\\
&& \hspace{-0.7cm}\Big((\mat S^{(j)})^{-1}+\mat R^\dagger\mat N^{-1}\mat R\Big)^{-1}\Big((\mat S^{(j)})^{-1/2} \mbi x_{\rm G_1}+\mat R^\dagger\mat N^{-1/2}\mbi x_{\rm G_2}\Big)\nonumber {,}
\end{eqnarray}
where $x_{\rm G_1}$ and $x_{\rm G_2}$ are two independent Gaussian variates. One can show by direct calculation that ${\mbi y}^{(j)}_{\sigma_{\rm WF}}$ has a covariance given by $\mbi\sigma^2_{\rm WF}$.
To stabilize the inversion \cite{2007ApJ...656..641E} suggest using the
following expression derived from the previous one by factorizing the
square-root of the power-spectrum
\begin{eqnarray}
\label{eq:Wsamp2}
 \lefteqn{{\mbi y}^{(j)}_{\sigma_{\rm WF}} =(\mat S^{(j)})^{1/2} \Big(\mbi 1 +(\mat S^{(j)})^{1/2}\mat R^\dagger\mat N^{-1}\mat R(\mat S^{(j)})^{1/2} \Big)^{-1}}\nonumber\\
&& \hspace{.cm}\Big(\mbi x_{\rm G_1}+(\mat S^{(j)})^{1/2} \mat R^\dagger\mat N^{-1/2}\mbi x_{\rm G_2}\Big) {.}
\end{eqnarray}
Accordingly, the reconstruction step can be done by solving the following set
of equations based on the signal-space representation of the Wiener-filter (eq~\ref{eq:WF2})
\begin{eqnarray}
\label{eq:Wsamp3}
 \lefteqn{{\mbi s}^{(j)}_{{\rm WF}} =(\mat S^{(j)})^{1/2} \Big(\mbi 1 +(\mat S^{(j)})^{1/2}\mat
 R^\dagger\mat N^{-1}\mat R(\mat S^{(j)})^{1/2} \Big)^{-1}}\nonumber\\
&& \hspace{.cm}(\mat S^{(j)})^{1/2}
 \mat R^\dagger\mat N^{-1}\mbi d {.}
\end{eqnarray}
This allows to perform the inversion for the fluctuating term and for the
reconstruction in one step.
An alternative way, permits us to use the data-space representation of the
Wiener-filter\footnote{Note, that the data-space representation for the covariance (eq.~\ref{app:WFvar2})
  is not appropriate, due to the inverse of the response operator (see appendix \ref{app:WIENER}).} by generating the fluctuations with a constrained
realization \citep[see][]{1987ApJ...323L.103B, 1991ApJ...380L...5H, 1993ApJ...415L...5G}
\begin{equation}
\label{eq:flucterm}
{\mbi y}^{(j)}_{\sigma_{\rm WF}} = {\mbi{\tilde{s}}}^{(j)}-\mat F_{\rm WF}{\mbi{ \tilde{d}}}^{(j)}{,}
\end{equation}
using two auxiliary Gaussian random fields $\tilde{s}$ and $\tilde{\epsilon}$ with zero mean and correlation $\langle{\mbi{\tilde{s}}}{\mbi{\tilde{s}}}^\dagger\rangle={\mat S}$
and
${\langle{\mbi{\tilde{\epsilon}}}{\mbi{\tilde{\epsilon}}}^\dagger\rangle}=\mat N$ respectively. Further we set ${\mbi{ \tilde{d}}}=\mat R{\mbi{\tilde{s}}}+\mbi{\tilde{\epsilon}}$. This method has the advantage that non-linear
reconstructions can be obtained with N-body simulations\footnote{Note, however, that a Gaussian constrained realization is good enough for power-spectrum estimation especially when one is interested in the traces of the linear regime, like the baryon acoustic oscillations, or the gravitational potential for the ISW-effect. Sampling with constrained N-body reconstructions requires a much deeper development, since the whole cosmological parameter space has to be scanned.}
 \citep[see][]{1998ApJ...492..439B}. It can be shown that the term in eq.~(\ref{eq:flucterm}) has the
appropriate Wiener covariance (see appendix~\ref{app:COV}). 
Each reconstruction step can then be done in one step with the direct Wiener representation by solving the following equations
\begin{equation}
\label{eq:Wsamp}
{\mbi s}^{(j)} = {\mbi{\tilde{s}}}^{(j)}+\mat F_{\rm WF}({\mbi{ {d}}}-{\mbi{ \tilde{d}}}^{(j)}){.}
\end{equation}

The power-spectrum can be sampled by an inverse gamma function, which we derive here for the case of the 3D power-spectrum \citep[see][for the analogous CMB case]{2004PhRvD..70h3511W} 
\begin{equation}
P({\mat S}\mid\mbi s)\propto P(\mat S)P(\mbi s\mid\mat S) {.}
\end{equation}
Assuming a Gaussian signal $\mbi s$ (see eq.~\ref{eq:priorgaus}) this yields
\begin{equation}
P({{ P}_{\rm S}(\mbi k)}\mid\mbi s^{(j)})\propto P({ P}_{\rm S}(\mbi k))\prod_{\mbi k}\frac{1}{\sqrt{{ P}_{\rm S}(\mbi k)}}\exp\Big({-\frac{|\mbi s^{(j)}(\mbi k)|^2}{2{ P}_{\rm S}(\mbi k)}}\Big) {,}
\label{eq:pslikel}
\end{equation}
with ${\mat S={\rm diag}({ P}_{\rm S}(\mbi k))}$.
The prior $P({ P}_{\rm S}(\mbi k))$ can be chosen to be flat (${P({ P}_{\rm S}(\mbi k))={\rm const}}$) or instead Jeffrey's prior can be used (${P({ P}_{\rm S}(\mbi k))\propto{ P}_{\rm S}(\mbi k)^{-1}}$), see section (\ref{subsubsec:jeff}) and appendix \ref{app:Jeff}.
Note, that the likelihood for the power-spectrum given by eq.~(\ref{eq:pslikel}) is clearly non-Gaussian.

\subsubsection[Joint signal and peculiar velocities estimation]{Joint signal and peculiar velocities estimation: \\redshift-distortions correction}
\label{sec:reds}

 We propose to sample the peculiar velocities in a MCMC fashion (see section \ref{sec:MCMC}), analogous to
 the case of the power-spectrum (see \cite{2004PhRvD..70h3511W} and section
 \ref{sec:PSEST}). We draw realizations of the matter field given the data, a
 power-spectrum and assumed galaxy peculiar velocities
\begin{equation}
{\mbi s^{(j+1)}}\sim P({\mbi s}\mid\mbi v^{(j)},\mat S,\mbi d){.} 
\end{equation}
The velocities are subsequently sampled too:
\begin{equation}
{\mbi v^{(j+1)}}\sim P({\mbi v}\mid\mbi s^{(j+1)}) {.} 
\end{equation}
In each step where we sample the peculiar velocity, the redshift-distortion can be corrected using
\begin{equation}
r^{(j+1)}=z - v_r^{(j+1)}
\end{equation}
We propose to sample the peculiar velocities from a PDF with a mean $\langle v\rangle_{\rm M}$ given by the linear theory $\mbi v_{\rm LT}$  and  a velocity dispersion $\sigma_{v}$ depending on the local value of the over-density,
\begin{equation}
P({\mbi v}\mid\mbi s^{(j)})\propto G\left(\mbi v-\langle{\mbi v}\rangle_{\rm M}(\mbi s^{(j)}),\sigma^2_{v}(\mbi s^{(j)})\right){,}
\end{equation}
where we have taken a Gaussian distribution, but this could be extended to other PDFs.

\section{Numerical method}
\label{sec:itinvreg}

In order to efficiently sample the joint PDF, as it is required in MCMC methods
(see section \ref{sec:MCMC}), fast inverse algorithms need to be considered to
regularize the solution. General iterative inverse methods scale as ${\cal O}
(n^3)$ since they imply matrix multiplications of a ${n\times n}$ matrix in an
iterative fashion (at most $n$-steps until convergence). This makes the study
of the joint PDFs as presented in section (\ref{sec:MCMC}), at a first glance,
un-feasible. However, a proper formulation of the problem in an operator
formalism allows treating the matrices as operators that have to be neither
calculated nor stored. Within this operator formalism, the inversion methods we
present here sped up to a scaling of ${{\cal O} (n\log_2 n)}$. We start with a
general formulation of iterative methods and subsequently present the different schemes that we have implemented in \textsc{argo}. 
Since a preconditioning treatment can dramatically enhance the performance of  iterative schemes (see our numerical experiments in section \ref{sec:codetesting}), we pay special attention to this point in the derivation of the different schemes.

\subsection[Iterative inverse and regularization methods]{Iterative inverse and regularization methods: a unified formulation of different linear methods}
\label{sec:itsche}  

Let us consider a region $D$ in the $n$-dimensional Euclidean space $E_{n}$ and
denote $L_2(D)$ the Hilbert space of all complex measurable square integrable
functions ${\int_D {\rm d}^n\mbi z |\mbi g|^2(\mbi z) <\infty}$ with inner
product \footnote{Here a Dirac type notation is introduced. It should not be confused with the ensemble average notation, which does not have a balk in-between.}
\begin{equation}
\langle \mbi g| \mbi s\rangle=\int_D {\rm d}^n\mbi z \,\overline{\mbi g(\mbi z)}\mbi s(\mbi z){,}
\label{eq:reg1}
\end{equation}
 and norm of ${\mbi g\in L_2(D)}$
\begin{equation}
||\mbi g||=\langle\mbi g| \mbi g\rangle^{1/2}{.}
\label{eq:reg2}
\end{equation}
Let $\Psi$ be a subspace of the Hilbert space $L_2(D)$ with the conditions that
every element ${\mbi\psi\in \Psi}$ must satisfy being smoothness, limit behavior at the boundary $D$, etc.
Let us now consider the linear operator $\mat A$, defined on the linear
manifold $\Psi$, and suppose that $\mat A$ is a positive definite, i.e.~${\langle \mat A\mbi\psi|\mbi\psi\rangle\ge0}$ \footnote{This expression can be written in matrix notation as ${\mbi\psi^\dagger\mat A\mbi\psi\ge0}$, where  ${\mbi\psi^\dagger}$ is the conjugate and transpose of the vector $\mbi\psi$.}
for all ${\mbi\psi\in\mbi\Psi}$. 
The kind of inverse problem we are interested in belongs to the stationary problems of the form 
\begin{equation}
\mat A\mbi\psi=\mbi f{,}
\label{eq:reg3}
\end{equation}
 since, for example, for the COBE-filter we have to invert ${\mat A{\langle{\mbi s}\rangle}_{\rm COBE}={\mat R}^\dagger{\mat N}^{-1}{\mbi d}}$, with \\${\mbi\psi={\langle{\mbi s}\rangle}_{\rm COBE}}$, ${\mat A={\mat R}^\dagger{\mat N}^{-1}{\mat R}}$ and ${\mbi f={\mat R}^\dagger{\mat N}^{-1}{\mbi d}}$, and for the Wiener-filtering we have \\${\mbi\psi=({\mat S}{\mat R}^\dagger)^{-1}{\langle{\mbi s}\rangle}_{\rm WF}}$, $\mat A=({\mat R}^\dagger\mat S{\mat R}+{\mat N})$ and ${\mbi f={\mbi d}}$.
Eq.~(\ref{eq:reg3}) has the same structure as eq.~(\ref{eq:data}), but without
a noise term. Hence, a regularization method is again required.

\subsubsection{Minimization of the quadratic form}
\label{sec:minim}

Another way of approaching the linear inverse problem is the minimization of a quadratic form given by
\begin{equation}
Q_{\mat A}(\mbi\psi)=\frac{1}{2}\langle \mat A\mbi\psi|\mbi\psi\rangle-\langle\mbi f|\mbi\psi\rangle+c{.}
\label{eq:reg16}
\end{equation}
The gradient of $Q_{\mat A}$ leads to
\begin{equation}
\frac{{\rm d}Q_{\mat A}}{{\rm d} \mbi\psi}(\mbi\psi)\equiv Q_{\mat A}'(\mbi\psi)=\mat A\mbi\psi-\mbi f{,}
\label{eq:reg17}
\end{equation}
assuming that the operator $\mat A$ is self-adjoint. Setting the gradient to zero, one obtains eq.~(\ref{eq:reg3}).
The surface defined by a quadratic form with a positive definite matrix $\mat A$ is shaped like a paraboloid bowl \citep[see e.g.][]{shewchuk}. This ensures the existence of a unique minimum or, equivalently, the convergence of appropriate algorithms. 

\subsubsection{Solution of the non-stationary problem: asymptotic regularization}
\label{sec:asymp}
Here, a unified framework for the regularization methods that we have implemented in \textsc{argo} is given based on the asymptotic regularization. Nevertheless, an original  Bayesian motivation to the asymptotic solution is presented in appendix \ref{app:bayinv}.

The stationary problem (eq.~\ref{eq:reg3})  can be replaced by a non-stationary equation, which relaxes to the equilibrium solution
\begin{equation}
\frac{\partial\mbi\psi}{\partial t}+ \mat A\mbi\psi=\mbi f{.}
\label{eq:reg4}
\end{equation}
We seek solutions of the form
\begin{equation}
\mbi\psi=\sum_l\psi_l {\mbi u}_l{,}
\label{eq:reg5}
\end{equation}
with a spectrum for the operator $\mat A$
\begin{equation}
\mat A{\mbi u}_l=\lambda_l {\mbi u}_l{.}
\label{eq:reg6}
\end{equation}
Expanding $\mbi f$ in this basis, yields 
\begin{equation}
\mbi f=\sum_l f_l {\mbi u}_l{.}
\label{eq:reg7}
\end{equation}
Then we get the following relations for the Fourier coefficients in the stationary case
\begin{equation}
\lambda_l \psi_l=f_l{,}
\label{eq:reg8}
\end{equation}
and for the non-stationary case
\begin{equation}
\frac{\partial\psi_l(t)}{\partial t}+\lambda_l \psi_l(t)=f_l, \,\, \psi_l(0)=0 {,}
\label{eq:reg9}
\end{equation}
which lead to the following solutions
\begin{equation}
\mbi\psi=\sum_l\frac{f_l}{\lambda_l}{\mbi u}_l{,}
\label{eq:reg10}
\end{equation}
and
\begin{equation}
\mbi\psi(t)=\sum_l\frac{f_l}{\lambda_l}(1-e^{-\lambda_l t}){\mbi u}_l{,}
\label{eq:reg11}
\end{equation}
for the stationary and non-stationary cases, respectively.
Since the spectrum of a positive definite operator $\mat A$ is real, ${\lambda_l>0}$, it follows that ${\lim_{t\to\infty}\mbi\psi |_{\rm non-stationary}=\mbi\psi |_{\rm stationary}}$.

The non-stationary problem can be solved using difference methods with respect to t 
\begin{equation}
\mbi\psi^{j+1}=\mbi\psi^{j}+\tau^j\mat M^j(\mbi f-\mat A\mbi\psi^{j}){,}
\label{eq:reg12}
\end{equation}
with $\{\mat M^j\}$ being a set of non-singular matrices \footnote{We implicitly generalized eq.~(\ref{eq:reg4}) to ${\partial\mbi\psi(t)/\partial t=\mat M(t)(\mbi f-\mat A\mbi\psi)}$, where the auxiliary matrix $\mat M$ is chosen to speed up convergence. } and $\{\tau_j\}$ being a sequence of real parameters. Here we concentrate on a constant, self-adjoint matrix $\mat M$.
Let us rewrite eq.~(\ref{eq:reg12}) as
\begin{equation}
\mbi\psi^{j+1}=\mbi\psi^{j}+\tau^j\mat M\mbi\xi^j{,}
\label{eq:reg13}
\end{equation}
with the residuals given by 
\begin{equation}
\mbi\xi^{j}=\mbi f-\mat A\mbi\psi^{j}{.}
\label{eq:reg14}
\end{equation}
The error vectors are defined as
\begin{equation}
\mbi\eta^{j}=\mbi\psi^{j}-\mbi\psi^*{,}
\label{eq:reg15}
\end{equation}
where ${\mbi\psi^*=\mat A^{-1}\mbi f}$ is the exact solution. 
The matrix $\mat M$ and the real number $\{\tau_j\}$ are chosen to speed up the convergence. $\mat M$ usually represents the preconditioning of eq.~(\ref{eq:reg12}) and $\tau_j$ can be interpreted as the time step (see appendix \ref{app:prec}), and is also called relaxation parameter.
Here truncation regularization occurs by quitting the iteration loop. Some
stopping rules are therefore required. In the case where no noise
regularization was conducted in the first step, they crucially define the noise
regularization. In the other cases, they mostly determine algorithmic
performance and accuracy. At this point we are interested in the regularization
for the inverse purpose, since we have already found expressions which
regularize the noise (e.g.~Wiener-filter, or MEM). However, the results
presented in section \ref{sec:codetesting} show that in some cases truncation leads to better results (see discussion in section \ref{sec:wind}). In the following sections, we will show how different iterative schemes are based on the general formula given by eq.~(\ref{eq:reg12}). It is worth mentioning that other methods that we do not discuss in this paper, like the algebraic reconstrcution technique \citep[ART, see ][]{gordon}, can  also be expressed through this formula.   

\subsubsection{Jacobi method}
\label{sec:jac}

The Jacobi iteration method splits the operator $\mat A$ in two matrices 
\begin{equation}
\mat A=\mat D+\mat B{,}
\label{eq:jac1}
\end{equation}
where $\mat D$ contains the diagonal elements of $\mat A$ and $\mat B$ contains the off-diagonal elements.
From eq.~(\ref{eq:reg3}) one  follows
\begin{equation}
\mbi\psi=\mat D^{-1}(\mbi f-\mat B\mbi\psi){.}
\label{eq:jac2}
\end{equation}
Substituting $\mat B$ by ${\mat A-\mat D}$ one gets the following iteration scheme
\begin{equation}
\mbi\psi^{j+1}=\mbi\psi^{j}+\mat D^{-1}(\mbi f-\mat A\mbi\psi^{j}){.}
\label{eq:jac3}
\end{equation}
The Jacobi method turns out to be a particular case of the iteration scheme
given by eq.~(\ref{eq:reg12}) with a preconditioning matrix given by ${\mat
  M=\mat D^{-1}}$ and ${\tau^j=1}$. This method can, must be optimized by increasing the timestep $\tau^j$ by a certain percentage if the solution converges and decreasing the timestep if the solution diverges.
An optimal timestep is hard to find, because the spectrum of the operator $\mat A$ has to be known (see appendix \ref{app:prec}).

\subsubsection{Steepest Descent method}
\label{sec:steep}

The steepest descent method searches the minimum of the quadratic form by choosing the direction in which $Q_{\mat A}$ decreases most rapidly.
This direction is given by the residual
\begin{equation}
-Q_{\mat A}'(\mbi\psi^j)=\mbi f-\mat A\mbi\psi^j=\mbi\xi^j{.}
\label{eq:SD1}
\end{equation}
The form of the iteration scheme is thus given by eq.~(\ref{eq:reg13}), with the length of the step in the direction of the residual given by $\tau^j$.
Steepest descent looks for the optimal length which minimizes the quadratic form with respect to $\tau^j$
\begin{equation}
0=\frac{{\rm d}Q_{\mat A}}{{\rm d} \tau^{j}}(\mbi\psi^{j+1})=\langle Q_{\mat A}'(\mbi\psi^{j+1})|\frac{{\rm d}\mbi\psi^{j+1}}{{\rm d} \tau^j}\rangle=\langle {\mbi\xi^{j+1}|\mat M\mbi\xi^{j}}\rangle{.}
\label{eq:SD2}
\end{equation}
This implies that subsequent searching directions must be orthogonal (say ${\mat M=\mat I}$).
Starting from this condition it is straightforward to derive the expression for $\tau^j$. It is only necessary to use the definition of residual for ${\mbi\xi^{j+1}}$ and substitute ${\mbi\psi^{j+1}}$ from eq.~(\ref{eq:reg13}). 
\begin{equation}
\tau^j=\frac{\langle \mbi\xi^{j} | \mat M{\mbi\xi^{j}}\rangle}{\langle \mat A\mat M\mbi\xi^{j}| \mat M{\mbi\xi^{j}}\rangle}{.}
\label{eq:SD3}
\end{equation}
Both the calculation of the factors $\tau^j$ and the residuals $\mbi\xi^j$ imply applying the operator $\mat A$, each time on different vectors. It is possible, however, to reduce the operation of $\mat A$ to the same vector for every iteration, but the residuals, must be calculated in a different way. Multiplying both sides of eq.~(\ref{eq:reg13}) by $-\mat A$ and adding $\mbi f$, one obtains the following relation for the residuals 
\begin{equation}
\mbi\xi^{j+1}=\mbi\xi^j-\tau^j\mat A\mat M\mbi\xi^j{.}
\label{eq:SD4}
\end{equation}
Notice that the vector ${\mat A\mat M\mbi\xi^j}$ already appears in the
expression for $\tau^j$, and consequently saves one operation. However, expression (\ref{eq:reg14}) has to be periodically used with the feedback of $\mbi\psi^{j}$, to avoid the accumulation of floating-point roundoff error.
The disadvantage of this method is that it ends up searching repeatedly in the same direction. This is especially severe when the quadratic form is highly deformed, which occurs when the matrix $\mat A$ deviates from the unity matrix. We will see, however, that steepest descent competes with any other method when the preconditioning is effective, and thus the stretched shape of the quadratic form is brought close to a spherical symmetric shape.
Preconditioning should not imply too many operations; that is the reason why
the inverse of the matrix, which contains only the diagonal elements of $\mat
A$, is usually taken for preconditioning. This will work especially fine when
the operator $\mat A$ is diagonally dominant, which in our case occurs when nearly full-sky data are available. 

\subsubsection{Krylov methods: Conjugate Gradients}
\label{sec:krylov}

To make the iteration scheme more efficient, Conjugate Gradients proposes to search each time in a different direction. This is achieved by imposing $\mat A$-orthogonality to two different (${i\not= j}$) searching vectors $\mbi\mu^i$ and $\mbi\mu^j$
\begin{equation}
\langle \mbi\mu^j|\mbi\mu^i\rangle_{\mat A}\equiv\langle \mat A\mbi\mu^j|\mbi\mu^i\rangle=0{,}
\label{eq:CG1}
\end{equation}
which are then said to be conjugated. In the preconditioned case, the searching vectors are multiplied by $\mat M$ so that the conjugacy has to be formulated in the following way: \\${\langle \mat M\mbi\mu^j|\mat M\mbi\mu^i\rangle_{\mat A}=0}$ (for ${i\not= j}$).
  
\begin{table*}
\rotatebox[]{0}{
\begin{tabular}{|c|c|c|c|c|} 
  \hspace{3cm} ${N}_l$        & $\langle\mbi\xi^{j+1}|\mat M\mbi\xi^{j+1}\rangle$ &  $\langle\mbi\xi^{j+1}|\mat M(\mbi\xi^{j+1}-\mbi\xi^{j})\rangle$ & $\langle\mbi\xi^{j+1}-\mbi\xi^{j}|\mat M(\mbi\xi^{j+1}-\mbi\xi^{j})\rangle$ & $-\langle\mat M\mbi\xi^{j+1}|\mat M\mbi\mu^{j}\rangle_{\mat A}$ \\ 
${D}_m$&&&$-\langle\mbi\xi^{j}|\mat M\mbi\xi^{j}\rangle$\\\hline
$\langle\mbi\xi^j|\mat M\mbi\xi^j\rangle$ & {\bf FR} & {\bf PR} & N3/D1 &--- \\ \hline
$\langle\mbi\mu^j|\mat M\mbi\xi^j\rangle$ & N1/D2 & N2/D2 & N3/D2  &--- \\ \hline
$-\langle\mbi\xi^j|\mat M(\mbi\xi^{j+1}-\mbi\xi^{j})\rangle$ & N1/D3 & N2/D3 & N3/D3  &--- \\ \hline
$-\langle\mbi\mu^j|\mat M(\mbi\xi^{j+1}-\mbi\xi^{j})\rangle$ & N1/D4 & {\bf HS} & N3/D4 &--- \\ \hline
$-(\langle\mbi\xi^{j+1}-\mbi\xi^{j}|\mat M(\mbi\xi^{j+1}-\mbi\xi^{j})\rangle$ & N1/D5 & N2/D5 & N3/D5 &--- \\ 
$-\langle\mbi\xi^{j+1}|\mat M\mbi\xi^{j+1}\rangle)$ &  &  \\ \hline
$\langle\mat M \mbi\mu^{j}|\mat M\mbi\mu^{j}\rangle_{\mat A}$ & --- & --- & --- & {\rm \bf EXP} \\ \hline
\end{tabular}
}
\caption{\label{tab:beta} Formulae for the $\beta$-factor: $\beta^{j+1}_{lm}=\frac{{N}_l}{{D}_m}$. Three of the methods are discussed in the literature: {\bf FR} (Fletcher-Reeves), {\bf PR} (Polak-Ribi\`ere, and {\bf HS} (Hestenes-Stiefels). The rest of the formulae are derived in this paper using equivalence relations derived in appendices \ref{app:CGorth}-\ref{app:CGFR}. The {\bf FR} and the {\bf PR} methods are tested against the {\bf EXP} algorithm in section (\ref{sec:codetesting}).}
\end{table*}

The iteration scheme is given by substituting the residuals in eq.~(\ref{eq:reg13}) by the new searching vectors $\{\mbi\mu^j\}$
\begin{equation}
\mbi\psi^{j+1}=\mbi\psi^{j}+\tau^j\mat M\mbi\mu^j{.}
\label{eq:CG2}
\end{equation}
By subtracting $\mbi\psi^*$ we obtain an equation for the errors,
\begin{equation}
\mbi\eta^{j+1}=\mbi\eta^{j}+\tau^j\mat M\mbi\mu^j{.}
\label{eq:CG3}
\end{equation}
Taking into account the relation between the residuals and the errors
\begin{equation}
\mbi\xi^{j+1}=-\mat A\mbi\eta^{j+1}{,}
\label{eq:CG4}
\end{equation}
we can derive the recurrent formula for the residuals
\begin{equation}
\mbi\xi^{j+1}=-\mat A(\mbi\eta^j+\tau^j\mat M\mbi\mu^j)=\mbi\xi^j-\tau^j\mat A\mat M\mbi\mu^j{.}
\label{eq:CG5}
\end{equation}
 Here again, expression (\ref{eq:reg14}) has to be used periodically with the feedback of $\mbi\psi^{j}$ to avoid the accumulation of floating-point roundoff error.
The optimal length of the step is found by minimizing the quadratic form
\begin{equation}
0=\frac{{\rm d}Q_{\mat A}}{{\rm d} \tau^{j}}(\mbi\psi^{j+1})=-\langle \mbi\xi^{j+1}| \mat M \mbi\mu^j\rangle=\langle \mbi\eta^{j+1}|\mat M\mbi\mu^j\rangle_{\mat A}{.}
\label{eq:CG6}
\end{equation}
Substituting expression (\ref{eq:CG3}) in (\ref{eq:CG6}) we then obtain
\begin{equation}
\tau^j=-\frac{\langle \mbi\eta^{j} | \mat M{\mbi\mu^j}\rangle_{\mat A}}{\langle \mat M\mbi\mu^j| \mat M{\mbi\mu^j}\rangle_{\mat A}}=\frac{\langle \mbi\xi^{j} | \mat M{\mbi\mu^j}\rangle}{\langle \mat M\mbi\mu^j| \mat M{\mbi\mu^j}\rangle_{\mat A}}{.}
\label{eq:CG7}
\end{equation}
It can be shown that this formula is equivalent to the following expression
\begin{equation}
\tau^j=\frac{\langle \mbi\xi^{j} |\mat M{\mbi\xi^{j}}\rangle}{\langle\mat M\mbi\mu^j| \mat M{\mbi\mu^j}\rangle_{\mat A}}{,}
\label{eq:CG8}
\end{equation}
using ${\langle \mbi\xi^j | \mat M\mbi\mu^j\rangle=\langle \mbi\xi^{j} |\mat M\mbi\xi^{j}\rangle}$ (see appendix \ref{app:CG}).

To generate $\mat A$-orthogonal searching vectors one could think of Gram-Schmidt-conjugation   
\begin{equation}
\mbi\mu^{j}=\mbi\xi^{j}+\sum^{j-1}_{k=0}\beta^{jk}\mbi\mu^k{.}
\label{eq:CG9}
\end{equation}
Here it was assumed that the residuals $\{\mbi\xi^{j}\}$ form a set of linearly independent vectors (see appendix \ref{app:CG}).
The expression for the factors $\beta^{jk}$ can be derived by calling $\mat A$-orthogonality in eq.~(\ref{eq:CG9})
\begin{eqnarray}
\langle \mat M\mbi\mu^{j}|\mat M\mbi\mu^i\rangle_{\mat A}&=&\langle \mat M\mbi\xi^{j}|\mat M\mbi\mu^i\rangle_{\mat A}+\sum^{j-1}_{k=0}\beta^{jk}\langle \mat M\mbi\mu^k|\mat M\mbi\mu^i\rangle_{\mat A}\nonumber\\
0&=&\langle \mat M\mbi\xi^{j}|\mat M\mbi\mu^i\rangle_{\mat A}+\beta^{ji}\langle \mat M\mbi\mu^i|\mat M\mbi\mu^i\rangle_{\mat A}{.}
\label{eq:eq}
\end{eqnarray}
One obtains the following formula for the factors
\begin{equation}
\beta^{ji}=-\frac{\langle \mat M\mbi\xi^{j}|\mat M\mbi\mu^i\rangle_{\mat A}}{\langle \mat M\mbi\mu^i|\mat M\mbi\mu^i\rangle_{\mat A}}{,}
\label{eq:CG10}
\end{equation}
where $i<j$ according to eq.~(\ref{eq:CG9})\footnote{Note that the sign of $\beta$ depends on the definition of the Gram-Schmidt conjugation. An alternative definition with the negation of the residuals would cancel the minus sign in eq.~(\ref{eq:CG10}). The sign of $\beta$ can be regarded as a free parameter.}.

This method seems to require too much memory, as apparently all previous searching vectors must be stored to calculate the new one. 
However, only one $\beta$-factor remains in the sum in eq.~(\ref{eq:CG9}), as we show in appendix \ref{app:CGFR}. Hence, Gram-Schmidt orthogonalization can be simplified to the following expression
\begin{equation}
\mbi\mu^{j+1}=\mbi\xi^{j+1}+\beta^{j+1}\mbi\mu^j{,}
\label{eq:CG11}
\end{equation}
where
\begin{equation}
\beta^{j+1}_{\rm EXP}\equiv\beta^{j+1}\equiv\beta^{j+1\,j}=-\frac{\langle \mat M\mbi\xi^{j+1}|\mat M\mbi\mu^{j}\rangle_{\mat A}}{\langle \mat M\mbi\mu^j|\mat M\mbi\mu^j\rangle_{\mat A}}{,}
\label{eq:CG112}
\end{equation}
with EXP meaning expensive, since the nominator of $\beta$ apparently requires
an extra $\mat A$ operation. This additional operation can be saved by taking
the vector $\mat A \mat M \mbi \mu^j$ from $\tau^j$ or  with alternative
methods (see table \ref{tab:beta} and appendix \ref{app:CG}), like the Fletcher-Reeves method \citep{fletcherreeves}
\begin{equation}
\beta^{j+1}_{\rm FR}=\frac{\langle \mbi\xi^{j+1}|\mat M\mbi\xi^{j+1}\rangle}{\langle \mbi\xi^{j}|\mat M\mbi\xi^{j}\rangle}{,}
\label{eq:CG12}
\end{equation}
the Polak-Ribi\'{e}re formula \citep{polakribiere}
\begin{equation}
\beta^{j+1}_{\rm PR}=\frac{\langle \mbi\xi^{j+1}|\mat M(\mbi\xi^{j+1}-\mbi\xi^j)\rangle}{\langle \mbi\xi^{j}|\mat M\mbi\xi^{j}\rangle}{,}
\label{eq:CG13}
\end{equation}
or the Hestenes-Stiefel expression \citep{hestenesstiefel}
\begin{equation}
\beta^{j+1}_{\rm HS}=-\frac{\langle \mbi\xi^{j+1}|\mat M(\mbi\xi^{j+1}-\mbi\xi^j)\rangle}{\langle \mbi\mu^{j}|\mat M(\mbi\xi^{j+1}-\mbi\xi^j)\rangle}{.}
\label{eq:CG14}
\end{equation}
However, $\beta_{\rm EXP}$ turns out to be a very efficient scheme, which behaves far more stably than the rest (see section \ref{sec:codetesting}). Since the $\beta$-formulae (eq.~\ref{eq:CG112}-\ref{eq:CG14}) are mathematically equivalent, one could think of combining them in a single scheme finding numerically different solutions.  However, this kind of hybrid scheme remains to be thouroughly studied. 

Formula (\ref{eq:CG11}) shows that new searching vectors are built from a linear combination of the current residual and the previous searching vector.
Since the subsequent residuals are given by the linear combination of the previous residual and the $\mat A$-operator applied to the searching vector, the manifold where the solution is being searched is spanned by the residuals and the so-called Krylov space. The latter is built by applying the $\mat A$ operator to the basis vector successively. In this manifold, curved quadratic forms appear to be spherical and thus the searching process becomes more effective.   
It is possible to derive the Conjugate Gradients method by minimizing the $\mat
A$-norm of the error: ${{\rm min}||\mbi\eta||_{\mat A}}$ \citep[see
e.g.][]{marchuk}. In this sense an optimal solution to the inverse problem can
be found even if no unique solution exists. Conjugate Gradients works, even if
the operator $\mat A$ is not a positive definite \citep[for a discussion see e.g.][]{shewchuk}.  
It can easily be shown that Conjugate Gradients converges at most in $n$-steps, with $n$ being the number of pixels/vector columns \citep[see e.g.][]{shewchuk}.

\subsection{Non-linear inverse methods}
\label{sec:nonlinear}

Non-linear inverse methods are especially required in reconstruction algorithms that do not assume a Gaussian distribution. The iterative method given in eq.~(\ref{eq:likelpois}), which makes use of a Poissonian likelihood, can alternatively be solved with the methods presented in this section. The same applies to the MEM, where zeros of the non-linear eq.~(\ref{eq:ME6}) have to be found.   

The generalization of the regularization methods to non-linear inverse problems is possible with methods like Tikhonov regularization as mentioned in section (\ref{sec:prior}) or like asymptotic regularization as will be shown below (a relation between both methods is shown in appendix \ref{app:bayinv}). However, the proofs of the convergence properties are different since the spectral theoretical foundation is missing here. We refer the reader to e.g.~\cite{95544}. 

Let us generalize eq.~(\ref{eq:reg3}) to non-linear equations of the form
\begin{equation}
\mat A(\mbi\psi)=\mbi f{,}
\label{eq:reg152}
\end{equation}
with $\mat A$ being a non-linear operator, and solve the non-linear and non-stationary equation given by
\begin{equation}
\frac{\partial\mbi\psi}{\partial t}+ \mat A(\mbi\psi)=\mbi f{,}
\label{eq:reg153}
\end{equation}
with the forward Euler method.
Discretizing the solution yields
\begin{equation}
\mbi\psi^{j+1}=\mbi\psi^{j}+\tau^j \mat T(\mbi\psi^{j})(\mbi f-\mat A(\mbi\psi^{j})){,}
\label{eq:reg154}
\end{equation}
with $\mat T$ being also a non-linear operator, typically given by $\nabla \mat
A^\dagger$ or $\nabla \mat A^{-1}$, though more complicated expressions exist (see the Levenberg-Marquardt method or the regularized Gauss-Newton method, \citealt{han97} or \citealt{bak92} and \citealt{bns97}, respectively). 

\subsubsection{Newton-Raphson method}
\label{sec:newt}

One of the most extended non-linear inverse methods is the so-called
Newton-Raphson method \citep[for an application in MEMs
see][]{1997MNRAS.290..313M, 1998MNRAS.300....1H}, which can easily be derived
by doing a Taylor expansion of the function under study and truncating it at
the first order
\begin{equation}
{\mbi \psi}^{j+1}={\mbi \psi}^{j}+(\nabla \mat A({\mbi \psi^j}))^{-1}(\mbi f-
\mat A({\mbi \psi}^j)){.}
\label{eq:New}
\end{equation}
This method requires the inverse of the gradient of $\mat A$, which for the cases we are interested in is the inverse of a Hessian matrix. 
Recalling the problem of finding extrema of a function as presented in section (\ref{sec:minim}) and taking into account eq.~(\ref{eq:SD1}), the previous equation can be rewritten as
\begin{equation}
{\mbi \psi}^{j+1}={\mbi \psi}^{j}-(\nabla\nabla Q_{\mat A}({\mbi \psi^j}))^{-1}\nabla Q_{\mat A}({\mbi \psi^j}){,}
\label{eq:New1}
\end{equation}
where $\nabla\nabla Q_{\mat A}\equiv\partial Q_{\mat A}/\partial\mbi\psi^l\partial\mbi\psi^m$ is the Hessian matrix of $Q_{\mat A}$. For a direct derivation of this equation, we require a Taylor expansion
until the second order of $Q_{\mat A}$, which is where the non-linearity arises.
The MEM can be solved (eq.~\ref{eq:ME6}) with expression (\ref{eq:New1}) by doing the substitutions: ${Q_{\mat A}\rightarrow Q^{\rm E}}$ and ${\mbi\psi^j\rightarrow\mbi s^j}$. Here the quantity $Q^{\rm E}$ is implicitly approximated by its quadratic expansion $Q_{\mat A}$.
Calculating the inverse of the Hessian ${(\nabla\nabla Q_{\mat A}({\mbi \psi^j}))^{-1}}$ implies solving a linear ill-posed problem in each iteration of the scheme (\ref{eq:New1}). Some solutions have been found to regularize this scheme, like the Levenberg-Marquardt method \citep[see][]{han97} or the regularized Gauss-Newton method \citep[see e.g.][]{bak92, bns97}.

\subsubsection{Landweber-Fridman method}
\label{sec:landw}

 Alternative algorithms to the above mentioned Newton-Raphson class of methods
 do not need to invert the Hessian matrix and can thus simultaneously speed up and stabilize the inversion process. The Landweber-Fridman algorithm belongs to the class of methods based on steepest descent
\begin{equation}
{\mbi \psi}^{j+1}={\mbi \psi}^{j}+(\nabla\mat A({\mbi \psi^j}))^\dagger(\mbi f-\mat A({\mbi \psi}^j)){.}
\label{eq:New3}
\end{equation}
Making the same substitutions as for eq.~(\ref{eq:New1}), we obtain
\begin{equation}
{\mbi \psi}^{j+1}={\mbi \psi}^{j}-(\nabla\nabla Q_{\mat A}({\mbi \psi^j}))^\dagger\mbi\nabla Q_{\mat A}({\mbi \psi^j}){.}
\label{eq:New34}
\end{equation}
Here just the adjoint of the Hessian must be taken $({\nabla\nabla Q_{\mat A}({\mbi \psi}^j))^\dagger}$.
For a convergence analysis of this method see \cite{hns95}.

\subsubsection{Non-linear Krylov methods}
\label{sec:nonlkry}

Another class of methods that do not require one to invert the Hessian matrix are the Krylov-based methods, which we have exposed in the previous section. The difference with respect to the linear case mainly resides in the calculation of the residuals $\mbi\xi^j$ and the step size $\tau^j$. The residuals are updated now by the negation of the gradient of the quadratic form that approximates the function under consideration ${\mbi\xi^j=-\nabla Q_{\mat A}(\mbi \psi^j)}$ (see eq.~\ref{eq:SD1}). The step size is given by 
\begin{equation}
\tau^j=-\frac{\hspace{-1.cm}\langle\nabla Q_{\mat A}(\mbi \psi^j)|\mat M\mbi \mu^j\rangle}{\hspace{.5cm}\langle\mat M\mbi\mu^{j}|\mat M\mbi\mu^j\rangle_{\nabla\nabla Q_{\mat A}(\mbi\psi^j)}}{.}
\label{eq:New5}
\end{equation}
The derivation of this expression (see appendix \ref{app:CGnonlinear}) is based on the second order Taylor expansion of $Q_{\mat A}$. That is why Krylov algorithms which use this formula are called Newton-Krylov methods. 
There are alternative expressions for the time step $\tau^j$ where the Hessian is approximated and does not need to be explicitly calculated, like those using a secant approximation. For various implementations of non-linear Krylov methods see, for example, \cite{shewchuk}.

\subsection{Operator formalism}
\label{sec:operators}

\begin{table*}
\rotatebox[]{0}{
\begin{tabular} {|c|c|c|c|c|c|c|c|c|c|c|c|c|c|}
& $\mat R$ &  $\mat R^\dagger$  &  $\mat S$ &  $\mat S^{-1}$ &  $\mat S^{-1/2}$ &  $\mat N$ &  $\mat N^{-1}$ &  $\mat S\mat R^\dagger$ &  $\mat R^\dagger\mat N^{-1}$&  $\mat R^\dagger\mat N^{-1/2}$ & $\mat R^\dagger\mat N^{-1}\mat R$& $\mat R\mat S\mat R ^\dagger$ \\ \hline
COBE    & & & & & & &X& &X& &X&   \\ \hline
WIENER  &X&X&X&X$^\#$&X$^\#$&X&X$^\#$&X& &X$^\#$&X$^\#$&X  \\ \hline
GAPMAP &X&X&X& & & & &X& & & &   \\ \hline
MEMG    &X&X& & & & &X& &X& &X&   \\ \hline
MEMP    &X&X& & & & & & & & & &   \\ \hline
\multicolumn{12}{|c|}{$^\#$ additional operators required for the signal-space representation (see eq.~\ref{eq:Wsamp})}\\
\hline
\end{tabular}
}
\caption{ \label{tab:oper} Operators in columns needed for the different estimators in rows, the COBE-filter (\ref{eq:COBE}), the Wiener-filter (\ref{eq:WF}), the GAPMAP estimator (\ref{eq:likelpois}), and the MEMs (sections \ref{sec:entr} \& \ref{sec:nonlinear}, and appendix \ref{app:MEM}). Note that the trivial diagonal matrices have been left out of this table. The first two estimators are linear estimators, whereas the rest are non-linear. MEMG and MEMP stand for the Maximum Entropy method with a Gaussian likelihood and with a Poissonian likelihood, respectively. Note that some of the operators have to be further inverted either directly, like $(\mat R^\dagger\mat N\mat R)^{-1}$ for the COBE-filter, or in combination with other operators, like $(\mat R\mat S\mat R ^\dagger+\mat N)^{-1}$ for the Wiener-filter. The methods presented in section (\ref{sec:itinvreg}) show how to do this implicitly by applying the operators in an iterative fashion.}
\end{table*}

The iterative methods presented so far require an operator formalism to become efficient. 
In this formalism, matrices should be represented in such a way that their action can be expressed as simple operations, like sums and multiplications. 
In order to achieve this, one has to carefully choose the adequate
representation, in which the individual matrix components are diagonal, though
the whole matrix may not be. 
In this section, we present the different operators under consideration (see table \ref{tab:oper}) in  k-space and real-space and discuss their optimal representation. In this way, we can take advantage of the fast Fourier-transform methods (FFTs) that scale as $n\log_2n$, with $n$ being the length of the arrays, and which ultimately determine the speed of the algorithm.

\subsubsection{Fourier-transform definitions and dimensionality of the problem}
\label{sec:dimen}

Let us introduce the following definitions of the $N_{\rm D}$-dimensional
forward and inverse Fourier-transforms just to make clear our notation
\begin{equation}
\hat{x}({\mbi{k})}\equiv{\rm FT}\big[x(\mbi r)\big]\equiv\int{\rm{d}}^{N_{\rm D}}{\mbi{r}} \,{\rm exp}({i}{\mbi k \cdot \mbi r})x({\mbi{r}}){,}
\label{eq:ft}
\end{equation}
and
\begin{equation}
x({\mbi{r}})\equiv{\rm IFT}\big[\hat{x}(\mbi k)\big]\equiv\int\frac{{\rm{d}}^{{N_{\rm D}}}{\mbi{k}}}{({\rm{2}\pi})^{{N}_{\rm D}}}\,{\rm exp}(-{i}{\mbi k \cdot \mbi r})\hat{x}({\mbi{k}}){,}
\label{eq:ift}
\end{equation}
 respectively. 

In general, the reconstruction problem has three spatial dimensions ($N_{\rm D}=3$), with the corresponding discrete array lengths for the real-space and k-space vectors given by $\mbi r=( {r_x},  {r_y},  {r_z})$ and $\mbi k=( {k_x},  {k_y},  {k_z})$. Each component has the following range: $r_x=\frac{L_x}{n_x}[0, n_x-1], r_y=\frac{L_y}{n_y}[0, n_y-1], r_z=\frac{L_z}{n_z}[0, n_z-1]$ and $k_x=\frac{2\pi}{L_x}[0, n_x-1], k_y=\frac{2\pi}{L_y}[0, n_y-1], k_z=\frac{2\pi}{L_z}[0, n_z-1]$, where the volume of the Universe under consideration is given by $V=L_x\times L_y\times L_z$ in [(Mpc/h)$^3$], and the box containing that volume is divided into $n=n_x\times n_y \times n_z$ cells, with $n$ being the length of the array $x$. 
In the following, we will treat the operators as being continuous. However, the discrete implementation can be derived in a straightforward way \citep[for a discussion on the relation between discrete and continuous representations see][]{2005astro.ph..6540M}.
Note that the methods presented here can be applied in arbitrary dimensions. The number of dimensions $N_{\rm D}$ is thus kept as a free parameter.

In our convention, vectors defined in real-space have plain notation ($x$) and
in k-space they are denoted with hats ($\hat{x}$). Matrices, however, have two
hats in k-space.  We represent convolutions with circles ``$\circ$'' and
multiplications with dots ``$\cdot$''. Due to the convolution theorem, where
convolutions are shown to be multiplications in the counter space, we can
either omit hats if they are present or include them if they are not, and replace
circles with dots and vice versa ``$\cdot\leftrightarrow\circ$'' to change from one representation to the other.
All the numerical iterative inversion schemes (see section \ref{sec:itinvreg}) of the different reconstruction algorithms (section \ref{sec:theory}) require only a small number of basic operators, listed in table (\ref{tab:oper}).
To show how the operators listed in table (\ref{tab:oper}) can efficiently be applied we derive their action on an arbitrary vector. 

\subsubsection{Data model: the response operator and its transpose}
\label{sec:respOp}

\begin{figure*}
\begin{eqnarray}
 \hat{\hat{{\mat R\mat S\mat R^\dagger}}} \{\hat{\mbi x}\} ({\mbi k})&=& \int \frac{{\rm d}^{N_{\rm D}}\mbi k'}{(2\pi)^{N_{\rm D}}}\langle\hat{ \alpha}(\mbi k)\overline{\hat{\alpha}(\mbi k')}\rangle_{({\mbi s,\mbi\epsilon}|\mbi p)} \{\hat{x}(\mbi k')\}\nonumber\\
&=& {{\hat{f}_{\rm B}(\mbi k)} \int \frac{{\rm d}^{N_{\rm D}}\mbi q}{(2\pi)^{N_{\rm D}}}{\hat{f}_{\rm SM}(\mbi k-\mbi q)} \int \frac{{\rm d}^{N_{\rm D}}\mbi q'}{(2\pi)^{N_{\rm D}}}{ { P_{\rm S}}(\mbi q')(2\pi)^{N_{\rm D}}\delta_{\rm D}(\mbi q-\mbi q') { \int \frac{{\rm d}^{N_{\rm D}}\mbi k'}{(2\pi)^{N_{\rm D}}}\overline{\hat{f}_{\rm SM}(\mbi k'-\mbi q')}\overline{\hat{f}_{\rm B}(\mbi k')}\{\hat{x}(\mbi k')\}}}}  \nonumber\\
&=& { \hat{f}_{\rm B}(\mbi k){ \int \frac{{\rm d}^{N_{\rm D}}\mbi q}{(2\pi)^{N_{\rm D}}}\hat{ f}_{\rm SM}(\mbi k-\mbi q) { {P_{\rm S}}(\mbi q) { \int \frac{{\rm d}^{N_{\rm D}}\mbi k'}{(2\pi)^{N_{\rm D}}}\overline{\hat{ f}_{\rm SM}(\mbi k'-\mbi q)} { \overline{\hat{f}_{\rm B}(\mbi k')}\cdot\{\hat{x}(\mbi k')\}}}}}} \nonumber\\
&=& \underbrace{ \hat{f}_{\rm B}(\mbi k)\underbrace{ \int \frac{{\rm d}^{N_{\rm D}}\mbi q}{(2\pi)^{N_{\rm D}}}\hat{ f}_{\rm SM}(\mbi k-\mbi q) \underbrace{ {P_{\rm S}}(\mbi q) \underbrace{ \int \frac{{\rm d}^{N_{\rm D}}\mbi k'}{(2\pi)^{N_{\rm D}}}{\hat{ f}_{\rm SM}(\mbi q-\mbi k')}\underbrace{ \overline{\hat{f}_{\rm B}(\mbi k')}\cdot\{\hat{x}(\mbi k')\}}_{\overline{\hat{f}_{\rm B}}\cdot\{\hat{x}\}}}_{{\hat{ f}_{\rm SM}}\circ\big[\overline{\hat{f}_{\rm B}}\cdot\{\hat{x}\}\big]}}_{{P_{\rm S}} \cdot \big[{ {\hat{ f}_{\rm SM}}\circ\big[\overline{\hat{f}_{\rm B}}\cdot\{\hat{x}\}}\big]\big]  } }_{\hat{f}_{\rm SM}\circ \big[ {P_{\rm S}}\cdot\big[  { {\hat{ f}_{\rm SM}}\circ\big[\overline{\hat{f}_{\rm B}}\cdot\{\hat{x}\}}\big]\big]\big]  } } \nonumber\\
&&\hspace{2cm} {\hat{f}_{\rm B}\cdot\big[\hat{f}_{\rm SM}\circ\big[ {P_{\rm S}} \cdot\big[ { {\hat{ f}_{\rm SM}}\circ\big[\overline{\hat{f}_{\rm B}}\cdot\{\hat{x}\}}\big]\big]\big]\big]}\nonumber\\
\label{eq:op15}
\end{eqnarray}

\begin{eqnarray}
 \hat{\hat{\mat R^\dagger\mat N_{\rm N}^{-1}\mat R}} \{\hat{\mbi x}\}({\mbi k})
 &=& { \big(\hat{f}_{\rm B}(\mbi k)  \int \frac{{\rm d}^{N_{\rm D}}\mbi q}{(2\pi)^{N_{\rm D}}} {\hat{f}_{\rm SM}(\mbi k-\mbi q)} \big)^\dagger\int \frac{{\rm d}^{N_{\rm D}}\mbi q'}{(2\pi)^{N_{\rm D}}}{ { P_{\rm N}}^{-1}(\mbi q')(2\pi)^{N_{\rm D}}\delta_{\rm D}(\mbi q-\mbi q')\hat{f}_{\rm B}(\mbi q') { \int \frac{{\rm d}^{N_{\rm D}}\mbi k'}{(2\pi)^{N_{\rm D}}}{\hat{f}_{\rm SM}(\mbi q'-\mbi k')}\{\hat{x}(\mbi k')\}}}}  \nonumber\\
 &=& { \int \frac{{\rm d}^{N_{\rm D}}\mbi q}{(2\pi)^{N_{\rm D}}}{\hat{f}_{\rm SM}(\mbi k-\mbi q)} \overline{\hat{f}_{\rm B}(\mbi q)}\int \frac{{\rm d}^{N_{\rm D}}\mbi q'}{(2\pi)^{N_{\rm D}}}{ { P_{\rm N}}^{-1}(\mbi q')(2\pi)^{N_{\rm D}}\delta_{\rm D}(\mbi q-\mbi q')\hat{f}_{\rm B}(\mbi q') { \int \frac{{\rm d}^{N_{\rm D}}\mbi k'}{(2\pi)^{N_{\rm D}}}{\hat{f}_{\rm SM}(\mbi q'-\mbi k')}\{\hat{x}(\mbi k')\}}}}  \nonumber\\
&=& \underbrace{ \int \frac{{\rm d}^{N_{\rm D}}\mbi q}{(2\pi)^{N_{\rm D}}} {\hat{f}_{\rm SM}(\mbi k-\mbi q)} \underbrace{ \overline{\hat{f}_{\rm B}(\mbi q)} \underbrace{ { P_{\rm N}}^{-1}(\mbi q) \underbrace{ \hat{f}_{\rm B}(\mbi q) \underbrace{ \int \frac{{\rm d}^{N_{\rm D}}\mbi k'}{(2\pi)^{N_{\rm D}}}{\hat{f}_{\rm SM}(\mbi q-\mbi k')}\{\hat{x}(\mbi k')\}}_{{f}_{\rm SM}\circ\{\hat{x}\}}}_{\hat{f}_{\rm B}\cdot\big[{f}_{\rm SM}\circ\{\hat{x}\}\big]}}_{{P_{\rm N}}^{-1}\cdot\big[ \hat{f}_{\rm B}\cdot\big[{f}_{\rm SM}\circ\{\hat{x}\}\big]\big]}}_{ \overline{\hat{f}_{\rm B}}\cdot\big[ {P_{\rm N}}^{-1} \cdot\big[{\hat{f}_{\rm B}}\cdot\big[\hat{ f}_{\rm SM}\circ\{\hat{x}\}\big]\big]\big] }}_{ \hat{f}_{\rm SM}\circ\big[\overline{\hat{f}_{\rm B}}\cdot\big[ {P_{\rm N}}^{-1} \cdot\big[{\hat{f}_{\rm B}}\cdot\big[\hat{ f}_{\rm SM}\circ\{\hat{x}\}\big]\big]\big]\big] } \nonumber\\
\label{eq:op25}
\end{eqnarray}

\begin{eqnarray}
 \hat{\hat{\mat R^\dagger\mat N_{\rm WN}^{-1}\mat R}} \{\hat{\mbi x}\}({\mbi k})
&=& \underbrace{ \int \frac{{\rm d}^{N_{\rm D}}\mbi q}{(2\pi)^{N_{\rm D}}} {\hat{f}_{\rm SM}(\mbi k-\mbi q)} \underbrace{ \overline{\hat{f}_{\rm B}(\mbi q)} \underbrace{ \int \frac{{\rm d}^{N_{\rm D}}\mbi q'}{(2\pi)^{N_{\rm D}}} { N_{\rm WN}}^{-1}(\mbi q-\mbi q') \underbrace{ \hat{f}_{\rm B}(\mbi q') \underbrace{ \int \frac{{\rm d}^{N_{\rm D}}\mbi k'}{(2\pi)^{N_{\rm D}}}{\hat{f}_{\rm SM}(\mbi q'-\mbi k')}\{\hat{x}(\mbi k')\}}_{{f}_{\rm SM}\circ\{\hat{x}\}}}_{\hat{f}_{\rm B}\cdot\big[{f}_{\rm SM}\circ\{\hat{x}\}\big]}}_{{ N_{\rm WN}}^{-1}\circ\big[ \hat{f}_{\rm B}\cdot\big[{f}_{\rm SM}\circ\{\hat{x}\}\big]\big]}}_{ \overline{\hat{f}_{\rm B}}\cdot\big[ {N_{\rm WN}}^{-1} \circ\big[{\hat{f}_{\rm B}}\cdot\big[\hat{ f}_{\rm SM}\circ\{\hat{x}\}\big]\big]\big] }}_{ \hat{f}_{\rm SM}\circ\big[\overline{\hat{f}_{\rm B}}\cdot\big[ {N_{\rm WN}}^{-1} \circ\big[{\hat{f}_{\rm B}}\cdot\big[\hat{ f}_{\rm SM}\circ\{\hat{x}\}\big]\big]\big]\big] } \nonumber\\
\label{eq:op252}
\end{eqnarray}

\caption{\label{fig:form} Here the action on an arbitrary vector $\hat{\mbi x}$
  of the most complex operators that appear in table (\ref{tab:oper}) is
  shown. The upper one is required for Wiener-filtering and represents the
  signal term in the covariance matrix of the data. The middle and  lower ones
  stand for the inverse of the ML variance (eq.~\ref{eq:COBEvar}) and are
  required for the COBE-filter, the MEMG and for sampling purposes with the
  Wiener-filter. The equations have to be read from right to left. The braces
  show the order in which the operations have to be done from top to
  bottom. One has to be very careful with the correct conjugation of the
  different functions. Note that, contrary to naiv expectations, the
  conjugation of the first selection function $\hat{f}_{\rm SM}$ to be applied
  in the upper operation disappears.   }
\end{figure*}

Let us first remember  the data model given in eq.~(\ref{eq:data1}), and suppose that the operator $\mat R_{\rm P}$ is given by a convolution in real-space with some blurring function $f_{\rm B}$
\begin{equation}
 d({\mbi{r}})\equiv \int{\rm{d}}^{{N}_{\rm D}}{{\mbi{r}'}}\,f_{\rm B}(\mbi r-\mbi r')f_{\rm S}({\mbi{r}'}) f_{\rm M}({\mbi{r}'}) s({\mbi{r}'})+ f_{\rm SF}({\mbi{r}})\epsilon_{\rm N}({\mbi{r}}) {.}
\label{eq:op1}
\end{equation}
The operator $\mat R$ acting on an arbitrary vector $\{x\}$ is thus given by
\begin{equation}
 \mat R\{\mbi{x}\} ({\mbi r}) \equiv \int{\rm{d}}^{{N_{\rm D}}}{{\mbi{r}'}}\,f_{\rm B}(\mbi r-\mbi r')f_{\rm S}({\mbi{r}'}) f_{\rm M}({\mbi{r}'})\{{x(\mbi r')}\}  {.}
\label{eq:op2}
\end{equation}
The selection function and the masks should conveniently be multiplied in real-space to save convolutions
\begin{equation}
 f_{\rm SM}({\mbi{r}})\equiv f_{\rm S}({\mbi{r}}) f_{\rm M}({\mbi{r}}){.}
\label{eq:op3}
\end{equation}
Accordingly, the same operation as in eq.~(\ref{eq:op2}) leads to 
\begin{eqnarray}
 \hat{\hat{\mat R}}\{\hat{\mbi{x}}\}({\mbi k})&=& \underbrace{ \hat{f}_{\rm B}(\mbi k) \underbrace{ \int \frac{{\rm d}^{N_{\rm D}}\mbi q}{(2\pi)^{N_{\rm D}}}\hat{ f}_{\rm SM}(\mbi k-\mbi q)\{\hat{{x}}(\mbi q)\} }_{\hat{f}_{\rm SM}\circ \{\hat{{x}}\} } } \\\nonumber
&&\hspace{1.5cm}\hat{f}_{\rm B}\cdot\big[\hat{f}_{\rm SM}\circ \{\hat{{x}}\} \big]  {,}
\label{eq:op4}
\end{eqnarray}
in k-space. 
Here we have introduced the operator notation in which the equations have to be read from right to left. The braces show the sequence in which the subsequent operations have to be performed in the algorithm.
The analogous operation for the adjoint $\mat R^\dagger$ can be derived from
the definition of the response operator in real space (see eq.~\ref{eq:op2})
leading to 
\begin{equation}
 {{{\mat R}}}^\dagger\{{\mbi{x}}\}({\mbi r})= f_{\rm S}({\mbi{r}}) f_{\rm M}({\mbi{r}})\int{\rm{d}}^{{N_{\rm D}}}{{\mbi{r}'}}\,f_{\rm B}(\mbi r'-\mbi r)\{{x(\mbi r')}\} {.}
\label{eq:op45}
\end{equation}
In k-space it yields
\begin{equation}
 {\hat{\hat{\mat R}}}^\dagger\{\hat{\mbi{x}}\}({\mbi k})= {\hat{f}_{\rm SM}}\circ \big[ \overline{\hat{f}_{\rm B}}\cdot\{\hat{\mbi{x}}\} \big] ({\mbi k}) {.}
\label{eq:op5}
\end{equation}
Note, that this expression can be naturally obtained by calculating the signal
term of the data-autocorrelation matrix (see the upper operator in  fig.~\ref{fig:form}).
In section~(\ref{sec:blurring}) we will consider a Gaussian smoothing of the signal, as could
happen through an observational process, where we test the deconvolution with
our scheme.
However, the blurring function that of main  interest in the matter-field
reconstruction is
given by the mass assignment function, or pixel window, which describes the
effect of representing point sources (such as galaxies) on a grid.
The most popular assignment functions are the nearest grid point (NGP), the
clouds-in-cell (CIC), and the triangular-shaped cloud functions (TSC) \citep[see][]{1981csup.book.....H}.

\subsubsection{Covariance matrix of the data}

The data model consists of two terms
\begin{equation}
 \alpha({\mbi{r}})= \int{\rm{d}}^{{N}_{\rm D}}{{\mbi{r}'}}\,f_{\rm B}(\mbi r-\mbi r')f_{\rm SM}({\mbi{r}'}) s({\mbi{r}'}){,}
\label{eq:op6}
\end{equation}
and
\begin{equation}
 \epsilon({\mbi{r}})= f_{\rm SF}({\mbi{r}}) \epsilon_{\rm N}({\mbi{r}}){.}
\label{eq:op7}
\end{equation}
The same quantities in k-space are given by
\begin{equation}
\hat{ \alpha}(\mbi k)=\hat{f}_{\rm B}(\mbi k)\int \frac{{\rm d}^{N_{\rm D}}\mbi q}{(2\pi)^{N_{\rm D}}}\hat{ f}_{\rm SM}(\mbi k-\mbi q)\hat{ s}(\mbi q){,}
\label{eq:op8}
\end{equation}
and
\begin{equation}
\hat{ \epsilon}(\mbi k)=\int \frac{{\rm d}^{N_{\rm D}}\mbi q}{(2\pi)^{N_{\rm D}}}\hat{ f}_{\rm SF}(\mbi k-\mbi q)\hat{ \epsilon}_{\rm N}(\mbi q){.}
\label{eq:op9}
\end{equation}
Consequently, the covariance matrix of the data is given by the following sum 
\begin{equation}
\langle\hat{d}(\mbi k)\overline{\hat{d}(\mbi k')}\rangle_{({\mbi s,\mbi\epsilon}|\mbi p)}=\langle\hat{ \alpha}(\mbi k)\overline{\hat{\alpha}(\mbi k')}\rangle_{{(\mbi s,\mbi\epsilon|\mbi p)}}+\langle\hat{ \epsilon}(\mbi k)\overline{\hat{\epsilon}(\mbi k')}\rangle_{(\mbi s,{\mbi\epsilon}|\mbi p)}{,}
\label{eq:op10}
\end{equation}
where we have assumed that the noise is uncorrelated to the signal, which is
consistent with our data model. Even though the structure function may be
correlated with the signal \\ ${\langle\hat{s}(\mbi k)\overline{\hat{f}_{\rm
      SF}(\mbi k')}\rangle_{({\mbi s, \mbi f_{\rm SF}}|\mbi p)}\neq0}$, the random noise part is not ${\langle\hat{s}(\mbi k)\overline{\hat{\epsilon}_{\rm N}(\mbi k')}\rangle_{({\mbi s,\mbi\epsilon}|\mbi p)}=0}$.
We will calculate the different terms of the data covariance matrix  and other related operators in the next sections.

\subsubsection{Covariance matrix of the data: the signal term}
\label{sec:signalterm}

Here it becomes necessary to choose the Fourier representation, since it is there that the signal-autocorrelation matrix appears to be diagonal in the form of a power spectrum (eq.~\ref{eq:op11}).
Taking into account statistical homogeneity for the signal $\mbi s$
\begin{equation}
\langle\hat{s}({\mbi{k}})\overline{\hat{s}({\mbi{k}'})}\rangle_{{(\mbi s|\mbi p)}}=(2\pi)^{N_{\rm D}}\delta_{\rm  D}({\mbi{k}}-{\mbi{k}'}){P_{\rm S}}({\mbi{k}'}){,}
\label{eq:op11}
\end{equation}
with $\delta_{\rm D}$ being the Dirac-delta function, we can derive the expression for the signal covariance matrix term
\begin{eqnarray}
\lefteqn{ \big(\hat{\hat{\mat R\mat S\mat R^\dagger}}\big)({\mbi k,\mbi k'})= \langle\hat{ \alpha}(\mbi k)\overline{\hat{\alpha}(\mbi k')}\rangle_{({\mbi s}|\mbi p)}}\\
&&=\hat{f}_{\rm B}(\mbi k)\int \frac{{\rm d}^{N_{\rm D}}\mbi q}{(2\pi)^{N_{\rm D}}}\hat{ f}_{\rm SM}(\mbi k-\mbi q){P_{S}}(\mbi q)\overline{\hat{ f}_{\rm SM}(\mbi k'-\mbi q)}\overline{\hat{f}_{\rm B}(\mbi k')}\nonumber\\
&&=\hat{f}_{\rm B}(\mbi k)\int \frac{{\rm d}^{N_{\rm D}}\mbi q}{(2\pi)^{N_{\rm D}}}\hat{ f}_{\rm SM}(\mbi k-\mbi q){P_{S}}(\mbi q){\hat{ f}_{\rm SM}(\mbi q-\mbi k')}\overline{\hat{f}_{\rm B}(\mbi k')}\nonumber{,}
\label{eq:op133}
\end{eqnarray}
For its action on a vector (see fig.~\ref{fig:form}), we get
\begin{equation}
 \hat{\hat{{\mat R\mat S\mat R^\dagger}}} \{\hat{\mbi x}\}({\mbi k}) ={\hat{f}_{\rm B}\cdot\big[\hat{f}_{\rm SM}\circ\big[ {P_{\rm S}} \cdot\big[ { {\hat{ f}_{\rm SM}}\circ\big[\overline{\hat{f}_{\rm B}}\cdot\{\hat{x}\}}\big]\big]\big]\big]}(\mbi k)
 {,}
\label{eq:op15}
\end{equation}
and consequently
\begin{eqnarray}
 \hat{\hat{\mat S\mat R^\dagger}} \{\hat{\mbi x}\}({\mbi{k}})&=&\int \frac{{\rm d}^{N_{\rm D}}\mbi k'}{(2\pi)^{N_{\rm D}}}\langle\hat{ s}({\mbi{k}})\overline{\hat d({\mbi{k'}})}\rangle_{(\mbi s|\mbi p)}\{\hat{x}({\mbi{k'}})\} \nonumber\\
\hspace{0cm}&=& \underbrace{ {P_{\rm S}}(\mbi k) \underbrace{ \int \frac{{\rm d}^{N_{\rm D}}\mbi k'}{(2\pi)^{N_{\rm D}}}{\hat{ f}_{\rm SM}(\mbi k-\mbi k')}\underbrace{ \overline{\hat{f}_{\rm B}(\mbi k')}\cdot\{\hat{x}(\mbi k')\}}_{\overline{\hat{f}_{\rm B}}\cdot\{\hat{x}\}}}_{{\hat{ f}_{\rm SM}}\circ\big[\overline{\hat{f}_{\rm B}}\cdot\{\hat{x}\}\big]}}  \nonumber\\
&&\hspace{0cm} \hspace{1.5cm}{ {P_{\rm S}} \cdot\big[ { {\hat{ f}_{\rm SM}}\circ\big[\overline{\hat{f}_{\rm B}}\cdot\{\hat{x}\}\big]\big]}}
 {.}
\label{eq:op16}
\end{eqnarray}
The inverse of the signal-autocorrelation matrix can be solved trivially in
Fourier-space: \\ ${\hat{\hat{\mat S}}^{-1} = {\rm diag}(P_{\rm S}(\mbi k)^{-1}})$. Hence, the inverse square root yields ${\hat{\hat{\mat S}}^{-1/2}={\rm diag}(P_{\rm S}(\mbi k)^{-1/2}})$.

\subsubsection{Covariance matrix of the data: the noise term}
\label{sec:noiseterm}

Here, we will consider the noise covariance matrix corresponding to the
definition of the likelihood. Note, that this expression is equivalent to the
noise term in eq.~(\ref{eq:op10}) if the noise structure function has no signal dependence
(see discussion in section \ref{sec:WF}).
We assume, analogous to the case of the signal, statistical homogeneity for $\epsilon_{\rm N}$
\begin{equation}
\langle\hat{\epsilon_{\rm N}}({\mbi{k}})\overline{\hat{\epsilon_{\rm N}}({\mbi{k}'})}\rangle_{{(\mbi \epsilon|\mbi p)}}=(2\pi)^{N_{\rm D}}\delta_{\rm  D}({\mbi{k}}-{\mbi{k}'}){ P_{\rm N}}({\mbi{k}'}){,}
\label{eq:op12}
\end{equation}
and then derive the expression for the noise covariance matrix
\begin{eqnarray}
\hat{\hat{N}}({\mbi k,\mbi k'})&=&\langle\hat{ \epsilon}(\mbi k)\overline{\hat{\epsilon}(\mbi k')}\rangle_{({\mbi\epsilon}|\mbi p)}\nonumber\\
&=&\int \frac{{\rm d}^{N_{\rm D}}\mbi q}{(2\pi)^{N_{\rm D}}}\hat{f}_{\rm SF}(\mbi k-\mbi q){P_{\rm N}}(\mbi q){\hat{ f}_{\rm SF}(\mbi q-\mbi k')}{.}
\label{eq:op14}
\end{eqnarray}
Its action on a vector yields
\begin{eqnarray}
\lefteqn{\hat{\hat{{\mat N}}} \{\hat{\mbi x}\}({\mbi k}) = \int \frac{{\rm d}^{N_{\rm D}}\mbi k'}{(2\pi)^{N_{\rm D}}}\langle\hat{ \epsilon}(\mbi k)\overline{\hat{\epsilon}(\mbi k')}\rangle_{({ \mbi\epsilon}|\mbi p)} \{\hat{x}(\mbi k')\}}\nonumber\\
&&= \underbrace{ \int \frac{{\rm d}^{N_{\rm D}}\mbi q}{(2\pi)^{N_{\rm D}}}\hat{f}_{\rm SF}(\mbi k-\mbi q) \underbrace{ { P_{\rm N}}(\mbi q) \underbrace{ \int \frac{{\rm d}^{N_{\rm D}}\mbi k'}{(2\pi)^{N_{\rm D}}}{\hat{f}_{\rm SF}(\mbi q-\mbi k')}\{\hat{x}(\mbi k')\}}_{{\hat{f}_{\rm SF}}\circ\{\hat{x}\}}}_{{P_{\rm N}} \cdot \big[{{\hat{f}_{\rm SF}}\circ\{\hat{x}\}}\big]  }}  \nonumber\\
&&  \hspace{2.5cm}{\hat{f}_{\rm SF}\circ \big[ {P_{\rm N}} \cdot \big[{{\hat{f}_{\rm SF}}\circ\{\hat{x}\}}\big]\big]}   
  {,}
\label{eq:op20}
\end{eqnarray}
In the case where there is no structure function, the noise-autocorrelation reduces to
\begin{equation}
\hat{\hat{{N}}}_{\rm N}({\mbi k, \mbi k'}) = (2\pi)^{N_{\rm D}}\delta_{\rm  D}({\mbi{k}}-{\mbi{k}'}){P_{\rm N}}(\mbi k') {.}
\label{eq:op17}
\end{equation}
Then, its action is given by
\begin{equation}
{\hat{\hat{{\mat N}}}_{\rm N}\{\hat{\mbi x}\}}({\mbi k})= {P_{\rm N}}\cdot \{\hat{x}\}(\mbi k){.}
\label{eq:op18}
\end{equation}
The corresponding inverse operation is 
\begin{equation}
{\hat{\hat{{\mat N}}}^{-1}_{\rm N}\{\hat{\mbi x}\}}({\mbi k})= {P_{\rm N}}^{-1}\cdot \{\hat{x}\}(\mbi k){.}
\label{eq:op19}
\end{equation}
Consequently, we obtain (see fig.~\ref{fig:form})
\begin{equation}
{ \hat{\hat{\mat R^\dagger\mat N_{\rm N}^{-1}\mat R}} \{\hat{\mbi x}\}({\mbi k}) }\hspace{0cm}= {{\hat{f}_{\rm SM}}\circ\big[ \overline{\hat{f}_{\rm B}}\cdot\big[{P_{\rm N}}^{-1} \cdot\big[{\hat{f}_{\rm B}}\cdot\big[ {\hat{ f}_{\rm SM}}\circ\{\hat{x}\}\big]\big]\big]\big]}(\mbi k)
 {,}
\label{eq:op25}
\end{equation}
and
\begin{equation}
{ \hat{\hat{{\mat R^\dagger\mat N_{\rm N}^{-1}}}} \{{\mbi x}\}({\mbi k}) }\hspace{0cm}= \hspace{0cm}{{\hat{f}_{\rm SF}}\circ\big[\overline{\hat{f}_{\rm B}}\cdot\big[ {P_{\rm N}}^{-1} \cdot \{\hat{x}\}\big]\big]}(\mbi k)
 {.}
\label{eq:op27}
\end{equation}
The inverse square root of $\hat{\hat{{\mat N}}}_{\rm N}$ can now be calculated and leads to 
\begin{equation}
\hat{\hat{{\mat N}}}_{\rm N}^{-1/2}({\mbi k})={\rm diag}({P_{\rm N}}^{-1/2}(\mbi k)){.}
\end{equation}
The operation $\hat{\hat{{\mat R^\dagger\mat N_{\rm N}^{-1/2}}}}\{\hat{\mbi x}\}$ can then be obtained by doing the following substitution $\hat{\hat{{\mat N}}}_{\rm N}^{-1}\rightarrow\hat{\hat{{\mat N}}}_{\rm N}^{-1/2}$ in eq.~(\ref{eq:op27})
\begin{equation}
{ \hat{\hat{{\mat R^\dagger\mat N_{\rm N}^{-1/2}}}} \{{\mbi x}\}({\mbi k}) }\hspace{0cm}= \hspace{0cm}{{\hat{f}_{\rm SF}}\circ\big[\overline{\hat{f}_{\rm B}}\cdot\big[ {P_{\rm N}}^{-1/2} \cdot \{\hat{x}\}\big]\big]}(\mbi k)
 {.}
\label{eq:op272}
\end{equation}
 We are especially interested in the case of white noise (${P_{\rm N}}={P_{\rm WN}}={ const}$) with a structure function (given by the Poissonian shot noise) 
\begin{equation}
\hat{\hat{{N}}}_{\rm WN}({\mbi k,\mbi k'}) = {P_{\rm WN}} \int \frac{{\rm d}^{N_{\rm D}}\mbi q}{(2\pi)^{N_{\rm D}}}\hat{f}_{\rm SF}(\mbi k-\mbi q)\overline{\hat{ f}_{\rm SF}(\mbi k'-\mbi q)}{.}
\label{eq:op212}
\end{equation}
The corresponding action yields
\begin{eqnarray}
\lefteqn{\hat{\hat{{\mat N}}}_{\rm WN}\{\hat{\mbi x}\}({\mbi k})} \nonumber\\
&&= \underbrace{ {P_{\rm WN}}   \underbrace{ \int \frac{{\rm d}^{N_{\rm D}}\mbi q}{(2\pi)^{N_{\rm D}}}\hat{f}_{\rm SF}(\mbi k-\mbi q)  \underbrace{ \int \frac{{\rm d}^{N_{\rm D}}\mbi k'}{(2\pi)^{N_{\rm D}}}{\hat{ f}_{\rm SF}(\mbi q-\mbi k')}\{\hat{ x}(\mbi k')\} }_{ { \hat{f}_{\rm SF} } \circ  \{ \hat{x} \} }  }_{ \hat{f}_{\rm SF} \circ \big[ { \hat{f}_{\rm SF} } \circ \{\hat{x}\} \big]} }\nonumber  {.} \\
&&{ P_{\rm WN} \cdot \big[ \hat{f}_{\rm SF} \circ \big[ { \hat{f}_{\rm SF} } \circ \{\hat{x}\} \big] \big]= {P_{\rm WN}}\cdot\big[\hat{f}_{\rm SF}^2\circ \{\hat{x}\}\big]}
\label{eq:op22}
\end{eqnarray}
It can be seen from this equation, that the preferential representation now is in real-space, where $\mat N$ is diagonal 
\begin{equation}
{N}_{\rm WN}({\mbi r,\mbi r'}) = \delta_{\rm  D}({\mbi{r}}-{\mbi{r}'}) {C_{\rm WN}}{f}_{\rm SF}^2 ({\mbi{r}'}){,}
\label{eq:op23}
\end{equation}
with ${C_{\rm WN}}={\rm IFT}\big[{P_{\rm WN}}\big]$ being a constant.
The inverse operation yields
\begin{equation}
{\mat N}^{-1}_{\rm WN} \{\mbi x\}({\mbi r}) = ({C_{\rm WN}}f_{\rm SF}^2)^{-1}\cdot\{{ x}\}(\mbi r) {.}
\label{eq:op24}
\end{equation}
Hence, the inverse square root yields 
\begin{equation}
{{{{N}}}_{\rm WN}^{-1/2}({{\mbi{r}}, {\mbi{r}'}})=\delta_{\rm  D}({\mbi{r}}-{\mbi{r}'}){C_{\rm WN}}^{-1/2}f_{\rm SF}^{-1}}(\mbi r){,}
\end{equation}
 and its action in k-space reads 
\begin{equation}
\hat{\hat{{\mat N}}}_{\rm WN}^{-1/2} \{\hat{\mbi x}\}({\mbi k})={P_{\rm WN}}^{-1/2}\cdot\big[\hat{f}_{\rm SF}^{-1}\circ \{\hat{ x}\}\big](\mbi k){.}
\end{equation}
Then we get (see fig.~\ref{fig:form})
\begin{equation}
{ \hat{\hat{\mat R^\dagger\mat N_{\rm WN}^{-1}\mbi R}} \{\hat{\mbi x}\}({\mbi k}) }\hspace{0cm}= \hspace{0cm}{{\hat{f}_{\rm SF}}\circ\big[\overline{\hat{f}_{\rm B}}\cdot\big[ \hat{\hat{N}}_{\rm WN}^{-1}\circ \big[{\hat{f}_{\rm B}}\cdot\big[ {\hat{ f}_{\rm SF}}\circ\{\hat{x}\}\big]\big]\big]\big]}(\mbi k)
 {,}
\label{eq:op26}
\end{equation}
and consequently
\begin{equation}
{ \hat{\hat{\mat R^\dagger\mat N_{\rm WN}^{-1}}} \{\hat{\mbi x}\}({\mbi k}) }\hspace{0cm}= \hspace{0cm}{{\hat{f}_{\rm SF}}\circ\big[\overline{\hat{f}_{\rm B}}\cdot\big[\hat{\hat{N}}_{\rm WN}^{-1}\circ \{\hat{x}\}\big]\big]}(\mbi k)
 {.}
\label{eq:op28}
\end{equation}
To calculate $\hat{\hat{{\mat R^\dagger\mat N_{\rm WN}^{-1/2}}}}\{\hat{\mbi x}\}$ one has to do the following substitution $\hat{\hat{{\mat N}}}_{\rm WN}^{-1}\rightarrow\hat{\hat{{\mat N}}}_{\rm WN}^{-1/2}$ in eq.~(\ref{eq:op28})
\begin{equation}
{ \hat{\hat{\mat R^\dagger\mat N_{\rm WN}^{-1/2}}} \{\hat{\mbi x}\}({\mbi k}) }\hspace{0cm}= \hspace{0cm}{{\hat{f}_{\rm SF}}\circ\big[\overline{\hat{f}_{\rm B}}\cdot\big[ \hat{\hat{N}}_{\rm WN}^{-1/2}\circ \{\hat{x}\}\big]\big]}(\mbi k)
 {.}
\label{eq:op282}
\end{equation}

In summary, we showed that the action of the different operators on a vector
required for the different reconstruction estimators (see table \ref{tab:oper})
can be calculated in a straightforward way, as an ordered series of products
and convolutions. Note that whenever we need to perform a convolution, we
change to the counter space representation with FFTs and do multiplications\footnote{In order to avoid aliasing effects one has to
  adequately perform {\it zero-padding} \citep[see
  e.g.][]{1992nrca.book.....P}.} there.
 

\section[Efficiency and quality validation of the inverse methods]{Efficiency and quality validation of the inverse methods with the Wiener-filter}
\label{sec:codetesting}

In this section the Wiener-filter implemented in \textsc{argo} is tested with the different linear inverse algorithms presented in the section of numerical methods (\ref{sec:itinvreg}) under several conditions determined by structured noise, blurring, selection function effects and windowing. 

The inverse methods that we test here are the Jacobi (J), the Steepest Descent
(SD), and several Krylov methods, like the Fletcher-Reeves ($\rm FR$), the
Polak-Ribi\`ere ($\rm PR$), and the {EXP} Conjugate
Gradients method (see section \ref{sec:krylov} and appendix \ref{app:CGFR}). This
scheme has not been previously discussed in the literature and turns out to be
very efficient as will be discussed below. Many other Krylov methods (see table
\ref{tab:beta}) can be built from simple equivalence relations, as we show in
appendix \ref{app:CG}. However, only the methods mentioned above are taken into
account here, as we consider them to be sufficiently representative.  The extra-regularization we propose with these Krylov methods converts the Wiener-filtering in a hybrid Tikhonov-Krylov space regularization method. In addition, we also test the Wiener-filter that uses hermitian redundancy as derived in appendix \ref{sec:mapeq}. We call the Wiener-filter defined by the mapping equation (\ref{eq:mapping}) the conjugated Wiener-filter (CJ), whereas the Wiener-filter defined by eq.~(\ref{eq:mapping2}) has no extra suffix.

\begin{figure*}
\centering
\resizebox{\hsize}{!}{\rotatebox[]{0}{\includegraphics[clip=true]{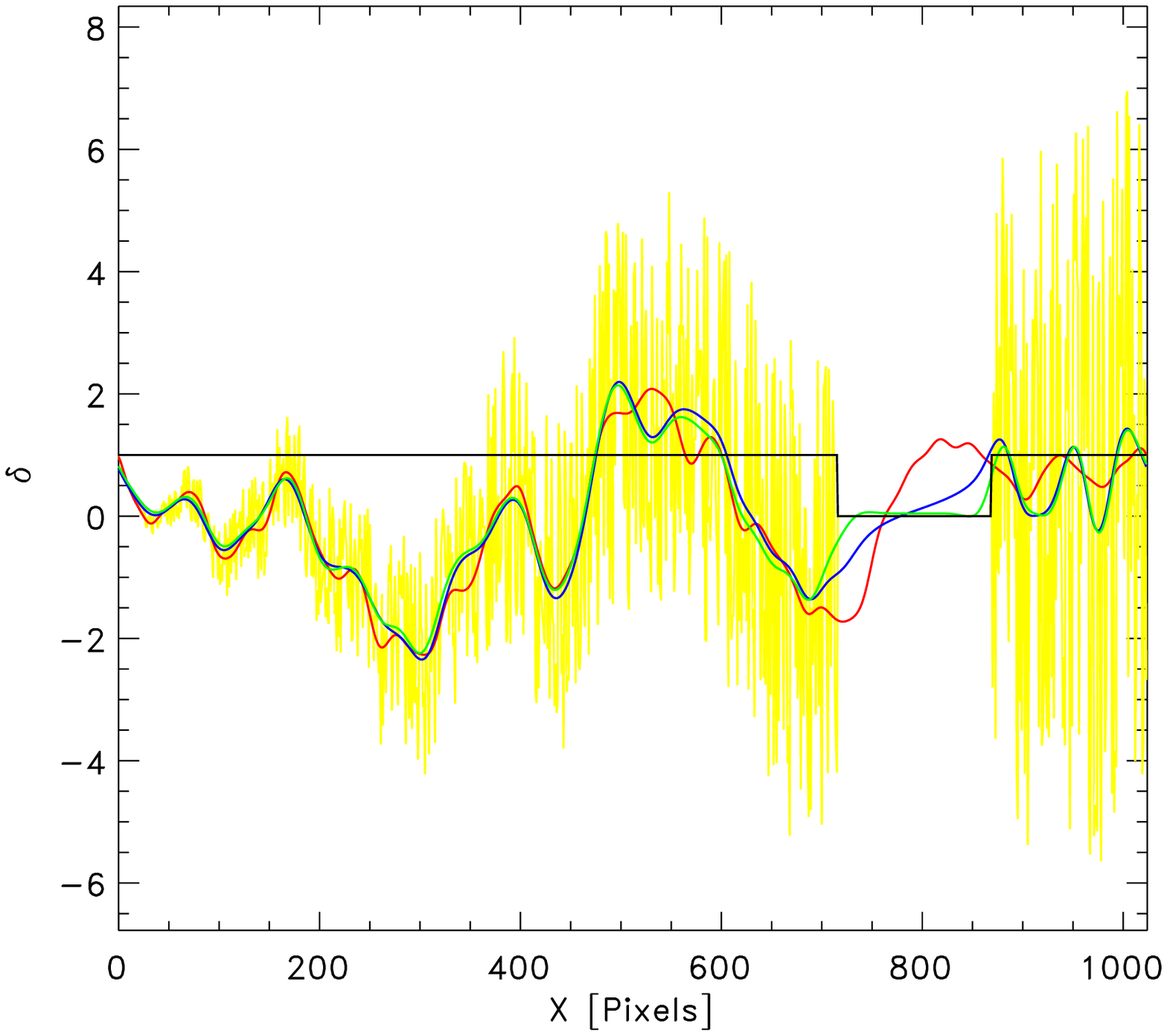}}
   \rotatebox[]{0}{\includegraphics[clip=true]{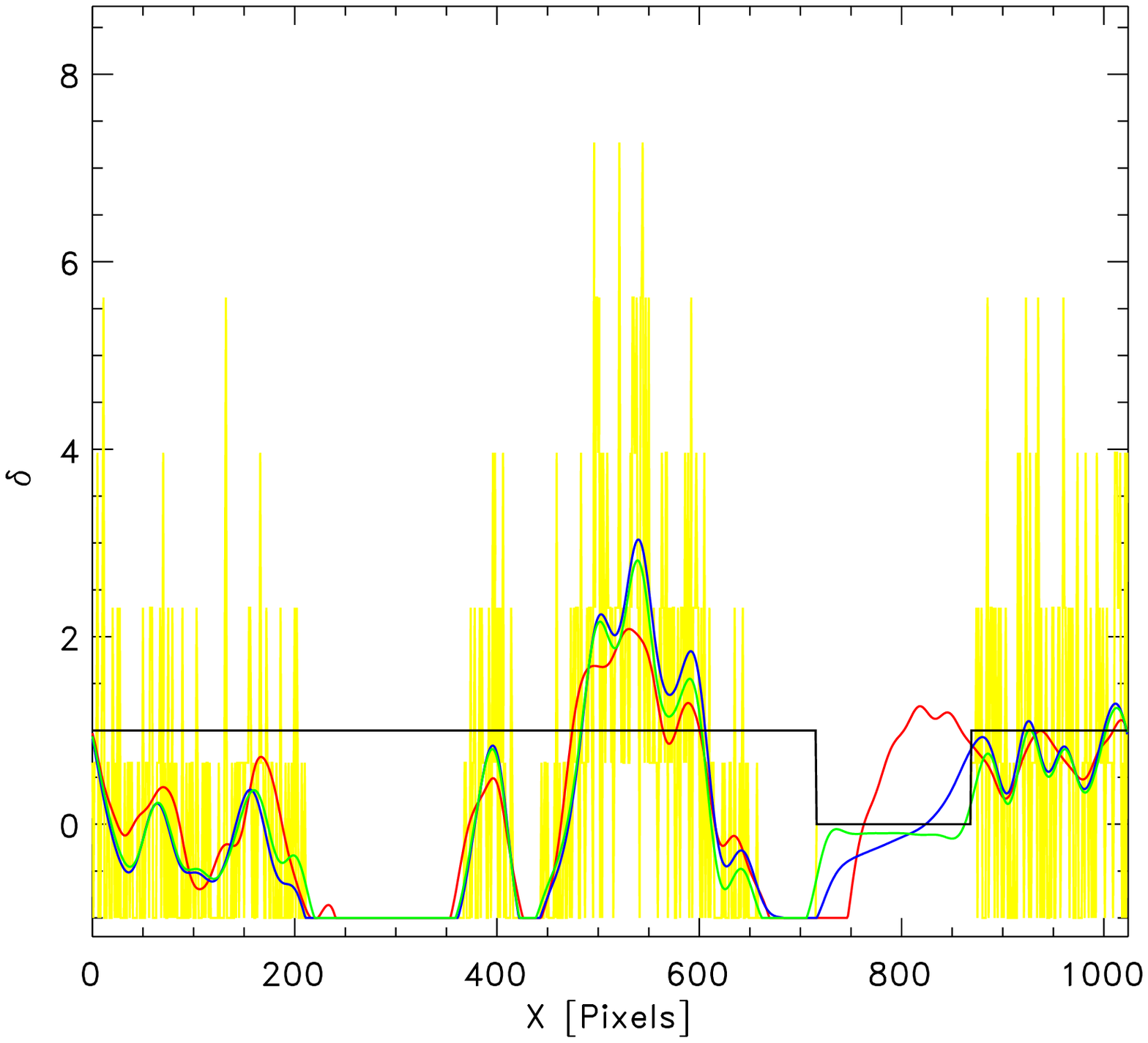}}}   
\caption[{\bf 1D Reconstruction with structured noise \&
    window}]{\label{delat1d}\small {\bf 1D Reconstruction with structured noise \&
    window:} {The left plot shows the reconstruction of a one-dimensional noisy
  signal. The red curve is the {\it true} underlying signal. The yellow lines
  represent the measured data in each grid cell. The data are windowed by a
  function given by the black line. A random noise with a structure function
  that increases with the distance with respect to the origin has been added to
  the {\it true} signal. The green and the blue lines show different
  reconstructions. In the blue case the windowing is formally treated, whereas
  in the green case the unseen region is modeled by a mean signal, which is
  zero in this case. We see that the unsampled region is estimated by
  the blue curve better than by the green curve, where the edge effects were
  neglected. The proper treatment of edge-effects gives even better
  results in the sampled regions close to the the borders of the unsampled
  regions. This improvement can clearly be seen in section (\ref{sec:wind}).} {\bf
  Poisson noise:} {In the right plot, two sampling processes are underlying
  the yellow signal. First the Gaussian random field that generates the red
  signal, which is then Poisson sampled leading to the yellow data. Again, the blue and the green curves
  represent the reconstructions with and without proper window treatment, respectively.}}  
\end{figure*}

With the aim of having full control over the synthetic data, we generate Gaussian random fields\footnote{ We use { GARFIELDS}: {\bf GA}ussian {\bf R}andom {\bf FIELDS}, a program we developed to generate Gaussian random fields from a given power spectrum. The method can be found in detail in \cite{2005astro.ph..6540M}.} with the \cite{1994MNRAS.267.1020P} formula for the power spectrum.  
The resulting {\it real} density field is denoted by $\delta_{\rm real}\equiv\delta_\rho$, and the reconstruction by $\delta_{\rm rec}\equiv\psi$.
The signals are discretized and arranged as vectors given by
$[k+n_z\times(j+n_y\times i)]$, where ${i\in[0,n_x-1]}$, ${j\in[0,n_y-1]}$, and
${k\in[0,n_z-1]}$. The algorithmic part of the reconstruction methods shown
in section (\ref{sec:itinvreg}) does not change with the dimensionality, but
solely the length of the vectors given by $n=n_x\times n_y\times n_z$ change  and
thus also the dimension of the involved matrices. The formulation of the matrices is
explained in detail in section (\ref{sec:operators}). The Fourier transforms
must be accordingly called with the  dimensions under consideration, which
occurs in \textsc{argo} by switching between the different FFTs given by FFTW\footnote{FFTW is a C subroutine library for computing fast discrete Fourier transforms in one or more dimensions of arbitrary input size and of both real and complex data: http://www.fftw.org/}. In addition, the power spectrum that is used for the reconstruction has to be set up with the corresponding length and the data have to be correctly rearranged to their original dimensions ($[i][j][k]\leftarrow[k+n_z\times(j+n_y\times i)]$) after their manipulation.  

\subsection{One-dimensional example}

We can see in fig.~(\ref{delat1d}) an example of a Gaussian realization in
one-dimension (red curve) that can represent a time-line. A structured noise
that increases with the distance ($f_{\rm SF}(r)\propto r$) and with a random
noise component ($\epsilon_{\rm WN}=G(0,1)$\footnote{$G(0,1)$: zero mean and
  variance 1.}) was added to
the signal. Finally a region was excluded simulating windowing effects. The
resulting curve was taken as the input signal (yellow curve). The
reconstruction given by \textsc{argo} is in blue and green, where the boundary effects
were considered in the first case, but not in the second. There the signal was
assumed to be zero in the UN-sampled region. We can see that the blue curve
better resembles the {\it real} signal guided by the trend at the
boundary. This effect is much larger in multiple dimensions as is shown in the
next section. 
In the right plot in fig.~(\ref{delat1d}), two sampling processes are underlying
  the yellow signal. First, the Gaussian random field that generates the red
  signal, which is then Poisson sampled thus leading to the yellow data. Again the blue and the green curves
  represent the reconstructions with and without proper window treatment,
  respectively. In this case, the blue curve also approaches the {\it true}
  signal better. 

\begin{figure*}
\begin{tabular}{cc}
\hspace{0.75cm}
\includegraphics[width=7cm]{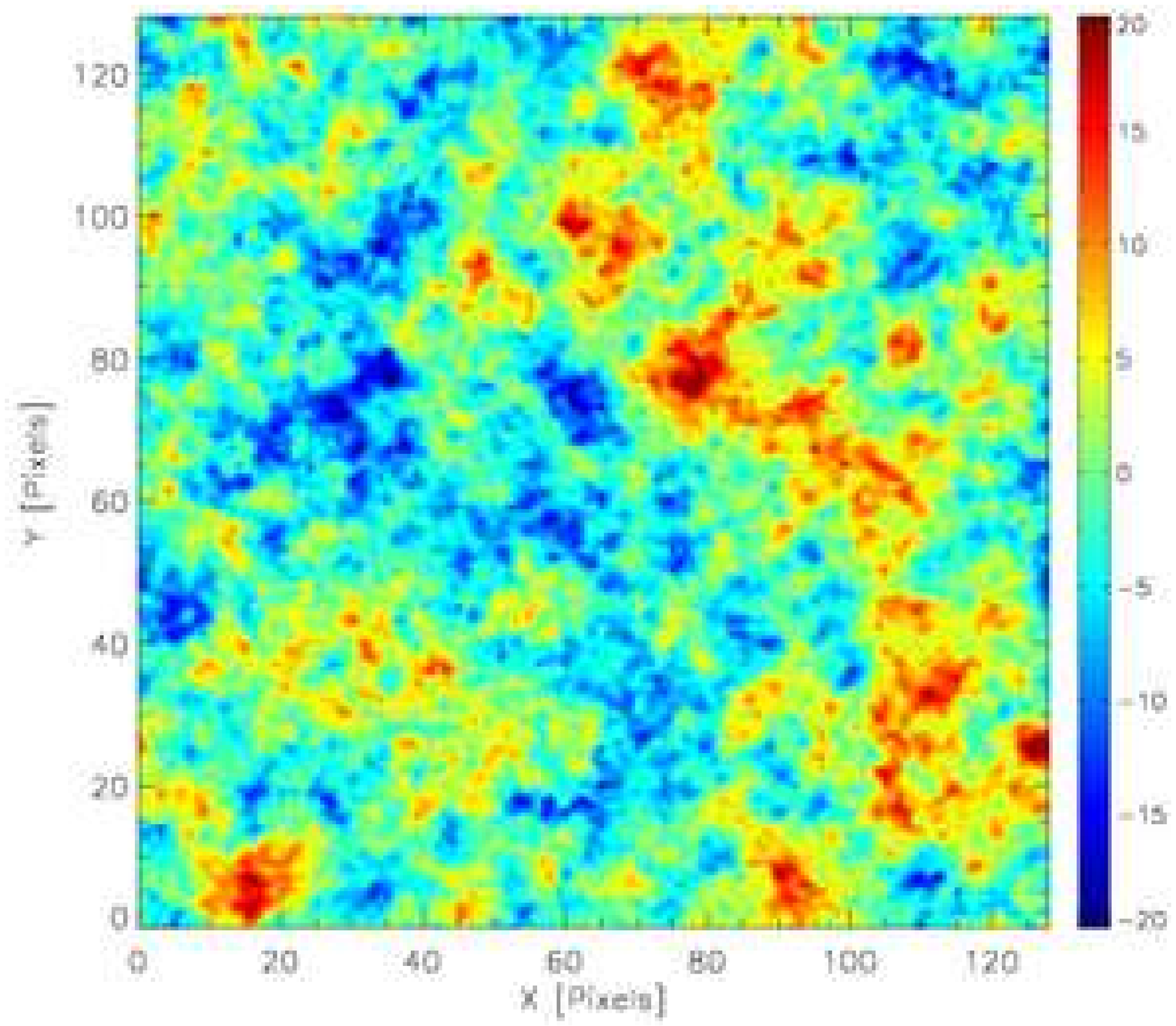}
\put(-190,0.5){{\Large\bf a}}
\hspace{0.5cm}
\includegraphics[width=7cm]{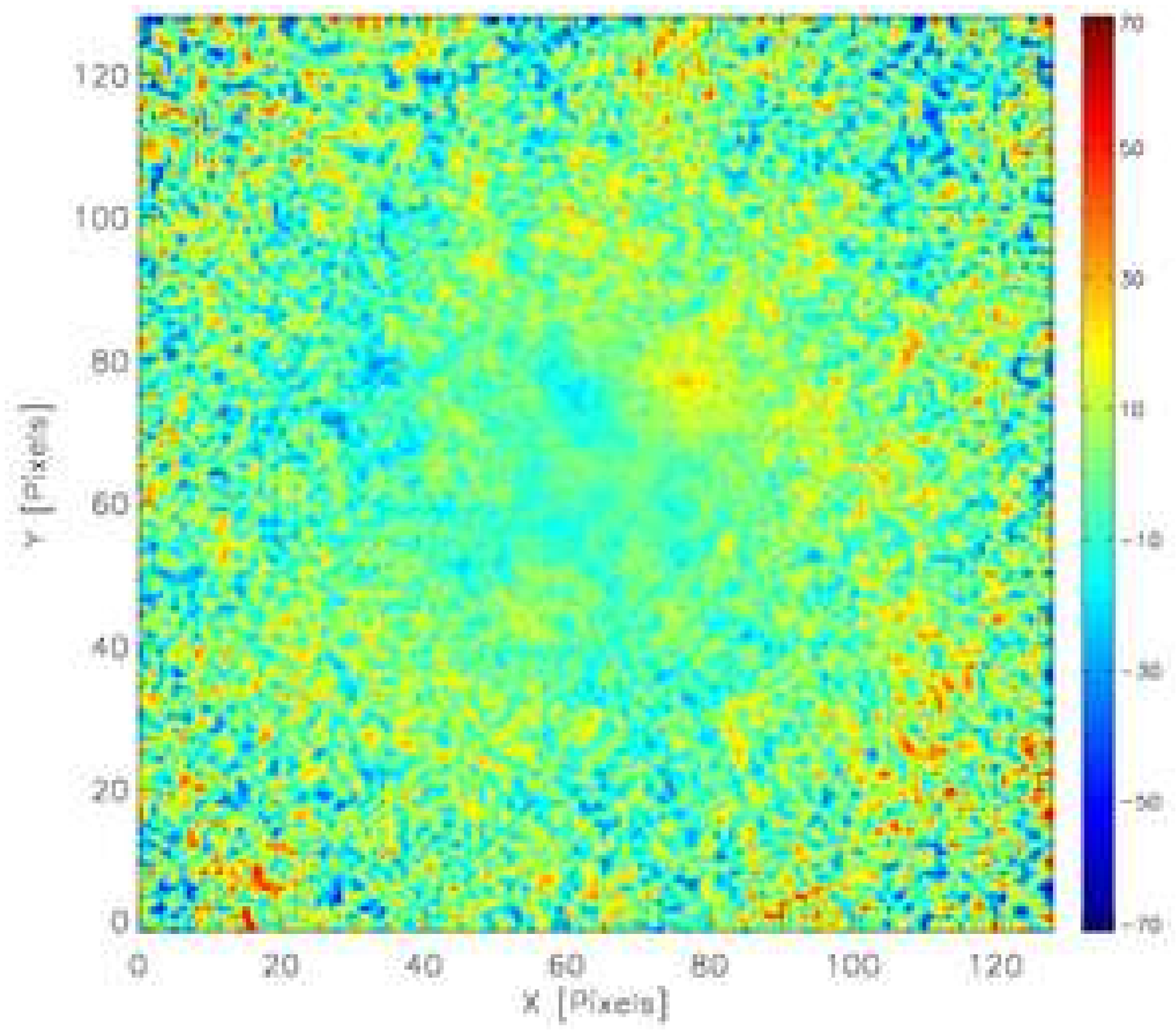}
\put(-190,0.5){{\Large\bf b}}
\\\\
\includegraphics[width=7cm]{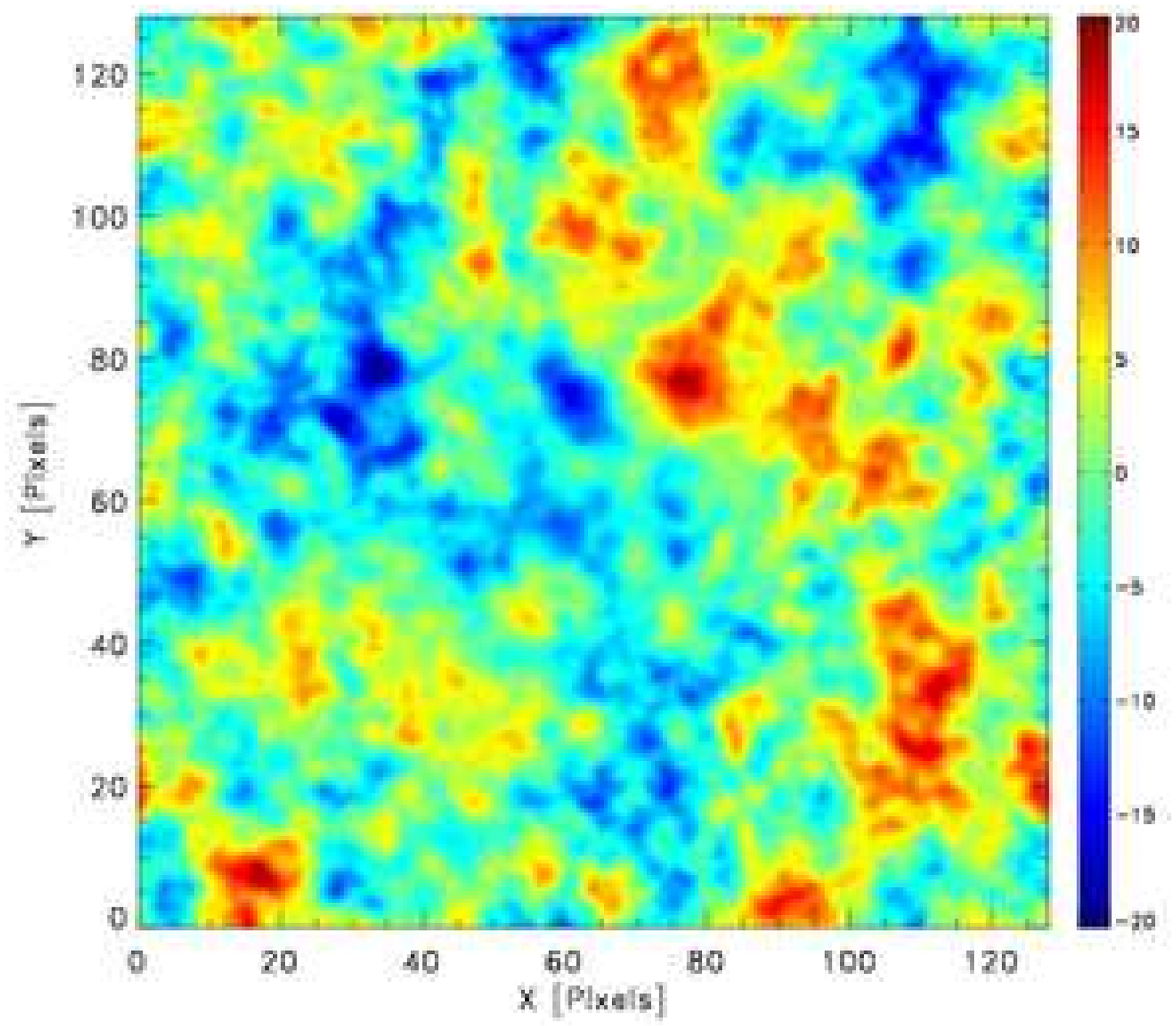}
\put(-190,0.3){{\Large\bf c}}

\hspace{0.2cm}
\includegraphics[width=6.5cm]{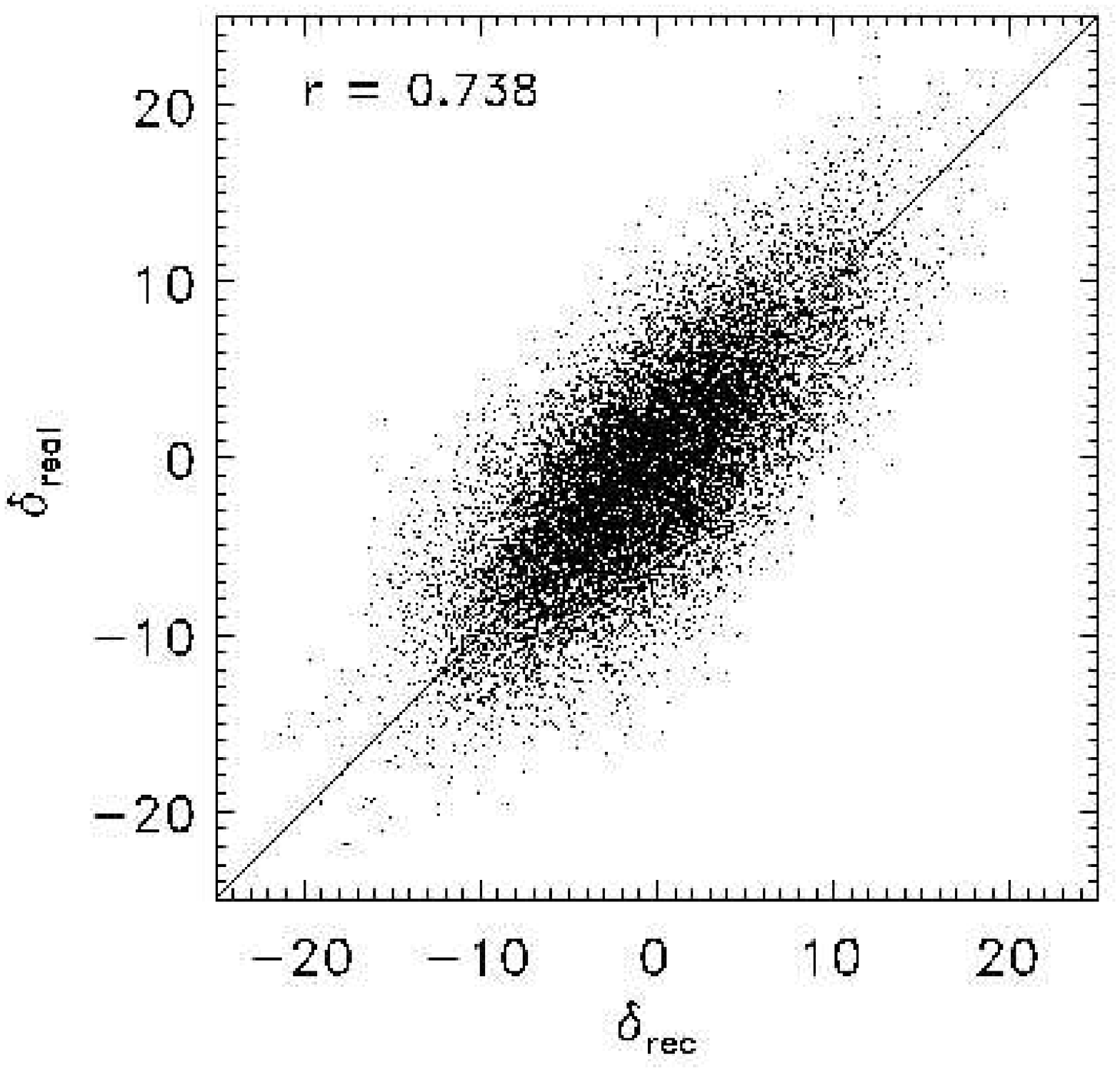}
\put(-160,0.){{\Large\bf d}}
\end{tabular}
\hspace{-0.5cm}
\caption{\small{\label{fig:2DNOW}{\bf Structured noise treatment:} The upper left picture
    shows the real signal. The upper right picture is the input signal, where
    some random noise that increases radially was added. Note that the scale of the colourbar changes from a maximum overdensity of 20 to 70. The lower left
    picture {\bf c} shows the reconstruction. The reconstructions using different
    numerical methods implemented in \textsc{argo} are indistinguishable. In the lower right image {\bf d}, the real density field is plotted against the reconstructed density field pixel by pixel without any smoothing. The numerical performance of this reconstrcution case is shown in the next figure.} }
\end{figure*}

\begin{figure*}
\begin{tabular}{cc}
\put(40.0,0.5){{\Large\bf a}}
\hspace{0.cm}
\includegraphics[width=7cm]{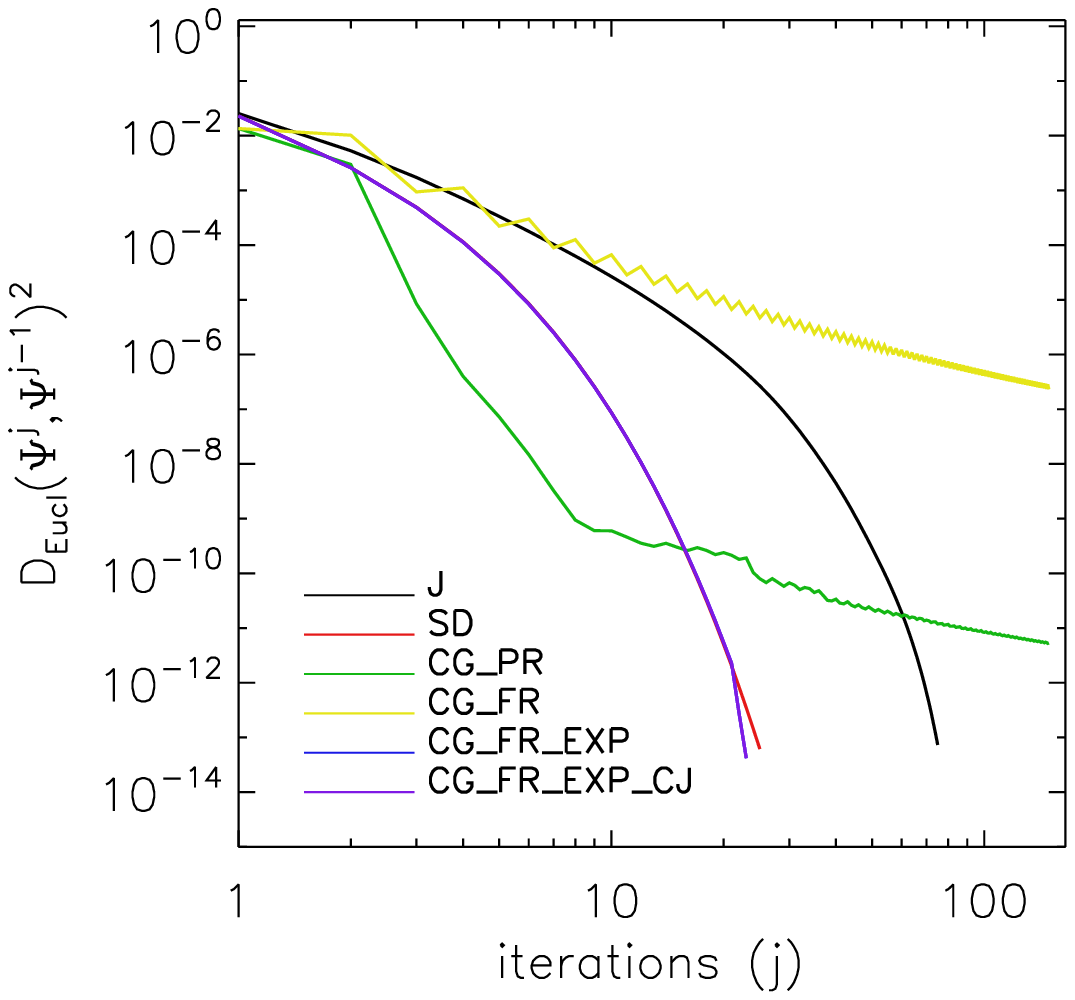}
\put(40,0.5){{\Large\bf b}}
\hspace{0.5cm}
\includegraphics[width=7cm]{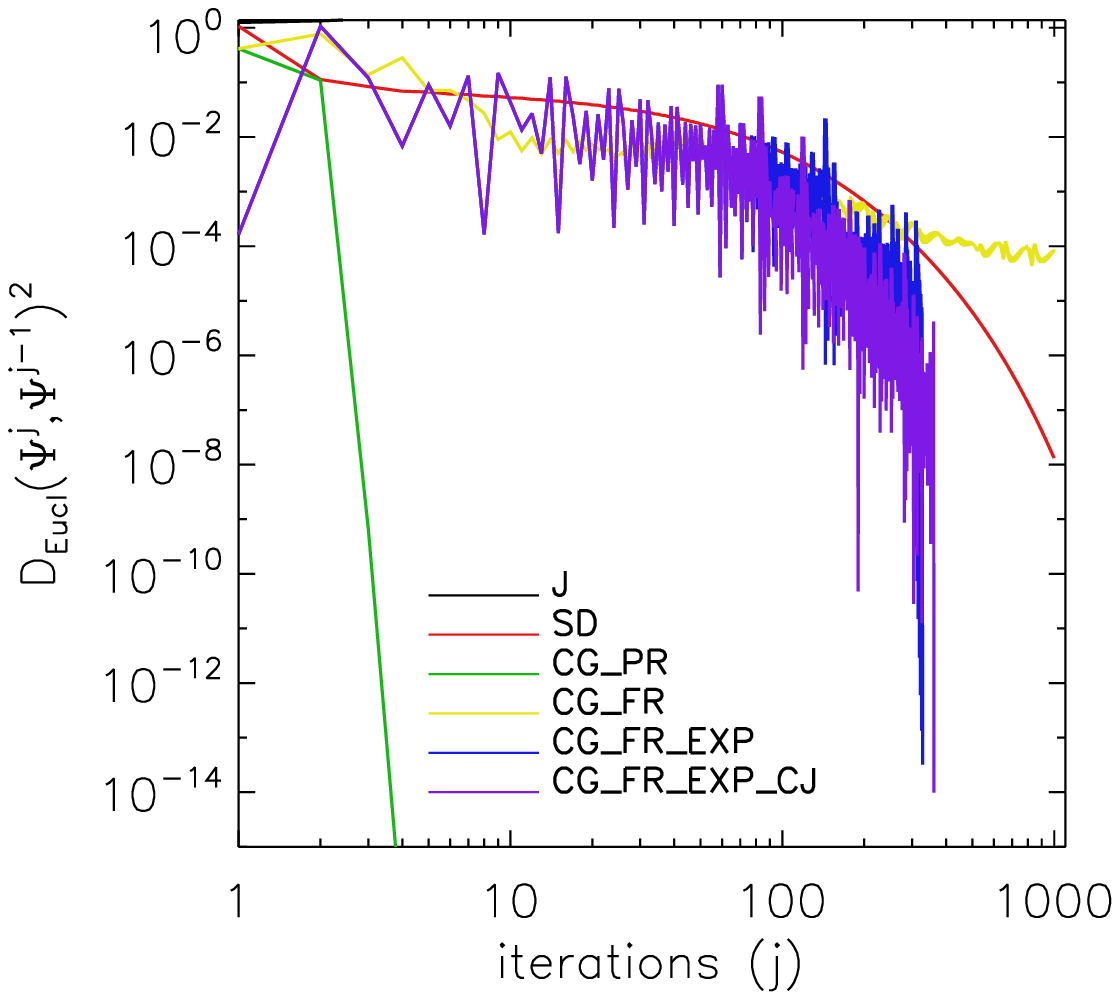}\\
\put(40.0,0.5){{\Large\bf c}}
\hspace{0.cm}
\includegraphics[width=7cm]{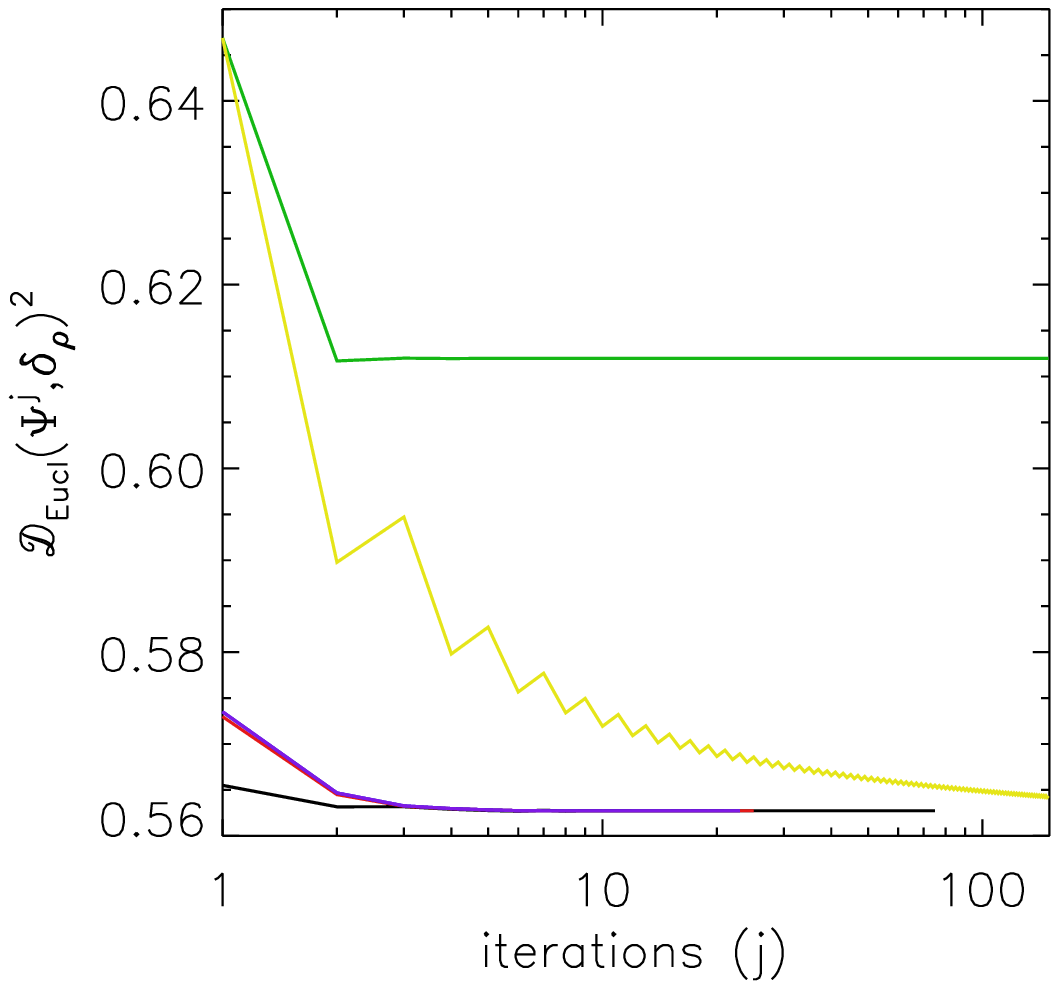}
\put(40,0.5){{\Large\bf d}}
\hspace{0.5cm}
\includegraphics[width=7cm]{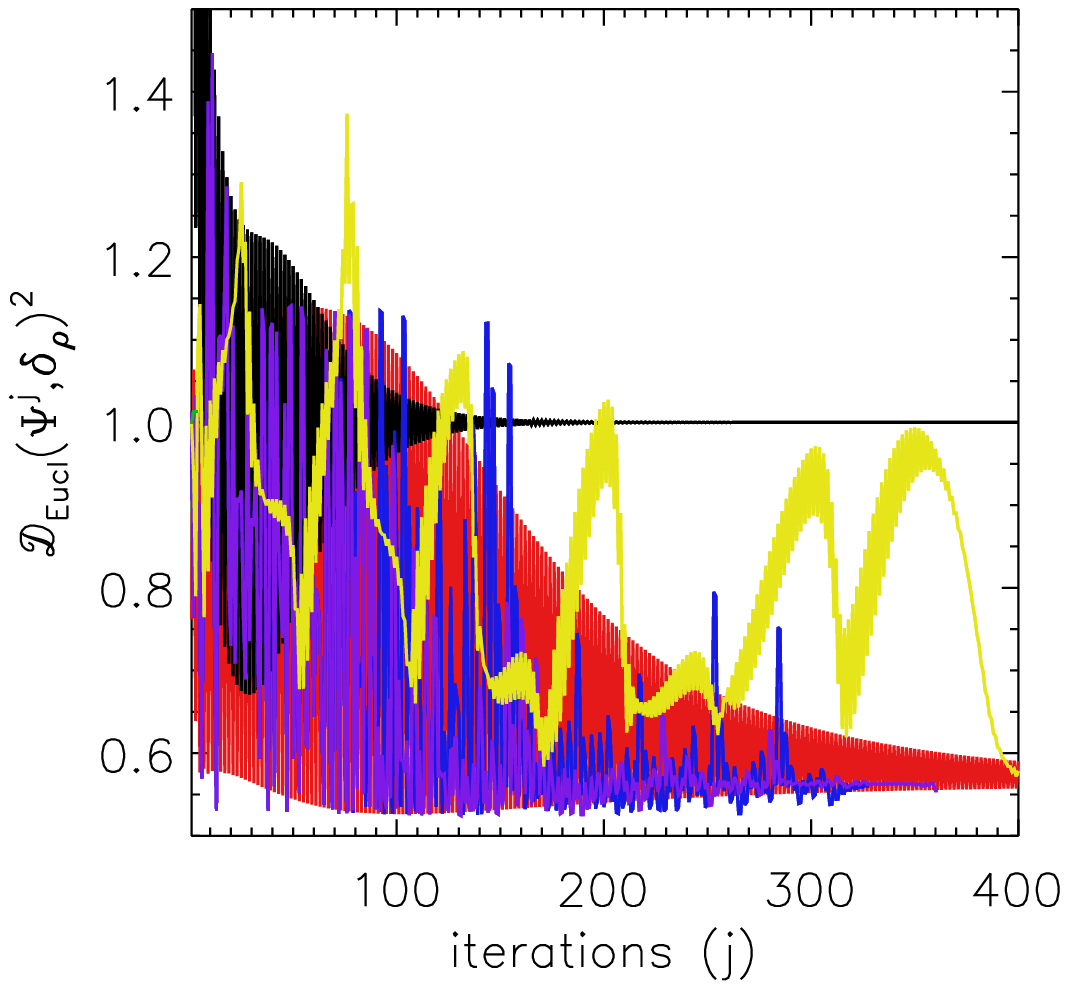}\\
\put(40.0,0.5){{\Large\bf e}}
\hspace{0.cm}
\includegraphics[width=7cm]{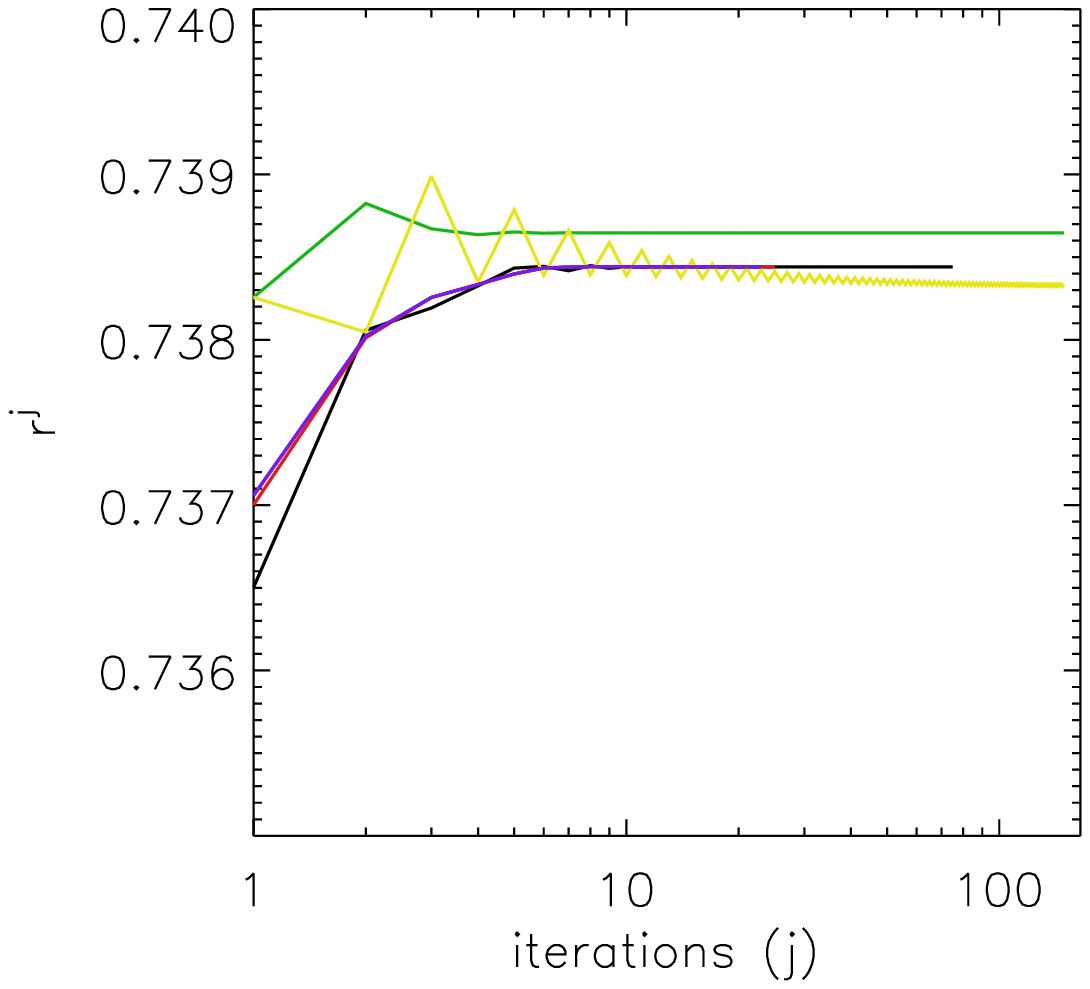}
\put(40,0.5){{\Large\bf f}}
\hspace{0.5cm}
\includegraphics[width=7cm]{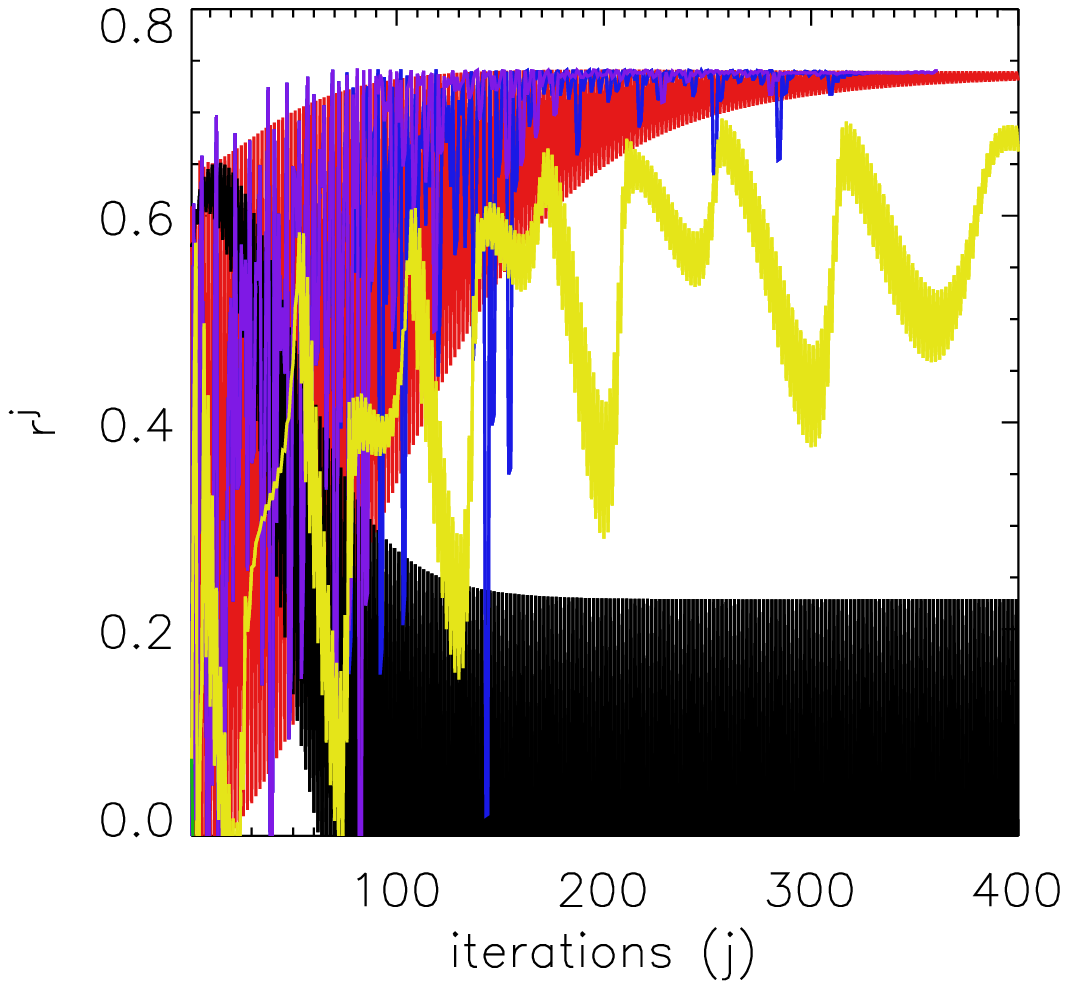}
\end{tabular}
\caption{\small{{\bf Numerical performance with and without preconditioning:}
    Here the convergence behaviour and the goodness of the reconstructions
    using different inversion algorithms can be seen. The pictures on the left
    show the methods using preconditioning, whereas the pictures on the right
    do not use preconditioning. The upper plots show the squared Euclidean distance
    between succesive reconstructions. The plots in the middle show the
    normalized Euclidean distance between the different reconstructions and the
    true signal. The lower plots show the evolution of the statistical
    correlation coefficient between reconstruction and signal. 
We see from panel {\bf c} and panel {\bf e} that after less than 10
iterations the reconstructions do not significantly improve with most of the
inversion algorithms. The different inversion algorithms used are: Jacobi (J),
Steepest Descent (SD), Conjugate Gradients (CG), Fletcher Reeves (FR), and
Polak Ribi\`ere (PR). We also tested a more expensive variant that uses
one additional operation of the involved matrix (EXP) and one other variant
(CJ), where a degree of freedom in the mapping equation for the Wiener-filter is used.}}
\label{fig:2DNOWSTAT}
\end{figure*}

\subsection{Multi-dimensional test cases}

\textsc{argo} has been implemented such that the global dimension $N_{\rm D}$ (see
section \ref{sec:dimen}), and even the  length in each  dimension ($n_x, n_y,
n_z$), can be chosen arbitrarily. Our tests in one-, two- and three dimensions show that the results do not differ qualitatively. The convergence behavior changes with the length of the arrays ($n=n_x\times n_y\times n_z$) as $n\log_2 n$ fully determined by the FFTs, as we showed in section (\ref{sec:itinvreg}).   
For the demonstration cases in this paper, we have selected the two-dimensional tests with $128\times128=16384$ pixels. 
However, three dimensional tests were also carried out leading to the same conclusions.

\subsubsection{Qualitative and quantitative measurement of the quality of the reconstruction }

To give a quantitative measurement of the quality of the reconstructions, we  define the correlation coefficient $r$ between the reconstructed and the real density field by
\begin{equation}
r\equiv\frac{\sum_i^n\delta_{\rho i}\psi_i}{\sqrt{\sum^n_i\delta_{\rho i}^2}\sqrt{\sum^n_j\psi_j^2}}{.}
\end{equation}
This statistical quantity is not very sensitive to the overall distribution and yields good values (close to unity) in some cases even with poor reconstructions (see section \ref{sec:selecfunc}).
The pixel to pixel plot of the {\it real} density field against the
reconstruction is highly informative because the scatter in the alignment of the pixels around the line of perfect correlation (45$^\circ$ slope) gives a qualitative goodness of the reconstruction.  
 In general, the quality of the recovered density map is better represented by the Euclidean distance between the real and the reconstructed signals. The ensemble average of this quantity can also be regarded as an action or loss function that leads to the Wiener-filter through minimization (see appendix \ref{sec:mapeq}).
Here we introduce the volume-averaged squared Euclidean distance\footnote{Note that ${\rm D}^2_{\rm Eucl}(\psi,\delta_\rho)=\frac{1}{V}{ D}^2_{\rm Eucl}(\psi,\delta_\rho)$.}
\begin{equation}
{\rm D}^2_{\rm Eucl}(\psi,\delta_\rho)\equiv\frac{1}{V} \int {\rm{d}}^{N_{\rm D}}{\mbi{r}}\,\Big[\psi({\mbi r})-  \delta_\rho({\mbi{r}})\Big]^2{,}
\label{eq:enseres}
\end{equation}
 with $V=L_x\times L_y\times L_z$.
 We further normalize the Euclidean distance through the following definition 
 \begin{equation}
{\mathcal D}^2_{\rm Eucl}(\psi,\delta_\rho)\equiv\frac{{\rm D}^2_{\rm Eucl}(\psi,\delta_\rho)}{{\rm D}^2_{\rm Eucl}(\psi_0,\delta_\rho)}{,}
\label{eq:ensres3}
\end{equation}
where $\psi_0$ is the zero vector.
We define the convergence tolerance criterion based on the squared Euclidean distance between subsequent reconstructions
\begin{equation}
{\rm tol_{crit}^{j+1}}\equiv{\rm D}^2_{\rm Eucl}(\psi^{j+1},\psi^j){.}
\label{eq:enseres}
\end{equation}
 We prefer this criterion with respect to the squared residuals $||\xi||^2$
 (see eq.~\ref{eq:reg14}) because all the tests show that no further statistical quality improvement in the reconstructions is reached after ${\rm tol_{crit}^{j+1}}$, as can be inferred from the correlation coefficients $r$ and the normalized squared  Euclidean distances ${\mathcal D}^2_{\rm Eucl}(\psi,\delta_\rho)$.  

\subsubsection{Numerical performance with and without preconditioning}

Here we analyze the convergence behavior of the different inverse schemes with
and without preconditioning. We start by considering a Gaussian random field
with some structured noise that increases radially and is modulated by a random noise component.  As a preconditioning expression, the diagonal part of the data covariance matrix is chosen, which is given by the sum of 
\begin{eqnarray}
 \lefteqn{\big(\hat{\hat{\mat R\mat S\mat R^\dagger}}\big)({\mbi k,\mbi k}) =}\nonumber\\ 
&&\hat{f}_{\rm B}(\mbi k)\int \frac{{\rm d}^{N_{\rm D}}\mbi q}{(2\pi)^{N_{\rm D}}}\hat{ f}_{\rm SM}(\mbi k-\mbi q){P_{S}}(\mbi q){\hat{ f}_{\rm SM}(\mbi q-\mbi k)}\overline{\hat{f}_{\rm B}(\mbi k)}\nonumber\\
&&=\underbrace{P_{\rm B}(\mbi k)\int \frac{{\rm d}^{N_{\rm D}}\mbi q}{(2\pi)^{N_{\rm D}}}P_{\rm SM}(\mbi k-\mbi q){P_{S}}(\mbi q)}\\
&&\hspace{1.5cm}{P_{\rm B}\cdot\big[P_{\rm SM}\circ P_{\rm S}\big]}\nonumber{,}
\label{eq:res5}
\end{eqnarray}
and
\begin{eqnarray}
\hat{\hat{N}}({\mbi k,\mbi k})&=&\int \frac{{\rm d}^{N_{\rm D}}\mbi q}{(2\pi)^{N_{\rm D}}}\hat{f}_{\rm SF}(\mbi k-\mbi q){P_{\rm N}}(\mbi q){\hat{ f}_{\rm SF}(\mbi q-\mbi k)} \nonumber\\
&=&\underbrace{\int \frac{{\rm d}^{N_{\rm D}}\mbi q}{(2\pi)^{N_{\rm D}}}P_{\rm SF}(\mbi k-\mbi q){P_{N}}(\mbi q)}\\
&&\hspace{1.5cm}{P_{\rm SF}\circ P_{\rm N}}\nonumber{,}
\end{eqnarray}
where we have used the following definitions: $P_{\rm B}\equiv||\hat{f}_{\rm B}||^2$, 
$P_{\rm SM}\equiv||\hat{f}_{\rm SM}||^2$ and $P_{\rm SF}\equiv||\hat{f}_{\rm SF}||^2$.
We can thus calculate the preconditioning matrix $\mat M$ required for the different schemes (section \ref{sec:itinvreg}) by just inverting each diagonal component.
The results summarized in figs.~(\ref{fig:2DNOW} and \ref{fig:2DNOWSTAT}) show important differences 
between the reconstructions done with (on the left side of fig.~\ref{fig:2DNOWSTAT}) and without (on the
right side of fig.~\ref{fig:2DNOWSTAT}) preconditioning. Some of the methods just speed up, like the
various EXP methods or the SD scheme. Others, however, are stabilized and
manage to converge to the solution only after preconditioning, like the J, the
FR and the CPR methods. Without preconditioning, the latter converges extremely
quickly  to a wrong solution. This is due to the fact that we did not impose
the following stabilization: $\beta_{\rm PR}={\rm max}(\beta_{\rm PR},0)$ in
this calculation \citep[see][for a discussion]{shewchuk}. However, our tests
show that upon imposing this stabilization the PR-method becomes significantly slower than the rest. On the other hand, the EXP-Krylov methods behave most stably and converge very quickly.
In the preconditioned case, we see that all methods converge to the same statistical result, as we can infer from the correlation coefficient $r$ and ${\mathcal D}^2_{\rm Eucl}(\psi,\delta_\rho)$, except for the PR scheme that yields slightly less optimal results (see the green line in comparison to the rest in panel {\bf c}).
We have tested preconditioning in the rest of the examples and could confirm  the results presented in this section.  Preconditioning turns out to be necessary to achieve fast algorithms.
In the next subsections we present results with a Poissonian distribution
(fig.~\ref{fig:2DPOIS}) and with blurring (fig.~\ref{fig:2DBLUR}). Their corresponding numerical
efficiency tests are shown in fig.~(\ref{fig:2DBPSTAT}). The same kind of studies are done
with a simulated selection function (fig.~\ref{fig:2DW4}) and with a mask
(fig.~\ref{fig:2DW1}). Their respective numerical behaviour can be seen in fig.~(\ref{fig:2DW14STAT}).

\subsubsection{Poissonian distribution}
\begin{figure*}
\begin{tabular}{cc}
\hspace{0.75cm}
\includegraphics[width=7cm]{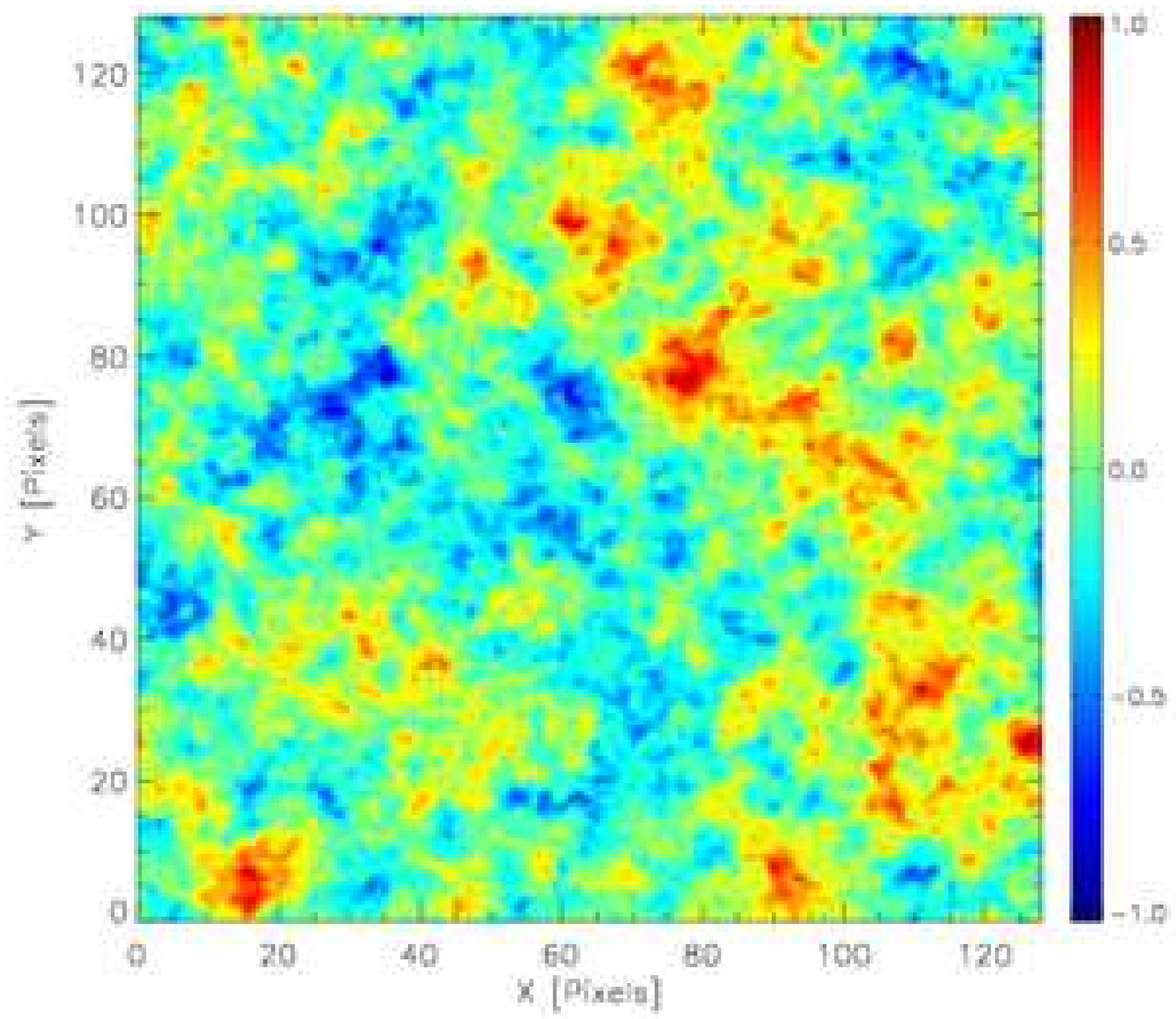}
\put(-190,0.5){{\Large\bf a}}
\hspace{0.5cm}
\includegraphics[width=7cm]{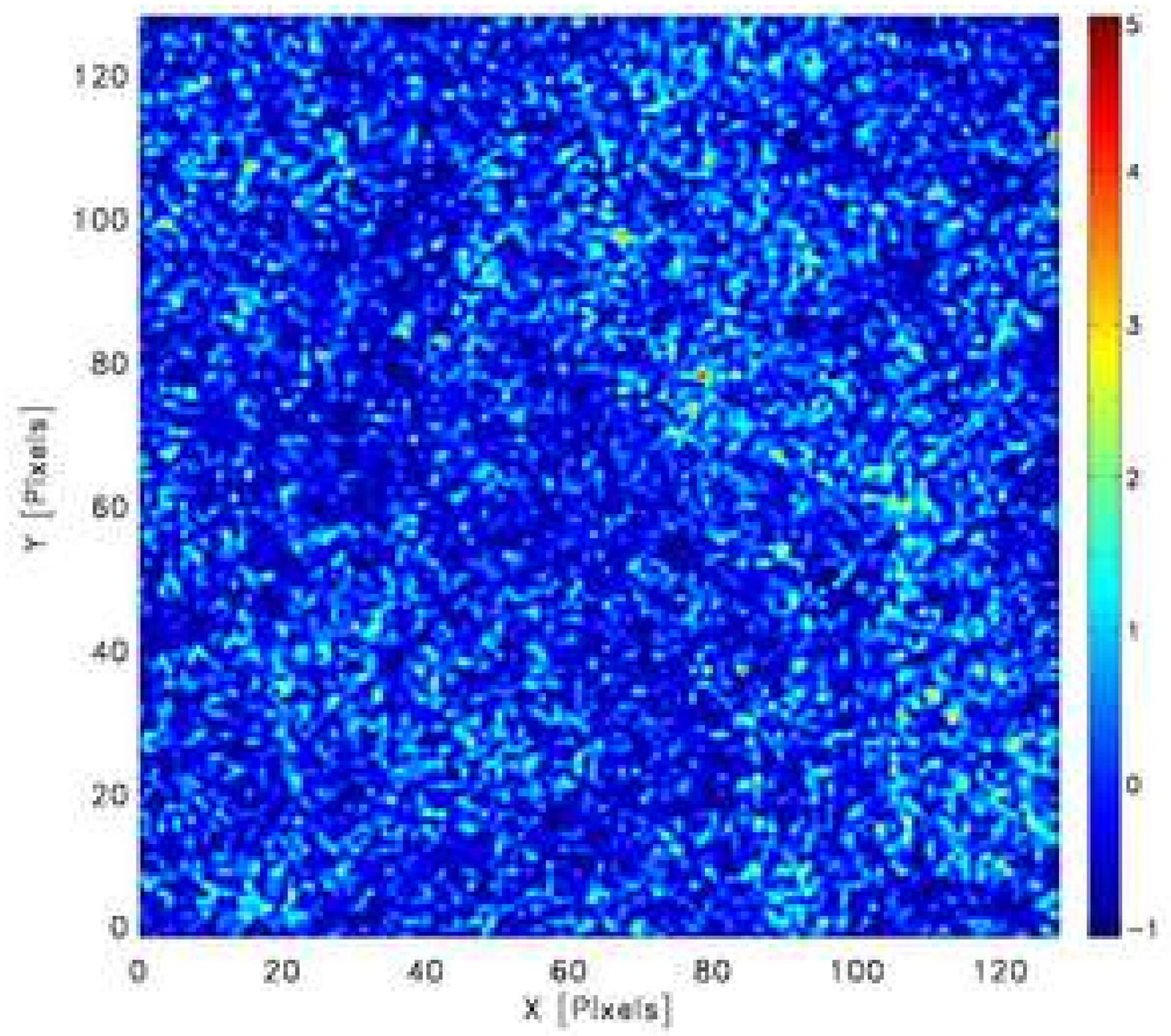}
\put(-190,0.5){{\Large\bf b}}\\\\
\includegraphics[width=7cm]{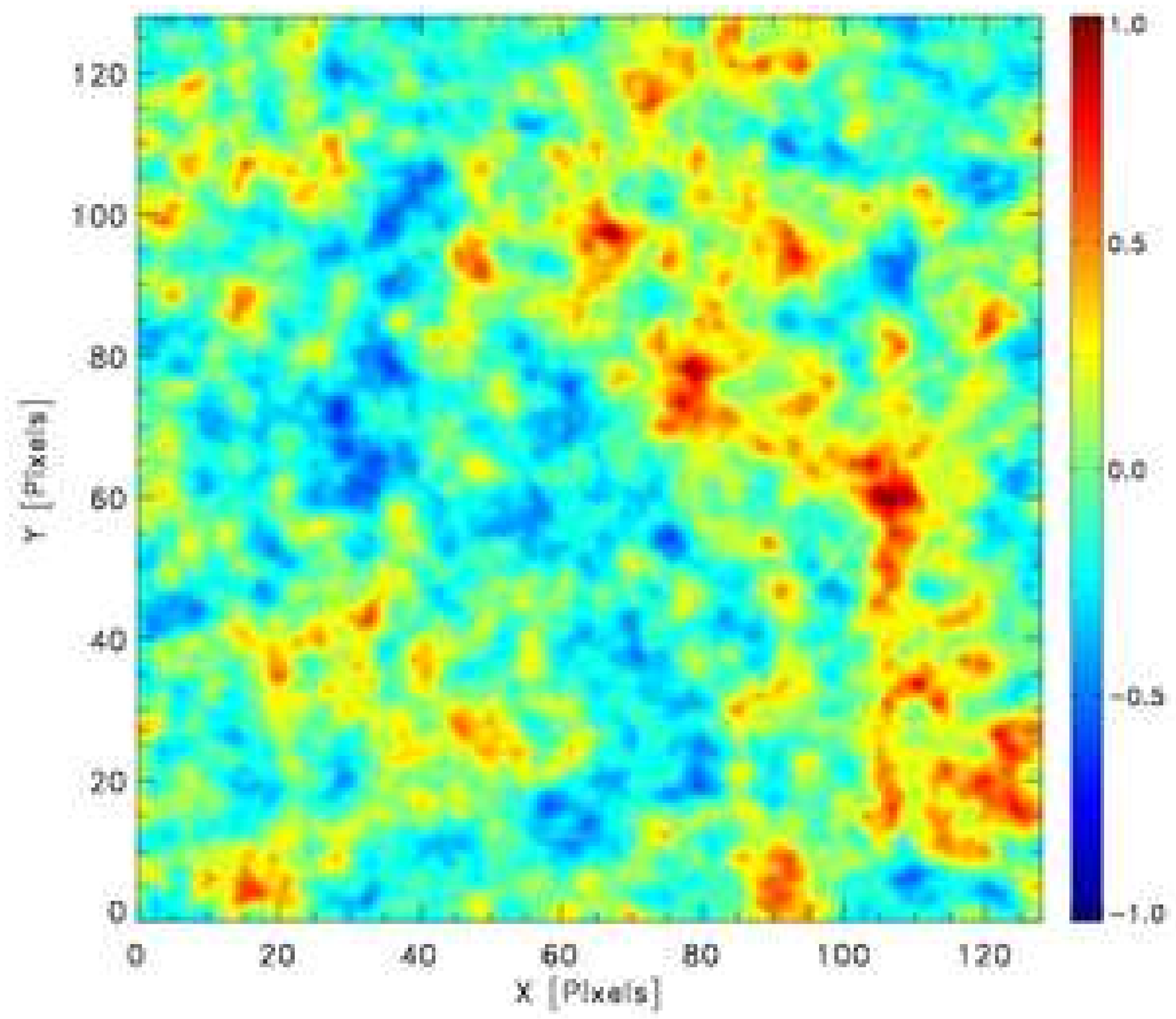}
\put(-190,0.3){{\Large\bf c}}
\hspace{0.2cm}
\includegraphics[width=6.5cm]{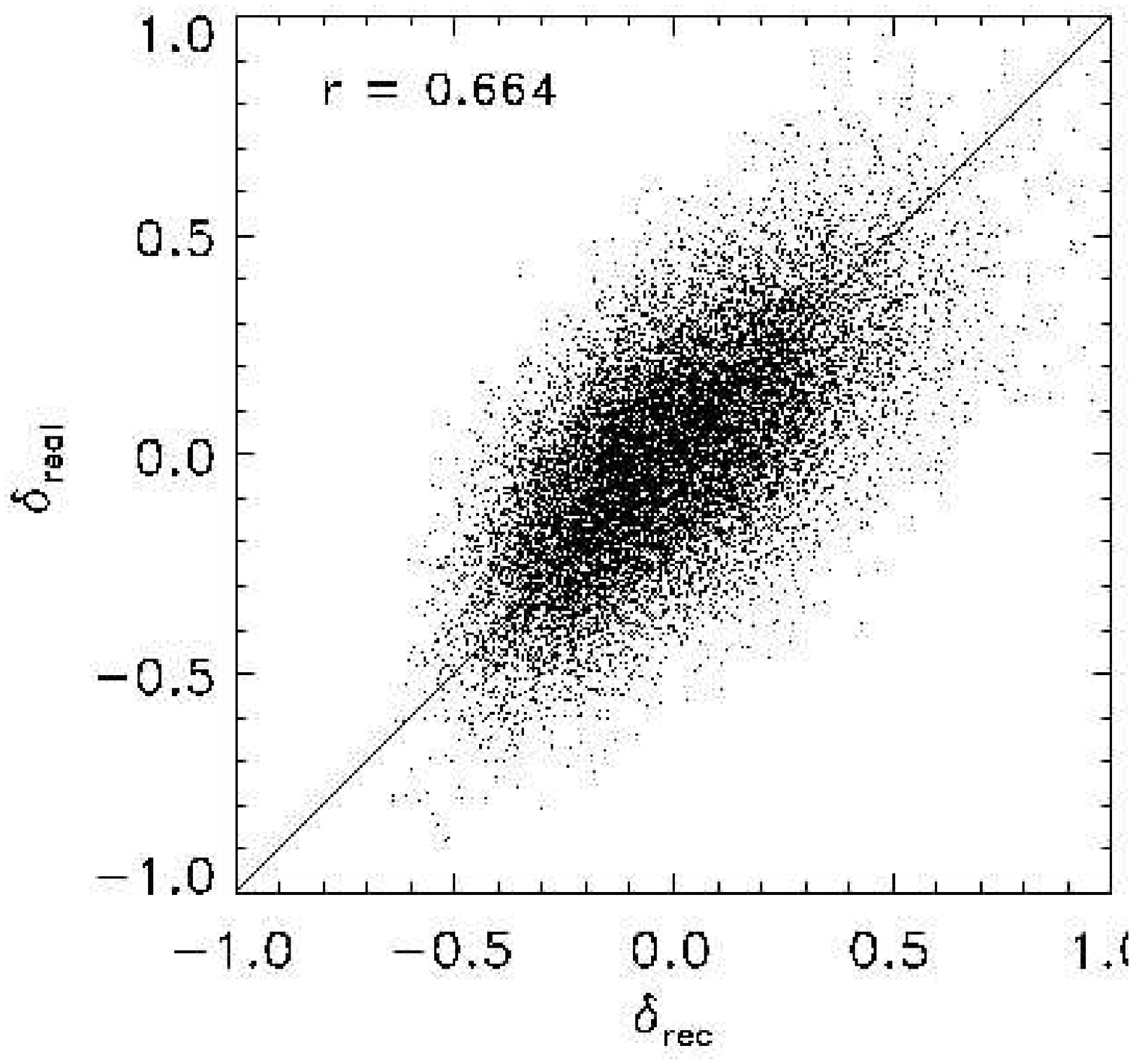}
\put(-160,0.){{\Large\bf d}}
\end{tabular}
\hspace{-0.5cm}

\caption{{\bf Poissonian noise:} Here two stochastic processes are underlying
  the input signal. First the Gaussian random field that generates the signal
  in panel {\bf a}, which is then Poisson sampled leading to the signal in
  panel {\bf b}. The reconstruction in panel {\bf c} is shown to be in good agreement with the underlying signal. The pixel values are correctly distributed as can be seen in panel {\bf d}. }
\label{fig:2DPOIS}
\end{figure*}

In this study case, we investigate the reconstruction of a Gaussian field based on a Poissonian distribution. This model is far from reality, where much more complex processes are known to occur (see discussion in section \ref{sec:datamodel}). However, we can model a non-Gaussian process in this way and test how good the Wiener-filter reconstruction works under such circumstances.
Here the assumed data model does not coincide with the one that has generated the data. However, the Poissonian noise can be modeled in the noise matrix of the Wiener-filtering through the structure function ${f}_{\rm S}$. 

The results presented in fig.~(\ref{fig:2DPOIS}) show very good agreement between the reconstruction and the {\it real} underlying density field (compare panels {\bf a} and {\bf c}). The convergence behaviour and statistical goodness is plotted in the left side of fig.~(\ref{fig:2DBPSTAT}), panels {\bf a}, {\bf c} and {\bf e}. There we can see that the FR and PR methods do not converge rapidly (see yellow and green curves in panel {\bf a}). On the contrary, the J, SD, and EXP schemes  are very efficient (panel {\bf c}) and lead to very similar results (panels {\bf c} and {\bf e}).

\subsubsection{Blurring effects: deconvolution}
\label{sec:blurring}
\begin{figure*}
\begin{tabular}{cc}
\includegraphics[width=7cm]{phiReal_2DPT1}
\put(-190,0.5){{\Large\bf a}}
\hspace{0.5cm}
\includegraphics[width=7cm]{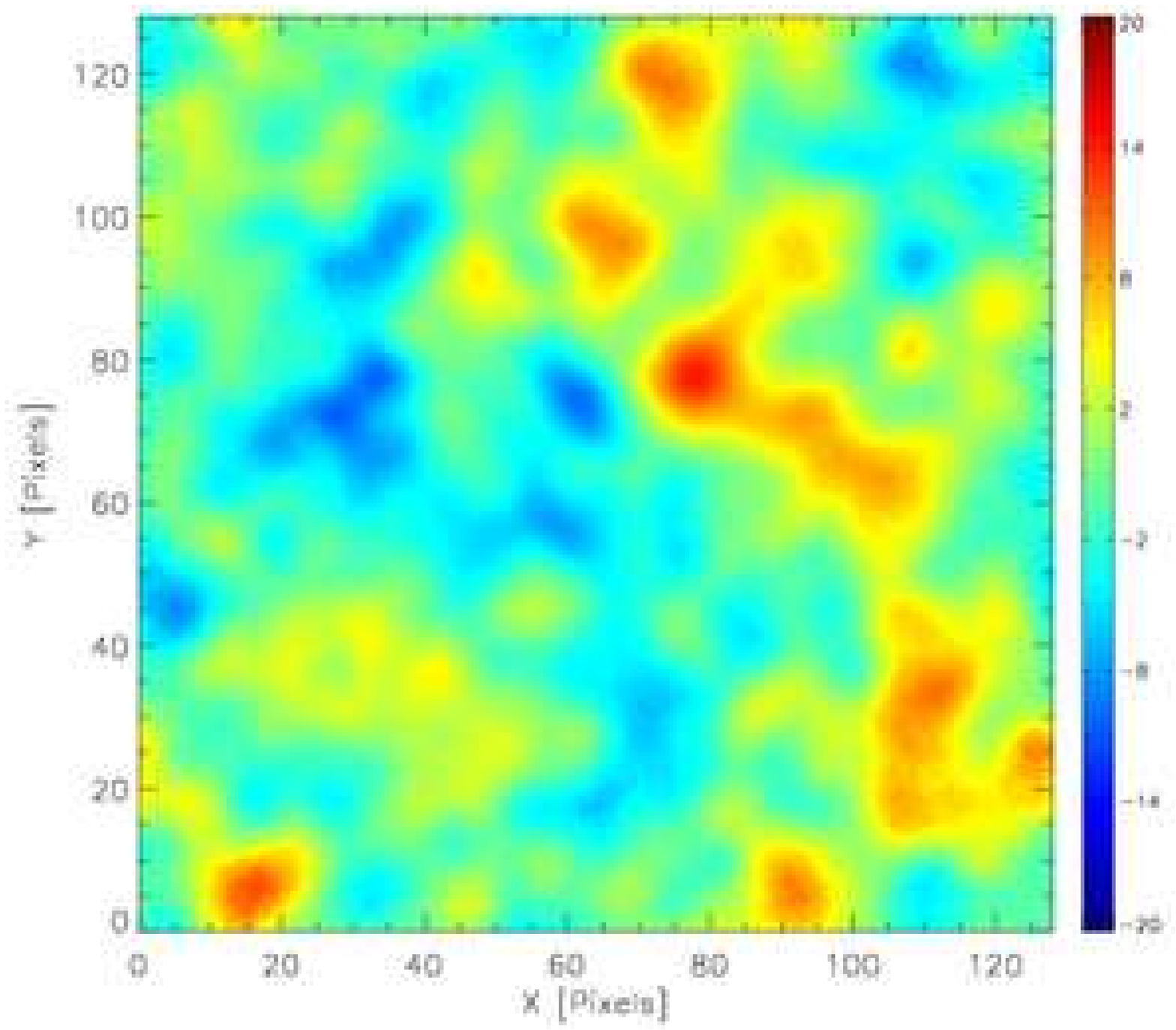}
\put(-190,0.5){{\Large\bf b}}\\\\
\includegraphics[width=7cm]{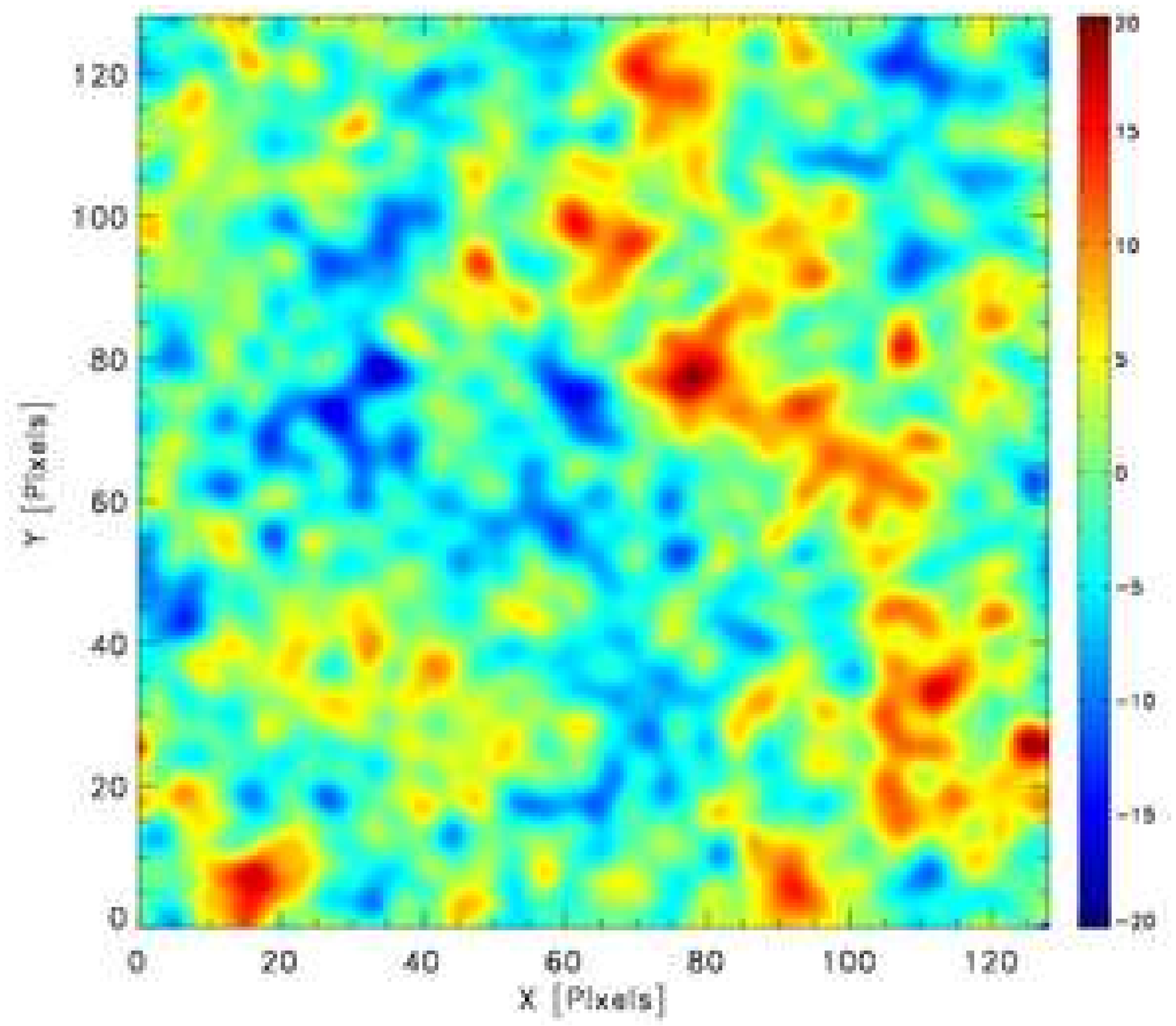}
\put(-190,0.5){{\Large\bf c}}
\hspace{0.5cm}
\includegraphics[width=7cm]{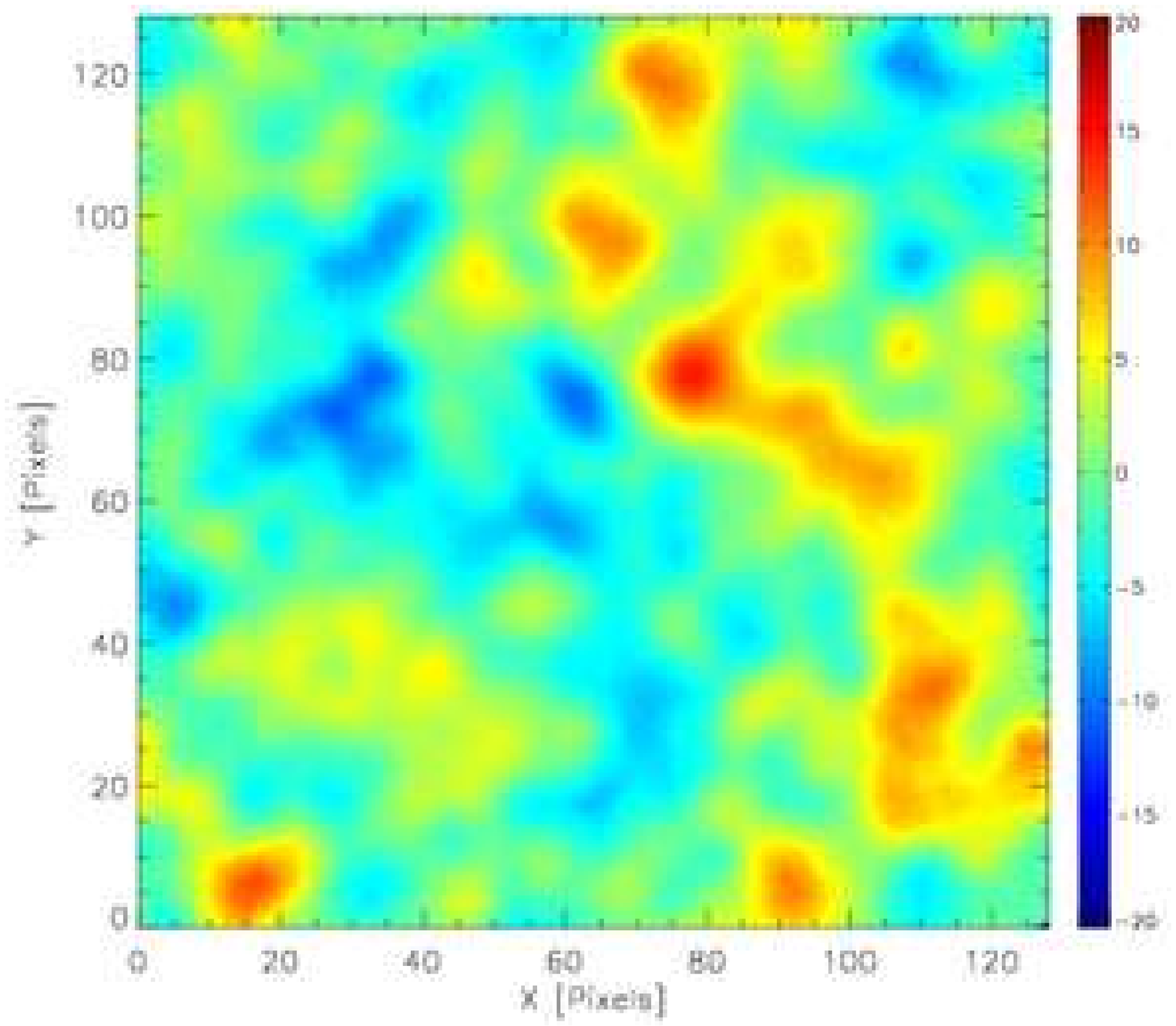}
\put(-190,0.5){{\Large\bf d}}
\end{tabular}

\begin{tabular}{ll}
\hspace{-.5cm}
\includegraphics[width=7cm]{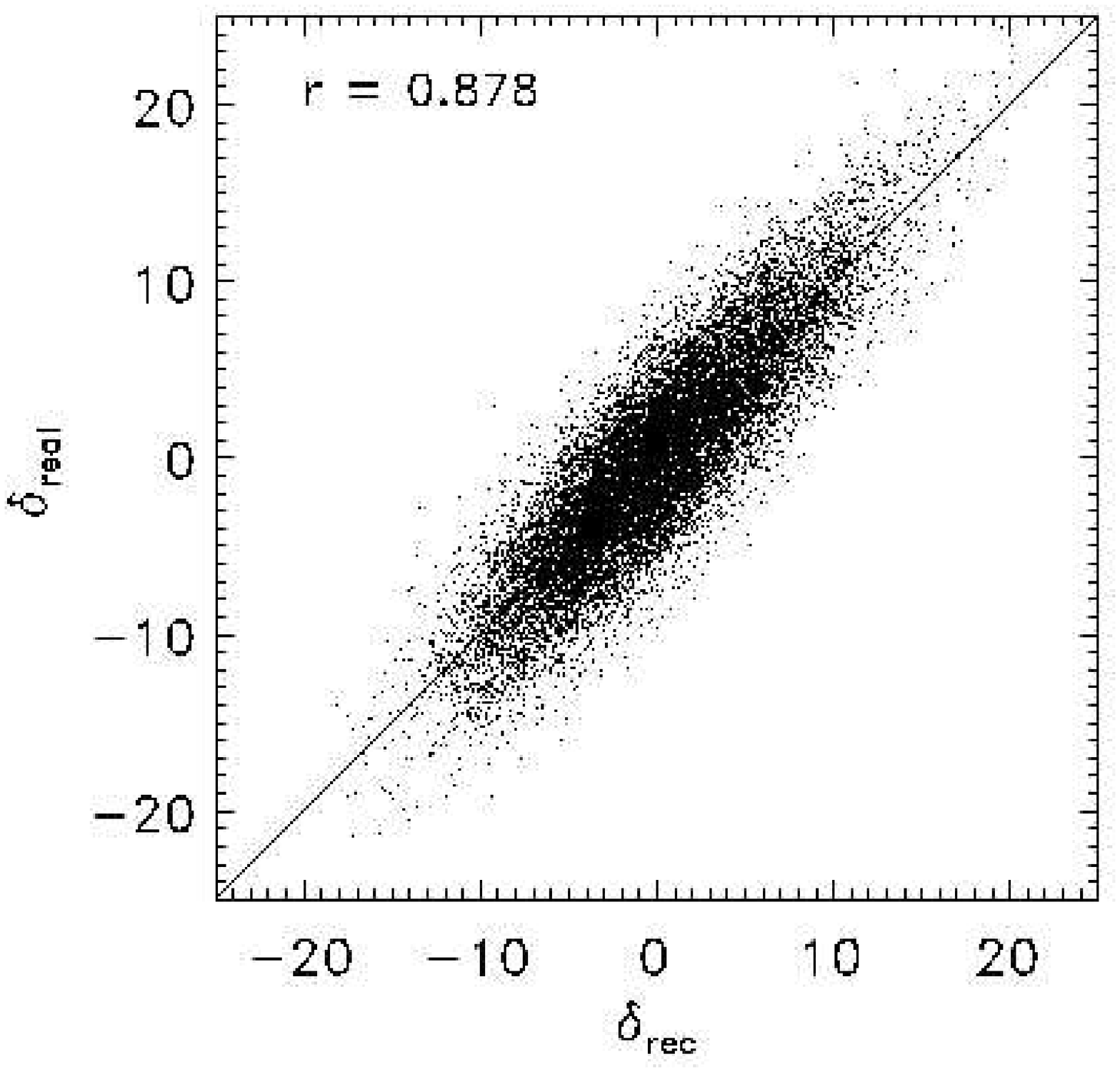}
\put(-180,0.5){{\Large\bf e}}
\hspace{0.5cm}
\includegraphics[width=7cm]{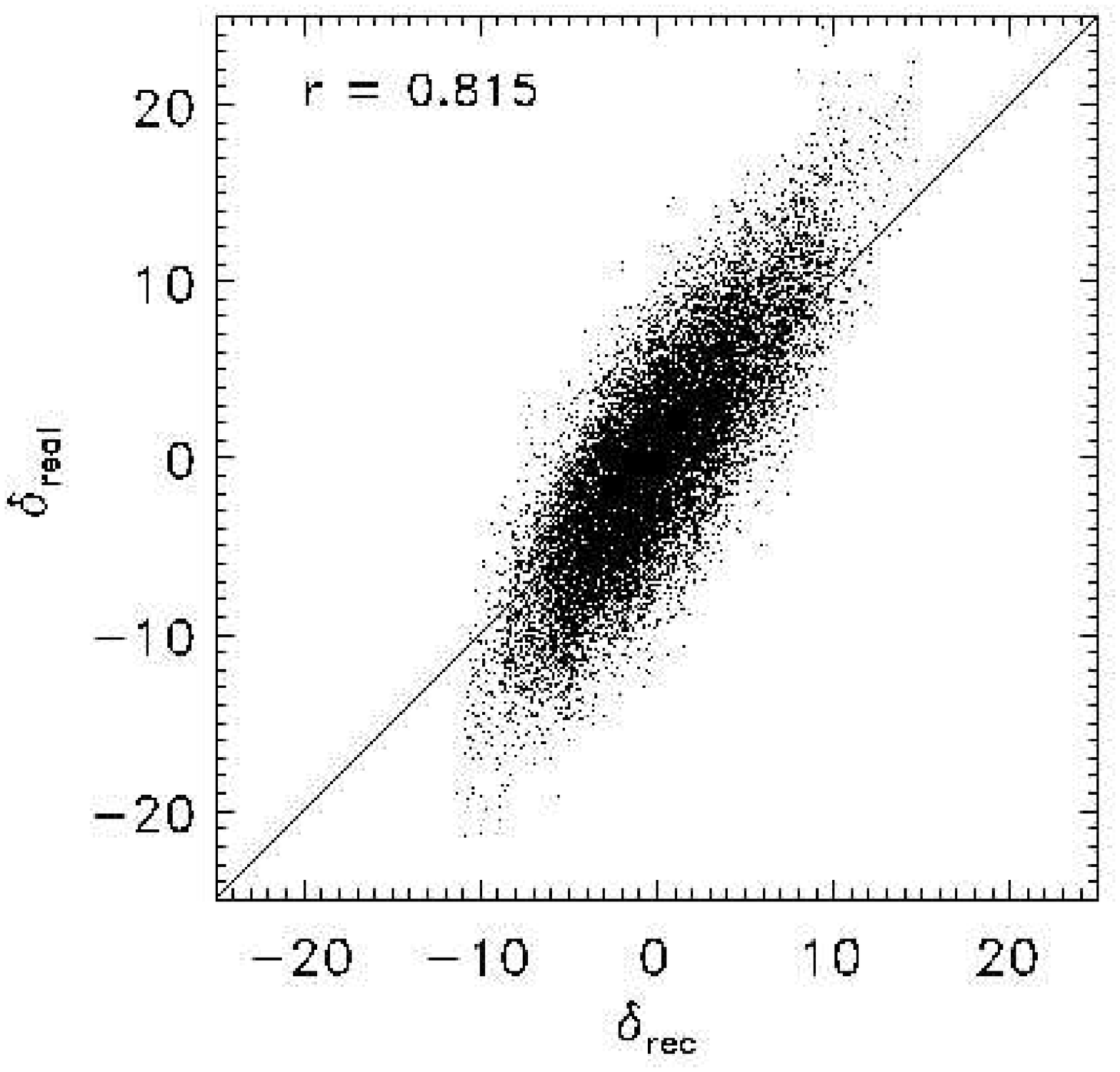}
\put(-180,0.5){{\Large\bf f}}
\end{tabular}
\hspace{0.5cm}

\caption{{\bf Blurring treatment:} Here the signal (panel {\bf a}) was
  convolved with a gaussian modeling blurring effects, as shown in panel {\bf
    b}. Some low noise with a structure function was added. Panel {\bf c} shows
  the deblurred result. Panel {\bf d} takes only the noise into account. We see in panel {\bf f} the correlation between the input signal and the {\it true} signal, because the noise is negligible. The correlation coefficient is thus very high, however, the alignment of the pixels in the plot is not correct. Overdensities and underdensities tend to be underestimated, which is consistent with the blurring effect. The reconstruction given in panel {\bf e} corrects this effect and consequently a higher correlation coefficient is achieved.}
\label{fig:2DBLUR}
\end{figure*}

\begin{figure*}
\begin{tabular}{cc}
\put(40.0,0.5){{\Large\bf a}}
\hspace{0.cm}
\includegraphics[width=7cm]{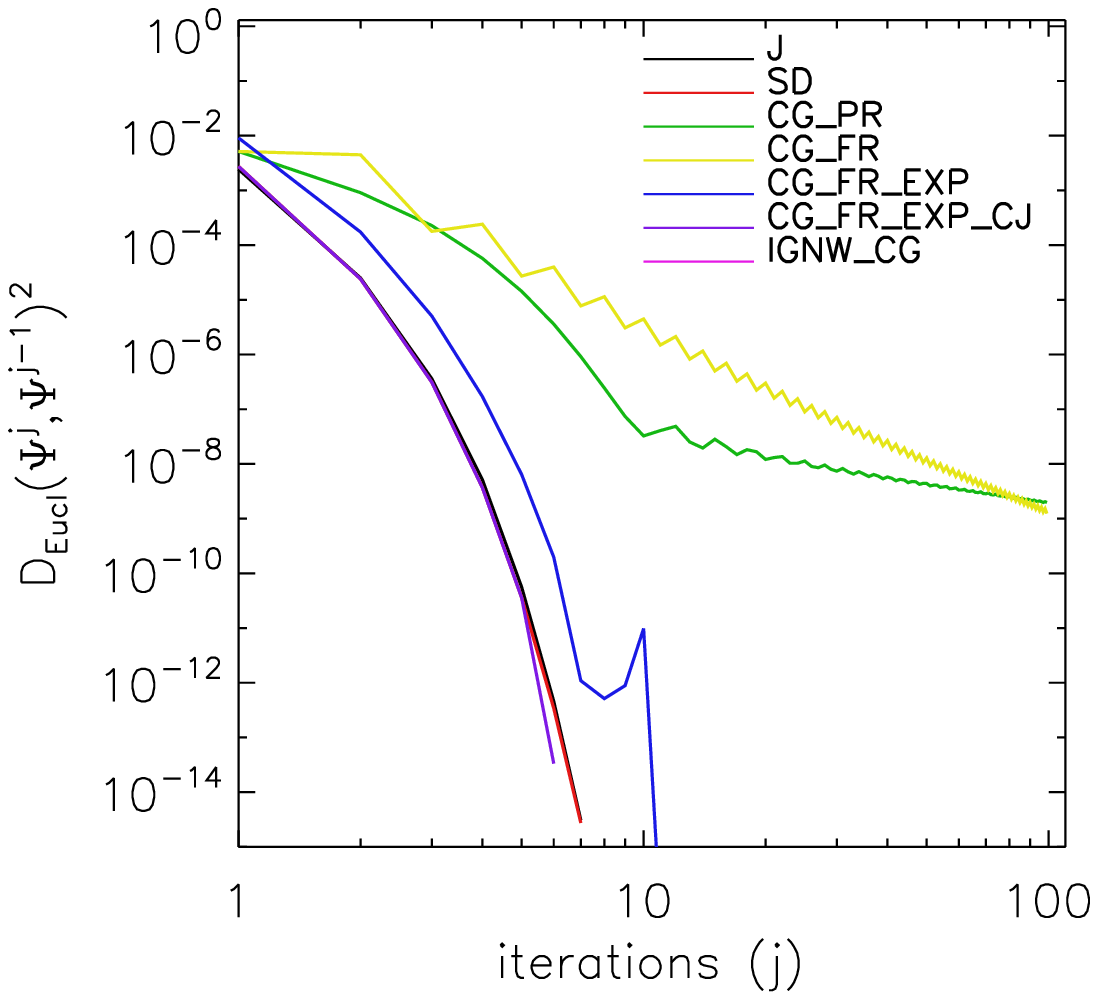}
\put(40,0.5){{\Large\bf b}}
\hspace{0.5cm}
\includegraphics[width=7cm]{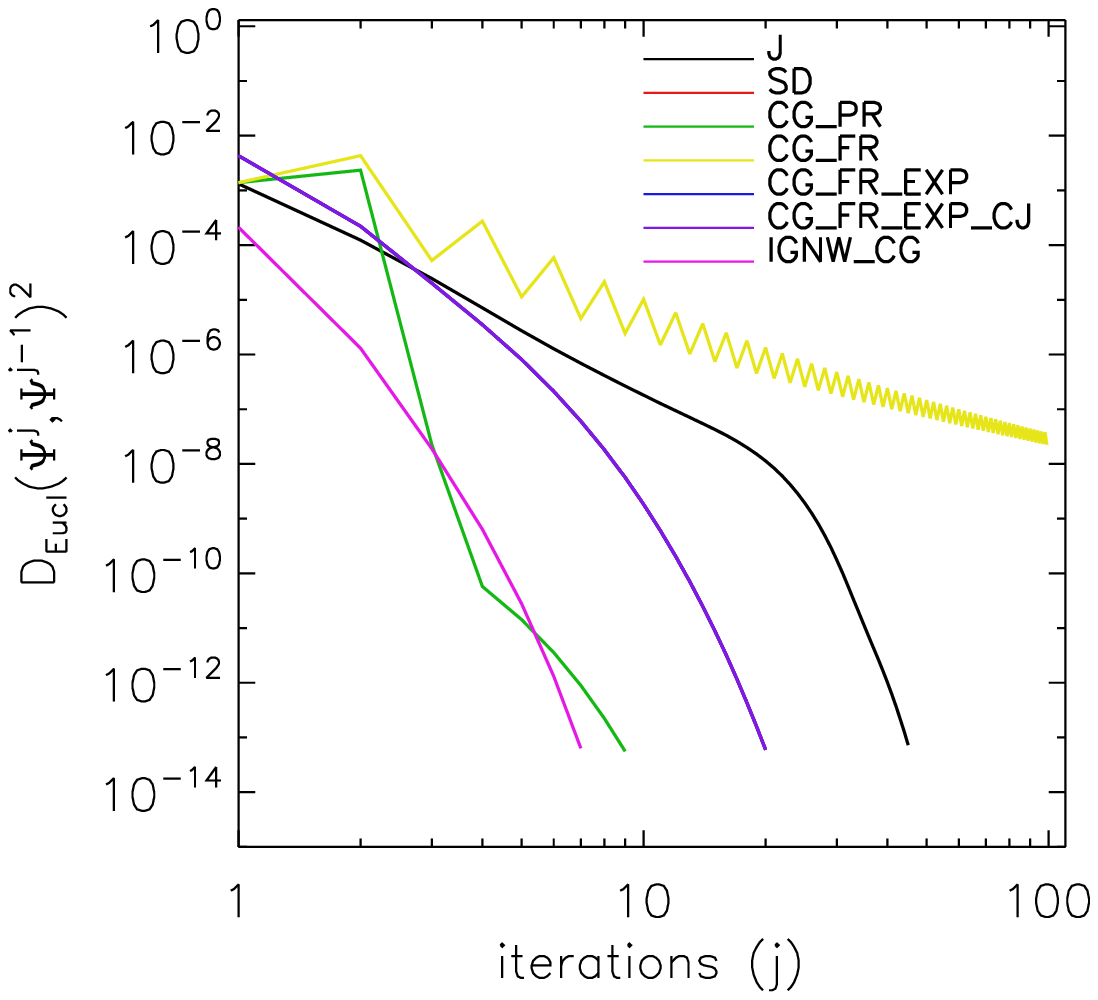}\\
\put(40.0,0.5){{\Large\bf c}}
\hspace{0.cm}
\includegraphics[width=7cm]{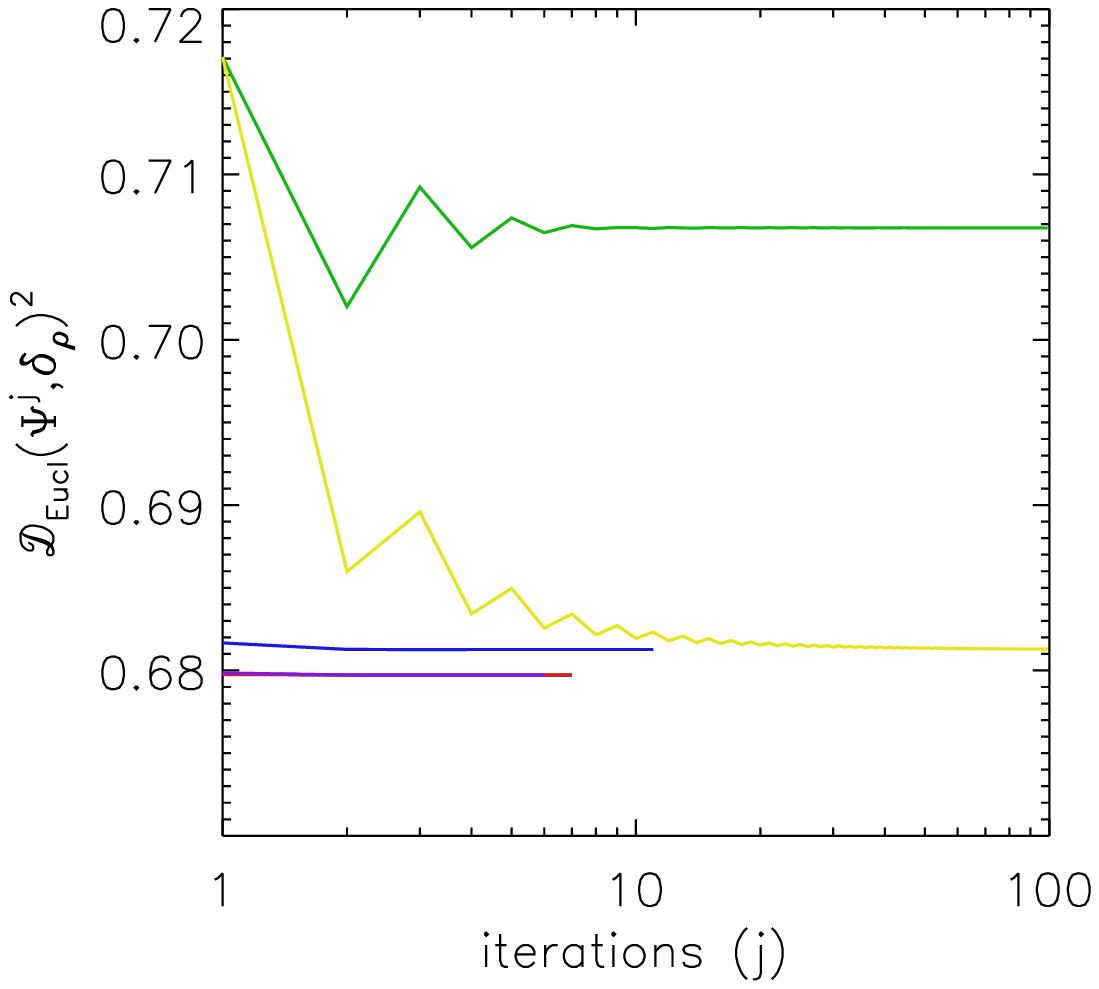}
\put(40,0.5){{\Large\bf d}}
\hspace{0.5cm}
\includegraphics[width=7cm]{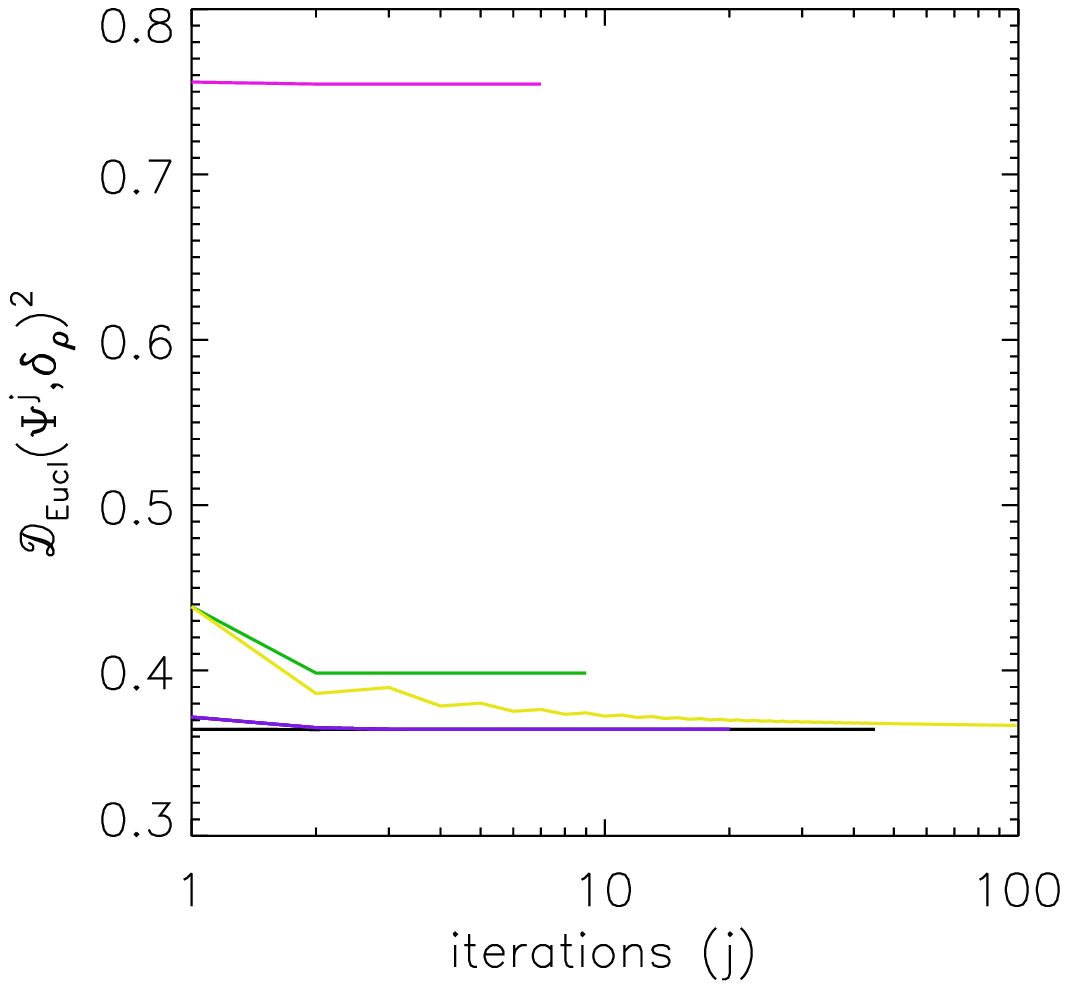}\\
\put(40.0,0.5){{\Large\bf e}}
\hspace{0.cm}
\includegraphics[width=7cm]{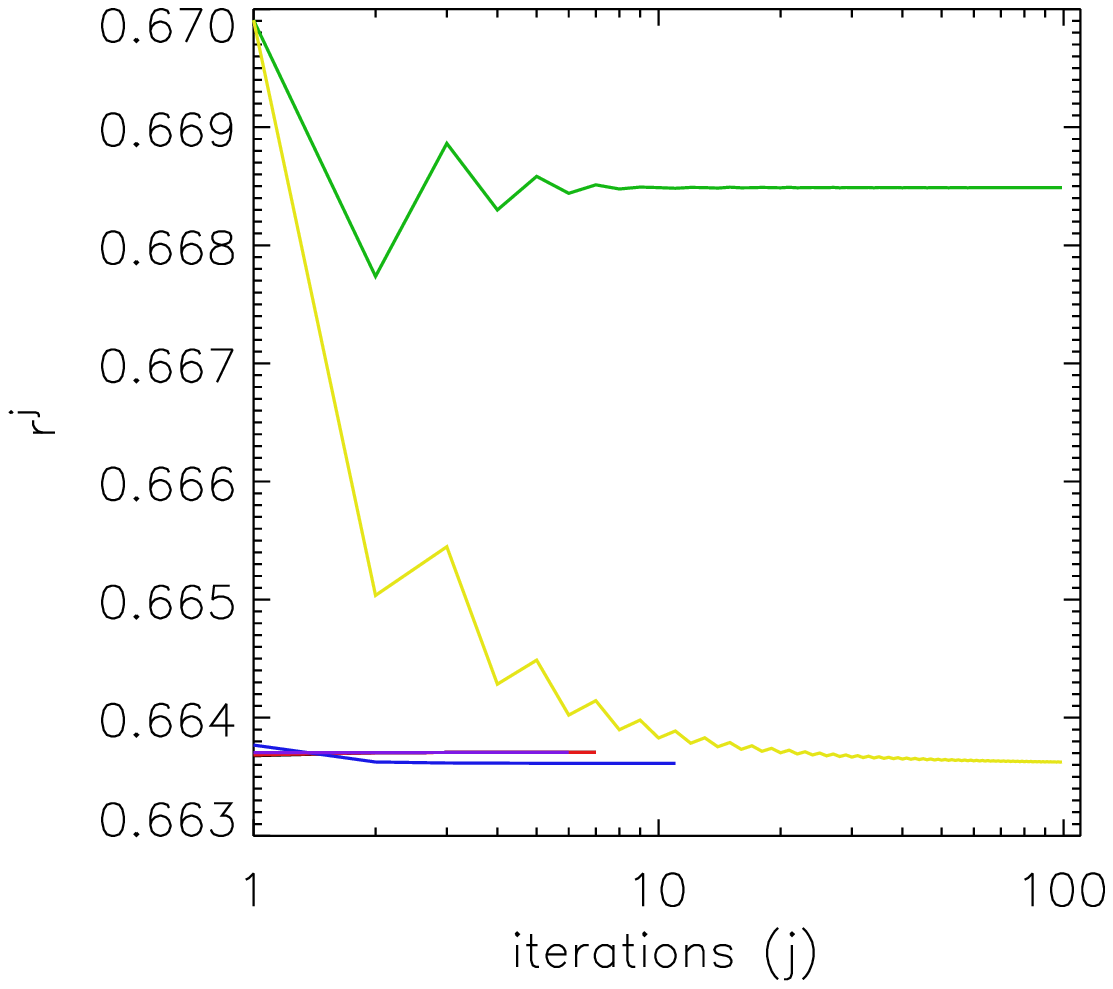}
\put(40,0.5){{\Large\bf f}}
\hspace{0.5cm}
\includegraphics[width=7cm]{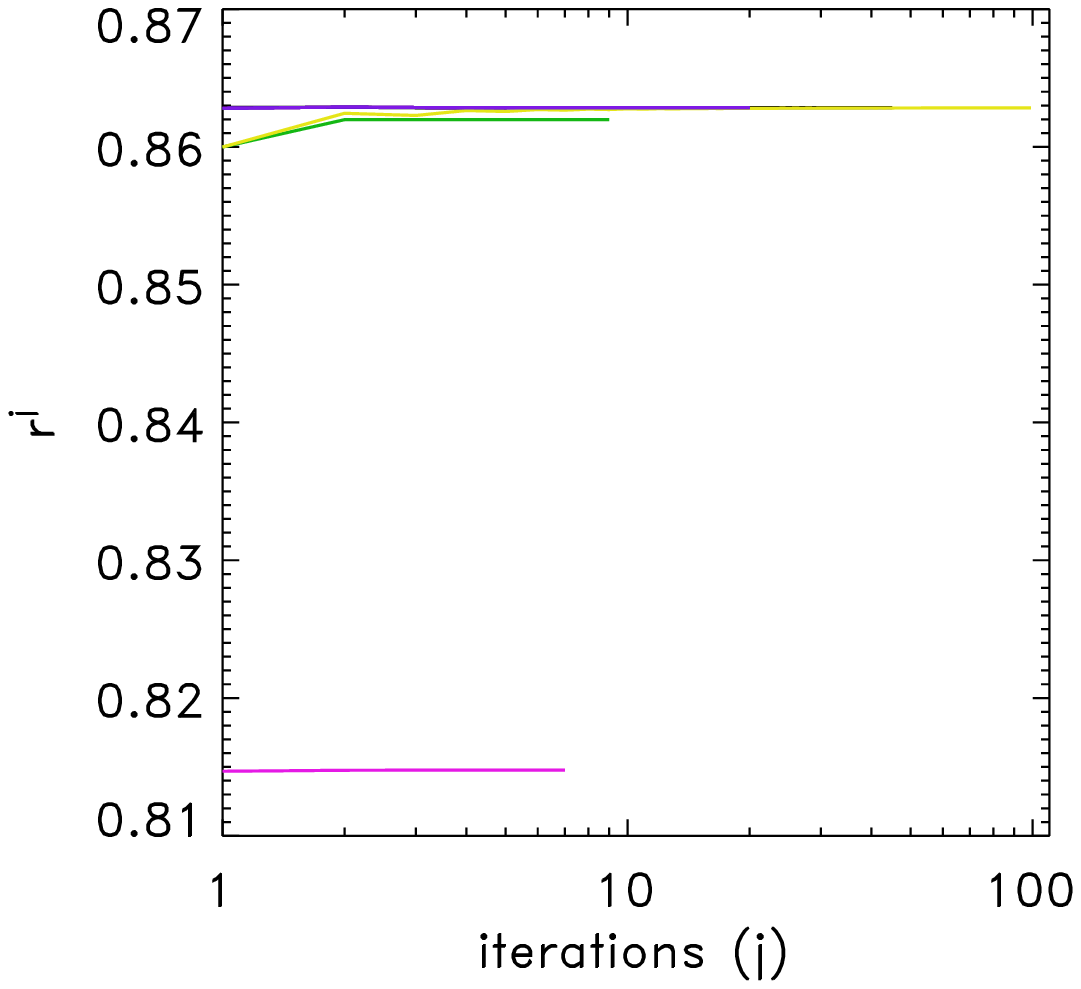}
\end{tabular}
\caption{ {\bf Poissonian noise and numerical performance (panels {\bf a}, {\bf
      c}, {\bf e}):} Here the convergence behaviour and quality of the
  reconstruction is comparable for the J, SD, EXP methods. The FR and PR
  schemes do not present a fast convergence (panel {\bf a}). Nevertheless, the
  FR scheme (yellow curve) seems to lead to the correct solution (panels {\bf
    c} and {\bf e}). The PR formula, on the contrary, stagnates at
  reconstructions that have much lower quality compared to the rest of the schemes.
 {\bf Blurring treatment and numerical performance (panels {\bf b}, {\bf d},
  {\bf f}):} In this study case, the EXP algorithm seems to work better than
  the rest of the schemes. Although the PR formula converges very rapidly
  (green curve in panel {\bf b}), it leads to a lower quality reconstruction
  (panels {\bf d} and {\bf f}). The FR scheme converges to the same solution as
  the J, SD, and EXP algorithms, however, with a slower convergence (yellow
  curve in panel {\bf b}). The J and SD methods have an overall good behaviour
  in this case, but still converge significantly slower than the EXP scheme
  (their convergence is identical black and red curves are overplotted).  The
  reconstruction considering just the noise is very poor, because the noise is negligible in this case (pink curves).
 }
\label{fig:2DBPSTAT}
\end{figure*}

\begin{figure*}
\begin{tabular}{cc}
\includegraphics[width=7cm]{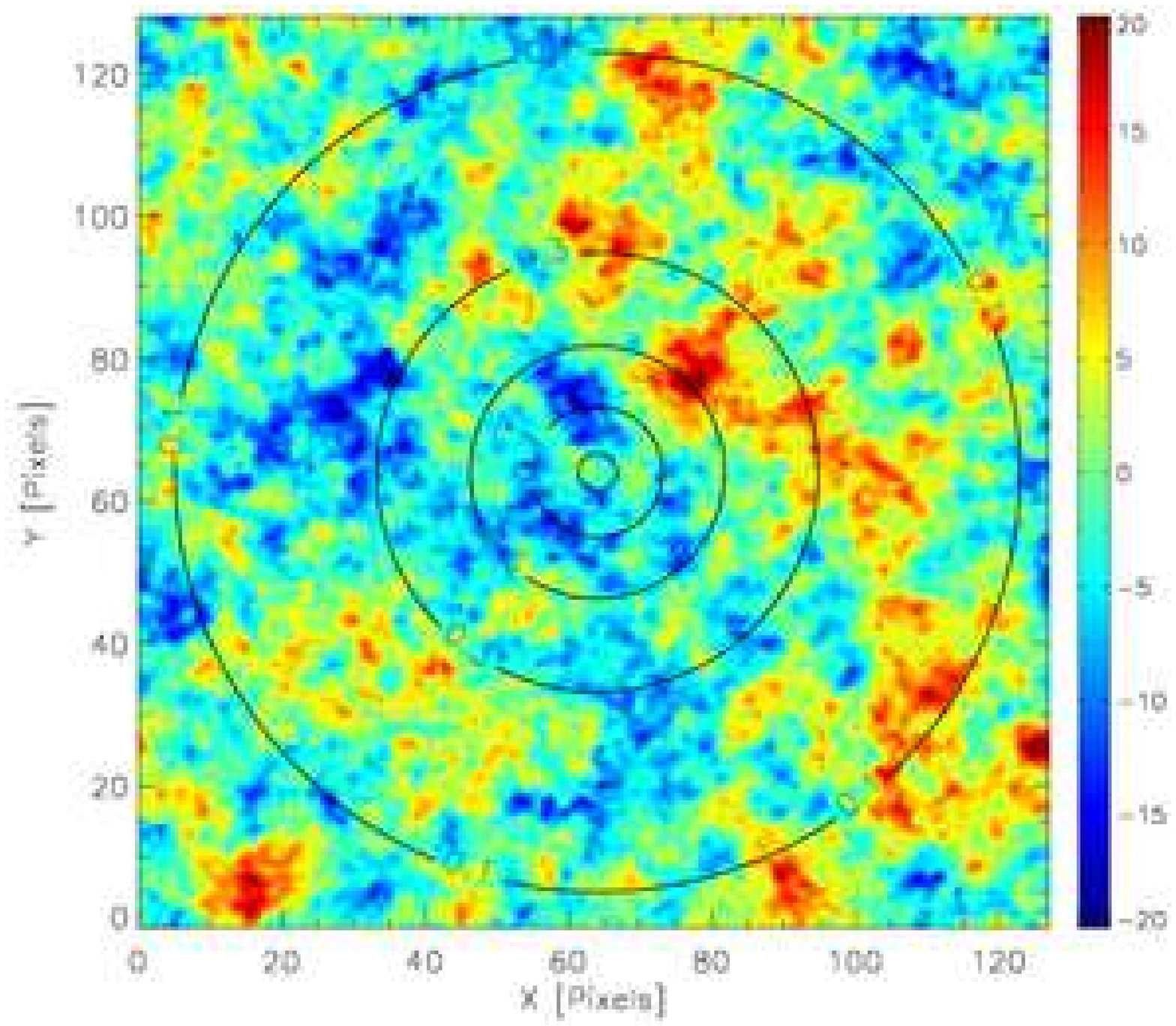}
\put(-190,0.5){{\Large\bf a}}
\hspace{0.5cm}
\includegraphics[width=7cm]{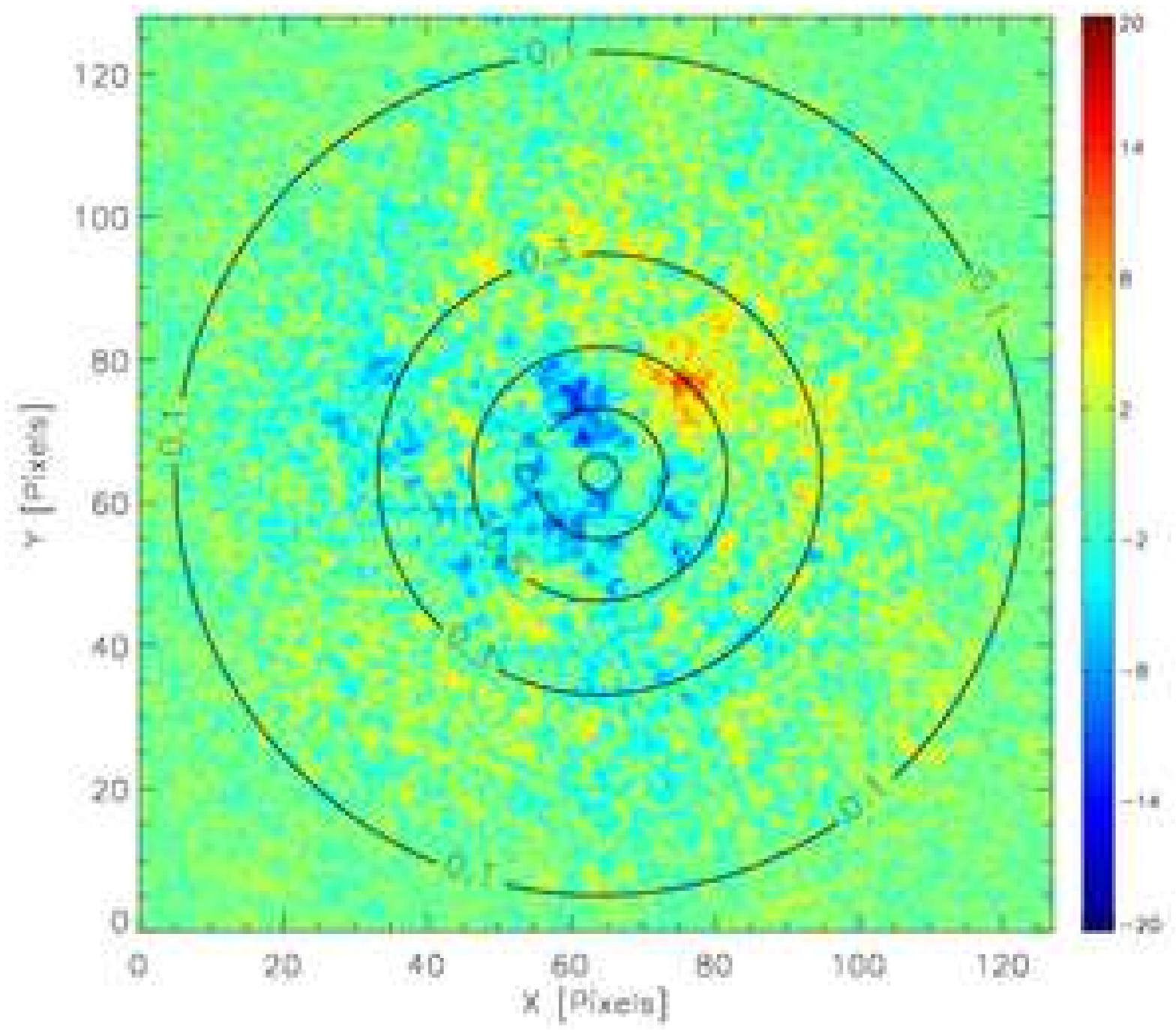}
\put(-190,0.5){{\Large\bf b}}
\\
\includegraphics[width=7cm]{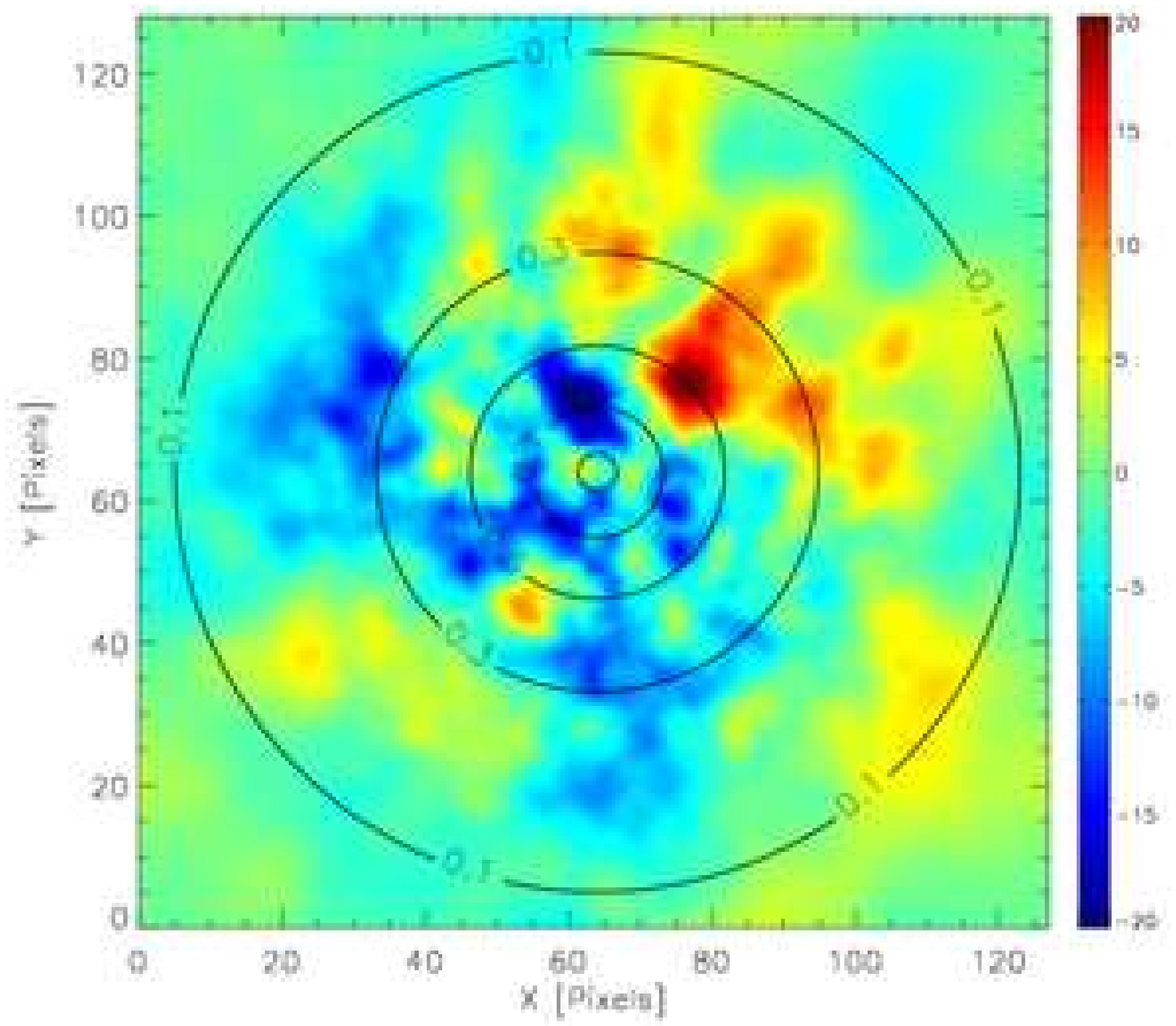}
\put(-190,0.5){{\Large\bf c}}
\hspace{0.5cm}
\includegraphics[width=7cm]{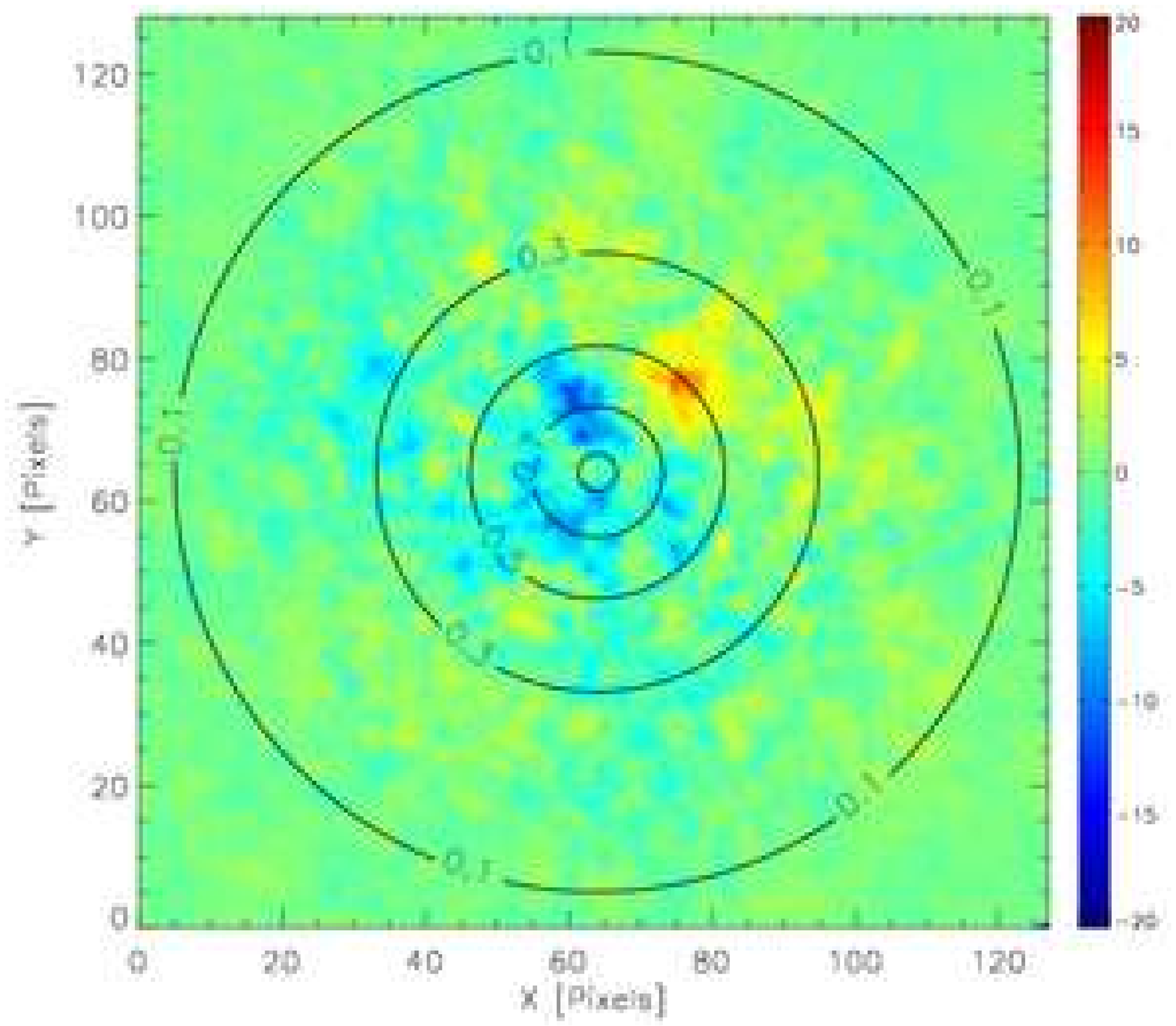}
\put(-190,0.5){{\Large\bf d}}
\end{tabular}

\begin{tabular}{ll}
\hspace{-.5cm}
\includegraphics[width=7cm]{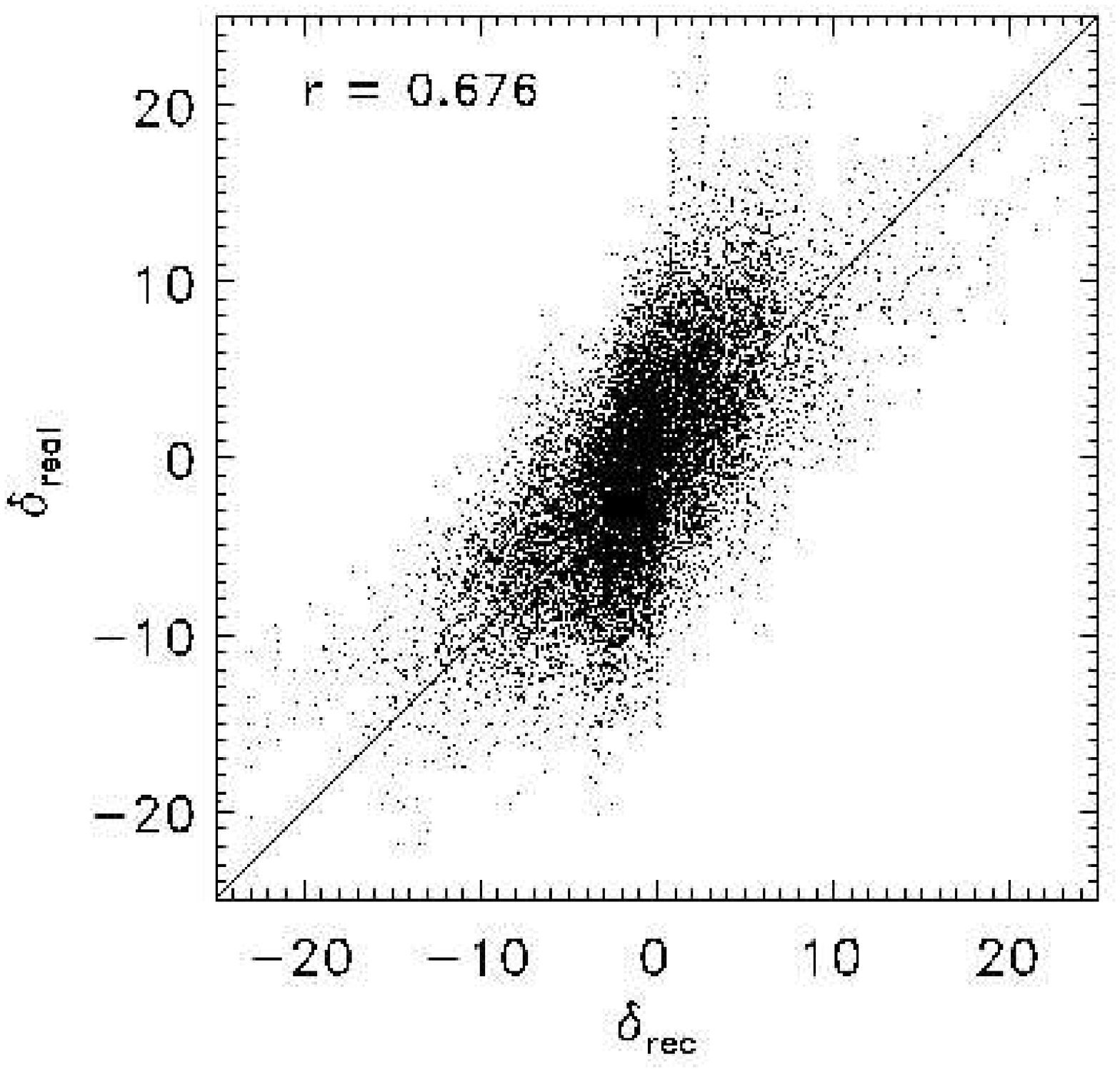}
\put(-180,0.5){{\Large\bf e}}
\hspace{0.5cm}
\includegraphics[width=7cm]{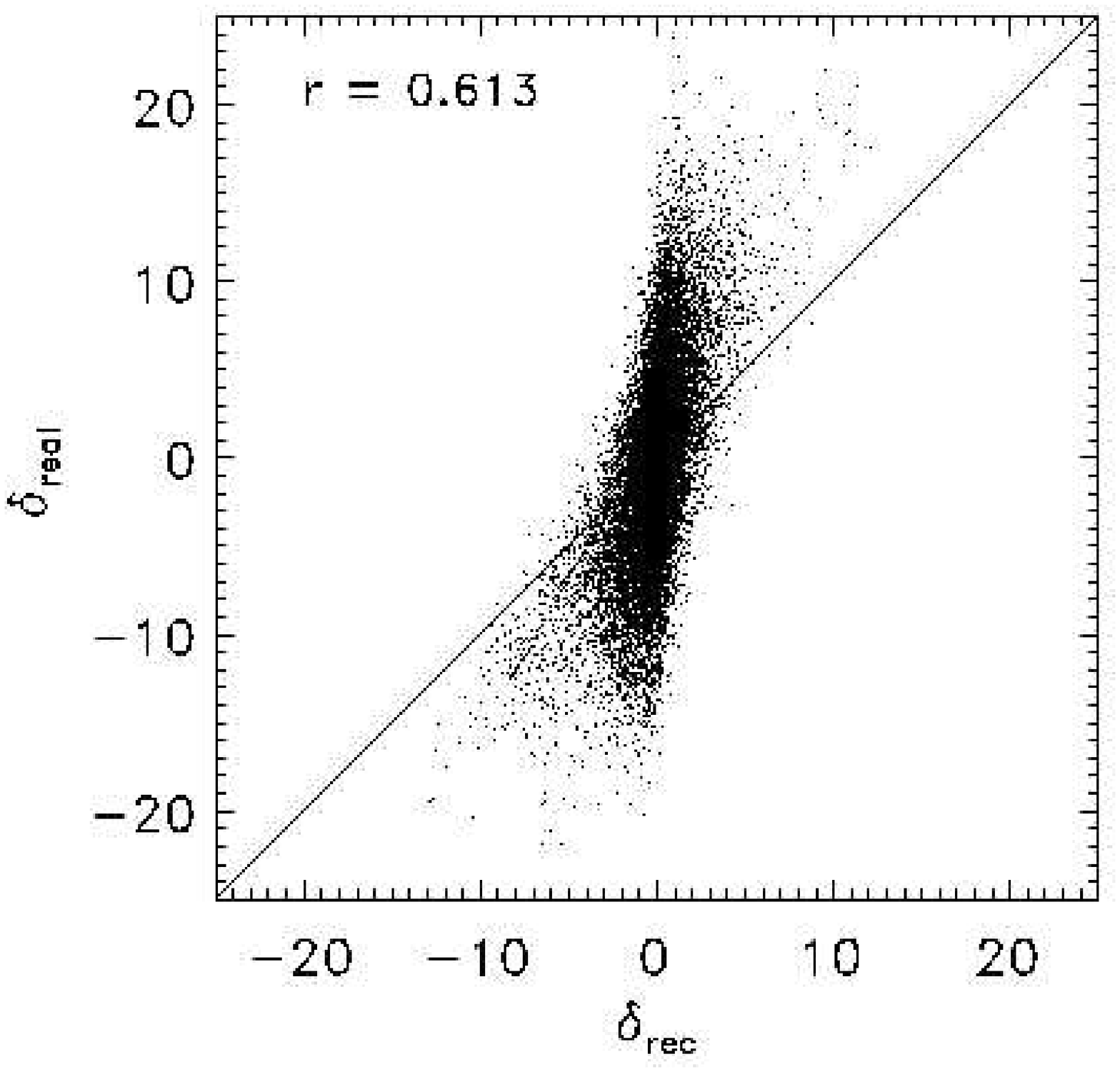}
\put(-180,0.5){{\Large\bf f}}
\end{tabular}

\hspace{0.5cm}
\caption{{\bf Selection function treatment:} Here selection function effects
  were simulated with a function that takes values between zero and one,
  decreasing exponentially in radial direction. The contours show different
  values of this function.  Panel {\bf a} shows the real density field. Panel
  {\bf b} shows the input data, where the true signal was multiplied in real
  space with the selection function and a radially increasing noise was
  added. The reconstruction and its correlation with the true signal are
  represented in panel {\bf c} and {\bf e}, respectively. The reconstruction
  ignoring selection effects by taking only the noise into account leads to
  panels {\bf d} and {\bf f}.  The reconstruction given in panel {\bf d} is
  very conservative and smooths the overdensities out due to noise
  supression. This leads to a high correlation coefficient, though the
  individual pixels are clearly not correctly aligned (panel {\bf f}). Panel
  {\bf c}, on the contrary, shows more structures that are enhanced due to
  consideration of the selection function effects. This correctly distributes
  the pixels, as can be seen in panel {\bf e}. The correlation coefficient
  seems to be significantly better than in panel {\bf f}, however, a better measure of the overall quality of the reconstruction can be seen in next figure.}
\label{fig:2DW4}
\end{figure*}

\begin{figure*}
\begin{tabular}{cc}
\includegraphics[width=7cm]{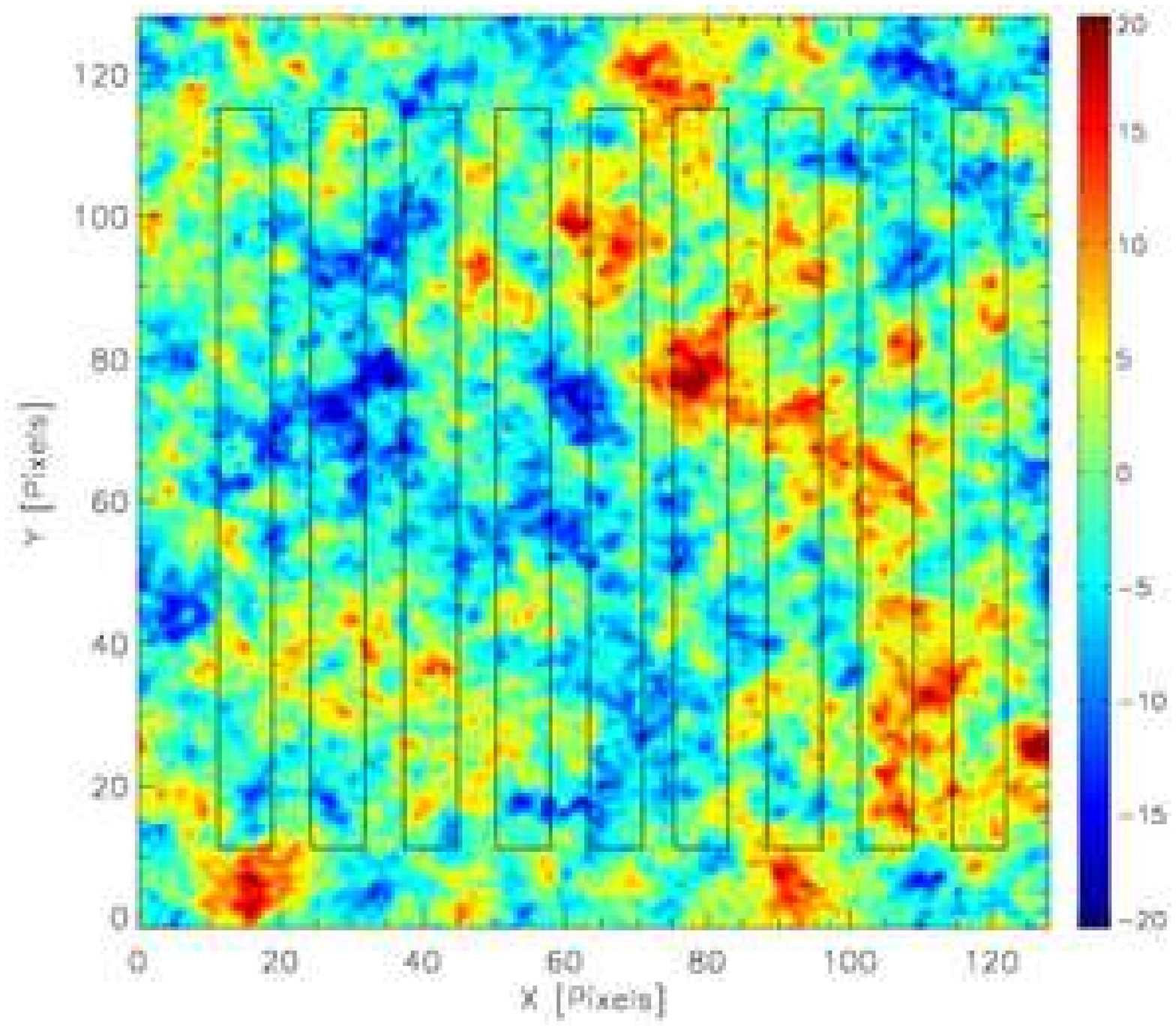}
\put(-190,0.5){{\Large\bf a}}
\hspace{0.5cm}
\includegraphics[width=7cm]{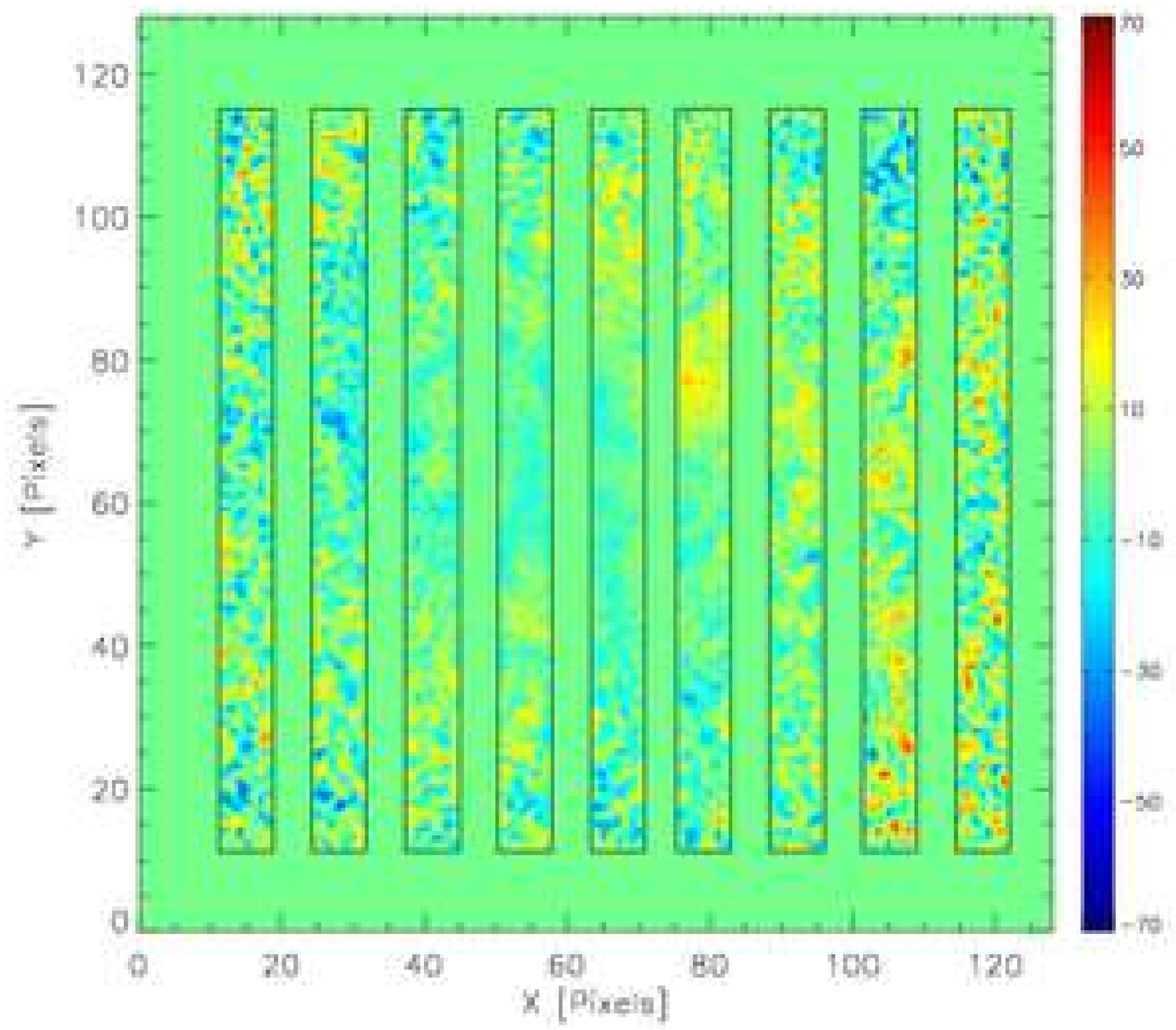}
\put(-190,0.5){{\Large\bf b}}
\\
\includegraphics[width=7cm]{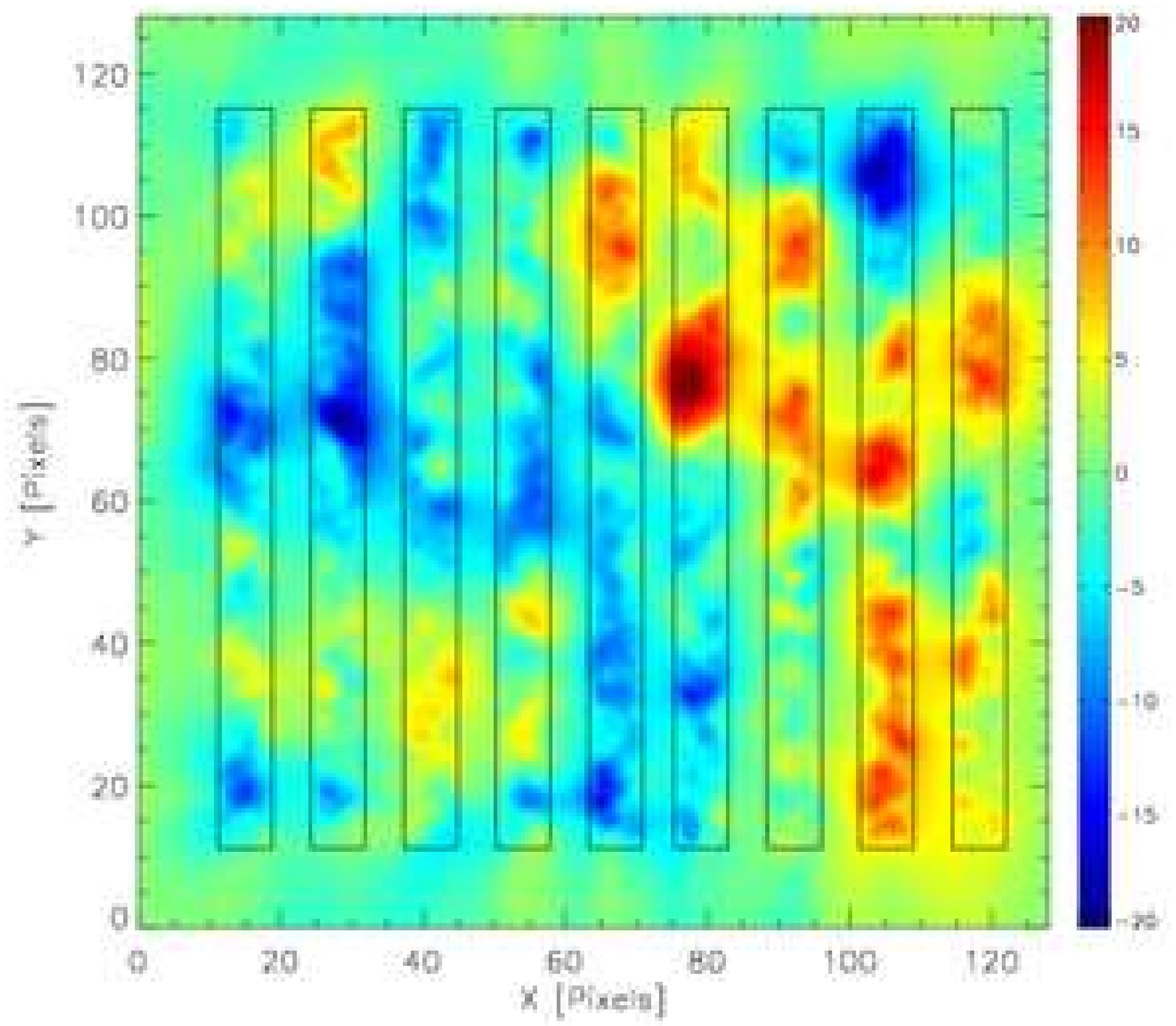}
\put(-190,0.5){{\Large\bf c}}
\hspace{0.5cm}
\includegraphics[width=7cm]{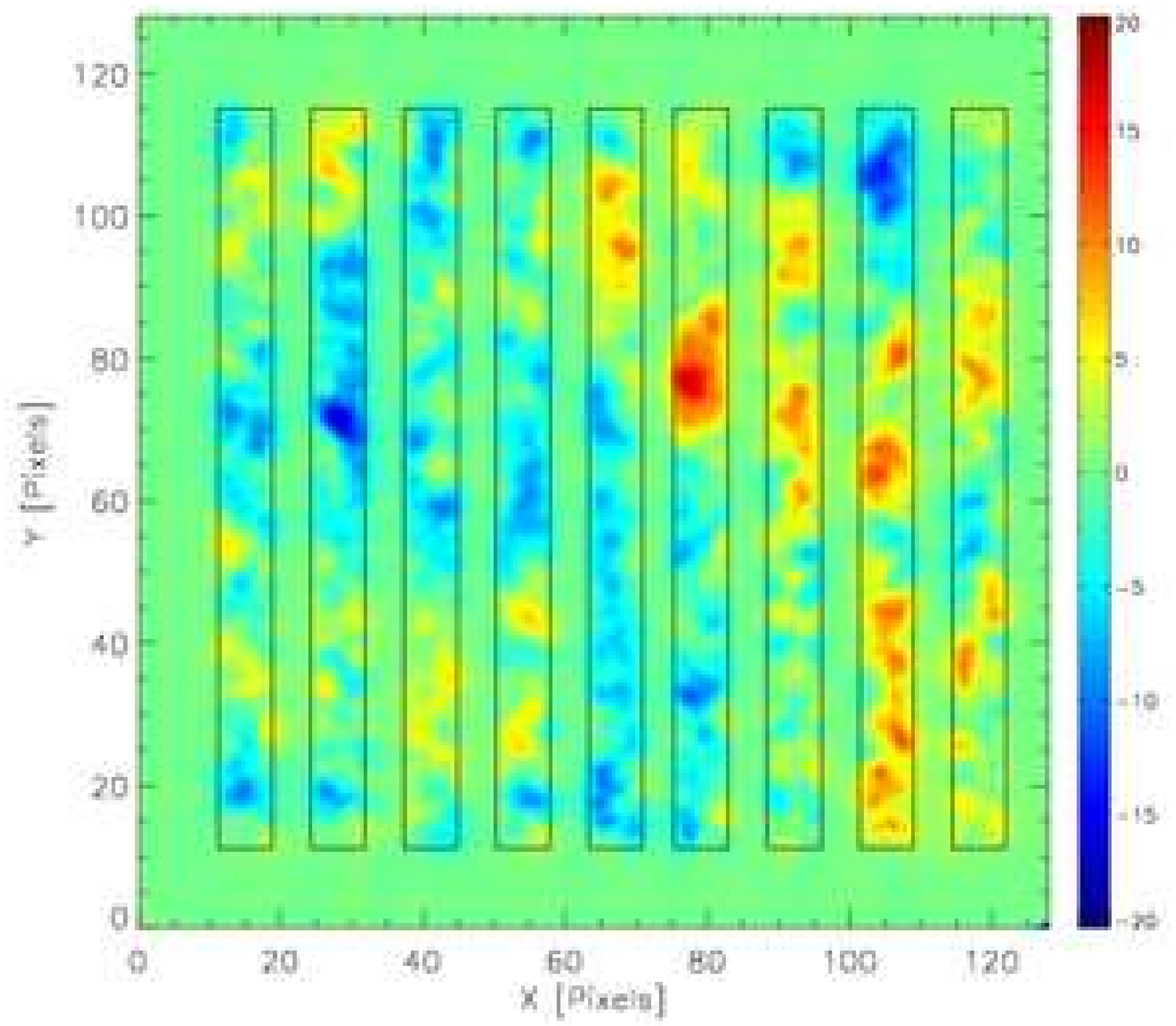}
\put(-190,0.5){{\Large\bf d}}
\end{tabular}

\begin{tabular}{ll}
\hspace{-0.5cm}
\includegraphics[width=7.0cm]{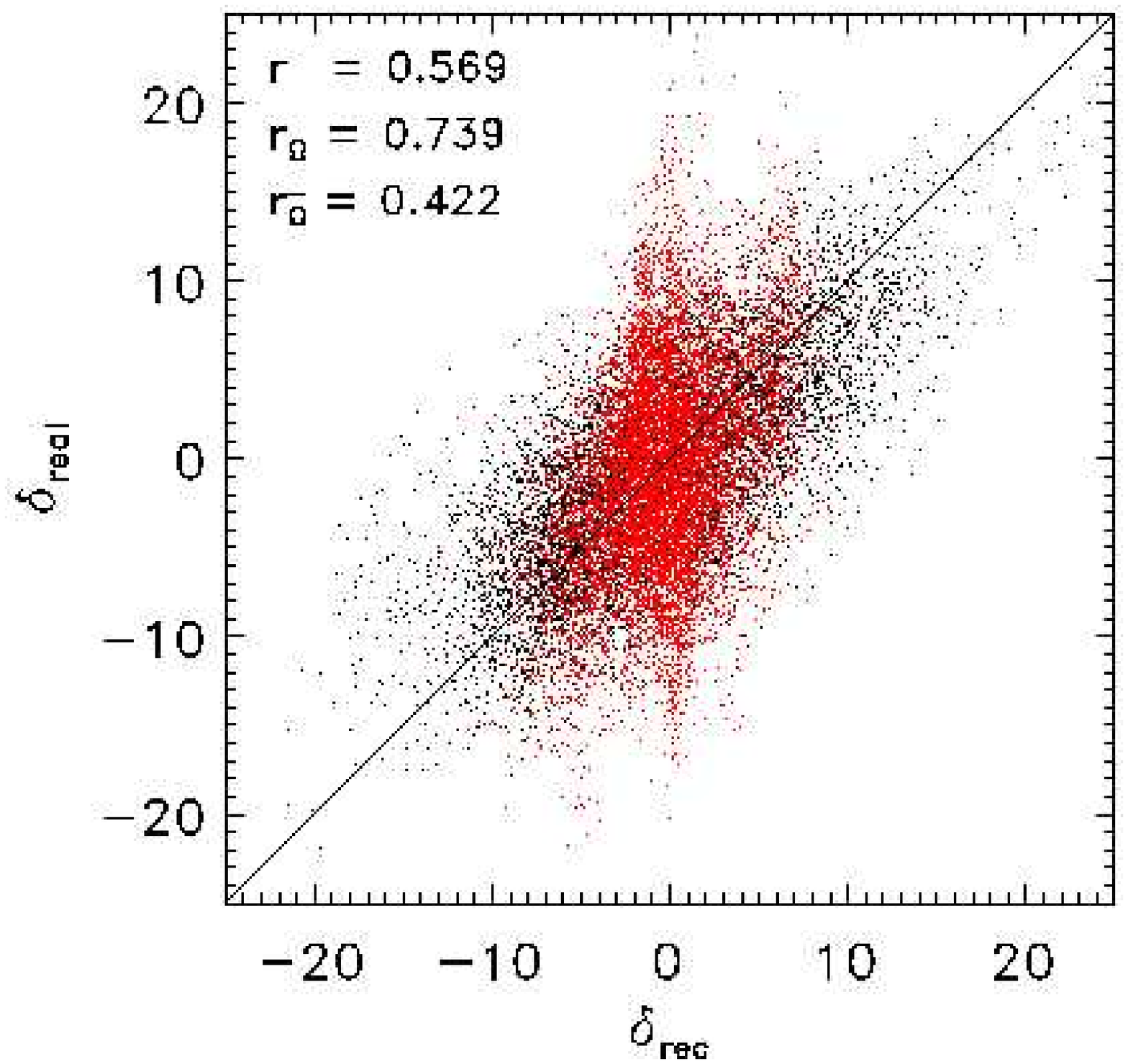}
\put(-180,0.5){{\Large\bf e}}
\hspace{0.5cm}
\includegraphics[width=7.0cm]{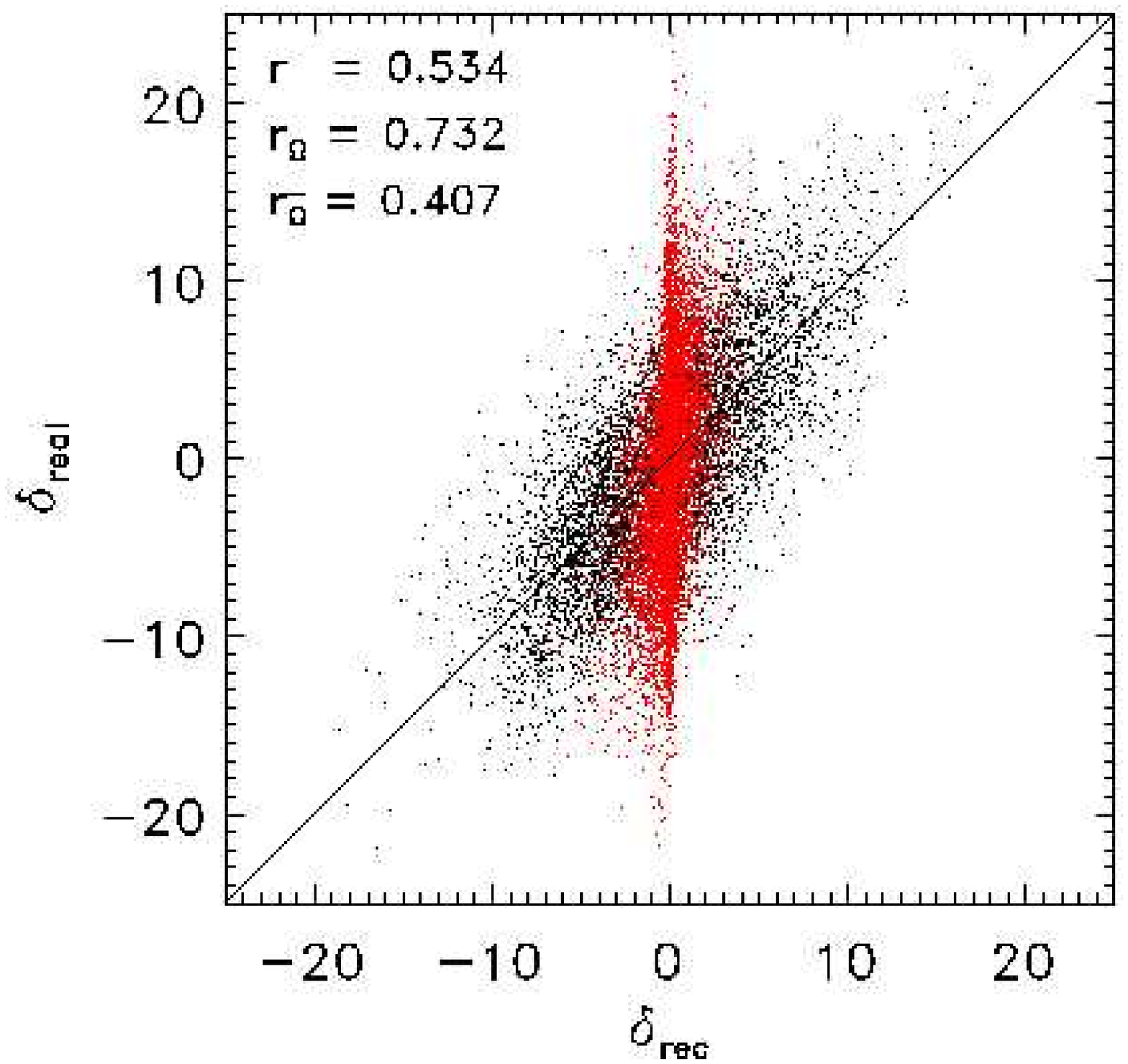}
\put(-180,0.5){{\Large\bf f}}
\end{tabular}

\hspace{1.cm}
\caption{\small{{\bf Windowing treatment:} Here the edge effects are shown in
    two dimensions. The true signal was multiplied by a windowing function that
    is one in the observed region ($\Omega$) and zero in the unknown region
    ($\bar{\Omega}$). The sampled regions are given by the vertical stripes. In addition, a radially
    increasing noise was added (see panel {\bf b}). Panel {\bf c} shows the
    reconstruction handling the edge effects. Panel {\bf d} represents the
    result taking only the noise into account. We see in panel {\bf c} how the information is
    propagated into the unsampled regions leading to a closer resemblance of the real
    signal, whereas the noise is just suppressed in panel {\bf d}. Panels {\bf e} and
    {\bf f} show the correlation coefficients for the whole reconstructed
    region, split into the sampled (black dots) and the unsampled regions
    (red dots). Note that the red dots are strongly aligned around the zero
    value in panel {\bf f}, whereas they are correctly spread in panel {\bf e},
    statistically representing the information propagation process mentioned above.   }}
\label{fig:2DW1}
\end{figure*}

\begin{figure*}
\begin{tabular}{cc}
\put(40.0,0.5){{\Large\bf a}}
\hspace{0.cm}
\includegraphics[width=7cm]{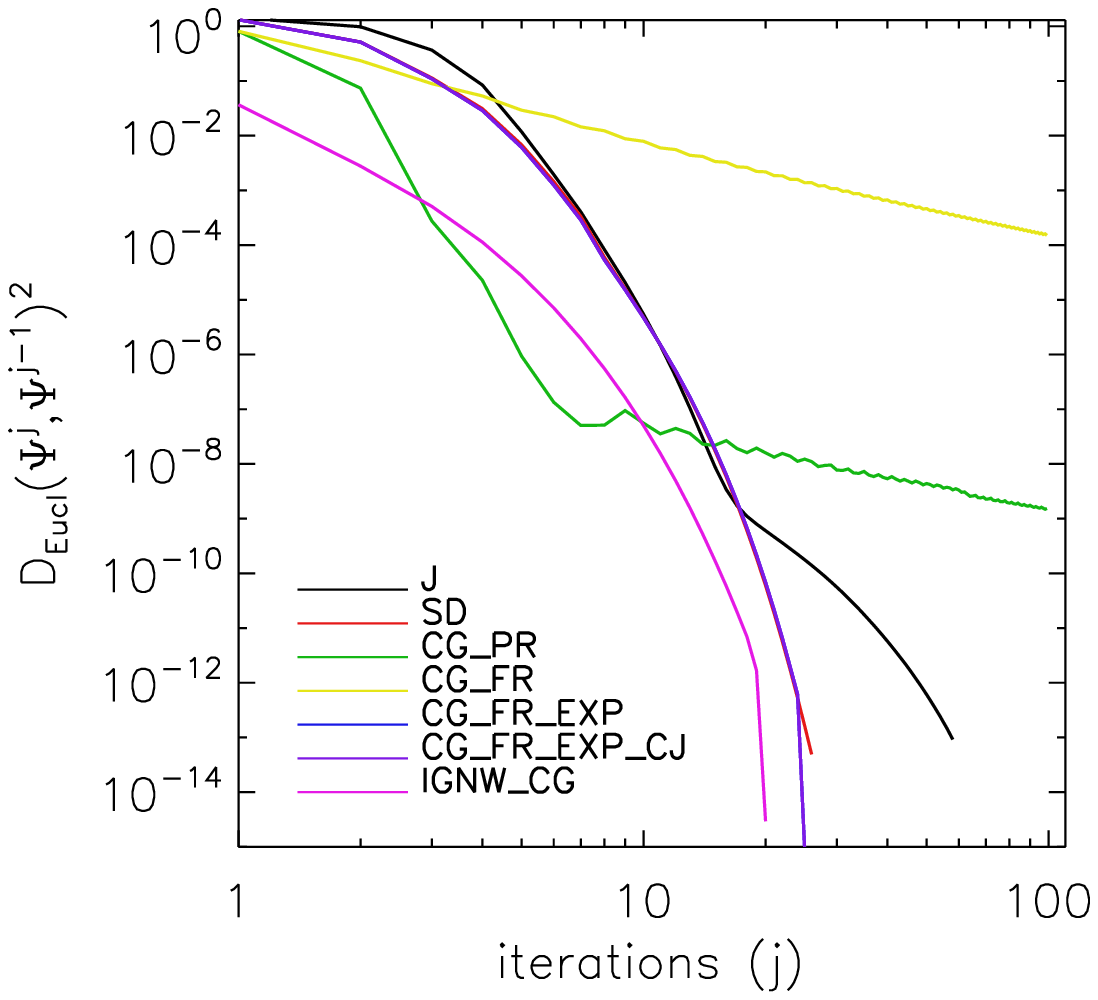}
\put(40,0.5){{\Large\bf b}}
\hspace{0.5cm}
\includegraphics[width=7cm]{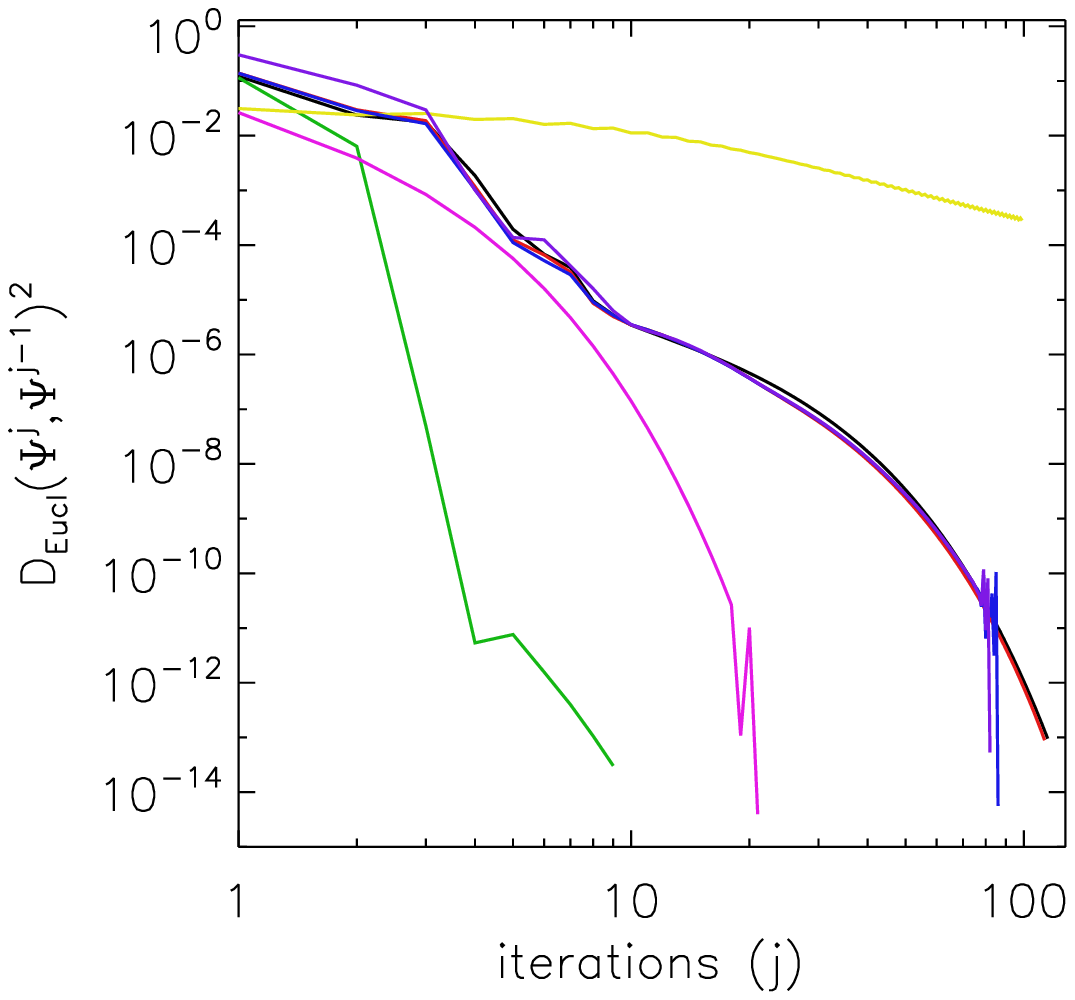}\\
\put(40.0,0.5){{\Large\bf c}}
\hspace{0.cm}
\includegraphics[width=7cm]{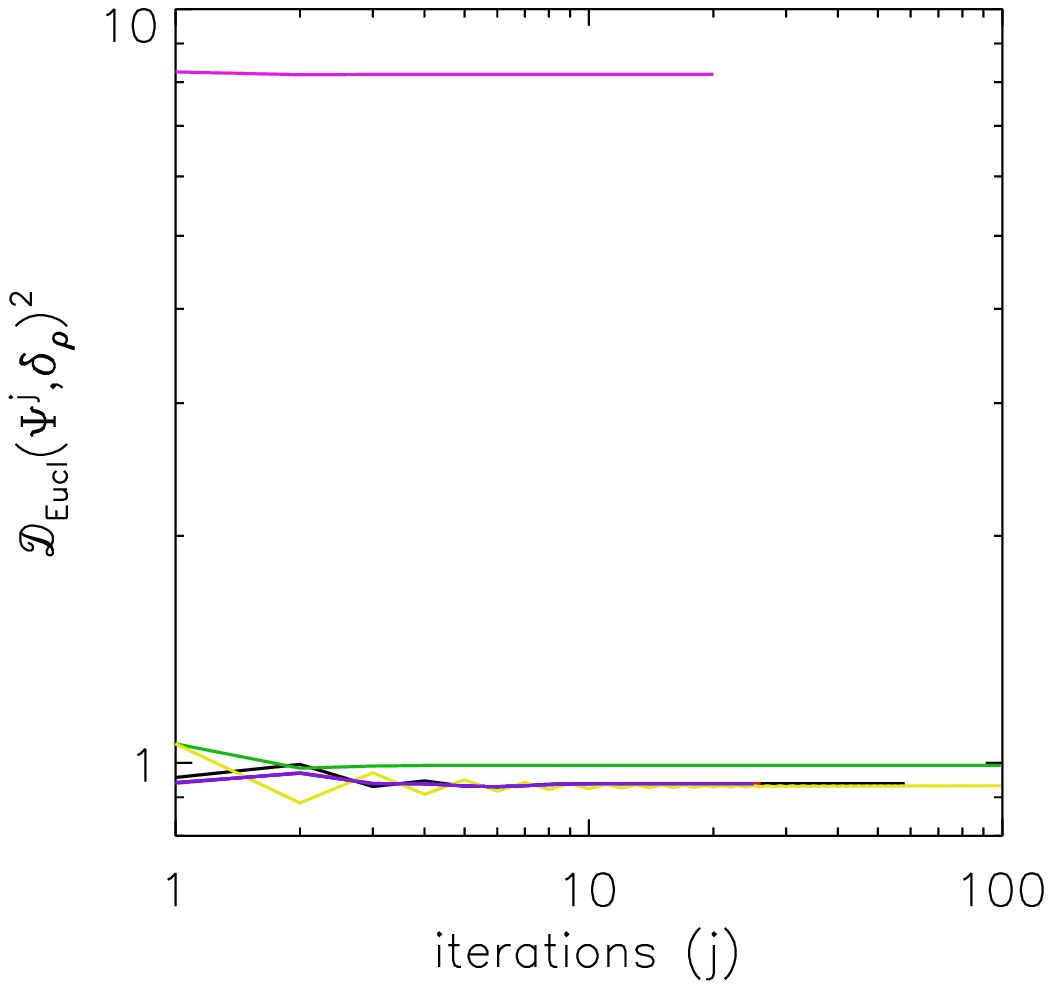}
\put(40,0.5){{\Large\bf d}}
\hspace{0.5cm}
\includegraphics[width=7cm]{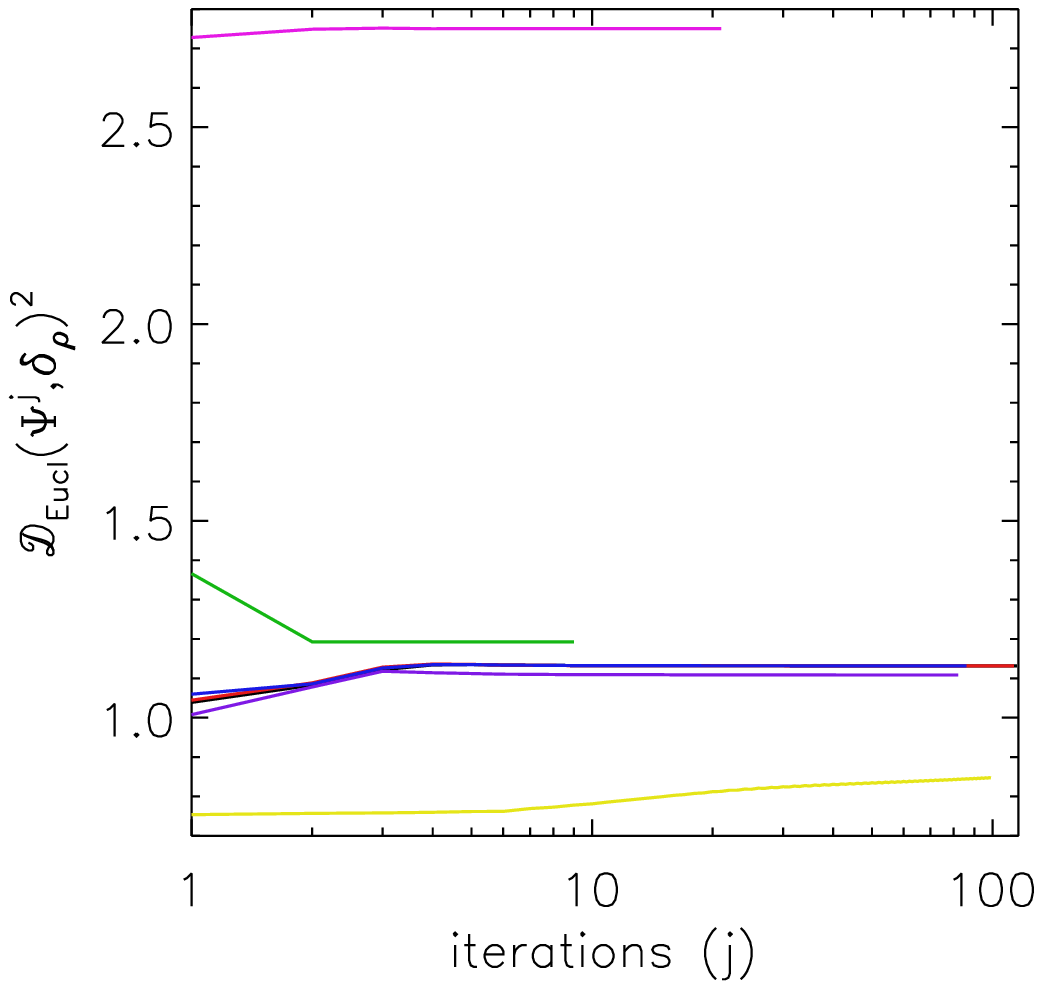}\\
\put(40.0,0.5){{\Large\bf e}}
\hspace{0.cm}
\includegraphics[width=7cm]{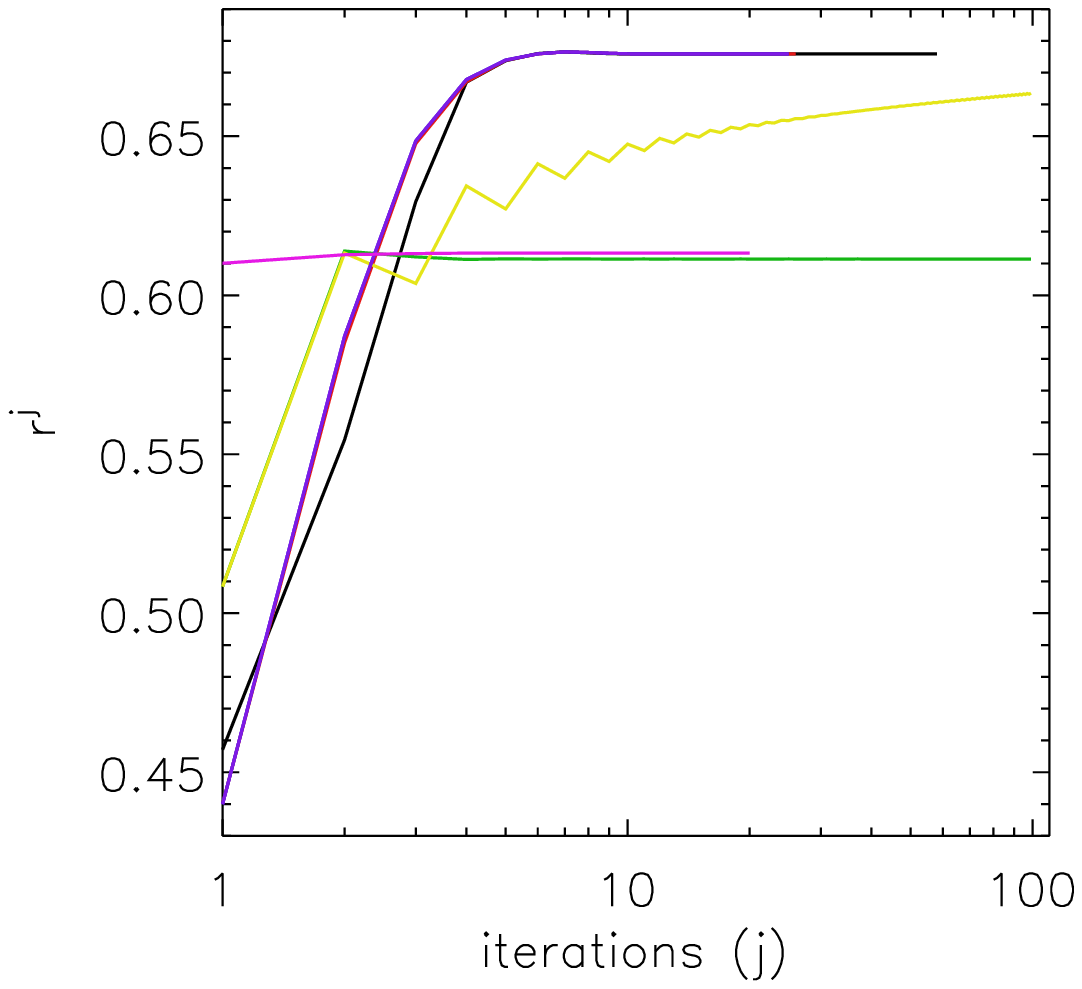}
\put(40,0.5){{\Large\bf f}}
\hspace{0.5cm}
\includegraphics[width=7cm]{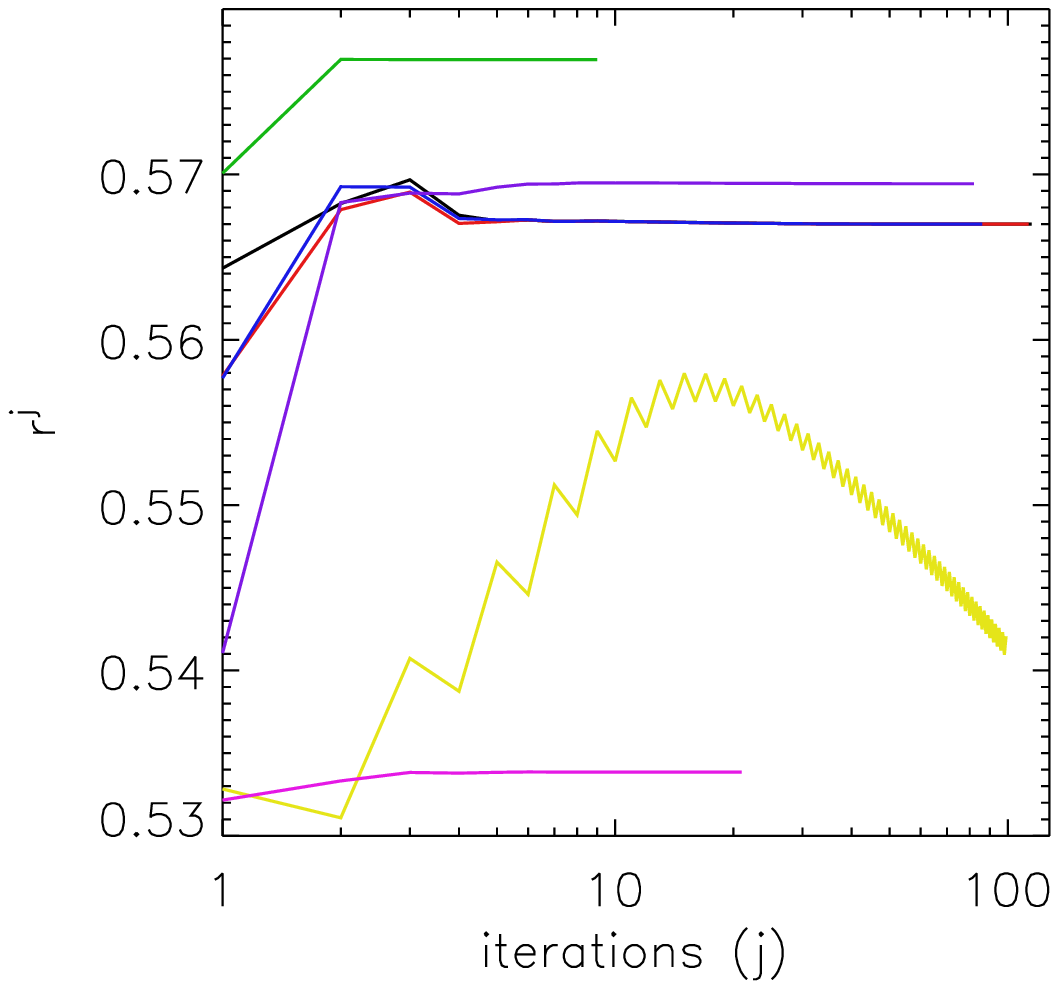}
\end{tabular}
\caption{{\bf Selection function treatment and numerical performance (panels a, c, e):} The same
  color coding is used as in fig.~(\ref{fig:2DNOWSTAT}) panel {\bf a}, except for 
  additional curve (represented in pink) that indicates the reconstruction in which
  the selection effects are ignored. Panel {\bf a} shows the squared Euclidean distance between subsequent reconstructions.  The squared Euclidean distance between the reconstruction and the true density field is plotted in panel {\bf c}, showing a huge difference between the reconstruction which takes only the noise into account and ignores the selection function and the rest of the methods.  Note that the
  statistical correlation r is also much better for the case where the selection
  effects are properly treated  (panel {\bf e}). One concludes from the three
  plots, that the SD and EXP methods (red, blue and violet curves) clearly
  converge faster to a more or equally optimal  solution in comparison with the rest of the methods. The J scheme shows a significantly slower convergence (black curve in panel {\bf a}). The PR algorithm stagnates at poorer reconstructions as can be seen from panel {\bf c} and {\bf e}. 
{\bf Windowing treatment and numerical performance (panels b, d, f):} In this
  case, the PR shows extremely good results: fast convergence (panel {\bf b})
  and a high correlation coeficient (panel {\bf d}). However, the Euclidean
  distance is slightly bigger than for the rest of the methods, except for the
  pink curve (ignoring windowing effects). The FR method is disastrous in this
  study case and diverges from the solution as can be seen in panel {\bf
  f}. The J, SD, and EXP methods show good and stable results. The J and SD algorithms give extremely similar results. Although their convergence behaviour is similar to the EXP schemes, the latter give slightly better results: smaller values for the Euclidean distance and higher values for the correlation coefficient (violet curves in panels {\bf d} and {\bf f}, respectively).}
\label{fig:2DW14STAT}
\end{figure*}

In this numerical experiment we tested the blurring effects by convolving the
density field with a Gaussian. The result is shown in fig.~(\ref{fig:2DBLUR}),
panel {\bf b}. We see how the small structures are smoothed out and only the
larger ones prevail. Some noise with a structure function was added to the
signal. However, the noise was kept low with the aim of investigating primarily
the blurring effect. The results of the reconstruction that considers only the noise does not change much with respect to the input signal, as can be expected. However, the extra-regularized Wiener-filtering deblurs the image applying eqs.~(\ref{eq:op133}) and (\ref{eq:op16}), and yields the figure shown in panel {\bf c}. 
We see how much of the small scale structure is restored and the peaks become enhanced. The correlation between this reconstruction and the original signal (panel {\bf e}) is significantly better than for the case where the blurring is ignored (panel {\bf f}). We can see in fig.~(\ref{fig:2DBPSTAT}) that the deconvolution algorithm is very fast for all the methods except for the FR-scheme.
The PR-method is the fastest, but it leads to slightly worse results (see the green curve in  panels {\bf c} and {\bf e}).
The EXP turns out to be more efficient than  the J and SD methods in this case. 

\subsubsection{Selection function effects}
\label{sec:selecfunc}

For this case we use a modified data model in which the selection function also affects the noise 
\begin{equation}
d=f_{\rm S}\cdot(s+f_{\rm SF}\cdot\epsilon_{\rm WN}){,}
\end{equation}
with $f_{\rm S}\in[0,1]$, simulating the fading strength of the signal with
increasing distance. The results are plotted in fig.~(\ref{fig:2DW4}), where
the structure of the signal can be seen to become indistinguishable in radial
direction (see panel {\bf b}). Taking only the noise into account leads to very
poor reconstructions (see panel {\bf d}). On the contrary, by also considering  the selection function effects, the structures are resolved even at contours where only 10 \% of the signal plus noise is left (see panel {\bf c}).
As can be appreciated in panels {\bf e} and {\bf f} there is an improvement in
the correlation between the {\it real} density field and the reconstructed
signal. Panel {\bf e} shows a higher correlation coefficient, but the quality
enhancement of the reconstruction can be seen better in the distribution of the
density values for each pixel. How the points are correctly spread along the diagonal line can be verified there. The longer Euclidean distance to the {\it real} density field shows the quantitative difference very clearly, by just comparing the pink curve with the rest (fig.~\ref{fig:2DW14STAT} and panel {\bf c}).   
It is worth mentioning that although the PR test seems to give a comparable
result to the calculation that ignores the selection function. The final
correlation coefficient in panel {\bf e} shows that the reconstructions actually strongly differ and panel {\bf c} shows that the quality of the recovered signal is notably better for the former experiment.   

In addition, we tested the same selection function affecting only the underlying signal with a model given by
\begin{equation}
d=f_{\rm S}\cdot s+f_{\rm SF}\cdot\epsilon_{\rm WN}{,}
\end{equation}
and obtained the same qualitative results.

\subsection{Windowing effects}
\label{sec:wind}

In this section we investigate the mask effects that introduce coupling between
different modes in Fourier-space so that the data covariance matrix is no longer diagonal. 
The input signal is given in panel {\bf b} of fig.~(\ref{fig:2DW1}). The noisy signal from panel {\bf b} in fig.~(\ref{fig:2DNOW}) was cut in stripes to simulate {\it observed} regions.
We compare two reconstructions here, the first one ignores windowing effects
given in panel {\bf d} and a second reconstruction employs the proper treatment
of the boundary through $f_{\rm M}$ in the algorithm (see eqs.~\ref{eq:op133}
and \ref{eq:op16}). The statistical correlation is given in panels {\bf e} and
{\bf f}, respectively. Our experiments show better results not only for the
latter reconstruction in the un-sampled region ($\overline\Omega$), represented
by the red dots in panels {\bf e} and {\bf f} in fig.~(\ref{fig:2DW1}), but
also in the sampled regions ($\Omega$). The global correlation $r$ is
significantly improved. Whereas the distribution of the black dots, the values
of the densities in the {\it observed} regions, does not apparently change, the
distribution of the un-sampled red dots clearly does. These are distributed
around the zero value for the case where windowing is ignored because a zero
signal is assumed by \textsc{argo} in the $\overline{\Omega}$ region. In contrast we see that the red dots are distributed along the diagonal line when edge effects are considered. This is equivalent to a propagation of the information to the un-sampled regions or the appropriate interpolation and extrapolation of signals.
Looking at the numerical performance in fig.~(\ref{fig:2DW14STAT}) reveals that
most of the methods behave very similarly, except for the PR and FR schemes
that deviate from the rest. The former converges rapidly to a good solution
that has a higher correlation (see green curve in panel {\bf f}), but a
slightly worse Euclidean distance to the {\it true} signal. The FR on the other
hand converges extremely slowly. The correlation coefficient is at a stage
where it becomes dramatically worse (see yellow curve in panel {\bf f}). The
smaller Euclidean distance is  no measure for the quality in this case because these low values can be achieved when the reconstruction is very conservative (closer to zero) and has no structure. Notice how many schemes start with better values for that distance measure (see panel {\bf d}).
The EXP methods converge faster and the CJ version leads  to even slightly better results (see violet curve in panels {\bf d} and {\bf f}).

It is also worth mentioning that the best reconstructions in terms of high correlation coefficients and low Euclidean distances to the underlying signal are achieved only after three iterations for the J, SD, and EXP methods, prior to numerical convergence. We furthermore tested \textsc{argo}  under extreme noise conditions in which the inversion diverges and produces density values that approach infinity. At early iterations, extremely good reconstructions were produced. These examples underline the regularization character of the inversion schemes under consideration in this paper. However, for the cases we are interested in, where the noise is mainly determined by the discrete sampling of galaxies, no additional stopping rules are required and the inversion algorithms can be run until full convergence.

\section{Summary and conclusions}
\label{sec:conclusion}

The goal of this work is to exploit the Bayesian formalism  to develop methods that reconstruct the underlying dark-matter distribution from the discrete sample of galaxies and their three-dimensional positions provided by galaxy redshift surveys.  
Such a general Bayesian analysis permits one to innovate methods and push this field forward to develop more accurate reconstruction algorithms. 

We show how a series of uncertainties demand a statistical approach (see figure \ref{fig:dag} and section \ref{sec:uncert}).  Some of the uncertainties
are intrinsic to the nature of the underlying signal (the dark matter) and have
a stochastic character, the cosmic variance.  Other uncertainties are intrinsic
to the nature of the observable (the galaxies) and lead to a kind of shot noise, galaxy-bias and redshift-distortions. Additional uncertainties, such as windowing, selection function effects and blurring effects, arise due to the observation process. 
The degeneracies that are produced by such uncertainties require regularization techniques, which should converge to optimal solutions.
We discuss the different Bayesian approaches specified through different options for the likelihood and the prior, and see how {\it natural} regularizations can be performed by the prior-choice (see section \ref{sec:prior}). 
Moreover, we see how the definition  of particular likelihoods and priors define classes of algorithms, each specific to a different problem approach (see table \ref{tab:rec}).

We develop new algorithms in this Bayesian framework which account for the discrete nature of a galaxy distribution by taking a Poissonian likelihood. This is done for the case of a Gaussian prior leading to the GAPMAP estimator (see section \ref{sec:gauspois} and appendix \ref{app:POISLIKE}) and for the case of an entropic prior (see section \ref{sec:entr} and appendix \ref{app:MEM}). The Maximum Entropy method is studied in detail as a non-informative prior, which does not assume a particular pattern for the underlying signal. This can be interesting when searching for intrinsic deviations from Gaussianity (see section \ref{sec:entr} and references therein). 

We extend the Wiener-filter (see section \ref{sec:WF} and appendix \ref{sec:mapeq}) and propose novel algorithms to do a joint estimation of the density field, its power-spectrum, and the peculiar velocities of the galaxies (see section \ref{sec:MCMC}). We also address the possibility of extending such work to determine  cosmological parameters and the bias between galaxies and dark matter. 

Such an aim requires a large number of repeated reconstructions, which can be only achieved with highly efficient inverse algorithms. We develop here the necessary numerical schemes in a preconditioned way for linear and non-linear inverse problems (see section \ref{sec:itinvreg} and appendix \ref{app:CG} \& \ref{app:prec}). Such iterative schemes acquire their real power only in an operator formalism, which we derive in detail for different Bayesian methods (see section \ref{sec:operators}).
A novel Krylov formula (see section \ref{sec:krylov} and appendix \ref{app:CG}) turns out to be superior in terms of performance and fidelity,  as we show in section (\ref{sec:codetesting}).

The novel \textsc{argo}-software package is presented in this paper. Different inverse schemes are tested with the Wiener-filter implemented in ARGO under several conditions determined by structured noise, blurring, selection function effects and windowing (see section \ref{sec:codetesting}).

We conclude that fast three-dimensional reconstructions of the large-scale structure scaling as $n\log_2 n$ (with $n$ being the total number of grid cells)  can be done with hybrid Wiener-Krylov iterative schemes under an operator formalism, which takes advantage of the speed of FFTs. 
This opens new horizons of possibilities, such as joint parameter and signal estimation, in the field of large-scale structure reconstruction.

It is our goal to apply such techniques to reconstruct the underlying density field, the power-spectrum and the peculiar velocities from galaxy surveys.  
Still, different problems, such as galaxy-bias studies,  have to be further analysed.
However, we are confident that such issues can be tackled from an information-theory approach.

\section*{Acknowledgements}

We acknowledge the valuable and encouraging discussions to Benjamin Wandelt,
Simon D.~M.~White, Jeremy Blaizot and  Hans-Martin Adorf.  We want to thank
especially Jens Jasche, Pilar Esquej and Sheridon Sauer for the careful proof-reading of this work. We are also very grateful to Martin Reinecke and Daniel Sauer for their computational advice.  F.~Kitaura thanks the International Max Planck Research School for supporting this project, the Faculty for Statistics at Penn State University for the statistics course in 2005, and especially Calyampudi R.~Rao and Tom Loredo for very enlightning lectures and conversations.

{\small
\bibliographystyle{mn2e}
\bibliography{lit}
}

\clearpage

\appendix

\section{The Wiener-filter as a Bayesian estimator}
\label{app:WF}

Let us recall eq.~(\ref{eq:WFBAY}) which comes from Bayes theorem assuming a
Gaussian prior and a Gaussian likelihood
\begin{eqnarray}
\lefteqn{P({\mbi s}\mid{\mbi d},\mbi p)}\nonumber\\
&&\propto {\rm exp}\left(-\frac{1}{2}\left[
    {\mbi s}^{\dagger}{\mat S}^{-1}{\mbi s}+({\mbi d-\mat R\mbi
      s})^{\dagger}{\mat N}^{-1}({\mbi d-\mat R \mbi s})\right]\right){.}
\end{eqnarray}
If we just look at the log-posterior distribution we have
\begin{eqnarray}
\lefteqn{\log P({\mbi s}\mid{\mbi d},\mbi p)\propto {\mbi s}^{\dagger}{\mat S}^{-1}{\mbi s}+({\mbi d-\mat R\mbi s})^{\dagger}{\mat N}^{-1}({\mbi d-\mat R \mbi s})} \\
&&\hspace{-0.cm}= {\mbi s}^{\dagger}{\mat S}^{-1}{\mbi s}+{\mbi s}^\dagger\mat R^\dagger{\mat N}^{-1}\mat R{\mbi s}-{\mbi s}^\dagger\mat R^\dagger{\mat N}^{-1}{\mbi d}-{\mbi d}^\dagger{\mat N}^{-1}\mat R\mbi s+{\mbi d}^\dagger{\mat N}^{-1}{\mbi d}\nonumber{.}
\label{app:post1}
\end{eqnarray}
We can combine the first two terms to one term: ${\mbi s}^{\dagger}(\mbi\sigma_{\rm WF}^2)^{-1}{\mbi s}$, with $(\mbi\sigma_{\rm WF}^2)^{-1}\equiv (\mat S^{-1}+\mat R^\dagger \mat N^{-1}\mat R)$.
Since we want to obtain a log-posterior of the form 
\begin{equation}
\log {P({\mbi s}\mid{\mbi d},\mbi p)}\propto ({\mbi s-\langle{\mbi s}\rangle_{\rm WF}})^{\dagger}{(\mbi\sigma_{\rm WF}}^2)^{-1}({\mbi s-\langle{\mbi s}\rangle_{\rm WF}}){,}
\label{app:post2}
\end{equation}
with $\langle{\mbi s}\rangle_{\rm WF}=\mat F_{\rm WF}\mbi d$, we can identify the third and the fourth term of eq.~(\ref{app:post1}) with the
corresponding terms in eq.~(\ref{app:post2})
\begin{equation}
-{\mbi s}^{\dagger}\mat R^\dagger{\mat N}^{-1}{\mbi d}=-{\mbi s}^{\dagger}(\mbi\sigma_{\rm WF}^2)^{-1}\mat F_{\rm WF}{\mbi d}{,}
\label{app:WFCOV1}
\end{equation}
and
\begin{equation}
-{\mbi d}^{\dagger}{\mat N}^{-1}\mat R{\mbi s}=-{\mbi d}^{\dagger}\mat F_{\rm WF}^\dagger(\mbi\sigma_{\rm WF}^2)^{-1}{\mbi s}{,}
\label{app:WFCOV2}
\end{equation}
respectively.
The remaining term depends only on the data and is thus factorized in the posterior distribution function as part of the evidence.
From both eq.~(\ref{app:WFCOV1}) and eq.~(\ref{app:WFCOV2}) we conclude that the Wiener-filter has the form 
\begin{equation}
\mat F_{\rm WF}=\mbi\sigma_{\rm WF}^2\mat R^\dagger{\mat N}^{-1}=(\mat S^{-1}+\mat R^\dagger \mat N^{-1}\mat R)^{-1}\mat R^\dagger{\mat N}^{-1}{.}
\end{equation}
This is the natural Bayesian representation in contrast to expression (\ref{eq:WF}),
 which is the outcome of a generalized LSQ approach (see appendix
 \ref{sec:mapeq} and discussion in section \ref{sec:WF}).
It can be shown that both expressions for the Wiener-filter are mathematically
equivalent (see appendix \ref{app:WIENER}).

\section{The mapping equation for the Wiener-filter in k-space}
\label{sec:mapeq}

Following the concept of minimum variance \citep[e.g.~][]{1992ApJ...398..169R, 1995ApJ...449..446Z}, we define an action given by the normalized volume integral of the square of the difference between the reconstruction ($\psi$) and the ensemble of different possible realizations of the density field ($s=\delta_\rho$)  
\begin{equation}
{\mathcal  A}=\langle \frac{1}{V} \int {\rm{d}}^{N_{\rm D}}{\mbi{r}}\,\Big[\psi({\mbi r})-  s({\mbi{r}})\Big]^2\rangle_{(\mbi s, \mbi\epsilon|\mbi p)}{.}
\label{eq:action}
\end{equation}
From the statistical point of view, the action $\mathcal A$ is the loss function that has to be minimized.
Note that this action can be expressed as the ensemble average of the squared Euclidean distance between the real density field $s$ and the reconstruction $\psi$
\begin{equation}
{\mathcal  A}=\frac{1}{V}\langle D^2_{\rm Eucl}(\psi,s)\rangle_{(\mbi s, \mbi \epsilon|\mbi p)}{.}
\label{eq:enseuc}
\end{equation}
Transforming expression (\ref{eq:action}) into Fourier space yields
\begin{eqnarray}
{\mathcal  A} &=& \frac{1}{V} \int \frac{ {\rm d}^{N_{\rm D}}{\mbi k}}{(2\pi)^{N_{\rm D}}} \,\Big[ \langle \hat{\psi}({\mbi k}) \overline{ \hat{\psi}({\mbi k}) }\rangle_{(\mbi s, \mbi\epsilon|\mbi p)} +  \langle\hat s({\mbi k})\overline{ \hat s({\mbi k})}\rangle_{(\mbi s, \mbi\epsilon|\mbi p)} \nonumber\\
&-& \langle\hat{\psi}({\mbi k})\overline{\hat s({\mbi k})}\rangle_{(\mbi s, \mbi\epsilon|\mbi p)} - \langle {\hat s({\mbi k})}\overline{\hat{\psi}({\mbi k})}\rangle_{(\mbi s, \mbi\epsilon|\mbi p)} \Big]{.}
\label{eq:eq}
\end{eqnarray}
Assuming a linear relation between the reconstruction $\psi$ and the data $ d$ 
\begin{equation}
\hat{\psi}({\mbi k})=\int \frac{{\rm{d}}^{{N}_{\rm D}}{\mbi k'}}{({\rm{2}\pi})^{{N}_{\rm D}}}\, \hat{ \hat{{ F}}}({\mbi k},{\mbi k'})\hat{ d}(\mbi{k'}){,}
\label{eq:eq}
\end{equation}
and statistical homogeneity
($\langle{\hat s({\mbi{k}})\hat s({\mbi{k'}})}\rangle_{(\mbi s, \mbi\epsilon|\mbi p)}=(2\pi)^{N_{\rm D}}\delta_{\rm
  D}({\mbi{k}}-{\mbi{k'}}){P_{\rm S}}({\mbi{k'}})$),
yields 
\begin{eqnarray}
\lefteqn{{\mathcal  A}=
 \frac{1}{V} \int \frac{ {\rm{d}}^{N_{\rm D}}{\mbi{k}} }{({\rm{2}\pi})^{{N_{\rm D}}}}  \int \frac{{\rm{d}}^{{N}_{\rm D}}{\mbi{k'}}}{({\rm{2}\pi})^{{N}_{\rm D}}}\Big[}\nonumber\\
&& \hat{ \hat{{ F}}}({\mbi{k}},{\mbi{k'}}) \int \frac{{\rm{d}}^{{N}_{\rm D}}{\mbi{q}}}{({\rm{2}\pi})^{{N}_{\rm D}}} \overline{\hat{ \hat{{ F}}}({\mbi{k}},{\mbi{q}})}  \langle\hat{ d}({\mbi{k'}}) \overline{ \hat{ d}({\mbi{q}})}\rangle_{(\mbi s, \mbi\epsilon|\mbi p)}\nonumber\\
&&+ (2\pi)^{N_{\rm D}}\delta_D({\mbi k- \mbi k'})\langle\hat
 s({\mbi{k'}})\overline{\hat s({\mbi{k'}})}\rangle_{(\mbi s, \mbi\epsilon|\mbi
 p)} \nonumber\\
&&  - \hat{ \hat{{ F}}}({\mbi{k}},{\mbi{k'}})\langle\hat{
 d}({\mbi{k'}})\overline{\hat s({\mbi{k}})}\rangle_{(\mbi s, \mbi\epsilon|\mbi
 p)}\nonumber\\
&& - \overline{ \hat{ \hat{{ F}}}({\mbi{k}},{\mbi{k'}})}\langle {\hat s({\mbi{k}})}\overline{\hat{ d}({\mbi{k'}})}\rangle_{(\mbi s, \mbi\epsilon|\mbi p)}\Big]{.}
\label{eq:maptable}
\end{eqnarray}
Now the action is minimized with respect to the linear operator, ${\frac{\delta\mathcal{A}}{\delta \hat{ \hat{ F}}}=0}$, to obtain the following mapping equation
\begin{equation}
\int \frac{{\rm{d}}^{{N}_{\rm D}}{\mbi{{q}}}}{({\rm{2}\pi})^{{N}_{\rm D}}}{ \hat{ \hat{{ F}}}({\mbi{k}},{\mbi{q}})}\langle\hat{ d}({\mbi{q}})\overline{\hat{ d}({\mbi{k'}})}\rangle_{(\mbi s, \mbi\epsilon|\mbi p)}=\langle\hat{ s}({\mbi{k}})\overline{\hat d({\mbi{k'}})}\rangle_{(\mbi s, \mbi\epsilon|\mbi p)}{.}
\label{eq:mapping}
\end{equation}
The desired filter can be thus expressed as the correlation matrix between the
signal and the data multiplied by the inverse of the autocorrelation matrix of the data \citep[see][]{1995ApJ...449..446Z}
\begin{equation}
{\mat F} =\langle{\mbi s} {\mbi d}^\dagger\rangle \langle{\mbi d}{\mbi d}^\dagger\rangle^{-1} {.}
\label{eq:mapping1}
\end{equation}
This filter is the LSQ estimator (see eq.~\ref{eq:WF3}). It is identical to the Wiener-filter in case the noise term
has no signal-dependent structure function or after applying an ensemble
average over all possible signals on the noise covariance matrix.
Note, that eq.~(\ref{eq:mapping}) allows us to substitute ${\mbi{k'}}$ by
${\mbi{-k'}}$, which is equivalent to the conjugation of
$\hat{ d}({\mbi{k'}})$ due to the hermitian redundancy of real numbers
\begin{equation}
\int \frac{{\rm{d}}^{{N}_{\rm D}}{\mbi{{q}}}}{({\rm{2}\pi})^{{N}_{\rm D}}}{ \hat{ \hat{{ F}'}}({\mbi{k}},{\mbi{q}})}\langle\hat{ d}({\mbi{q}}){\hat{ d}({\mbi{k'}})}\rangle_{(\mbi s, \mbi\epsilon|\mbi p)}=\langle\hat{ s}({\mbi{k}}){\hat d({\mbi{k'}})}\rangle_{(\mbi s, \mbi\epsilon|\mbi p)}{.}
\label{eq:mapping2}
\end{equation}
 The linear operator one obtains in this way is
different, but fulfills the same requirements. We compare both cases in section (\ref{sec:codetesting}).
Let us see how one would apply such a filter. The covariance matrix of the data is given by
\begin{equation}
\langle\hat{d}(\mbi k){\hat{d}(\mbi k')}\rangle_{({\mbi s,\mbi\epsilon}|\mbi p)}=\langle\hat{ \alpha}(\mbi k){\hat{\alpha}(\mbi k')}\rangle_{{(\mbi s, \mbi\epsilon|\mbi p)}}+\langle\hat{ \epsilon}(\mbi k){\hat{\epsilon}(\mbi k')}\rangle_{({\mbi s, \mbi\epsilon}|\mbi p)}{,}
\end{equation}
and its action on some vector by
\begin{eqnarray}
\lefteqn{\int \frac{{\rm d}^{N_{\rm D}}\mbi k'}{(2\pi)^{N_{\rm D}}}\langle\hat{ \alpha}(\mbi k){\hat{\alpha}(\mbi k')}\rangle_{({\mbi s,\mbi\epsilon}|\mbi p)} \{\hat{x}(\mbi k')\}}\nonumber\\
&&={\hat{f}_{\rm B}\cdot\big[\hat{f}_{\rm SM}\circ\big[ {P_{\rm S}} \cdot\big[ { \overline{\hat{ f}_{\rm SM}}\circ\big[\overline{\hat{f}_{\rm B}}\cdot\{\hat{x}\}}\big]\big]\big]\big] (\mbi k)}{,}
\end{eqnarray}
and
\begin{equation}
\int \frac{{\rm d}^{N_{\rm D}}\mbi k'}{(2\pi)^{N_{\rm D}}}\langle\hat{ \epsilon}(\mbi k){\hat{\epsilon}(\mbi k')}\rangle_{({\mbi s, \mbi\epsilon}|\mbi p)} \{\hat{x}(\mbi k')\}= {\hat{f}_{\rm SF}\circ \big[ {P_{\rm N}} \cdot \big[{\overline{\hat{f}_{\rm SF}}\circ\{\hat{x}\}}\big]\big] (\mbi k)}   {.}
\end{equation}
The correlation matrix between the data and the signal applied to that vector yields
\begin{equation}
\int \frac{{\rm d}^{N_{\rm D}}\mbi k'}{(2\pi)^{N_{\rm D}}}\langle\hat{ s}({\mbi{k}}){\hat d({\mbi{k'}})}\rangle_{(\mbi s, \mbi \epsilon|\mbi p)}\{\hat{x}({\mbi{k'}})\} ={ {P_{\rm S}} \cdot\big[ { \overline{\hat{ f}_{\rm SM}}\circ\big[\overline{\hat{f}_{\rm B}}\cdot\{\hat{x}\}\big]\big](\mbi k)}}{.}
\end{equation}
We see that the difference with respect to the operations derived in section (\ref{sec:operators}) resides in the conjugation of certain functions.

\section{Data-space and signal-space representations for the Wiener-filter}
\label{app:WIENER}

Here we show the equivalence between the data-space and the signal-space representations for
the Wiener-filter (see section~\ref{sec:WF}).
In a first approach, we start assuming that the inverse of the response operator exists
($ \mat R^{-1}$). Then after some operations the equivalence can be shown for both the Wiener-filter
\begin{eqnarray}
{\mat F}_{\rm WF}&=&{(\mat S^{-1} +\mat R^\dagger \mat N^{-1}\mat R)^{-1}\mat R^\dagger\mat N^{-1}}{,}\nonumber\\
&=&(\mat S^{-1} (\mat R)^{-1}\mat N+\mat R^\dagger)^{-1}{,}\nonumber\\
&=&{\mat S \mat R^\dagger(\mat R \mat S \mat R^\dagger+\mat N)^{-1}}{,}
\label{app:WF1}
\end{eqnarray}
and the covariance
\begin{eqnarray}
{\mbi\sigma}^2_{\rm WF}&=&({\mat S^{-1}+\mat R^\dagger\mat N^{-1}\mat R})^{-1}{,}\nonumber\\
&=&{\mat S \mat R^\dagger(\mat R +\mat R\mat S\mat R^\dagger\mat N^{-1}\mat R})^{-1}{,}\nonumber\\
&=&{\mat S \mat R^\dagger(\mat R \mat S \mat R^\dagger+\mat N)^{-1}}\mat N
(\mat R^\dagger)^{-1} {.}
\label{app:WFvar2}
\end{eqnarray}
Note that the covariance given by eq.~(\ref{app:WFvar2}) has limited practical use, since
it requires the inverse of the response operator $\mat R$, which is in general
a singular matrix.
To find a data-space representation for the covariance one has to introduce the concept of constrained realizations (see section~\ref{sec:PSEST} and appendix~\ref{app:COV}).
 In order to find a general proof for the equivalence between the data-space and the signal-space representation of the Wiener-filter, we have to look at the residuals
\begin{eqnarray}
\lefteqn{{\mbi\sigma_{\rm WF}^2}=\langle\mbi r\mbi r^\dagger\rangle=\langle ({\mbi{{s}}}-\mat F_{\rm WF}{\mbi{ {d}}})({\mbi{{s}}}-\mat F_{\rm WF}{\mbi{ {d}}})^\dagger \rangle}\\
&&= \mat S-\mat S\mat R^\dagger{\mat F_{\rm WF}}^\dagger-\mat F_{\rm WF}\mat
R\mat S+\mat F_{\rm WF}(\mat R\mat S\mat R^\dagger+\mat N){\mat F_{\rm WF}}^\dagger\nonumber{,}
\label{app:directWF}
\end{eqnarray}
where we have done the substitution: $\mbi d=\mat R\mbi s+\mbi \epsilon$ and
$\langle \mbi s\mbi \epsilon^\dagger\rangle=0$.
The first two terms lead to the Wiener covariance, as we show here
\begin{eqnarray}
\mat S-\mat S\mat R^\dagger{\mat F_{\rm WF}}^\dagger&=&(\mat S({\mbi\sigma_{\rm WF}^2})^{-1} -\mat S\mat R^\dagger{\mat F_{\rm WF}}^\dagger({\mbi\sigma_{\rm WF}}^2)^{-1}){\mbi\sigma_{\rm WF}^2}\nonumber\\
&=& \big(\mat S(\mat S^{-1}+\mat R^\dagger\mat N^{-1}\mat R)-\mat S\mat R^\dagger\mat N^{-1}\mat R\big){\mbi \sigma_{\rm WF}^2}\nonumber\\
&=& {\mbi \sigma_{\rm WF}^2}{,}
\label{app:directWF2}
\end{eqnarray}
where we have used the signal-space relation obtained in section~(\ref{app:WF}): $\mat F_{\rm WF}=\mbi\sigma_{\rm WF}^2\mat R^\dagger{\mat N}^{-1}$.
Consequently, the last two terms of eq.~(\ref{app:directWF}) have to cancel out
\begin{eqnarray}
0&=&-\mat F_{\rm WF}\mat R\mat S+\mat F_{\rm WF}(\mat R\mat S\mat R^\dagger+\mat N){\mat F_{\rm WF}}^\dagger\nonumber\\
0&=&\mat F_{\rm WF}(-\mat R\mat S+(\mat R\mat S\mat R^\dagger+\mat N){\mat F_{\rm WF}}^\dagger){.}
\end{eqnarray}
Now we take the transpose and conjugate of the last equation and factorize
the data correlation matrix out (which is always invertible, since the noise
covariance matrix is invertible)
\begin{equation}
0=({\mat F_{\rm WF}}-\mat S\mat R^\dagger(\mat R\mat S\mat R^\dagger+\mat N)^{-1})(\mat R\mat S\mat R^\dagger+\mat N)\mat F_{\rm WF}^\dagger{.}
\label{app:dirinv}
\end{equation}
The last equation motivates the data-space representation of the Wiener-filter without performing least
squares, i.e.~without demanding the Filter to be optimal ($\partial\mbi\sigma_{\rm WF}^2/\partial\mat F_{\rm WF}=0$), which is already imposing some
regularity condition on $\mat F_{\rm WF}$. Note that we also obtain the trivial zero solution ($\mat F_{\rm
  WF}=0$), which is equivalent to $\mat R=0$ or $\mat N=\infty$ with covariance
$\mbi\sigma^2=\mat S$. Since the data-space and the signal-space representation have the
same null-spaces eq.~(\ref{app:dirinv}) already proves the
equivalence between the data-space and the signal-space representations for the Wiener-filter. Nevertheless, let us directly test this equivalence
\begin{eqnarray}
\mat S\mat R^\dagger(\mat R\mat S\mat R^\dagger+\mat N)^{-1}&\stackrel{?}{=}&\mbi \sigma_{\rm
  WF}^2\mat R^\dagger\mat N^{-1}\nonumber\\
\mat S\mat R^\dagger&\stackrel{?}{=}&\sigma_{\rm  WF}^2\mat R^\dagger\mat N^{-1}(\mat R\mat S\mat R^\dagger+\mat N)\nonumber\\
\mat R\mat S&\stackrel{?}{=}&(\mat R\mat S\mat R^\dagger+\mat N)\mat N^{-1}\mat R\mbi \sigma_{\rm  WF}^2\nonumber\\
\mat R\mat S&\stackrel{?}{=}&\mat R\mat S\mat R^\dagger\mat N^{-1}\mat R\mbi\sigma_{\rm  WF}^2+\mat R\mbi\sigma_{\rm  WF}^2\nonumber\\
\mat R\mat S (\mbi\sigma_{\rm  WF}^2)^{-1}&\stackrel{?}{=}&\mat R\mat S\mat R^\dagger\mat N^{-1}\mat R+\mat R\nonumber\\
\mat R\mat S (\mat S^{-1}+\mat R^\dagger\mat N^{-1}\mat R)&\stackrel{?}{=}&\mat R\mat S\mat R^\dagger\mat N^{-1}\mat R+\mat R\nonumber\\
\mat R+\mat R\mat S\mat R^\dagger\mat N^{-1}\mat R&\stackrel{?}{=}&\mat R\mat S\mat R^\dagger\mat N^{-1}\mat R+\mat R{.}
\end{eqnarray}
Since the left-hand-side is equal to the right-hand-side both representations are
equivalent. Note that we did not assume the response operator to be
invertible. We solely demanded that the inverse of the signal and of the noise
covariance matrices can be built ($\exists \mat S^{-1}$ and $\exists \mat N^{-1}$). This implies that the covariance matrix and the inverse of the data
autocorrelation matrix exist ($\exists (\mat S^{-1}+\mat R^\dagger\mat N^{-1}\mat R)^{-1}$ and $\exists (\mat R\mat S\mat R^\dagger+\mat N)^{-1}$), as
we required in our proof.

\section{Covariance of a constrained realization}
\label{app:COV}

Following \cite{1991ApJ...380L...5H, 1993ApJ...415L...5G,
  1998ApJ...492..439B} we can generate a synthetic realization with
\begin{equation}
{\mbi y} = {\mbi{\tilde{s}}}-\mat F_{\rm WF}{\mbi{ \tilde{d}}}{,}
\end{equation}
If the following relations hold\footnote{Note that the realization does not
  need to be Gaussian distributed, but just fulfill these requirements.}:
$\langle{\mbi{\tilde{s}}}{\mbi{\tilde{s}}}^\dagger\rangle={\mat S}$,
${\langle{\mbi{\tilde{\epsilon}}}{\mbi{\tilde{\epsilon}}}^\dagger\rangle}=\mat N$ and ${\langle{\mbi{\tilde{s}}}{\mbi{\tilde{\epsilon}}}^\dagger\rangle}=0$
then we obtain
\begin{eqnarray}
\lefteqn{\langle {\mbi y}{\mbi y}^\dagger \rangle=\langle ({\mbi{\tilde{s}}}-\mat F_{\rm WF}{\mbi{ \tilde{d}}})
({\mbi{\tilde{s}}}-\mat F_{\rm WF}{\mbi{ \tilde{d}}})^\dagger \rangle}\\
&&= \mat S-\mat S\mat R^\dagger{\mat F_{\rm WF}}^\dagger-\mat F_{\rm WF}\mat
R\mat S+\mat F_{\rm WF}(\mat R\mat S\mat R^\dagger+\mat N){\mat F_{\rm WF}}^\dagger\nonumber
\end{eqnarray}
We can identify these terms with eq.~(\ref{app:directWF}).
Thus, following relation is fulfilled
\begin{equation}
{\langle {\mbi y}}{\mbi y}^\dagger \rangle=\langle\mbi r\mbi r^\dagger\rangle={\mbi \sigma_{\rm WF}^2}{.}
\end{equation}

\section{GAPMAP: MAP with a Gaussian prior and a Poissonian likelihood}
\label{app:POISLIKE}

Remember ${P({\mbi s}\mid {\mbi d, \mbi p})\propto {{\cal L}({\mbi d}\mid {\mbi s, \mbi p})P({\mbi s\mid \mbi p})}}$ to be extremized.
First we write the log-likelihood taking the logarithm of eq.~(\ref{eq:likepois})
\begin{eqnarray}
\lefteqn{\log{\cal L}({\mbi s}\mid{\mbi d}, \mbi p) = \sum_i\Big[-(\mat R \mbi s')_i-{ c}_i+ d'_{i}\log\Big((\mat R \mbi s')_i+{ c}_i\Big)}\nonumber\\
&&-\log( d'_{i}!){\Big]}{.}
\label{ap:pois1}
\end{eqnarray}
Then we differentiate with respect to the signal to yield
\begin{equation}
{\frac{\partial\log{\cal L}({\mbi s}\mid{\mbi d}, \mbi p)}{\partial { s}_k}}= \sum_i\Big[{ R}_{ik}b\overline{n_{\rm g}}\Big(-1+ (\sum_j{ R}_{ij}{ s}'_{j}+{ c}_i)^{-1}{ d}'_{i}\Big)\Big]\nonumber{.}
\end{equation}
The same exercise for the Gaussian prior leads to
\begin{equation}
\frac{\partial\log{ P}({\mbi s}\mid \mbi p)}{\partial { s}_k}= -\sum_j{ S}^{-1}_{kj}{ s}_j{.}\end{equation}
 Now we demand ${0=\partial\log{P}({\mbi s}\mid{\mbi d}, \mbi p)/\partial { s}_k}$ to get an equation for the MAP estimator. After applying ${\mat S}$ to the equation we obtain
\begin{eqnarray}
\lefteqn{{ s}_k^{j}=}\\
&&\hspace{-1.25cm}{\sum_i\sum_l{\Big[}{ S}_{kl}{ R}_{il}b\overline{n_{\rm g}}}\hspace{0.cm}{\Big(}-1+\Big(\sum_m{ R}_{im}\overline{n_{\rm g}}(1+b{ s}_{m}^j)+{ c}_i\Big)^{-1}{{ d}'_{i}}{\Big)}{\Big]}{.}\nonumber
\end{eqnarray}
Adding the index $j+1$ and $j$ to $\mbi s$ on lhs and rhs respectively, an iteration scheme is formed
\begin{eqnarray}
\lefteqn{{ s}_k^{j+1}=}\\
&&\hspace{-1.25cm}\sum_i\sum_l{\Big[}{ S}_{kl}{ R}_{il}b\overline{n_{\rm g}}}{\Big(}-1+\Big(\sum_m{ R}_{im}\overline{n_{\rm g}}(1+b{ s}_{m}^j)+{ c}_i\Big)^{-1}{{ d}'_{i}}{\Big)\Big]{.}\nonumber
\end{eqnarray}
Let us simplify this algorithm for positive signals $\mbi s'$ in matrix notation
\begin{equation}
{\mbi s}'^{j+1}=\overline{\mbi s}'^2{ \mat S}{\mat R}^\dagger{\Big[}-\vec{1}+{\rm diag}({\mat R}{\mbi s}'^j+{\mbi c})^{-1}{{\mbi d}'}{\Big]+\overline{\mbi s}'}{,}
\end{equation}
where we made following substitutions $b\rightarrow1$ and $\overline{\mbi n}_{\rm g}\rightarrow\overline{\mbi s}'$, with $\overline{\mbi s}'$ being the average of the positive signal.

\section{Poissonian maximum likelihood}
\label{app:POISLIKEML}

The context in which the Richardson-Lucy algorithm is applied has positive
intensity signals  and the kernel $\mat R$ in eq.~(\ref{eq:fred}) is understood
as a blurring function that can be expressed  mathematically as a convolution with the {\it true} signal $\mbi s$.
We will further assume no background (${\mbi c=0}$) so that the log-likelihood of eq.~(\ref{eq:likepois}) can be written as
\begin{equation}
\log{\cal L}({\mbi s}'\mid{\mbi d}', \mbi p) = \sum_i{\Big[}-(\mat R \mbi s')_i+d'_{i}\log(\mat R \mbi s')_{i}-\log( d'_{i}!){\Big]}{,}
\label{ap:poisML}
\end{equation}
differentiating with respect to the signal yields
\begin{equation}
0=\frac{\partial\log{\cal L}({\mbi s}'\mid{\mbi d}', \mbi p)}{\partial { s'}_{k}}= \sum_i\Big[{ R}_{ik}\Big(-1+ ({\mat R}{\mbi s}')_i^{-1}{ d'}_{i}\Big)\Big]{.}
\end{equation}
We can multiply this equation with the signal $\mbi s'$ and make an iterative method which coincides with Richardson-Lucy algorithm
\begin{equation}
\mbi s'^{j+1}={\rm diag}\Big({\mat R}^\dagger{\rm diag}({\mat R}{\mbi s}'^j)^{-1}{\mbi d}'\Big)\mbi s'^j{,}
\end{equation}
with ${\mat R^\dagger\vec{1}=\vec{1}}$ due to the convolution operation.

\section{COBE-filter}
\label{app:COBE}

We briefly show here that the COBE-filter is an unbiased estimator only and only if the response matrix is invertible.
\begin{eqnarray}
\langle\langle{\mbi s}\rangle_{\rm COBE}\rangle_{(\mbi d|\mbi s, \mbi p)}&=&\langle({\mat R^\dagger\mat N^{-1}\mat R})^{-1}\mat R^\dagger\mat N^{-1} {\mbi d}\rangle_{(\mbi d|\mbi s, \mbi p)}\nonumber\\
&=&({\mat R^\dagger\mat N^{-1}\mat R})^{-1}\mat R^\dagger\mat N^{-1} \langle{\mat R\mbi s +\mbi \epsilon}\rangle_{(\mbi d|\mbi s, \mbi p)}\nonumber\\
&=&({\mat R^\dagger\mat N^{-1}\mat R})^{-1}\mat R^\dagger\mat N^{-1} {\mat R}{\mbi s}\nonumber\\
&=&{\mbi s} {\rm ,\,\,if\,\,} \mat R {\rm \,\,is \,invertible.}
\end{eqnarray}

\section{Linear filters need to be invertible to conserve information}
\label{app:FISHER}

The Fisher information matrix ${\mat J}$ for a Gaussian distribution\footnote{Here a Gaussian likelihood is assumed, but the result does not rely on the Gaussianity of the data \cite[see e.g.][]{1998ApJ...503..492S}.} with zero mean and covariance matrix ${\mat C}$ calculated by \cite{1996ApJ...465...34V} has the form
\begin{equation}
\label{eq:fisher}
{\mat J}_{ij}=\frac{1}{2}{\rm tr}\,\left({\mat G}_i {\mat G}_j\right) {,}
\end{equation}
with
\begin{equation}
{\mat G}_{i}={\mat C}^{-1} {\mat C}_{,i} {,}
\end{equation}
where the comma notation ${\mat C}_{,i}$ stands for the derivative with respect to the parameter $\theta_i$: ${\rm d}{\mat C}/{\rm d}\theta_i$.
Following \cite{1997ApJ...480L..87T}, we calculate the Fisher information matrix ${\mat J}$ for the filtered and un-filtered signal. 
Let us assume a linear filter $\mat L$, which provides us with an estimator of the signal 
\begin{equation}
{\langle{\mbi s}\rangle_{\rm L}}\equiv{\mat L} {\mbi d} {.}
\end{equation}
The correlation matrix of the estimator yields
\begin{equation}
{\mat C}^{\rm est}=\langle{\langle{\mbi s}\rangle_{\rm L}}{\langle{\mbi s}\rangle^\dagger_{\rm L}}\rangle_{(\mbi s,\mbi\epsilon|\mbi p)}={\mat L}^\dagger\left(\mat R\mat S \mat R^\dagger +\mat N\right){\mat L} {.}
\end{equation}
We get then
\begin{eqnarray}
{\mat C}^{\rm est}_{,i} & = & {\mat L}^\dagger\left(\mat R \mat S_{,i} \mat R^\dagger\right){\mat L} {,}\\
\label{eq:g}
{\mat G}_{i}^{\rm est} & = & \tilde{{\mat L}}\left(\mat R \mat S \mat R^\dagger+\mat N\right)^{-1}\tilde{{\mat L}}^{\dagger}  {\mat L}^\dagger\left(\mat R \mat S_{,i} \mat R^\dagger\right){\mat L}{,}
\end{eqnarray}
where we have denoted the approximate inverse of $\mat L$ as $\tilde{\mat L}$.
Doing the same for the data yields
\begin{eqnarray}
{\mat C}^{\rm data}&=&\langle{{\mbi d}}{{\mbi d}^\dagger}\rangle_{(\mbi s,\mbi\epsilon|\mbi p)}=\left(\mat R\mat S \mat R^\dagger +\mat N\right) {,}\\
{\mat C}^{\rm data}_{,i} & = & \mat R\mat S_{,i} \mat R^\dagger  {,}\\
{\mat G}_{i}^{\rm data} & = & \left(\mat R\mat S \mat R^\dagger +\mat N\right)^{-1} \left(\mat R\mat S_{,i} \mat R^\dagger  \right){.}
\end{eqnarray}
If we now insert expression (\ref{eq:g}) in the Fisher matrix (\ref{eq:fisher}), we get
\begin{eqnarray}
{\mat J}_{ij}^{\rm est}&=&\frac{1}{2}{\rm tr}\,\left({\mat G}_i^{\rm est}{\mat G}_j^{\rm est}\right)\\
\label{ap:fest}
&=&\frac{1}{2}{\rm tr}\,\left(\tilde{\mat L}{\mat C}^{\rm data -1}\tilde{\mat L}^\dagger{\mat L}^\dagger{\mat C}_{,i}^{\rm data}{\mat L}\tilde{\mat L}{\mat C}^{\rm data -1}\tilde{\mat L}^\dagger{\mat L}^\dagger{\mat C}_{,j}^{\rm data}{\mat L}\right)\nonumber{.}
\end{eqnarray}
In general, this will differ from the Fisher matrix of the data.
If we assume, however, that the linear operator is invertible ($\exists \mat L^{-1}$), then eq.~(\ref{ap:fest}) reduces to
\begin{equation}
{\mat J}_{ij}^{\rm est}=\frac{1}{2}{\rm tr}\,\left({\mat L}^{-1}{\mat G}_i^{\rm data}{\mat G}_j^{\rm data}{\mat L}\right){.}
\end{equation}
Invoking that the trace of a product of matrices is invariant under cyclic permutations, we see that
\begin{equation}
{\mat J}_{ij}^{\rm est}=\frac{1}{2}{\rm tr}\,\left({\mat G}_i^{\rm data}{\mat G}_j^{\rm data}\right)={\mat J}_{ij}^{\rm data} {.}
\end{equation}
This shows the result that any linear invertible filter conserves information, regardless of the parameters that one wants to estimate.  However, one should be careful with this statement because linear filters are, in general, not invertible unless the data and signal space have the same dimension, the noise is non-zero for any frequency, and the $\mat R$- and $\mat S$-matrices are invertible. Usually the data and signal space will differ and the $\mat R$-matrix will not be exactly invertible.

\section{Jeffrey's prior for the 3-dimensional power spectrum}
\label{app:Jeff}

Let us start by assuming a Gaussian likelihood\footnote{Note that the likelihood for ${ P}_{\rm S}(\mbi k)$ is the prior for $\mbi s$.} 
\begin{equation}
P(\mbi s\mid{{ P}_{\rm S}(\mbi k)})\propto \prod_{\mbi k}\frac{1}{\sqrt{{ P}_{\rm S}(\mbi k)}}\exp\Big({-\frac{| s(\mbi k)|^2}{2{ P}_{\rm S}(\mbi k)}}\Big) {.}
\end{equation}
The log-likelihood is then given by
\begin{equation}
\log\Big(P(\mbi s\mid{{ P}_{\rm S}(\mbi k)})\Big)\propto \sum_{\mbi k}\Big[\log\Big({{{ P}_{\rm S}(\mbi k)}}\Big){+\frac{| s(\mbi k)|^2}{{ P}_{\rm S}(\mbi k)}}\Big] {.}
\end{equation}
We now need the second derivatives of the log-likelihood with respect to the parameter ${ P}_{\rm S}$ 
\begin{equation}
\frac{\partial^2}{\partial{{ P}_{\rm S}(\mbi k)}^2}\log\Big(P(\mbi s\mid{{ P}_{\rm S}(\mbi k)})\Big)\propto \Big[-\frac{1}{{{ P}^2_{\rm S}(\mbi k)}}{+\frac{2|\ s(\mbi k)|^2}{{ P}^3_{\rm S}(\mbi k)}}\Big] {.}
\end{equation}
The next step consists of calculating the Fisher information by performing the
integral \\${\int{\rm d}s P( s\mid{{ P}_{\rm S}(\mbi k)})}$ on the above quantity, which is equivalent to performing the following ensemble average (see section \ref{sec:stat})
\begin{equation}
J({ P}_{\rm S}(\mbi k))=\langle\frac{\partial^2}{\partial{{ P}_{\rm S}(\mbi k)}^2}\log\Big(P(\mbi s\mid{{ P}_{\rm S}(\mbi k)})\Big)\rangle_{(\mbi s|\mbi p)}\propto \frac{1}{{{ P}^2_{\rm S}(\mbi k)}} {,}
\end{equation}
where we have taken into account that ${ P}_{\rm S}(\mbi k)=\langle| s(\mbi k)|^2\rangle_{(\mbi s|\mbi p)}$.
Finally the square-root of the Fisher information leads to Jeffrey's prior
\begin{equation}
P({P_{\rm S}}(\mbi k))=\sqrt{J({ P}_{\rm S}(\mbi k))}\propto {P_{\rm S}}(\mbi k) ^{-1}{.}
\end{equation}
Following \cite{2004PhRvD..70h3511W} we can argue in a more intuitive way that $P({P_{\rm S}}(\mbi k))\propto {P_{\rm S}}(\mbi k)^{-1}$ is a solution to a measure invariant under scale transformations of the form \\$P({P_{\rm S}}(\mbi k)){\rm d}{P_{\rm S}}(\mbi k)=P(\alpha{P_{\rm S}}(\mbi k))\alpha{\rm d}{P_{\rm S}}(\mbi k)$ (here we have generalized this result to the 3-dimensional power spectrum).

\section{MEM with Gaussian and Poissonian likelihoods}
\label{app:MEM}

The quantity to maximize is given by
\begin{equation}
Q^{\rm E}({\mbi s}\mid \mbi p)=\alpha S^{\rm E}({\mbi s}\mid \mbi p) + \log{\cal L}({\mbi s}\mid{\mbi d},\mbi p){.}
\end{equation}

After some calculations we see that the gradient of the entropy for PADs is 
\begin{equation}
\nabla S^{\rm E}_+({\mbi s}'\mid \mbi p)_i=-\log\left(\frac{s'_{i}}{m_i}\right){,}
\end{equation}
and for positive and negative distributions
\begin{equation}
\nabla S^{\rm E}_\pm({\mbi s}\mid \mbi p)_i=-\log\left(\frac{w_i +s_i}{m_i}\right){.}
\end{equation}
We took into account that ${\partial w_i/\partial s_j=s_i/w_i\delta_{ij}}$.
It is then more straightforward to calculate the $S^{\rm E}$ curvature for PADs
\begin{equation}
\nabla\nabla S^{\rm E}_+({\mbi s}'\mid \mbi p)=-{\rm diag}({\mbi s}')^{-1}{,}
\end{equation}
and for positive and negative distributions,
\begin{equation}
\nabla\nabla S^{\rm E}_\pm({\mbi s}\mid \mbi p)=-{\rm diag}({\mbi w})^{-1}{.}
\end{equation}
Analogously, we calculate the gradient of the $\log{\cal L}({\mbi s}\mid{\mbi d})$ for the Gaussian case valid for positive ($\mbi s'$)  and positive and negative signals ($\mbi s_\pm$)
\begin{equation}
\nabla \log{\cal L}_{\rm G}({\mbi s}\mid{\mbi d},\mbi p)_i = -\frac{1}{2}\nabla \chi^2({\mbi s})_i=-\left(\mat R^\dagger \mat N^{-1}({\mat R \mbi s -\mbi d})\right)_i{,}
\end{equation}
and the corresponding curvature
\begin{equation}
\nabla\nabla \log{\cal L}_{\rm G}({\mbi s}\mid{\mbi d},\mbi p) = -\frac{1}{2}\nabla\nabla \chi^2({\mbi s})=-\mat R^\dagger \mat N^{-1}\mat R{.}
\end{equation}
The Poissonian case leads to
\begin{eqnarray}
\lefteqn{\nabla \log{\cal L}_{\rm P}({\mbi s}\mid{\mbi d},\mbi p)_i}\nonumber\\
&&= b\overline{n_{\rm g}}\sum_k\Big[{ R}_{ki}\Big(-1+ (\sum_j{ R}_{kj}{ s'}_{j}+{ c}_k)^{-1}{ d'}_{k}\Big)\Big]\nonumber\\
&&= b\overline{n_{\rm g}}\Big[{\mat R}^\dagger\Big(-\vec{1}+ {\rm diag}\Big(({\mat R}{\mbi s}')+{\mbi c}\Big)^{-1}{\mbi d}'\Big)\Big]_i\nonumber{,}
\end{eqnarray}
and
\begin{eqnarray}
\lefteqn{\nabla\nabla \log{\cal L}_{\rm P}({\mbi s}\mid{\mbi d},\mbi p)_{ij}}\nonumber\\&&= -b^2\overline{n_{\rm g}}^2 \sum_k \Big[ { R}_{ki}(\sum_l{ R}_{kl}{ s'}_{l}+{ c}_k)^{-2} { R}_{kj} { d'}_{k} \Big] \nonumber\\
&&= -b^2\overline{n_{\rm g}}^2\Big[{\mat R}^\dagger\Big( {\rm diag}\Big(({\mat R}{\mbi s}')+{\mbi c}\Big)^{-2}{\mat R}^\dagger{\mbi d}'\Big)\Big]_{ij}\nonumber{.}
\end{eqnarray}
Note that when dealing with over-density fields one should do the following substitution: ${\mbi s'}_{i}=\overline{n_{\rm g}}(1+b{\mbi s}_i)$ in the last two expressions.

Summing up, we have the following gradient of $Q^{\rm E}$ for PADs 
\begin{equation}
\nabla Q^{\rm E}_+({\mbi s}'\mid\mbi p)_i=-\alpha\log\left(\frac{s'_{i}}{m_i}\right)+\nabla\log{\cal L}({\mbi s}'\mid{\mbi d},\mbi p)_i{,}
\label{ap:dQminus}
\end{equation} 
and for positive and negative distributions
\begin{equation}
\nabla Q^{\rm E}_\pm({\mbi s}\mid \mbi p)_i=-\alpha\log\left(\frac{w_i -s_i}{m_i}\right)+\nabla\log{\cal L}({\mbi s}\mid{\mbi d},\mbi p)_i{,}
\end{equation}
and the corresponding curvatures
\begin{equation}
\nabla\nabla Q^{\rm E}_+({\mbi s}'\mid \mbi p)= -\alpha {\rm diag}({\mbi s}')^{-1}+\nabla\nabla\log{\cal L}({\mbi s}'\mid{\mbi d},\mbi p){,}
\end{equation}
\begin{equation}
\nabla\nabla Q^{\rm E}_\pm({\mbi s}\mid \mbi p)= -\alpha {\rm diag}({\mbi w})^{-1}+\nabla\nabla\log{\cal L}({\mbi s}\mid{\mbi d},\mbi p){.}
\label{ap:ddQminus}
\end{equation}
The corresponding likelihood (Gaussian or Poissonian) has to be inserted in each of the expressions for the gradient or curvature of $Q^{\rm E}$. 
For the choice of an optimal regularization constant $\alpha$ see e.g.~\cite{1997MNRAS.290..313M} and \cite{1998MNRAS.300....1H}.

\section{Krylov methods: Conjugate Gradients}
\label{app:CG}

\subsection{Orthogonality between the residuals and the searching vectors}
\label{app:CGorth}

Eq.~(\ref{eq:CG6}) tells us that each error vector $\mbi\eta^{j+1}$ is $\mat A$-orthogonal to the previous searching vector $\mat M\mbi\mu^j$. Since all different searching vectors $\mat M\mbi\mu^i$ are $\mat A$-orthogonal to each other by construction, and the error vectors are given by the linear combination of the previous error vector and the previous searching vector (eq.~(\ref{eq:CG3})), it follows that each error vector $\mbi\eta^{j+1}$ is $\mat A$-orthogonal to all previous searching vectors $\mbi\mu^i$, i.e.~for $i\le j$,
\begin{equation}
\langle \mbi\eta^{j+1}|\mat M\mbi\mu^i \rangle_{\mat A} = 0{.}
\label{eq:CGorth0}
\end{equation}
Using eq.~(\ref{eq:CG4}) we can write eq.~(\ref{eq:CGorth0}) as
\begin{equation}
\langle \mbi\xi^{j+1}|\mat M\mbi\mu^i \rangle = 0{,}
\label{eq:CGorth01}
\end{equation}
being $i\le j$.

\mat Applying the inner product between the searching vectors $\mat M\mbi\mu^i$ and the recurrent formula for the residuals (eq.~\ref{eq:CG5}), we get
\begin{equation}
\langle \mbi\xi^{j+1}|\mat M\mbi\mu^i \rangle = \langle \mbi\xi^j|\mat M\mbi\mu^i \rangle-\tau^j\langle\mat M\mbi\mu^j|\mat M\mbi\mu^i \rangle_{\mat A}{.}
\label{eq:CGorth1}
\end{equation}
For $i\not= j$ this equation reduces to
\begin{equation}
\langle \mbi\xi^{j+1}|\mat M\mbi\mu^i \rangle = \langle \mbi\xi^j|\mat M\mbi\mu^i \rangle{.}
\label{eq:CGorth2}
\end{equation}
From eq.~(\ref{eq:CGorth01}) and eq.~(\ref{eq:CGorth2}) we conclude that for $i<j$,
\begin{equation}
\langle \mbi\xi^j|\mat M\mbi\mu^i \rangle=0{.}
\label{eq:CGorth3}
\end{equation}

\subsection{The set of residuals as a basis of linearly independent vectors}
\label{app:CGli}

Taking the Gram-Schmidt orthogonalization scheme (eq.~\ref{eq:CG9}) and multiplying it with the residuals, we obtain
\begin{equation}
\langle \mbi\xi^{i}|\mat M\mbi\mu^j \rangle = \langle \mbi\xi^i|\mat M\mbi\xi^j \rangle+\sum_{k=0}^{j-1}\beta^{kj}\langle \mbi\xi^i|\mat M\mbi\mu^k \rangle{.}
\label{eq:CGli1}
\end{equation}
Using the result obtained in the appendix \ref{app:CGorth} (eq.~\ref{eq:CGorth3}), one shows the orthogonality (strictly orthogonal, if $\mat M=\mat I$) between any different residuals (for $i\not=j$)\footnote{This result is at first glance only valid for $i<j$. However, with the additional requirement that the matrix $\mat M$ be self-adjoint, the generalization to $i\not=j$ is trivial.}
\begin{equation}
\langle \mbi\xi^{i}|\mat M\mbi\xi^j \rangle = 0{.}
\label{eq:CGli2}
\end{equation}
For $i=j$ by combining (\ref{eq:CGorth3}) and (\ref{eq:CGli1}) we get  the relation we used in equation (\ref{eq:CG8})
\begin{equation}
\langle \mbi\xi^{i}|\mat M\mbi\mu^i \rangle = \langle \mbi\xi^{i}|\mat M\mbi\xi^i \rangle{.}
\label{eq:CGli3}
\end{equation}

\subsection[Formulae for the different Krylov methods]{Formulae for the $\beta$-factor}
\label{app:CGFR}

From the scalar product between eq.~(\ref{eq:CG5}) and the residual $\mbi\xi^i$
\begin{equation}
\langle \mbi\xi^{j+1}|\mat M\mbi\xi^i \rangle = \langle \mbi\xi^j|\mat M\mbi\xi^i \rangle-\tau^j\langle \mat M\mbi\mu^j|\mat M\mbi\xi^i \rangle_{\mat A}{,}
\label{eq:CGFR0}
\end{equation}
it is clear that the $\beta$-factors are all zero except for one. Notice that the denominator in $\beta$, given by $\langle \mat M\mbi\mu^j|\mat M\mbi\xi^i \rangle_{\mat A}$ cancels out if neither $i=j+1$ nor $i=j$. The latter is excluded according to the definition of $\beta$ (see eqs.~\ref{eq:CG9} and \ref{eq:CG10}). Gram-Schmidt orthogonalization thus simplifies to eq.~(\ref{eq:CG11}), with 
\begin{equation}
\beta^{j+1}_{\rm EXP}=-\frac{\langle \mat M\mbi\xi^{j+1}|\mat M\mbi\mu^{j}\rangle_{\mat A}}{\langle \mat M\mbi\mu^j|\mat M\mbi\mu^j\rangle_{\mat A}}{.}
\label{eq:CGFR1}
\end{equation}
   Other expressions for this factor can be derived by replacing $i=j+1$ in eq.~(\ref{eq:CGFR0})
\begin{equation}
\langle \mat M\mbi\mu^j|\mat M\mbi\xi^{j+1} \rangle_{\mat A} = -\frac{1}{\tau^{j}}\langle \mbi\xi^{j+1}|\mat M\mbi\xi^{j+1} \rangle{.}
\label{eq:CGFR2}
\end{equation}
Substituting this expression in eq.~(\ref{eq:CG10}) and using the formula for
$\tau^j$ (eq.~\ref{eq:CG8}) one obtains the Fletcher-Reeves equation
\begin{equation}
\beta^{j+1}_{\rm FR}=\frac{\langle \mbi\xi^{j+1}|\mat M\mbi\xi^{j+1}\rangle}{\langle \mbi\xi^j|\mat M\mbi\xi^j\rangle}{.}
\label{eq:CGFR3}
\end{equation}
Polak-Ribi\`eres formula can now be obtained  trivially by taking
expression (\ref{eq:CGli2}) into account.
Let us do an invariant operation by adding ${-\langle \mbi\xi^{j+1}|\mat M\mbi\xi^{j}\rangle}$ to the nominator in Fletcher-Reeves formula
\begin{equation}
{\langle \mbi\xi^{j+1}|\mat M\mbi\xi^{j+1}\rangle-\langle \mbi\xi^{j+1}|\mat M\mbi\xi^{j}\rangle}=\langle \mbi\xi^{j+1}|\mat M(\mbi\xi^{j+1}-\mbi\xi^{j})\rangle{,}
\label{eq:CGFR4}
\end{equation}
which immediately leads to Polak-Ribi\`eres expression
\begin{equation}
\beta^{j+1}_{\rm PR}=\frac{\langle \mbi\xi^{j+1}|\mat M(\mbi\xi^{j+1}-\mbi\xi^{j})\rangle}{\langle \mbi\xi^j|\mat M\mbi\xi^j\rangle}{.}
\label{eq:CGFR5}
\end{equation}
In order to get Hestenes-Stiefels formula one has to consider eqs.~(\ref{eq:CGli3}) and (\ref{eq:CGorth3}) in the denominator of $\beta_{\rm PR}$
\begin{equation}
{\langle \mbi\xi^j|\mat M\mbi\xi^j\rangle}={\langle \mbi\mu^j|\mat M\mbi\xi^{j}\rangle-\langle \mbi\mu^j|\mat M\mbi\xi^{j+1}\rangle}={\langle \mbi\mu^j|\mat M(\mbi\xi^{j}-\mbi\xi^{j+1})\rangle}{,}
\label{eq:CGFR6}
\end{equation}
resulting in the following expression
\begin{equation}
\beta^{j+1}_{\rm HS}=-\frac{\langle\mbi\xi^{j+1}|\mat M(\mbi\xi^{j+1}-\mbi\xi^{j})\rangle}{\langle\mbi\mu^j|\mat M(\mbi\xi^{j+1}-\mbi\xi^{j})\rangle}{.}
\label{eq:CGFR7}
\end{equation}
Due to the relations derived in this appendix other equivalent formulae for $\beta$ (summarized in table \ref{tab:beta}) can be found, which differ in their numerical behavior. Note that from the 16 possible schemes presented here, only 3 are discussed in the literature.

\subsection{Preconditioned non-linear time step}
\label{app:CGnonlinear}

The function under consideration is expanded until the second order around $\tau^j\mat M\mbi\mu^j$ according to eq.~(\ref{eq:CG2})
\begin{eqnarray}
\lefteqn{Q_{\mat A}(\mbi \psi^j+\tau^j\mat M\mbi\mu^j)}\\
&&\hspace{-.5cm}\simeq Q_{\mat A}(\mbi\psi^j)+\tau^j\langle\nabla Q_{\mat A}(\mbi\psi^j)|\mat M\mbi\mu^j\rangle +\frac{\tau^{j2}}{2}\langle\mat M\mbi\mu^j|\mat M\mbi\mu^j\rangle_{\nabla\nabla Q_{\mat A}(\mbi\psi^j)}\nonumber{.}
\end{eqnarray}
Then the derivative with respect to the searching vector is done to find the extremum
\begin{eqnarray}
\lefteqn{\frac{\rm d}{{\rm d}\tau^j}Q_{\mat A}(\mbi \psi^j+\tau^j\mat M\mbi\mu^j)}\nonumber\\
&&\simeq \langle\nabla Q_{\mat A}(\mbi\psi^j)|\mat M\mbi\mu^j\rangle +\tau^{j}\langle\mat M\mbi\mu^j|\mat M\mbi\mu^j\rangle_{\nabla\nabla Q_{\mat A}(\mbi\psi^j)}{.}\hspace{0.8cm}
\end{eqnarray}
By setting this equation to zero, one finds an expression for the time step
\begin{equation}
\tau^j=-\frac{\hspace{-1cm}\langle\nabla Q_{\mat A}(\mbi \psi^j)|\mat M\mbi \mu^j\rangle}{\hspace{0.5cm}\langle\mat M\mbi\mu^{j}|\mat M\mbi\mu^j\rangle_{\nabla\nabla Q_{\mat A}(\mbi\psi^j)}}{.}
\label{eq:New5}
\end{equation}
Note that the last equation can be rewritten using relation (\ref{eq:CGli3}) as 
\begin{equation}
\tau^j=-\frac{\hspace{0.cm}\langle\nabla Q_{\mat A}(\mbi \psi^j)|\mat M \nabla Q_{\mat A}(\mbi \psi^j) \rangle}{\hspace{0.8cm}\langle\mat M\mbi\mu^{j}|\mat M\mbi\mu^j\rangle_{\nabla\nabla Q_{\mat A}(\mbi\psi^j)}}{.}
\label{eq:New6}
\end{equation}

\section{Bayes, Tikhonov, asymptotic regularization and learning algorithms}
\label{app:bayinv}

We want to solve eq.~(\ref{eq:reg3}) from a Bayesian perspective.
Let us assume a Gaussian likelihood with covariance $\mat I$
\begin{equation}
{{\cal L}(\mbi \psi\mid \mbi f,\mbi p)=G({\mbi f}-\mat A \mbi \psi, \mat I)}{,}
\end{equation}
 which is a fair assumption in the absence of noise (eq.~(\ref{eq:reg3}) is equivalent to eq.~(\ref{eq:data}) without noise, ${\mbi\epsilon=0}$). Let us further assume a Gaussian prior around a prior solution ${\mbi \psi}^*$ with covariance $\tau\tilde{\mat M}^{-1}$
\begin{equation}
{P(\mbi \psi\mid\mbi p)=G(\mbi \psi-{\mbi \psi}^*, \tau\tilde{\mat M}^{-1})}{.}
\end{equation}
 We can now calculate the MAP which coincides in this case with the mean of the posterior.
Let us look at the quantity given by the $\log$-posterior PDF
\begin{equation}
||\mbi f-\mat A\mbi \psi||^2+\tau||\mbi \psi- {\mbi \psi}^*||^2_{\tilde{\mat M}} {,}
\end{equation}
which is a generalization of Tikhonov regularization.
Minimizing the negative $\log$-posterior yields the following equation for the Bayesian estimator $\langle{\mbi \psi}\rangle_{\rm B}$
\begin{equation}
\mat A^\dagger(\mat A\langle{\mbi\psi}\rangle_{\rm B}-\mbi f) +\tau^{-1}\tilde{\mat M}(\langle{\mbi \psi}\rangle_{\rm B} -{\mbi \psi}^*) =0 {.}
\end{equation}
 If we now choose ${\tilde{\mat M}=\mat A^\dagger\mat M^{-1}}$ ($\mat M$ is an invertible matrix) we get
\begin{equation}
\mat A^\dagger\left( \mat M^{-1}({\mbi \psi}^*- \langle{\mbi\psi}\rangle_{\rm B})+\tau(\mbi f-\mat A\langle{\mbi\psi}\rangle_{\rm B}) \right) =0 {,}
\end{equation}
This equation will be fulfilled if the following equality holds
\begin{equation}
\label{eq:bayma2}
\langle{\mbi \psi}\rangle_{\rm B}= {\mbi\psi}^*+\tau\mat M(\mbi f-\mat A\langle{\mbi\psi}\rangle_{\rm B}){.}
\end{equation}
The estimator $\langle{\mbi\psi}\rangle_{\rm B}$ for the solution to the
inverse problem (eq.~(\ref{eq:reg3})) is expressed in eq.~(\ref{eq:bayma2}) as
the prior solution ${\mbi\psi}^*$ plus a correction term given by the residual
$\mbi f-\mat A\langle{\mbi\psi}\rangle_{\rm B}$. Since only the residual based
on the prior solution is known, the following substitution must be done on the
right-hand-side (rhs) $\langle\mbi \psi\rangle_{\rm B}\rightarrow \psi^*$ leading to 
\begin{equation}
\langle{\mbi \psi}\rangle_{\rm B}\simeq {\mbi\psi}^*+\tau\mat M(\mbi f-\mat A{\mbi\psi}^*){.}
\end{equation}
This can be interpreted as an iterative scheme, in which the estimator is the
update $j+1$ ($\langle{\mbi\psi}\rangle_{\rm B}\rightarrow\mbi\psi^{j+1}$ on
the left-hand-side (lhs)) of the estimator at the previous step $j$ (${\mbi\psi}^*\rightarrow\mbi\psi^{j}$ on the rhs)
\begin{equation}
{\mbi \psi}^{j+1}= \mbi \psi^{j}+\tau\mat M(\mbi f-\mat A\mbi\psi^{j}){.}
\end{equation}
In this way, we have found the general iterative method (eq.~\ref{eq:reg12}) derived with the asymptotic regularization in section (\ref{sec:asymp}).
From the Bayesian point of view, this scheme could be interpreted as a learning algorithm, in which the estimator of the solution to the inverse problem is calculated from the prior solution and becomes itself the prior solution for the subsequent iteration.

\section{Preconditioning}
\label{app:prec}

We can enhance the convergence of the iteration methods by multiplying the matrix we want to invert by another matrix that is close to its inverse
\begin{equation}
\mat M\mat A\mbi\psi=\mat M\mbi f{,}
\label{eq:prec1}
\end{equation}
with ${\mat M\sim \mat A^{-1}}$.
Let us show this by deriving eq.~(\ref{eq:reg12}) in a different way.
We can invert $\mat M\mat A$ using the Neumann expansion for the inverse of an operator 
\begin{equation}
\mbi\psi=(\mat M\mat A)^{-1}\mat M\mbi f=\sum_{i=0}^\infty(\mat I-\mat M\mat A)^i\mat M\mbi f{.}
\label{eq:prec2}
\end{equation}
This iteration scheme will converge if ${||\mat I-\mat M\mat A||<1}$.
Let us introduce the following notation
\begin{equation}
\mbi\psi\equiv\sum_{i=0}^\infty\mbi\psi[i]{,}
\label{eq:prec3}
\end{equation}
\begin{equation}
\mbi\psi^j\equiv\sum_{i=0}^j\mbi\psi[i]{,}
\label{eq:prec4}
\end{equation}
with 
\begin{equation}
\mbi\psi[i]\equiv(\mat I-\mat M\mat A)^i\mat M\mbi f{.}
\label{eq:prec45}
\end{equation}
It follows that
\begin{equation}
\mbi\psi[i+1]=(\mat I-\mat M\mat A)\mbi\psi[i]{,}
\label{eq:prec5}
\end{equation}
and summing over $i$ we get
\begin{equation}
\sum_{i=0}^j\mbi\psi[i+1]=\sum_{i=0}^j\mbi\psi[i]-\sum_{i=0}^j\mat M\mat A\mbi\psi[i]{.}
\label{eq:prec6}
\end{equation}
Manipulating the indices, we see that
\begin{equation}
\sum_{i=0}^j\mbi\psi[i+1]=\sum_{i=0}^{j+1}\mbi\psi[i]-\mbi\psi[0]{.}
\label{eq:prec7}
\end{equation}
Combining the last two equations we obtain eq.~(\ref{eq:reg12})\footnote{The iteration time step $\tau$ has been absorbed here in the matrix $\mat M$.}
\begin{equation}
\mbi\psi^{j+1}=\mbi\psi^{j}+\mat M(\mbi f-\mat A\mbi\psi^j){,}
\label{eq:prec8}
\end{equation}
with
\begin{equation}
\mbi\psi[0]=\mbi\psi^{0}=\mat M\mbi f{.}
\label{eq:prec9}
\end{equation}
The meaning of the preconditioning matrix $\mat M$ is clear when we look at
eq.~(\ref{eq:prec2}). There it can be seen that a much more rapid convergence
is obtained if ${(\mat I-\mat M\mat A)}$ is close to zero, that is if $\mat M$ is close to the inverse of $\mat A$.

\end{document}